\catcode`\@=11					



\font\fiverm=cmr5				
\font\fivemi=cmmi5				
\font\fivesy=cmsy5				
\font\fivebf=cmbx5				

\skewchar\fivemi='177
\skewchar\fivesy='60


\font\sixrm=cmr6				
\font\sixi=cmmi6				
\font\sixsy=cmsy6				
\font\sixbf=cmbx6				

\skewchar\sixi='177
\skewchar\sixsy='60


\font\sevenrm=cmr7				
\font\seveni=cmmi7				
\font\sevensy=cmsy7				
\font\sevenit=cmti7				
\font\sevenbf=cmbx7				

\skewchar\seveni='177
\skewchar\sevensy='60


\font\eightrm=cmr8				
\font\eighti=cmmi8				
\font\eightsy=cmsy8				
\font\eightit=cmti8				
\font\eightbf=cmbx8				

\skewchar\eighti='177
\skewchar\eightsy='60


\font\ninei=cmmi9
\font\ninesy=cmsy9

\skewchar\ninei='177
\skewchar\ninesy='60


\font\tenrm=cmr10				
\font\teni=cmmi10				
\font\tensy=cmsy10				
\font\tenex=cmex10				
\font\tenit=cmti10				
\font\tensl=cmsl10				
\font\tenbf=cmbx10				
\font\tentt=cmtt10				
\font\tenss=cmss10				
\font\tensc=cmcsc10				
\font\tenbi=cmmib10				

\skewchar\teni='177
\skewchar\tenbi='177
\skewchar\tensy='60

\def\tenpoint{\ifmmode\err@badsizechange\else
	\textfont0=\tenrm \scriptfont0=\sevenrm \scriptscriptfont0=\fiverm
	\textfont1=\teni  \scriptfont1=\seveni  \scriptscriptfont1=\fivemi
	\textfont2=\tensy \scriptfont2=\sevensy \scriptscriptfont2=\fivesy
	\textfont3=\tenex \scriptfont3=\tenex   \scriptscriptfont3=\tenex
	\textfont4=\tenit \scriptfont4=\sevenit \scriptscriptfont4=\sevenit
	\textfont5=\tensl
	\textfont6=\tenbf \scriptfont6=\sevenbf \scriptscriptfont6=\fivebf
	\textfont7=\tentt
	\textfont8=\tenbi \scriptfont8=\seveni  \scriptscriptfont8=\fivemi
	\def\rm{\tenrm\fam=0 }%
	\def\it{\tenit\fam=4 }%
	\def\sl{\tensl\fam=5 }%
	\def\bf{\tenbf\fam=6 }%
	\def\tt{\tentt\fam=7 }%
	\def\ss{\tenss}%
	\def\sc{\tensc}%
	\def\bmit{\fam=8 }%
	\rm\setparameters\setbaselines\fi}


\font\twelverm=cmr12				
\font\twelvei=cmmi12				
\font\twelvesy=cmsy10	scaled\magstep1		
\font\twelveex=cmex10	scaled\magstep1		
\font\twelveit=cmti12				
\font\twelvesl=cmsl12				
\font\twelvebf=cmbx12				
\font\twelvett=cmtt12				
\font\twelvess=cmss12				
\font\twelvesc=cmcsc10	scaled\magstep1		
\font\twelvebi=cmmib10	scaled\magstep1		

\skewchar\twelvei='177
\skewchar\twelvebi='177
\skewchar\twelvesy='60

\def\twelvepoint{\ifmmode\err@badsizechange\else
	\textfont0=\twelverm \scriptfont0=\eightrm \scriptscriptfont0=\sixrm
	\textfont1=\twelvei  \scriptfont1=\eighti  \scriptscriptfont1=\sixi
	\textfont2=\twelvesy \scriptfont2=\eightsy \scriptscriptfont2=\sixsy
	\textfont3=\twelveex \scriptfont3=\tenex   \scriptscriptfont3=\tenex
	\textfont4=\twelveit \scriptfont4=\eightit \scriptscriptfont4=\sevenit
	\textfont5=\twelvesl
	\textfont6=\twelvebf \scriptfont6=\eightbf \scriptscriptfont6=\sixbf
	\textfont7=\twelvett
	\textfont8=\twelvebi \scriptfont8=\eighti  \scriptscriptfont8=\sixi
	\def\rm{\twelverm\fam=0 }%
	\def\it{\twelveit\fam=4 }%
	\def\sl{\twelvesl\fam=5 }%
	\def\bf{\twelvebf\fam=6 }%
	\def\tt{\twelvett\fam=7 }%
	\def\ss{\twelvess}%
	\def\sc{\twelvesc}%
	\def\bmit{\fam=8 }%
	\rm\setparameters\setbaselines\fi}


\font\fourteenrm=cmr12	scaled\magstep1		
\font\fourteeni=cmmi12	scaled\magstep1		
\font\fourteensy=cmsy10	scaled\magstep2		
\font\fourteenex=cmex10	scaled\magstep2		
\font\fourteenit=cmti12	scaled\magstep1		
\font\fourteensl=cmsl12	scaled\magstep1		
\font\fourteenbf=cmbx12	scaled\magstep1		
\font\fourteentt=cmtt12	scaled\magstep1		
\font\fourteenss=cmss12	scaled\magstep1		
\font\fourteensc=cmcsc10 scaled\magstep2	
\font\fourteenbi=cmmib10 scaled\magstep2	

\skewchar\fourteeni='177
\skewchar\fourteenbi='177
\skewchar\fourteensy='60

\def\fourteenpoint{\ifmmode\err@badsizechange\else
	\textfont0=\fourteenrm \scriptfont0=\tenrm \scriptscriptfont0=\sevenrm
	\textfont1=\fourteeni  \scriptfont1=\teni  \scriptscriptfont1=\seveni
	\textfont2=\fourteensy \scriptfont2=\tensy \scriptscriptfont2=\sevensy
	\textfont3=\fourteenex \scriptfont3=\tenex \scriptscriptfont3=\tenex
	\textfont4=\fourteenit \scriptfont4=\tenit \scriptscriptfont4=\sevenit
	\textfont5=\fourteensl
	\textfont6=\fourteenbf \scriptfont6=\tenbf \scriptscriptfont6=\sevenbf
	\textfont7=\fourteentt
	\textfont8=\fourteenbi \scriptfont8=\tenbi \scriptscriptfont8=\seveni
	\def\rm{\fourteenrm\fam=0 }%
	\def\it{\fourteenit\fam=4 }%
	\def\sl{\fourteensl\fam=5 }%
	\def\bf{\fourteenbf\fam=6 }%
	\def\tt{\fourteentt\fam=7}%
	\def\ss{\fourteenss}%
	\def\sc{\fourteensc}%
	\def\bmit{\fam=8 }%
	\rm\setparameters\setbaselines\fi}




\newdimen\rp@
\newcount\@basestretchnum
\newskip\@baseskip
\newskip\headskip
\newskip\footskip


\def\setparameters{\rp@=.1em
	\headskip=24\rp@
	\footskip=\headskip
	\delimitershortfall=5\rp@
	\nulldelimiterspace=1.2\rp@
	\scriptspace=0.5\rp@
	\abovedisplayskip=10\rp@ plus3\rp@ minus5\rp@
	\belowdisplayskip=10\rp@ plus3\rp@ minus5\rp@
	\abovedisplayshortskip=5\rp@ plus2\rp@ minus4\rp@
	\belowdisplayshortskip=10\rp@ plus3\rp@ minus5\rp@
	\normallineskip=\rp@
	\lineskip=\normallineskip
	\normallineskiplimit=0pt
	\lineskiplimit=\normallineskiplimit
	\jot=3\rp@
	\setbox0=\hbox{\the\textfont3 B}\p@renwd=\wd0
	\skip\footins=12\rp@ plus3\rp@ minus3\rp@
	\skip\topins=0pt plus0pt minus0pt}


\def\setbaselines{\maxdepth=4\rp@\baselinestretch=\@basestretchnum}


\def\baselinestretch{\afterassignment\@basestretch\@basestretchnum}
\def\@basestretch{%
	\@baseskip=12\rp@ \divide\@baseskip by1000
	\normalbaselineskip=\@basestretchnum\@baseskip
	\baselineskip=\normalbaselineskip
	\bigskipamount=\the\baselineskip
		plus.25\baselineskip minus.25\baselineskip
	\medskipamount=.5\baselineskip
		plus.125\baselineskip minus.125\baselineskip
	\smallskipamount=.25\baselineskip
		plus.0625\baselineskip minus.0625\baselineskip
	\setbox\strutbox=\hbox{\vrule height.708\baselineskip
		depth.292\baselineskip width0pt }}



\def\makeheadline{\vbox to0pt{\baselinestretch=1000
	\vskip-\headskip \vskip1.5pt
	\line{\vbox to\ht\strutbox{}\the\headline}\vss}\nointerlineskip}

\def\makefootline{\baselineskip=\footskip\line{\the\footline}}

\def\big#1{{\hbox{$\left#1\vbox to8.5\rp@ {}\right.\n@space$}}}
\def\Big#1{{\hbox{$\left#1\vbox to11.5\rp@ {}\right.\n@space$}}}
\def\bigg#1{{\hbox{$\left#1\vbox to14.5\rp@ {}\right.\n@space$}}}
\def\Bigg#1{{\hbox{$\left#1\vbox to17.5\rp@ {}\right.\n@space$}}}


\mathchardef\alpha="710B
\mathchardef\beta="710C
\mathchardef\gamma="710D
\mathchardef\delta="710E
\mathchardef\epsilon="710F
\mathchardef\zeta="7110
\mathchardef\eta="7111
\mathchardef\theta="7112
\mathchardef\iota="7113
\mathchardef\kappa="7114
\mathchardef\lambda="7115
\mathchardef\mu="7116
\mathchardef\nu="7117
\mathchardef\xi="7118
\mathchardef\pi="7119
\mathchardef\rho="711A
\mathchardef\sigma="711B
\mathchardef\tau="711C
\mathchardef\upsilon="711D
\mathchardef\phi="711E
\mathchardef\chi="711F
\mathchardef\psi="7120
\mathchardef\omega="7121
\mathchardef\varepsilon="7122
\mathchardef\vartheta="7123
\mathchardef\varpi="7124
\mathchardef\varrho="7125
\mathchardef\varsigma="7126
\mathchardef\varphi="7127
\mathchardef\imath="717B
\mathchardef\jmath="717C
\mathchardef\ell="7160
\mathchardef\wp="717D
\mathchardef\partial="7140
\mathchardef\flat="715B
\mathchardef\natural="715C
\mathchardef\sharp="715D


\def\err@badsizechange{%
	\immediate\write16{--> Size change not allowed in math mode, ignored}}

\baselinestretch=1000
\tenpoint

\catcode`\@=12					
\catcode`\@=11
\expandafter\ifx\csname @iasmacros\endcsname\relax
	\global\let\@iasmacros=\par
\else	\immediate\write16{}
	\immediate\write16{Warning:}
	\immediate\write16{You have tried to input iasmacros more than once.}
	\immediate\write16{}
	\endinput
\fi
\catcode`\@=12



\def\singlespace{\baselineskip=\normalbaselineskip}

\def\doublespace{\baselineskip=2\normalbaselineskip}


\def\nonarrower{\advance\leftskip by-\parindent
	\advance\rightskip by-\parindent}


\def\boxit#1{\vbox{\hrule\hbox{\vrule\kern3pt
	\vbox{\kern3pt#1\kern3pt}\kern3pt\vrule}\hrule}}

\def\hence{\leavevmode\hbox{\bf .\raise5.5pt\hbox{.}.} }

\def\dalemb#1#2{{\vbox{\hrule height.#2pt
	\hbox{\vrule width.#2pt height#1pt \kern#1pt \vrule width.#2pt}
	\hrule height.#2pt}}}
\def\gtorder{\mathrel{\raise.3ex\hbox{$>$}\mkern-14mu
             \lower0.6ex\hbox{$\sim$}}}
\def\ltorder{\mathrel{\raise.3ex\hbox{$<$}\mkern-14mu
             \lower0.6ex\hbox{$\sim$}}}

\newdimen\fullhsize
\newbox\leftcolumn
\def\twoup{\hoffset=-.5in \voffset=-.25in
  \hsize=4.75in \fullhsize=10in \vsize=6.9in
  \def\fullline{\hbox to\fullhsize}
  \let\lr=L
  \output={\if L\lr
        \global\setbox\leftcolumn=\columnbox\global\let\lr=R \advancepageno
      \else \doubleformat \global\let\lr=L\fi
    \ifnum\outputpenalty>-20000 \else\dosupereject\fi}
  \def\doubleformat{\shipout\vbox{
    \fullline{\box\leftcolumn\hfil\columnbox}\advancepageno}}
  \def\columnbox{\leftline{\vbox{\makeheadline\pagebody\makefootline}}}
  \tolerance=1000 }
\twelvepoint
\bigskip
\bigskip
\centerline{\fourteenpoint 
Statistical Dynamics of Global Unitary Invariant Matrix Models}
\centerline{\fourteenpoint as Pre-Quantum Mechanics}
\vskip 1in
{\singlespace{\centerline{\rm Stephen L. Adler}
\centerline{\rm Institute for Advanced Study}
\centerline{\rm Einstein Drive, Princeton, NJ 08540} 
\centerline{\rm adler@ias.edu}   }
\vskip   4in
\centerline{\copyright~~~ Stephen L. Adler 2002    }

\vfill
\eject
\centerline{\fourteenpoint  Contents}
\bigskip

\leftline{~~~~Introduction and Overview\dotfill 4}
\bigskip

\leftline{1.~~~Trace Dynamics:  The Classical Lagrangian and Hamiltonian}
\leftline{~~~~~~Dynamics of Matrix Models\dotfill 6}
\bigskip

\leftline{2.~~~Additional Generic Conserved Quantities\dotfill 16}
\bigskip

\leftline{3.~~~Trace Dynamics Models With Global Supersymmetry${}^*$\dotfill 
30}
\leftline{~~~3A.~~The Wess-Zumino Model\dotfill 30}
\leftline{~~~3B.~~The Supersymmetric Yang-Mills Model\dotfill 34}
\leftline{~~~3C.~~The Matrix Model for M Theory\dotfill 37}
\leftline{~~~3D.~~Superspace Considerations and Remarks\dotfill 40}
\bigskip

\leftline{4.~~~Statistical Mechanics of Matrix Models\dotfill 42}
\leftline{~~~4A.~~The Liouville Theorem\dotfill 43}
\leftline{~~~4B.~~The Canonical Ensemble\dotfill 47}
\leftline{~~~4C.~~The Microcanonical Ensemble\dotfill 55}
\leftline{~~~4D.~~Gauge Fixing in the Partition Function${}^*$\dotfill 60}
\leftline{~~~4E.~~Reduction of the Hilbert Space Modulo $i_{\rm eff}$\dotfill 
68}
\bigskip

\leftline{5.~~~Ward Identities and the Emergence of Quantum}
\leftline{~~~~~~Field Dynamics\dotfill 78}
\leftline{~~~5A.~~The General Ward Identity\dotfill 79}
\leftline{~~~5B.~~Variation of the Source Terms\dotfill 84}
\leftline{~~~5C.~~Approximations/Assumptions Leading to the Emergence}
\leftline{~~~~~~~~~~of Quantum Theory\dotfill 88}
\leftline{~~~5D.~~Restrictions on the Underlying Theory Implied}
\leftline{~~~~~~~~~~by Further Ward Identities\dotfill 97}
\leftline{~~~5E.~~Derivation of the Schr\"odinger Equation\dotfill 107}
\leftline{~~~5F.~~Brownian Motion Corrections to Schr\"odinger Dynamics}
\leftline{~~~~~~~~~~and the Emergence of the Probability Interpretation
\dotfill 113}
\bigskip

\leftline{6.~~~Discussion and Outlook\dotfill 125}
\bigskip

\leftline{Acknowledgements\dotfill 130}
\bigskip

\leftline{Appendices\dotfill 131}
\leftline{~~~Appendix A: Modifications in Real and Quaternionic}
\leftline{~~~~~~~~~~~~~~~~~~~~~Hilbert Space\dotfill 131}
\leftline{~~~Appendix B: Algebraic Proof of the Jacobi Identity for the}
\leftline{~~~~~~~~~~~~~~~~~~~~~Generalized Poisson Bracket\dotfill 133}
\leftline{~~~Appendix C: Symplectic Structures in Trace Dynamics\dotfill 137}
\leftline{~~~Appendix D: Gamma Matrix Identities for Supersymmetric}
\leftline{~~~~~~~~~~~~~~~~~~~~~Trace Dynamics Models\dotfill 141}
\leftline{~~~Appendix E: Trace Dynamics Models with Operator} 
\leftline{~~~~~~~~~~~~~~~~~~~~~Gauge Invariance\dotfill 145}
\leftline{~~~Appendix F: Properties of Wightman Functions Needed for}
\leftline{~~~~~~~~~~~~~~~~~~~~~Reconstruction of Local Quantum Field Theory
\dotfill 148}     
\leftline{~~~Appendix G: Proof of Reduction with Born Rule Probabilities
\dotfill 151}
\leftline{~~~Appendix H: ``No Go'' Theorem for Heisenberg}
\leftline{~~~~~~~~~~~~~~~~~~~~~Picture Reduction\dotfill 154}
\leftline{~~~Appendix I: Phenomenology of Energy Driven} 
\leftline{~~~~~~~~~~~~~~~~~~~~~Stochastic Reduction\dotfill 156}
\bigskip

\leftline{~~~References\dotfill 168}
\vfill

\eject

\doublespace      
\centerline{Introduction and Overview}
Quantum mechanics is our most successful physical theory.  It underlies 
our very detailed understanding of atomic physics, chemistry, and 
nuclear physics, and the many technologies to which physical systems    
in these regimes give rise.  
Additionally, relativistic quantum mechanics is the basis for  
the standard model of elementary particles,  
which very successfully gives a partial unification of 
the forces operating at the atomic, nuclear, and 
subnuclear levels. 

However, from its inception the probabilistic nature of 
quantum mechanics, and the fact that ``quantum measurements'' require the   
intervention of non-quantum mechanical ``classical systems'', has led to 
speculations by many physicists, mathematicians, and philosophers of science 
that quantum mechanics may be incomplete.  In an opposing camp, many others 
in these communities have attempted  
to provide an interpretational foundation in which quantum mechanics remains 
a complete and self-contained system.   The debate continues, and has 
spawned an enormous literature.  

We shall not attempt any  historical review, although specific aspects 
of the discussions of the past decades will make their appearance where 
relevant to our exposition.  Rather, we shall turn to a statement of the   
purpose of this book.  Our aim is to make the case that {\it  quantum 
mechanics is not a complete theory, but rather is an emergent 
phenomenon arising from the 
statistical mechanics of matrix models that have a 
global unitary invariance}.  We use ``emergent'' here in the sense that  
it is used in condensed matter, molecular dynamics,  
and complex systems theory, where higher 
level phenomena (phonons, superconductivity, fluid mechanics,...) are seen 
to arise or ``emerge'' as the expressions, in appropriate dynamical 
contexts, of an underlying dynamics that at first glance shows little  
resemblance to these phenomena.  
Initial ideas in this direction were developed 
by the author and collaborators in a number of papers dealing with the 
properties of what we termed ``generalized quantum dynamics'' or, in the  
terminology that we shall use in this exposition,``trace dynamics.''  The 
purpose of this book is to give a comprehensive review of this earlier work, 
with a number of significant additions and modifications that bring the 
project closer to its goal.  We shall also relate our proposal to  
a substantial body of 
literature on stochastic modifications of the Schr\"odinger 
equation, which we believe provides the low energy phenomenology for the 
pre-quantum dynamics that we develop here.    

Certain sections of this book are more technical and, although included for  
completeness, are not essential to follow the 
main line of development;  these are marked with an asterisk (*) in the 
section head.  The exposition of the text is set within the framework of 
complex Hilbert space, but many of the ideas carry over to a statistical 
dynamics of 
matrix models in  real or quaternionic Hilbert space where, as in the case of  
complex Hilbert space, an emergent 
complex quantum mechanics  still results.    
Modifications necessary in order for 
the results derived in the text to apply in 
real and quaternionic Hilbert space are sketched in Appendix A.    
Discussions of other topics needed to keep our treatment self-contained
are given in the remaining appendices.  In particular, a survey of the 
properties of the energy-driven stochastic Schr\"odinger equation is given 
in Appendices G, H, and I, and our notational conventions are reviewed in 
the introductory paragraphs preceding Appendix A.    
\vfill
\eject

\twelvepoint
\doublespace
\pageno=6
\centerline{1.~~~Trace Dynamics: The Classical 
Lagrangian and Hamiltonian Dynamics of Matrix Models}
\bigskip

We begin by setting up a classical Lagrangian and Hamiltonian dynamics  
for matrix models.  We shall assume  finite dimensional matrices, 
although ultimately an extension to the infinite dimensional case may 
be needed.  
Let $B_1$ and $B_2$ be two $N \times N$ matrices with matrix elements that 
are even grade elements of a Grassmann algebra over the 
complex numbers, and let Tr be the ordinary 
matrix trace, which obeys the cyclic property 
$${\rm Tr} B_1 B_2 = \sum_{m,n}(B_1)_{mn}(B_2)_{nm} 
=\sum_{m,n} (B_2)_{nm}(B_1)_{mn}= {\rm Tr}B_2B_1~~~.
\eqno(1.1a)$$
Similarly, let $\chi_1$ and $\chi_2$ be two $N \times N$ matrices with 
matrix elements that are odd grade elements of a Grassmann algebra 
over the complex numbers, 
which anticommute rather than commute, so that the cyclic property for these 
takes the form 
$${\rm Tr}\chi_1 \chi_2 = \sum_{m,n} (\chi_1)_{mn}(\chi_2)_{nm}
=-\sum_{m,n}(\chi_2)_{nm}(\chi_1)_{mn}=-{\rm Tr}\chi_2\chi_1~~~.
\eqno(1.1b)$$
Since the even and odd grade elements of a Grassmann algebra over the 
complex numbers commute, one has a final bilinear cyclic identity 
$${\rm Tr} B \chi ={\rm Tr} \chi B~~~.\eqno(1.1c)$$
We shall refer to the Grassmann even and Grassmann odd matrices $B, \chi$ 
as being of bosonic and fermionic type, respectively.  

The cyclic/anticyclic properties of Eqs.~(1.1a-c) are the basic 
identities from which further cyclic properties can be derived.  
{}For example, from the basic bilinear identities  
one immediately derives the trilinear cyclic identities 
$$\eqalign{
{\rm Tr} B_1[B_2,B_3]=&{\rm Tr}B_2[B_3,B_1]={\rm Tr}B_3[B_1,B_2]~~~,\cr
{\rm Tr} B_1\{B_2,B_3\}=&{\rm Tr}B_2\{B_3,B_1\}={\rm Tr}B_3\{B_1,B_2\}~~~,\cr
{\rm Tr} B \{\chi_1,\chi_2\}=&{\rm Tr}\chi_1 [\chi_2,B]
={\rm Tr} \chi_2[\chi_1,B]~~~,  \cr
{\rm Tr} \chi_1\{B,\chi_2\}=&{\rm Tr}\{\chi_1,B\}\chi_2
={\rm Tr}[\chi_1,\chi_2]B ~~~, \cr
{\rm Tr} \chi [B_1,B_2] =& {\rm Tr}B_2 [\chi,B_1]=
{\rm Tr} B_1 [B_2,\chi]~~~,\cr
{\rm Tr} \chi \{B_1,B_2\}=&{\rm Tr} B_2 \{\chi,B_1\}
={\rm Tr}B_1\{B_2,\chi\}~~~,\cr
{\rm Tr} \chi_1 \{\chi_2,\chi_3\}=&{\rm Tr} \chi_2\{\chi_3,\chi_1\}
={\rm Tr} \chi_3 \{\chi_1,\chi_2\}~~~,\cr
{\rm Tr} \chi_1 [\chi_2,\chi_3]=&{\rm Tr} \chi_2 [\chi_3,\chi_1]
={\rm Tr} \chi_3 [\chi_1,\chi_2]~~~,\cr
}\eqno(1.2) $$
which are used repeatedly in trace dynamics calculations.  In these  
equations, and throughout the text, $[X,Y]\equiv XY-YX$ denotes a matrix 
commutator, and $\{X,Y\}=XY+YX$ a matrix anticommutator. 

The basic observation of trace dynamics [1] is that given the trace of a 
polynomial $P$ constructed from noncommuting 
matrix or operator variables (we 
shall use the terms ``matrix'' and ``operator'' interchangeably in the 
following discussion), 
one can {\it define} a derivative of the complex number ${\rm Tr} P$ with 
respect to 
an operator variable ${\cal O}$ by varying and then cyclically permuting 
so that in each term the factor $\delta {\cal O}$ stands on the right. 
This gives the fundamental definition 
$$\delta {\rm Tr P}={\rm Tr} {\delta {\rm Tr}P \over \delta {\cal O}} 
\delta {\cal O}~~~,\eqno(1.3a)$$
or in the condensed notation that we shall use throughout this book, in 
which ${\bf P} \equiv {\rm Tr}P$, 
$$\delta {\bf P}   = {\rm Tr} {\delta {\bf P} \over \delta {\cal O}}  
\delta {\cal O}~~~.\eqno(1.3b)$$
Note that the derivative $\delta {\bf P} / \delta {\cal O}$ thus   
defined is an operator.   In general we will take ${\cal O}$ to be 
either of bosonic or fermionic (but not of mixed) type, and we will 
construct ${\bf P}$ to always be an even grade element of the Grassmann 
algebra.  With these restrictions, $\delta {\bf P} / \delta {\cal O}$
will be of the same type as ${\cal O}$, that is, either both will be 
bosonic or both will be fermionic.  

The definition of Eq.~(1.3b) has the important property that 
if $\delta {\bf P}$ vanishes for arbitrary 
variations $\delta {\cal O}$ of the same type as ${\cal O}$, 
then the operator derivative 
$\delta {\bf P}/ \delta {\cal O}$ must vanish.  To see this, 
let us expand $\delta {\bf P}/\delta {\cal O}$ in the form
$${\delta {\bf P} \over \delta {\cal O}}=\sum_n C_n K_n~~~,\eqno(1.4a)$$ 
with  the $K_n$ distinct Grassmann monomials that are all c-numbers (i.e.,  
multiples of the $N \times N$  unit matrix), and with 
the $C_n$ complex matrix    
coefficients that are unit elements in the Grassmann algebra.  
Let us choose $\delta {\cal O}$ to be an infinitesimal $\epsilon$ times 
$C_p^{\dagger}$, with $\epsilon$ a real number when ${\cal O}$ is bosonic, 
and with $\epsilon$ a Grassmann element not appearing in $K_p$ when 
${\cal O}$ is fermionic. (There must be at least one such element, or 
else $K_p$ would make an identically vanishing contribution to Eq.~(1.3b),  
and could not appear in the sum in Eq.~(1.4a).)
We then have 
$$0=\sum_n {\rm Tr} C_p^{\dagger} C_n  K_n \epsilon  ~~~,\eqno(1.4b)$$
and since the coefficients of all distinct Grassmann monomials must vanish 
separately, we have in particular
$$0={\rm Tr} C_p^{\dagger} C_p~~~.\eqno(1.4c)$$ 
This implies the vanishing of the matrix coefficient $C_p$, and   
letting $p$ range over all index 
values appearing in the sum in Eq.~(1.4a), we conclude that 
$${\delta {\bf P} \over \delta {\cal O}}=0~~~.\eqno(1.4d)$$

When ${\cal O}$ is bosonic, a useful extension of the above result 
states that the vanishing of $\delta {\bf P}$  for all  self-adjoint  
variations $\delta {\cal O}$, or for all  anti-self-adjoint  
variations $\delta {\cal O}$, still implies the vanishing of $\delta{\bf P} 
/\delta{\cal O}$.  To prove this, split each $C_n$ in Eq.~(1.4a) into 
self-adjoint and anti-self-adjoint parts, $C_n=C_n^{sa} + C_n^{asa}$, with 
$C_n^{sa}=C_n^{sa \dagger}$ and $C_n^{asa}=-C_n^{asa\dagger}$. For 
self-adjoint $\delta {\cal O}$, Eq.~(1.1a) implies that ${\rm Tr} C_n^{sa}
\delta {\cal O}$ is real, and ${\rm Tr} C_n^{asa} \delta {\cal O}$ is 
imaginary, and so the vanishing of $\delta{\bf P}$ implies that both of these 
traces must 
vanish separately.  Taking $\delta {\cal O} = C_p^{sa}$ then implies 
the vanishing of $C_p^{sa}$, while taking $\delta {\cal O} =i C_p^{asa}$ then 
implies the vanishing of $C_p^{asa}$. A similar argument, with the role 
of reals and imaginaries interchanged, applies to the case in which 
$\delta {\cal O}$ is restricted to be anti-self-adjoint.  

In our applications, we shall often consider trace functionals ${\bf P}$ 
that are real.  These have the important property 
that when ${\cal O}$ is 
a self-adjoint bosonic operator, then $\delta {\bf P}/ \delta {\cal O}$
is also self-adjoint.  To prove this, we make a self-adjoint 
variation $\delta {\cal O}$, and use the reality of ${\bf P}$ to write 
$$\eqalign{
0\equiv&{\rm Im} {\rm Tr} \delta {\bf P} \propto {\rm Tr} \left[
{\delta {\bf P} \over \delta {\cal O}}  \delta{\cal O} 
-(\delta{\cal O})^{\dagger} \left({\delta {\bf P} 
\over \delta {\cal O}}\right)^{\dagger}\right] \cr
=& {\rm Tr}  \delta{\cal O} 
\left[{\delta {\bf P} \over \delta {\cal O}}  
-\left( {\delta {\bf P} \over \delta {\cal O}} \right)^{\dagger} \right] 
~~~.\cr
}\eqno(1.5)$$
This implies, by the extension given in the preceding paragraph, 
that the anti-self-adjoint part of 
$\delta {\bf P}/\delta {\cal O}$ must vanish.  Similarly, when ${\bf P}$ is   
real and $\delta {\cal O}$ is anti-self-adjoint, then  
$\delta {\bf P}/\delta {\cal O}$ is also anti-self-adjoint.  

We can now proceed to use the apparatus just described to set up a 
Lagrangian and Hamiltonian dynamics for matrix models.   Let 
$L[\{q_r\},\{\dot q_r\}]$ be a Grassmann even polynomial function  
of the bosonic or fermionic operators $\{q_r\}$ and their time 
derivatives $\{\dot q_r\}$, which are all assumed to obey the 
cyclic relations of Eqs.~(1.1a-c) and (1.2) under the trace.  
{}From $L$, we form the {\it trace 
Lagrangian} 
$${\bf L}[\{q_r\},\{\dot q_r\}]={\rm Tr}L[\{q_r\},\{\dot q_r\}]~~~,
\eqno(1.6a)$$ 
and the corresponding {\it trace action} 
$${\bf S}=\int dt {\bf L}~~~.\eqno(1.6b)$$   
We shall assume that the trace action is real valued, which requires 
that $L$ be self-adjoint up to a possible total time derivative and/or  
a possible term with vanishing trace, such as a commutator. That is, we  
require 
$$L-L^{\dagger}= {d\over dt} \Delta_1 +[\Delta_2, \Delta_3]~~~,\eqno(1.6c)$$
with $\Delta_{1,2,3}$ arbitrary.  
Requiring that the trace action be stationary with respect to variations 
of the $q_r$'s that preserve their bosonic or fermionic type, and using the 
definition of Eq.~(1.3b), we get 
$$0=\delta {\bf S} = \int dt \sum_r \left(
{\delta {\bf L} \over \delta q_r }\delta q_r  
+{\delta {\bf L} \over \delta \dot q_r }\delta \dot q_r \right)~~~,  
\eqno(1.7a)$$
or after integrating by parts in the second term and discarding surface 
terms, 
$$0=\delta {\bf S} = \int dt \sum_r \left(
{\delta {\bf L} \over \delta q_r }\delta q_r  
-{d \over dt} {\delta {\bf L} \over \delta \dot q_r }  \right) \delta q_r 
~~~.  \eqno(1.7b)$$
{}For this to hold for general same-type operator 
variations $\delta q_r$, the  
coefficient of $\delta q_r$ in Eq.~(1.7b) must vanish for all $t$, 
giving the operator Euler-Lagrange equations 
$${\delta {\bf L} \over \delta q_r} -
{d \over dt} {\delta {\bf L} \over \delta \dot q_r} =0~~~.\eqno(1.7c)$$
Because, by the definition of Eq.~(1.3b), we have 
$$\left( {\delta {\bf L} \over \delta q_r} \right)_{ij}=
{\partial {\bf L} \over \partial (q_r)_{ji} }~~~,\eqno(1.8)$$
for each $r$ the single Euler-Lagrange equation of Eq.~(1.7c) 
is equivalent 
to the $N^2$ Euler-Lagrange equations obtained by regarding ${\bf L}$ 
as a function of the $N^2$ matrix element variables $(q_r)_{ji}$.  (For  
future reference, we note that the identity of Eq.~(1.8) still holds 
when ${\bf L}$ is replaced by a general complex 
valued trace functional ${\bf A}$.)     

Let us now define the momentum operator $p_r$ conjugate to $q_r$ by 
$$p_r \equiv {\delta {\bf L} \over \delta \dot q_r}~~~.\eqno(1.9a)$$
Since the Lagrangian is Grassmann even, $p_r$ is of the 
same bosonic or fermionic type as $q_r$.  We can now introduce a trace  
Hamiltonian ${\bf H}$ by analogy with the usual definition,  
$${\bf H}={\rm Tr}\sum_rp_r \dot q_r - {\bf L}~~~.\eqno(1.9b)$$
In correspondence with Eq.~(1.8), the matrix elements 
$(p_r)_{ij}$ of the 
momentum operator $p_r$ just correspond to the momenta canonical to the 
matrix element variables $(q_r)_{ji}$.  
Performing general same-type operator variations, and using the 
Euler-Lagrange 
equations, we find from Eq.~(1.9b) that the trace Hamiltonian {\bf H} is a 
trace functional of the operators $\{q_r\}$ and $\{p_r\}$, 
$${\bf H}= {\bf H}[\{q_r\},\{p_r\}]~~~,\eqno(1.10a)$$
with the operator derivatives 
$${\delta {\bf H} \over \delta q_r}=-\dot p_r~,~~~
{\delta {\bf H} \over \delta p_r}=\epsilon_r \dot q_r~,~~~\eqno(1.10b)$$
with $\epsilon_r=1(-1)$ according to whether $q_r,p_r$ are 
bosonic (fermionic).  

Letting {\bf A} and {\bf B} be two bosonic trace functionals of the operators 
$\{q_r\}$ and $\{p_r\}$, it is convenient to define the {\it generalized 
Poisson bracket} 
$$\{{\bf A}, {\bf B} \}={\rm Tr} \sum_r \epsilon_r \left(
{\delta {\bf A} \over \delta q_r}{\delta {\bf B} \over \delta p_r}
-{\delta {\bf B} \over \delta q_r} {\delta {\bf A} \over \delta p_r} \right)
~~~.\eqno(1.11a)$$
Then using the Hamiltonian form of the equations of motion, one readily 
finds that for a general bosonic trace functional ${\bf A}[\{q_r\},\{p_r\}]$, 
the time derivative is given by 
$${d \over dt} {\bf A}={\partial {\bf A} \over \partial t} +
\{ {\bf A}, {\bf H} \}~~~;\eqno(1.11b)$$
in particular, letting {\bf A} be the trace Hamiltonian {\bf H}, which has 
no explicit time dependence when the Lagrangian has no explicit 
time dependence, and using 
the fact that the generalized Poisson bracket is antisymmetric in its 
arguments, it follows that the time derivative of {\bf H} vanishes,  
$${d \over dt} {\bf H}=0~~~.\eqno(1.12)$$

An important property of the generalized Poisson bracket is that it 
satisfies the Jacobi identity, 
$$\{ {\bf A},\{ {\bf B},{\bf C} \}\}    
+\{ {\bf C},\{ {\bf A},{\bf B} \}\}    
+\{ {\bf B},\{ {\bf C},{\bf A} \}\} =0~~~.\eqno(1.13a)$$   
This can be proved algebraically in a basis independent way [2] (see 
Appendix B), and can 
also be proved [3] by inserting a complete set of intermediate states into 
the trace on the right of Eq.~(1.11a) and using the 
complex valued analogs of Eq.~(1.8), 
giving
$$\eqalign{
\{{\bf A}, {\bf B} \}
=& \sum_{m,n,r} \epsilon_r \left[
\left({\delta {\bf A} \over \delta q_r}\right)_{mn}
\left({\delta {\bf B} \over \delta p_r}\right)_{nm}
-\left({\delta {\bf B} \over \delta q_r}\right)_{mn} 
\left({\delta {\bf A} \over \delta p_r} \right)_{nm}    \right] \cr
=& \sum_{m,n,r} \epsilon_r \left[
{\partial {\bf A} \over \partial (q_r)_{nm}}
{\partial {\bf B} \over \partial (p_r)_{mn}}
-{\partial {\bf B} \over \partial (q_r)_{nm}} 
{\partial {\bf A} \over \partial (p_r)_{mn}} \right] ~~~.\cr
}\eqno(1.13b)$$
In the second line of Eq.~(1.13b), the generalized Poisson bracket has 
been reexpressed as a sum of 
classical Poisson brackets in which   
the matrix elements of $q_r,~p_r$ are the classical variables, and the   
Jacobi identity of Eq.~(1.13a) then follows from the Jacobi identity 
for the classical Poisson bracket.  
As a consequence of the Jacobi identity, if ${\bf Q_1}$ and ${\bf Q_2}$ are 
two conserved charges, that is if
$$0={d \over dt} {\bf Q_1} = \{ {\bf Q_1}, {\bf H} \}~,~~~ 
  0={d \over dt} {\bf Q_2} = \{ {\bf Q_2}, {\bf H} \}~,~~~ 
  \eqno(1.14)$$ 
then their generalized Poisson bracket $\{ {\bf Q_1}, {\bf Q_2} \}$ 
also has a vanishing generalized Poisson bracket with {\bf H}, and is 
conserved.  This has the consequence that Lie algebras of symmetries can  
be represented as Lie algebras of trace functionals under the generalized 
Poisson bracket operation.   More generally, the Jacobi identity implies 
that trace dynamics has an underlying symplectic geometry that is 
preserved by the time evolution generated by the total trace Hamiltonian [4], 
in analogy with corresponding symplectic structures in classical dynamics. 
(See Appendix C.)  

It will be useful at this point to introduce a compact notation for the 
operator phase space variables, which emphasizes the symplectic structure. 
Let us introduce the notation  $x_1=q_1~,~x_2=p_1~,~x_3=q_2~,~x_4=p_2~,~...
,x_{2D-1}=q_D~,~  x_{2D}=p_D$, where by 
convention we list all of the bosonic variables   
before all of the fermionic ones in the $2D$-dimensional phase space 
vector $x_r$.  The generalized Poisson bracket of Eq.~(1.11a) can now 
be rewritten as 
$$\{{\bf A}, {\bf B}\}={\rm Tr} \sum_{r,s=1}^{2D}  \left(
{\delta {\bf A} \over \delta x_r} \omega_{rs}
{\delta {\bf B} \over \delta x_s} \right)
~~~,\eqno(1.15a)$$
and the operator Hamiltonian equations of Eq.~(1.10b) can be compactly 
rewritten as 
$$\dot x_r=\sum_{s=1}^{2D} \omega_{rs} {\delta {\bf H} \over \delta x_s}
~~~.\eqno(1.15b)$$
The numerical matrix $\omega_{rs}$ that appears here is given by 
$$\omega={\rm diag}(\Omega_B,...,\Omega_B,\Omega_F,...,\Omega_F)~~~, 
\eqno(1.16a)$$ 
with the $2 \times 2$ bosonic and fermionic matrices $\Omega_B$ and 
$\Omega_F$ given respectively by 
$$\Omega_B=\left(\matrix{0 & 1 \cr -1 & 0 \cr}\right)~,~~~  
  \Omega_F=-\left(\matrix{ 0 & 1 \cr 1 & 0 \cr}\right)~~~.  \eqno(1.16b)$$ 
It is easy to verify that the matrix $\omega$ obeys the properties   
$$\eqalign{
(\omega^2)_{rs}=&-\epsilon_r \delta_{rs}~,~~~
\omega_{sr}=-\epsilon_r \omega_{rs}=-\epsilon_s\omega_{rs}~~~,\cr
(\omega^4)_{rs}=&\delta_{rs}~,~~~\sum_r \omega_{rs} 
\omega_{rt}=\sum_r \omega_{sr} \omega_{tr}
=\delta_{st}~~~.\cr
}\eqno(1.17)$$
Henceforth, as in Eq.~(1.17), we shall not explicitly 
indicate the range of the summation 
indices; the index $r$ on $q_r~,~p_r$  will be understood to have an upper 
summation limit of $D$, while the index $r$ on $x_r$ will be understood to 
have an upper limit of $2D$.  

Using this compact notation one can formally integrate the 
trace dynamics equations of 
motion.  Let $j_r$ be a source matrix of the same bosonic or fermionic 
type as $x_r$, and let us define 
$${\bf X}_r={\rm Tr}j_r x_r~~~,\eqno(1.18a)$$ 
so that 
$${\delta {\bf X}_r \over \delta x_t}= \delta_{rt} j_r~~~.\eqno(1.18b)$$
Then the Hamiltonian equations of motion of Eq.~(1.15b) can 
be rewritten [5] as 
$$\eqalign{
\dot{\bf X}_r=&{\rm Tr} j_r \dot x_r = {\rm Tr} \sum_t \delta_{rt} j_r 
\dot x_t \cr
=&{\rm Tr} \sum_{s,t} {\delta {\bf X}_r \over \delta x_t} \omega_{ts} 
{\delta {\bf H} \over \delta x_s} = \{ {\bf X}_r, {\bf H} \} = 
-\{  {\bf H}, {\bf X}_r \}~~~,\cr 
}\eqno(1.18c)$$
which expresses them as generalized Poisson brackets with the trace 
Hamiltonian.  We can now formally integrate the equation of motion for 
${\bf X}_r(t)$ by writing
$$\eqalign{
{\bf X}_r(t)=&\exp(-\{ {\bf H},...\} t) {\bf X}_r(0) 
\exp(\{ {\bf H},...\} t) \cr
=&{\bf X}_r(0)- t \{{\bf H},{\bf X}_r(0)\} + {1\over 2} t^2 
\{ {\bf H}, \{ {\bf H},{\bf X}_r(0) \}\}
-{1\over 6} t^3 \{ {\bf H}, \{ {\bf H}, \{ {\bf H},{\bf X}_r(0) \}\}\}+...
~~~.\cr
}\eqno(1.19a)$$
In general, the matrix dynamics specified by Eqs.~(1.15b) and (1.19a) is not 
unitary, in other words, Eq.~(1.19a) is {\it not} equivalent to an  
evolution of the form
$$x_r(t)=U^{\dagger}(t) x_r(0)  U(t)~~~,\eqno(1.19b) $$
for some unitary $U(t)$.  
\vfill
\eject
\centerline{2.~~~Additional Generic Conserved Quantities}
\bigskip

We have seen in the previous section that the trace Hamiltonian 
${\bf H}$ is always a conserved quantity in the dynamics of matrix models. 
In this section we introduce two structural restrictions on the form of 
the trace Hamiltonian (or Lagrangian), which lead to two further generic 
conserved quantities, one a trace quantity ${\bf N}$ analogous to the 
fermion number operator in field theory, the other an 
operator $\tilde C$ that is reminiscent of the 
canonical commutator-anticommutator structure of field theory. 

Although we shall allow the trace Hamiltonian to have 
arbitrary polynomial dependences on the bosonic variables, let us for  
the moment 
restrict the fermionic structure to have the bilinear form found in all  
renormalizable quantum field theory models, by taking ${\bf H}$ to have 
the form 
$${\bf H}={\rm Tr} \sum_{r,s,F} (p_r q_s B_{1rs} + p_r B_{2rs} q_s) 
+ {\rm purely~bosonic}~~~.\eqno(2.1a)$$
Here the subscript $F$ indicates a sum over only the fermionic operator 
phase space variables, and  $B_{1,2}$ are general polynomials in the 
bosonic variables.   From Eq.~(2.1a) and the Hamilton equations 
of Eq.~(1.10b),  we have for fermionic $r$ 
$$\eqalign{
\dot p_s=&-{\delta {\bf H} \over \delta q_s} = 
-\sum_{r,F}(B_{1rs} p_r + p_r B_{2rs})~~~,\cr
\dot q_r=&-{\delta {\bf H} \over \delta p_r} = 
\sum_{s,F}(q_s B_{1rs} + B_{2rs} q_s)~~~.\cr 
}\eqno(2.1b)$$
Let us now define the trace quantity ${\bf N}$ by 
$${\bf N}={1\over 2}i{\rm Tr} \sum_{r,F}[q_r,p_r]
=i {\rm Tr} \sum_{r,F}q_rp_r =-i {\rm Tr} \sum_{r,F}p_rq_r~~~.
\eqno(2.2a)$$
Then for the time derivative of ${\bf N}$ we have, from the second of  
the three equivalent forms of ${\bf N}$, 
$$\dot {\bf N}= i{\rm Tr} \sum_{r,F} (\dot q_r p_r + q_r \dot p_r)~~~,
\eqno( 2.2b)$$
which on substituting the fermion equations of motion of Eq.~(2.1b) 
becomes 
$$\dot {\bf N}= i{\rm Tr} \sum_{r,s,F} [B_{2rs},q_sp_r]=0~~~.\eqno(2.3)$$ 
Thus, ${\bf N}$ is a conserved trace quantity when the trace Hamiltonian 
has the bilinear fermionic structure of Eq.~(2.1a).    Inverting the 
Legendre transformation of Eq.~(1.9b), the corresponding bilinear 
fermionic trace Lagrangian is 
$${\bf L}={\rm Tr} \sum_{r,F} p_r \dot q_r  
-{\rm Tr} \sum_{r,s,F} (p_r q_s B_{1rs} + p_r B_{2rs} q_s)  + 
{\rm purely~bosonic}~~~.\eqno(2.4a)$$
In order for the kinetic part of 
$L$ to be self-adjoint up to a total time derivative, 
we assign adjointness properties of the fermionic variables according to 
$$q_r=\psi_r~,~~~ p_r=i\psi_r^{\dagger} ~~~,\eqno(2.4b)$$
which gives 
$$\eqalign{
(p_r \dot q_r)^{\dagger}=&(i\psi_r^{\dagger} \dot \psi_r)^{\dagger} ~~~\cr
=&-i \dot \psi_r^{\dagger} \psi_r  ~~~\cr
=& i \psi_r^{\dagger}\dot \psi_r -i {d \over dt} 
(\psi_r^{\dagger}\psi_r) ~~~\cr
=&p_r \dot q_r + {\rm total~time~derivative}~~~,\cr
}\eqno(2.4c)$$
as needed.  (A more general construction of the fermionic kinetic Lagrangian,  
and a correspondingly more general assignment of adjointness properties of 
the fermionic variables, will be taken up at the end of this section.)
Substituting Eq.~(2.4b) into Eq.~(2.4a), 
the fermionic trace 
Lagrangian takes the form 
$${\bf L}={\rm Tr} \sum_{r,F} i \psi_r^{\dagger} \dot \psi_r  
-{\rm Tr} \sum_{r,s,F}i (\psi_r^{\dagger} \psi_s B_{1rs} 
+ \psi_r^{\dagger} B_{2rs} \psi_s)  + 
{\rm purely~bosonic}~~~.\eqno(2.4d)$$
Correspondingly, substituting Eq.~(2.4b) into Eq.~(2.2a) for ${\bf N}$ 
we get
$${\bf N}=-{1\over 2} {\rm Tr} \sum_{r,F}[\psi_r,\psi_r^{\dagger}]
=-{\rm Tr} \sum_{r,F}\psi_r\psi_r^{\dagger} 
={\rm Tr} \sum_{r,F}\psi_r^{\dagger}\psi_r~~~,\eqno(2.4e)$$
showing that, since ${\bf N}$ is the trace of a self-adjoint quantity, it 
is real when the fermionic adjointness properties are assigned as in   
Eq.~(2.4b).  

The resemblance of ${\bf N}$ to a fermion number operator suggests that it 
will be conserved even when ${\bf H}$ is not bilinear, as long as each 
monomial in ${\bf H}$ has equal numbers of fermionic operators 
$p_r = i\psi_r^{\dagger}$ and $q_s = \psi_s$, with any values of 
the mode indices $r,s$.  This is indeed the case, and can be seen as 
follows.  Let ${\bf H}_{n_q,n_p}$ be a monomial term in ${\bf H}$ containing 
exactly $n_q$ factors of fermionic $q$'s, and $n_p$ factors 
of fermionic $p$'s, 
with any values of the  indices $r,...$ labeling fermionic degrees of 
freedom. Then by a simple counting argument (an application of Euler's  
theorem for homogeneous functions) we have
$$\eqalign{
{\rm Tr} \sum_{r,F} {\delta {\bf H}_{n_q,n_p} \over \delta q_r} q_r 
=& n_q {\bf H}_{n_q,n_p}~~~,\cr
{\rm Tr} \sum_{r,F} {\delta {\bf H}_{n_q,n_p} \over \delta p_r} p_r  
=& n_p {\bf H}_{n_q,n_p}~~~.\cr 
}\eqno(2.5a)$$
Denoting by  $\dot {\bf N}_{n_q,n_p}$   the contribution of 
${\bf H}_{n_q,n_p}$ to $\dot {\bf N}$ , we have by use of  
Eqs.~(1.10b) and (2.2b), 
$$\eqalign{
\dot {\bf N}_{n_q,n_p}
=&-i{\rm Tr}\sum_{r,F} \left[
{\delta {\bf H}_{n_q,n_p} \over \delta p_r} p_r  
+q_r {\delta {\bf H}_{n_q,n_p} \over \delta q_r} \right] \cr
=&-i{\rm Tr}\sum_{r,F} \left[
{\delta {\bf H}_{n_q,n_p} \over \delta p_r} p_r  
-{\delta {\bf H}_{n_q,n_p} \over \delta q_r}q_r \right] \cr
=&-i (n_p-n_q) {\bf H}_{n_q,n_p}~~~.\cr
}\eqno(2.5b)$$
Hence if ${\bf H}$ is constructed solely from monomials which have 
equal numbers of fermionic $\psi$'s and $\psi^{\dagger}$'s, so that 
$n_q=n_p$ for all monomial terms in ${\bf H}$, then the trace quantity  
${\bf N}$ remains conserved.  This gives the most general structural 
restriction on ${\bf H}$ leading to conservation of ${\bf N}$.

As a second structural specialization, let us restrict the class of matrix 
models under consideration to those in which the {\it only} non-commuting 
matrix quantities are the Lagrangian dynamical variables $q_r,\dot q_r$, or 
their Hamiltonian equivalents $q_r,p_r$.  In other words, we shall assume 
that the trace Lagrangian and Hamiltonian are constructed from the dynamical 
variables using only c-number complex coefficients, excluding 
the more general 
case in which fixed matrix coefficients are used.  With this restriction,   
we shall show that 
there is a generic conserved operator 
$$\tilde C \equiv \sum_{r,B}[q_r,p_r]-\sum_{r,F}
\{q_r, p_r\}=\sum_{r,s}x_r \omega_{rs}x_s~~~,\eqno(2.6)$$
with subscripts $B,F$ denoting respectively sums over bosonic and fermionic 
operator phase space variables.  
The existence of the conserved quantity $\tilde C$ was first 
discovered by Millard [6] under the more
restrictive assumption of a bosonic theory with a 
Weyl ordered (i.e., symmetrized) Hamiltonian, but was 
soon seen to hold [7,8] under the less restrictive conditions assumed here. 

When the Lagrangian and Hamiltonian are constructed using only c-number 
fixed coefficients, there is a global unitary invariance which preserves   
the adjointness properties of the dynamical variables 
(in the sense that   
writing $y_r=U^{\dagger} x_r U$, then 
$y_r^{\dagger}=U^{\dagger} x_r^{\dagger} U$.) That is, if there are no  
fixed matrix coefficients, then 
the trace Lagrangian obeys 
$${\bf L}[\{U^{\dagger} q_r U\},\{U^{\dagger}\dot q_r U\}]=
{\bf L}[\{q_r\},\{\dot q_r\}]~~~,\eqno(2.7a)$$
and the trace Hamiltonian correspondingly obeys 
$${\bf H}[\{ U^{\dagger}q_rU\},\{U^{\dagger}p_rU\}]={\bf H}[\{q_r\},\{p_r\}]
~~~,\eqno(2.7b)$$
with $U$ a constant unitary $N \times N$ matrix.  
Let us now find the 
conserved Noether current corresponding to this global unitary invariance. 
Setting $U=\exp\Lambda$, with $\Lambda$ an anti-self-adjoint bosonic 
generator matrix, 
and expanding to first order in $\Lambda$, Eq.~(2.7b) implies that 
$${\bf H}[\{q_r -[\Lambda,q_r]\}, \{p_r - [\Lambda,p_r]\}]
={\bf H}[\{q_r\},\{p_r\}] ~~~.\eqno(2.8a)$$
But applying the definition of the variation of a trace functional given 
in Eq.~(1.3b), Eq.~(2.8a) becomes 
$${\rm Tr} \sum_r \left(- {\delta {\bf H} \over \delta q_r}
[\Lambda,q_r] - {\delta {\bf H} \over \delta p_r} [\Lambda,p_r] \right) 
=0~~~,\eqno(2.8b)$$
which by use of the bilinear cyclic identities of Eqs.~(1.1a,b) yields  
$${\rm Tr}  \Lambda 
\sum_r \left( 
{\delta {\bf H} \over \delta q_r}  q_r -\epsilon_r q_r {\delta {\bf H} 
\over \delta q_r} 
+{\delta {\bf H} \over \delta p_r}  p_r -\epsilon_r p_r {\delta {\bf H} 
\over \delta p_r} \right)=0.~~~~\eqno(2.8c)$$
Since the generator $\Lambda$ is an arbitrary anti-self-adjoint $N \times N$ 
matrix, the matrix that multiplies it in Eq.~(2.8c) must 
vanish, giving the matrix  identity 
$$ 
\sum_r \left( 
{\delta {\bf H} \over \delta q_r}  q_r-\epsilon_r q_r {\delta {\bf H} 
\over \delta q_r} 
+{\delta {\bf H} \over \delta p_r}  p_r -\epsilon_r p_r {\delta {\bf H} 
\over \delta p_r} \right)=0.~~~~\eqno(2.9a)$$
But now substituting the Hamilton equations 
of Eq.~(1.10b), Eq.~(2.9a) takes 
the form 
$$\eqalign{
0=&\sum_r \left( 
-\dot p_r  q_r +\epsilon_r q_r \dot p_r 
+\epsilon_r \dot q_r p_r - p_r \dot q_r \right)  \cr
=&{d \over dt}\sum_r  \left( -p_r q_r + \epsilon_r q_r p_r \right)   \cr
=& {d \over dt}\left( \sum_{r,B}[q_r,p_r]-\sum_{r,F}
\{q_r, p_r\} \right)~~~,\cr 
}\eqno(2.9b)$$
completing the demonstration of the conservation of $\tilde C$.  Assigning 
adjointness properties to the fermionic variables as in Eq.~(2.4b), 
and taking the bosonic variables $q_r$ to be self-adjoint 
(or anti-self-adjoint), which for a real  
trace Lagrangian implies that the corresponding bosonic $p_r$ are 
respectively self-adjoint 
(or anti-self-adjoint), we 
see that the conserved operator $\tilde C$ is anti-self-adjoint, 
$$\tilde C=-\tilde C^{\dagger}~~~.\eqno(2.10a)$$
(A more general  
adjointness structure for $\tilde C$, corresponding to an alternative  
assignment of fermion adjointness properties, will be discussed shortly.)     
Also, from the bilinear cyclic identities of Eqs.~(1.1a,b), we see that 
$\tilde C$ is traceless, 
$${\rm Tr}\tilde C =0~~~.\eqno(2.10b)$$

Corresponding to the fact that $\tilde C$ is the conserved Noether current 
in any matrix model 
with a global unitary invariance, it is easy to see [5,7] that 
$\tilde C$ can be used to construct the generator of global unitary 
transformations of the Hilbert space basis.  Consider the trace functional  
$${\bf G}_{\Lambda} ={\rm Tr} \Lambda \tilde C~~~,\eqno(2.11a)$$
with $\Lambda$ a fixed bosonic anti-self-adjoint operator, which can be 
rewritten, using cyclic 
invariance of the trace, as 
$${\bf G}_{\Lambda}=-{\rm Tr}\sum_r[\Lambda,p_r]q_r 
={\rm Tr}\sum_r p_r [\Lambda ,q_r]~~~.\eqno(2.11b)$$
Hence for the variations of $p_r$ and $q_r$ induced 
by using ${\bf G}_{\Lambda}$ as 
canonical generator, which by definition (see Eq.~(2.13a) below) 
have a structure analogous to the 
Hamilton equations of Eq.~(1.10b), we get     
$$\eqalign{
\delta p_r\equiv&-{\delta {\bf G}_{\Lambda} \over \delta q_r} 
=[\Lambda,p_r]~~~,\cr
\delta q_r\equiv&\epsilon_r {\delta {\bf G}_{\Lambda} \over \delta p_r} 
=[\Lambda,q_r]~~~.\cr 
}\eqno(2.11c)$$ 
Comparing with Eqs.~(2.7b) and (2.8a), we see that these have 
just the form of an infinitesimal global unitary transformation.  

The generalized Poisson bracket of the trace generators of two infinitesimal 
global unitary transformations 
${\bf G}_{\Lambda}$ and ${\bf G}_{\Sigma}$ can be computed [5] by combining 
Eq.~(2.11c) with the definition of the bracket in Eq.~(1.11a), with 
the result 
$$\{ {\bf G}_{\Lambda}, {\bf G}_{\Sigma} \}=
{\bf G}_{[\Lambda,\Sigma]}~~~.\eqno(2.12a) $$
Hence the Lie algebra of the generators ${\bf G}_{\Lambda}$ under the 
generalized Poisson bracket is isomorphic to the algebra of the matrices 
$\Lambda$ under commutation.  Equation (2.12a), which is an analog 
of the  ``current algebra'' group properties 
of integrated charges in quantum field theory, can be 
generalized [5] 
to an analog of the local current algebra of quantum field theory as follows. 
Let us write 
$$\eqalign{
\tilde C=&\sum_r \tilde C_r~~~,\cr
\tilde C_r\equiv &\epsilon_rq_rp_r-p_rq_r~~~,\cr
}\eqno(2.12b)$$
and let us define a ``local'' trace generator ${\bf G}_{\Lambda r}$ by 
$${\bf G}_{\Lambda r}={\rm Tr} \Lambda \tilde  C_r~~~.\eqno(2.12c)$$
Then a straightforward calculation, similar to that leading to Eq.~(2.12a), 
shows that 
$$\{ {\bf G}_{\Lambda r}, {\bf G}_{\Sigma s} \} = 
\delta_{rs} {\bf G}_{[\Lambda,\Sigma]r}~~~.\eqno(2.12d)$$

In addition to the canonical generators for global unitary transformations 
given in Eq.~(2.11c), we can also define general canonical 
transformations.  Letting ${\bf G}= {\rm Tr} G$, with $G$ self-adjoint but 
otherwise arbitrary, a general infinitesimal canonical transformation is 
defined by 
$$\eqalign{
p_r^{\prime} -p_r \equiv \delta p_r=&-{\delta {\bf G}\over \delta q_r} 
~~~,\cr
q_r^{\prime}-q_r \equiv \delta q_r=&\epsilon_r {\delta {\bf G}\over 
\delta p_r}~~~,\cr 
 }\eqno(2.13a)$$ 
which is the natural extension to trace dynamics of an
infinitesimal canonical transformation in
classical mechanics.  In terms of the symplectic variables $x_r$ introduced  
in Sec.~1, Eq.~(2.13a) can be written in the compact form 
$$x_r^{\prime}-x_r \equiv \delta x_r = \sum_s \omega_{rs} 
{ \delta {\bf G} \over \delta x_s} ~~~. \eqno(2.13b)$$

Letting ${\bf A} \equiv {\bf A}[\{ x_r \}]$ 
be an arbitrary trace functional, we find immediately that to first
order under a canonical transformation, 
$$\eqalign{
{\bf A}^{\prime} \equiv& {\bf A} [\{ x_r^{\prime} \}] \cr
=&{\bf A} +{\bf Tr} \sum_r {\delta {\bf A} \over \delta x_r}
\delta x_r \cr 
=&{\bf A} +{\bf Tr} \sum_{r,s} {\delta {\bf A} \over \delta x_r} 
\omega_{rs}  { \delta {\bf G} \over \delta x_s} \cr
=&{\bf A} +\{ {\bf A}, {\bf G} \}~~~, \cr
} \eqno(2.14a)$$
that is,
$$\delta {\bf A} \equiv 
{\bf A}^{\prime}-{\bf A}= \{ {\bf A}, {\bf G} \}~~~.\eqno(2.14b)$$
Comparing Eq.~(2.13b) with Eq.~(1.15b), we see that when {\bf G} 
is taken as ${\bf H}dt$, with {\bf H}
the trace Hamiltonian and $dt$ an infinitesimal time step, 
then $\delta x_r =\dot x_r dt$ gives the small change in $x_r$ 
resulting from
the dynamics of the system over that time step.  So as expected,   
the Hamiltonian 
dynamics of the system is a special case of a canonical transformation.  

Let us now consider canonical transformations with generators ${\bf G}$
that are global unitary invariant, that is, that are constructed from the 
$\{x_r\}$ using only c-number fixed coefficients.  Global unitary 
invariance implies that these generators 
obey 
$$\{{\bf G},{\bf G}_{\Lambda}\}=0~~~,\eqno(2.15)$$
with ${\bf G}_{\Lambda}$ the global unitary generator of Eq.~(2.11a).
But using Eq.~(2.14b), Eq.~(2.15) has the alternate interpretation 
that ${\bf G}_{\Lambda}$ is invariant 
under a canonical transformation ${\bf G}$ that is global unitary invariant, 
and since the anti-self-adjoint matrix $\Lambda$ is arbitrary, this implies 
that $\tilde C$ is invariant under any canonical transformation with a 
global unitary invariant generator.  (This could also have been deduced by 
a calculation in direct analogy with Eqs.~(2.7a) through (2.9b).)
This invariance group of $\tilde C$ has the following significance.  
Consider a Poincar\'e invariant trace dynamics field theory with a 
global unitary invariant trace Lagrangian, examples of which are given in 
Sec. 3 below.  In such a theory, the Poincar\'e generators will be trace 
functionals which are also global unitary invariant (that is, if the 
Lagrangian 
involves only c-number fixed coefficients, this property carries over to 
the trace energy momentum tensor and to the trace Poincar\'e generators), 
and so we can conclude from the above discussion of canonical invariance 
that $\tilde C$ is Poincar\'e invariant.  This will be seen explicitly in 
the examples given in Sec. 3, and will play a role in our later analysis 
of the emergence of quantum behavior from the statistical dynamics of 
global unitary invariant matrix models.  

{}For each phase space variable $q_r,p_r$, let us 
define the {\it classical 
part} $q_r^c, p_r^c$ and the noncommutative remainder $q_r^{\prime}, 
p_r^{\prime}$, by
$$\eqalign{
q_r^c=&{1 \over N}{\rm Tr} q_r~,~~~ p_r^c={1 \over N} {\rm Tr} p_r~~~,\cr
q_r^{\prime}=&q_r-q_r^c~,~~~p_r^{\prime}=p_r-p_r^c~~~~,\cr
}\eqno(2.16a)$$
so that bosonic $q_r^c,p_r^c$ are $c$-numbers, fermionic $q_r^c,p_r^c$ 
are Grassmann $c$-numbers, and the remainders are traceless,
$${\rm Tr} q_r^{\prime}={\rm Tr}p_r^{\prime}=0~~~.\eqno(2.16b)$$
Then since $q_r^c,p_r^c$ commute (anticommute) with $q_s^{\prime},
p_s^{\prime}$ for $r,s$ both bosonic (fermionic), we see that the classical 
parts of the phase space variables make no contribution to $\tilde C$, and 
Eq.~(2.6) can be rewritten as 
$$\tilde C=\sum_{r,B}[q_r^{\prime},p_r^{\prime}]
-\sum_{r,F}\{q_r^{\prime},p_r^{\prime}\}~~~.\eqno(2.16c)$$
Thus $\tilde C$ depends only on the non-commutative 
parts of the matrix phase space variables.

We conclude this section by showing that the argument for the adjointness  
properties of  $\tilde C$ can be generalized, allowing for 
the possibility that $\tilde C$ can have a component which is self-adjoint, 
when we allow a more general assignment of fermionic adjointness 
properties than that of Eq.~(2.4b).  
Let us consider a trace Lagrangian which has 
a fermionic kinetic term of the form 
$${\bf L}_{\rm kin}={\rm Tr}\sum_{r,s} \psi_r^{\dagger} A_{rs}\dot \psi_s
~~~,\eqno(2.17a)$$
with $A_{rs}$ for each $r,s$ an $N\times N$ matrix.  This trace Lagrangian 
will be real, up to a total time derivative, provided that the set of 
matrices $A_{rs}$ obeys 
$$A_{rs}=-A_{sr}^{\dagger}~~~.\eqno(2.17b)$$
If we identify $q_s =\psi_s$, then the corresponding canonical momentum is 
$$p_s = {\delta {\bf L} \over \delta \dot q_s} = \sum_r 
\psi_r^{\dagger} A_{rs}~~~,\eqno(2.17c)$$
and the kinetic Lagrangian takes the form 
$${\bf L}_{\rm kin}={\rm Tr} \sum_s p_s \dot q_s~~~,\eqno(2.17d)$$ 
which is clearly global unitary invariant as a function of the 
phase space variables $x_r$, even though ${\bf L}$ was not global unitary 
invariant when expressed in terms of the original variables $\psi_r,\,
\psi_r^{\dagger}$.  Let us now suppose that the remaining terms in ${\bf L}$ 
also have the property that they are global unitary invariant when expressed 
in terms of the phase space variables $x_r$; then the trace Hamiltonian 
${\bf H}$ will also be global unitary invariant.  
The argument of Eqs.~(2.7b) through (2.9b), which does not make use of 
the adjointness assignment of Eq.~(2.4b), then implies that 
$\tilde C$  of Eq.~(2.6) is still conserved.

To see when the possibility 
of a self-adjoint component of $\tilde C$ can be realized, we consider the 
fermionic part $\tilde C_F$, for which we have
$$ \eqalign{
\tilde C_F=&-\sum_{s,F} \{q_s,p_s\}= -\sum_{r,s}
[\psi_s \psi_r^{\dagger} A_{rs} 
+ \psi_r^{\dagger} A_{rs} \psi_s ]~~~,\cr
\tilde C_F^{\dagger} =& -\sum_{r,s} [ A_{rs}^\dagger \psi_r \psi_s^{\dagger} 
+\psi_s^{\dagger} A_{rs}^{\dagger} \psi_r]\cr
=& \sum_{r,s}[A_{sr} \psi_r \psi_s^{\dagger} 
+\psi_s^{\dagger} A_{sr} \psi_r]~~~,\cr
}\eqno(2.18a)$$
where in getting the final line we have used the condition of Eq.~(2.17b).   
Adding the two equations, we get 
$$\tilde C_F + \tilde C_F^{\dagger} = -\sum_{r,s} 
[\psi_s \psi_r^{\dagger}, A_{rs}]~~~,\eqno(2.18b)$$ 
showing that when the right hand side of Eq.~(2.18b) is nonzero, the 
operator $\tilde C$ is no longer anti-self-adjoint.  As a direct check 
that the commutator on the right hand side 
of Eq.~(2.18b) is self-adjoint, we have 
$$\eqalign{
(\sum_{r,s}  [\psi_s \psi_r^{\dagger}, A_{rs}] )^{\dagger} 
=&\sum_{r,s} [A_{rs}^\dagger, \psi_r \psi_s^{\dagger}] 
=-\sum_{r,s} [A_{sr},  \psi_r \psi_s^{\dagger}] \cr
=& \sum_{r,s} [  \psi_r \psi_s^{\dagger} ,A_{sr}]
= \sum_{r,s}  [\psi_s \psi_r^{\dagger}, A_{rs}] ~~~.\cr
}\eqno(2.18c)$$
When $A_{rs}$ is a $c$-number for all  
$r,s$, then the right hand side of Eq.~(2.18b) vanishes, and $\tilde C$ 
is anti-self-adjoint.  Thus, with the $c$-number choice 
$$A_{rs} =i\delta_{rs}~~~,\eqno(2.18d)$$ 
which trivially satisfies the condition of Eq.~(2.17b)  
and corresponds to the fermion kinetic structure and adjointness 
assignment used in Eqs.~(2.4a,b), we recover our earlier conclusion 
that $\tilde C$ is anti-self-adjoint.  Throughout most of this book we shall  
use this simple choice of $A_{rs}$, and we shall see that 
an anti-self-adjoint $\tilde C$  
naturally leads to an emergent quantum dynamics.   However, in Section 5F, 
where we consider stochastic corrections to the Schr\"odinger equation, 
we shall allow for the possibility that $\tilde C$ can have a self-adjoint 
part as well.  

As a very simple example of a nontrivial trace 
action that has a conserved $\tilde C$ that is 
not anti-self-adjoint, consider 
$${\bf L} = {\rm Tr} [\psi^{\dagger} A(\dot \psi  +\psi B) 
+{1\over 2}\dot B^2 ]~~~,\eqno(2.19a)$$
with $A=-A^{\dagger}$  a fixed matrix and with  
$B=-B^{\dagger}$  an  
anti-self-adjoint bosonic operator.    
Then with $q_F=\psi~,~~ p_F=\psi^{\dagger}A,$   and  $q_B=B~,~~
p_B=\dot B$,   
the corresponding trace Hamiltonian is 
$${\bf H} =  {\rm Tr}[ -p_F q_F q_B + {1 \over 2} p_B^2 ]
~~~,\eqno(2.19b)$$
which is a global unitary invariant  function of its arguments.   
Thus the operator 
$$\tilde C= [q_B,p_B] -\{q_F,p_F\}~~~\eqno(2.20a)$$ 
is conserved, as can be checked explicitly by use of the operator 
equations of motion 
$$\eqalign{
\dot q_F=&-q_F q_B~,~~\dot p_F=q_B p_F~~~,\cr
\dot q_B=&p_B~,~~\dot p_B=p_F q_F~~~.\cr
}\eqno(2.20b)$$
However, when $A$ is not a $c$-number, the calculations 
of Eqs.~(2.18b,c) show that 
$\tilde C$ has a piece that is self-adjoint,   
$$\tilde C + \tilde C^{\dagger}=[A,\psi \psi^{\dagger}]
=[A,q_Fp_FA^{-1}]~~~.\eqno(2.21a)$$
The anti-self-adjoint and self-adjoint parts of $\tilde C$ are separately 
conserved, as we readily verify from the equations of motion of Eq.~(2.21), 
$${d\over dt}(q_Fp_F)=\dot q_F p_F+q_F \dot p_F= 0 ~~~.\eqno(2.21b)$$

We conclude this discussion by exploring the connection between time reversal 
noninvariance and the appearance of a self-adjoint piece in $\tilde C$. 
Let us define the anti-unitary time reversal transformation ${\cal T}$, in  
analogy with the standard definition for fermion fields, by 
$$\eqalign{
{\cal T} i {\cal T}^{-1} =& -i~~~,\cr
{\cal T} \psi_r(t) {\cal T}^{-1} =&\sum_s U_{rs} \chi_s(-t)~~~,\cr
{\cal T} \psi^{\dagger}_r(t) {\cal T}^{-1} =& \sum_s U_{rs}^* \chi_s^{\dagger}
(-t)=\sum_s U_{sr}^{\dagger} \chi_s^{\dagger}(-t)~~~.\cr 
}\eqno(2.22a)$$
Here $U$ is a $c$-number unitary matrix, and we use the notation $\chi$ 
instead of $\psi$ on the right hand side as a reminder that under a 
linear superposition of the $\psi_r$ with complex coefficients, one 
obtains an antilinear superposition of the corresponding $\chi_s$, that is, 
a superposition with complex conjugated coefficients.  
With this definition, we find 
that the transformation of the expression appearing in Eq.~(2.17a) for 
the fermionic kinetic energy is 
$$\eqalign{
&{\cal T}\left(\sum_{r,s} \psi^{\dagger}_r(t) A_{rs} \partial_t \psi_s(t) 
\right) {\cal T}^{-1}   \cr
=&\sum_{m,n,r,s} \chi^{\dagger}_r(-t) U_{rm}^{\dagger} A_{mn}^* U_{ns} 
\partial_t \chi_s(-t) ~~~.\cr
}\eqno(2.22b)$$
Hence the kinetic action $\int dt {\bf L}_{\rm kin}$ will be form-invariant 
under time reversal if
$$A_{rs}=-\sum_{m,n} U_{rm}^{\dagger} A_{mn}^* U_{ns}~~~,\eqno(2.22c)$$
which is satisfied by the simplest choice $A_{rs}=i\delta_{rs}$ of 
Eq.~(2.18a).  If we instead take the somewhat more general choice 
$A_{rs}=A \delta_{rs}$, with $A=-A^{\dagger}$, then since $U$ is a 
$c$-number Eq.~(2.22c) simplifies to 
$$A=-A^*=A^T~~~,\eqno(2.22d)$$
with the superscript $T$ denoting the matrix transpose.  
Hence in this case, which is the relevant one for a Dirac spinor 
in an irreducible group representation,  
with $r$ a composite index 
labeling the Lorentz spinor and the group representation 
indices,  
time reversal noninvariance 
occurs when $A \not= A^T$.  This implies that $A$ cannot be a $c$-number,  
with the consequence that $[A,\sum_r \psi_r\psi_r^{\dagger}] \not= 0$, 
and thus $\tilde C$ 
necessarily has a self-adjoint part.  For more complicated index structures 
of $A_{rs}$, time reversal invariance does not necessarily imply the 
presence of a self-adjoint piece in $\tilde C$, although both can 
simultaneously be present.  
 
\vfill \eject

\twelvepoint
\doublespace
\pageno=30
\overfullrule=0pt
\centerline{3.~~~Trace Dynamics 
Models With Global Supersymmetry $^*$}
\bigskip
In the preceding sections, the degrees of freedom have been indexed by 
a discrete index $r$.  As is familiar in quantum field theory, 
when dealing with a continuum spacetime system with coordinates $\vec x$ 
and time $t$, the index $r$ labels infinitesimal boxes which fill out the 
coordinate space.  When the trace Lagrangian $\bf L$ 
has spacetime symmetries, there 
will be additional conserved trace quantities reflecting the presence 
of these symmetries.  For example, in a Poincar\'e invariant system, the 
trace Hamiltonian $\bf H$ and a trace momentum $\vec{\bf P}$ form a 
four vector ${\bf P}^{\mu}$, which appears as the $0\mu$ component of a 
trace stress-energy tensor ${\bf T}_{\nu \mu}$.  If the system is 
additionally scale invariant, the Lorentz trace of the trace stress-energy 
tensor vanishes, i.e. ${\bf T}_{\mu}^{\mu}=0$  
(with repeated spacetime indices 
summed over).  If the system has a global supersymmetry, there will be a 
conserved trace supersymmetry current with a time independent trace 
supercharge ${\bf Q}_{\alpha}$, which together with the trace four momentum 
obeys the Poincar\'e supersymmetry algebra under the generalized Poisson 
bracket of Eq.~(1.11a).   In this section we shall illustrate 
some of these 
comments with three concrete examples, the trace dynamics versions [9,10] 
of the Wess-Zumino model, the supersymmetric Yang-Mills model, and the 
so-called ``matrix model for $M$-theory''.  These three examples are worked    
out using component field methods; we close with a short discussion of a
superspace approach, and of the obstacles facing the construction of a trace  
dynamics theory with local supersymmetry.

\medskip
\centerline{3A.~~~The Wess-Zumino Model}
\medskip
We begin with the trace dynamics transcription of the 
Wess-Zumino model.  We follow the notational conventions of West [11], 
except that we normalize the fermion terms in the action differently, and 
always use the Majorana representation for the Dirac gamma matrices, in   
which $\gamma^{1,2,3}$ are real symmetric and $\gamma^0\,,~i\gamma^5$ are 
real skew-symmetric.  
(For useful cyclic identities satisfied by this representation of the 
$\gamma$ matrices, see the text and Appendix 
of [9].)    The trace Lagrangian for the Wess-Zumino model is
$$\eqalign{
{\bf L}=&\int d^3x {\rm Tr}\big( -{1\over 2} (\partial_{\mu}A)^2 -{1 \over 2} 
(\partial_{\mu} B)^2-\bar{\chi} \gamma^{\mu} \partial_{\mu} \chi + 
{1 \over 2} F^2 + {1 \over 2} G^2   \cr
&-m (AF+BG - \bar{\chi} \chi)   \cr             
&-\lambda[(A^2-B^2)F +G\{A,B\} 
-2 \bar{\chi}(A-i\gamma_5 B) \chi] \big)~~~,\cr
}\eqno(3.1)$$
with $A,B,F,G$ self-adjoint $N \times N$ matrices and with  
$\chi$ a Grassmann 4-component column vector spinor, each spin component 
of which is a self-adjoint Grassmann $N \times N$ matrix.  
The notation $\bar{\chi}$ is defined by $\bar{\chi}=\chi^T \hat\gamma^0$, 
with the transpose $T$ acting only on the Dirac spinor structure, 
so that $\chi^T$ is the 4-component row vector spinor constructed from the 
same $N \times N$ matrices that appear in $\chi$,  and $\hat\gamma^0$ is 
an abbreviation for $i \gamma^0$.  The numerical parameters $\lambda$ and 
$m$ are respectively the coupling constant and mass.  Equation (3.1) is 
identical in appearance to the usual Wess-Zumino model Lagrangian, except 
that we have explicitly symmetrized the term $G\{A,B\}$; symmetrization of 
the other terms is automatic (up to total derivatives that do not contribute 
to the action) by virtue of the cyclic property of the trace.  

Taking operator variations of Eq.~(3.1) by using the 
recipe of Eq.~(1.3b), the 
Euler-Lagrange equations of Eq.~(1.7c) take the form
$$\eqalign{
\partial^2A=&mF+\lambda (\{A,F\}+\{B,G\}-2\bar{\chi} \chi)~~~,\cr
\partial^2B=&mG+\lambda(-\{B,F\}+\{A,G\}+2i\bar{\chi}\gamma_5\chi)~~~,\cr
\gamma^{\mu}\partial_{\mu}\chi=
&m\chi+\lambda(\{A,\chi\}-i\{B,\gamma_5\chi\})~~~,\cr
{}F=&mA+\lambda(A^2-B^2)~~~,\cr
G=&mB+ \lambda\{A,B\}~~~.  \cr
}\eqno(3.2)$$
Transforming to Hamiltonian form, the canonical momenta of Eq.~(1.9a) are 
$$\eqalign{
p_{\chi}=&-\bar{\chi} \gamma^0=i\chi^T~~~,  \cr
p_A=&\partial_0 A~~~, \cr
p_B=&\partial_0 B~~~, \cr 
}\eqno(3.3)$$
and the trace Hamiltonian is given by 
$$\eqalign{
{\bf H}=&\int d^3x {\rm Tr} \big({1\over 2}[p_A^2+p_B^2+(\vec \nabla A)^2
+(\vec \nabla B)^2] -ip_{\chi} \hat \gamma^0 \vec \gamma 
\cdot \vec \nabla \chi   \cr
+&{1\over 2}(F^2+G^2) 
-m\bar{\chi} \chi+i \lambda p_{\chi} 
\hat\gamma^0\{(A-i\gamma_5B),\chi\} \big)~~~, \cr
}\eqno(3.4a)$$
in which $F$ and $G$ are understood to be the functions of $A$ and $B$ 
given by the final two lines of Eq.~(3.2), and where we have taken care   
to write {\bf H} so that it is manifestly symmetric 
in the identical quantities $p_{\chi}$ and $i\chi^T$.
The trace three-momentum $\vec{\bf P}$ is given by 
$$\vec{\bf P}=-\int d^3x {\rm Tr} (p_A \vec\nabla A+p_B \vec \nabla B
+p_{\chi} \vec \nabla \chi)~~~,\eqno(3.4b)$$
while the conserved trace quantity ${\bf N}$ of Eq.~(2.4e) and the 
conserved operator $\tilde C$ of Eq.~(2.6) are given respectively by 
$$\eqalign{
{\bf N}=&\int d^3x {\rm Tr} \chi^T \chi~~~,\cr
\tilde C=&\int d^3x ([A,p_A]+[B,p_B]-\{\chi,p_{\chi} \})  ~~~,\cr
}\eqno(3.5)$$
with a contraction of the spinor indices in the final term on the 
second line of Eq.~(3.5) understood.  
Equations (3.4a,b) are clearly formed from the usual field theoretic 
expressions for the Hamiltonian and three-momentum by taking the trace, 
and symmetrizing factors where this is not already implicit from the cyclic 
properties of the trace.  Exactly the same procedure can be used to form 
the full trace energy-momentum tensor ${\bf T}_{\nu\mu}$.  

Let us now perform a supersymmetry variation of the fields given by 
$$\eqalign{
\delta A=&\bar \epsilon \chi ~~,~~~\delta B=i \bar \epsilon 
\gamma_5 \chi~~~,  \cr
\delta \chi=&{1\over 2}[F+i\gamma_5 G+\gamma^{\mu}\partial_{\mu}
(A+i\gamma_5 B)]
\epsilon~~~,  \cr
\delta F=&\bar \epsilon \gamma^{\mu} \partial_{\mu} \chi~~,~~~
\delta G=i \bar \epsilon \gamma_5 \gamma^{\mu} \partial_{\mu} \chi~~~,\cr
}\eqno(3.6)$$
with $\epsilon$ a $c$-number Grassmann spinor (i.e., a four component spinor, 
the spin components of which are $1 \times 1$ Grassmann matrices).
Substituting Eq.~(3.6) into the trace Lagrangian of Eq.~(3.1), 
a lengthy 
calculation shows that when $\epsilon$ is constant, the variation of {\bf L} 
vanishes.   The calculation parallels that done in the conventional 
$c$-number Lagrangian case, except that the cyclic properties of the trace 
and cyclic identities obeyed by the Majorana representation $\gamma$ 
matrices [9] (see Appendix D) 
are used extensively in place of commutativity/anticommutativity of the 
fields. 
When $\epsilon$ is not constant, the variation of {\bf L} is given by 
$$\eqalign{
\delta {\bf L}=&\int d^3x {\rm Tr}(\bar J^{\mu} 
\partial_{\mu} \epsilon)~~~,  \cr
\bar J^{\mu}=&-\bar \chi \gamma^{\mu}
\big[(\gamma^{\nu}\partial_{\nu}+m)
(A+i\gamma_5 B)+\lambda(A^2-B^2+i\gamma_5\{A,B\}) \big]~~~,   \cr
}\eqno(3.7a)$$
which identifies the trace supercharge ${\bf Q}_{\alpha}$ as 
$$\eqalign{
{\bf Q}_{\alpha}\equiv&\int d^3x {\rm Tr} \bar J^0 \alpha \cr
=&\int d^3x {\rm Tr}{1 \over 2}(p_{\chi}+i\chi^T)
\big[(\gamma^{\nu}\partial_{\nu}+m)
(A+i\gamma_5 B)+\lambda(A^2-B^2+i\gamma_5\{A,B\}) \big]\alpha ~~~,   \cr
}\eqno(3.7b)$$
with $\alpha$ a  $c$-number Grassmann spinor, 
where we have again taken care to express ${\bf Q}_{\alpha}$ 
symmetrically in the identical quantities $p_{\chi}$ and $i\chi^T$. It  
is straightforward to check, using the equations of motion and the 
cyclic identity, that 
$\bar{\bf J}^{\mu}\equiv {\rm Tr} \bar J^{\mu}$ is a 
conserved trace supercurrent, which implies 
that the trace supercharge is time independent.    

It is now straightforward (but tedious) to check the closure of the 
supersymmetry algebra under 
the generalized Poisson bracket of Eq.~(1.11a), which for the 
Hamiltonian dynamics of the Wess-Zumino model gives
$$\eqalign{
\{ {\bf Q}_{\alpha},{\bf Q}_{\beta} \}=&{\rm Tr}\big[
{\delta {\bf Q}_{\alpha} \over \delta A} {\delta {\bf Q}_{\beta} 
\over \delta p_A}+
{\delta {\bf Q}_{\alpha} \over \delta B} {\delta {\bf Q}_{\beta} 
\over \delta p_B}
-{\delta {\bf Q}_{\alpha} \over \delta \chi} 
{\delta {\bf Q}_{\beta} 
\over \delta p_{\chi}}
-\big(\alpha \leftrightarrow \beta\big) 
\big] \cr
=&\bar\alpha \gamma^0 \beta {\bf H}
-\bar \alpha \vec \gamma \beta \cdot \vec {\bf P} ~~~,\cr
}\eqno(3.8a)$$
with {\bf H} and $\vec {\bf P}$ the trace Hamiltonian and three-momentum
given above.  
It is also easy to check that ${\bf Q}_{\epsilon}$ plays the role of the 
generator of supersymmetry transformations for the dynamical variables 
$A,B,\chi$ under the generalized Poisson bracket, since we readily find 
(for constant Grassmann even parameters $a,b$ and Grassmann odd 
parameter $c$)
$$\{ {\rm Tr}(aA+bB+c\chi), {\bf Q}_{\epsilon}\}=
{\rm Tr}(a \delta A+ b \delta B+c \delta \chi)~~~,\eqno(3.8b)$$
with $\delta A, \delta B, \delta \chi$ the supersymmetry variations 
given by Eq.~(3.6) above, after 
elimination of the auxiliary fields $F,G$ by their equations of motion.

\medskip
\centerline{3B.~~~The Supersymmetric Yang-Mills Model}
\medskip

As a second example of a trace dynamics model with global 
supersymmetry, we discuss 
supersymmetric Yang-Mills theory.  
We start from the trace Lagrangian 
$${\bf L}=\int d^3x {\rm Tr}\big[ {1\over 4g^2} F^2_{\mu\nu} -\bar \chi
\gamma^{\mu}D_{\mu} \chi +{1\over 2} D^2\big]~~~,\eqno(3.9a)$$
with the field strength $F_{\mu\nu}$ and covariant derivative $D_{\mu}$ 
constructed from the gauge potential $A_{\mu}$ according to 
$$\eqalign{
{}F_{\mu\nu}=&\partial_{\mu}A_{\nu}-\partial_{\nu}A_{\mu}
+[A_{\mu},A_{\nu}]~~~, \cr
D_{\mu}{\cal O}=&\partial_{\mu}{\cal O}+[A_{\mu},{\cal O}] \cr
\Rightarrow&D_{\mu}F_{\nu\lambda}+D_{\nu}F_{\lambda\mu}+D_{\lambda}F_{\mu\nu}             
=0  ~~~.\cr
}\eqno(3.9b)$$
In Eq.~(3.9b), the potential components $A_{\mu}$ are each an 
anti-self-adjoint, and 
the auxiliary field $D$ a self-adjoint, 
$N \times N$ matrix, and each spinor component of  
$\chi$ is a self-adjoint Grassmann $N \times N$ matrix.
The Euler-Lagrange equations of motion are 
$$\eqalign{
D=&0~~~,   \cr
\gamma^{\mu}D_{\mu} \chi=&0~~~,   \cr
D_{\mu}F^{\mu\nu}=&2g^2\bar \chi \gamma^{\nu} \chi ~~~;\cr
}\eqno(3.10a)$$
as usual for a gauge system, the $\nu=0$ component of Eq.~(3.10a) is not  
a dynamical evolution equation, but rather the constraint
$$D_{\ell}F^{\ell 0}=2g^2 \bar \chi \gamma^0 \chi~~~.\eqno(3.10b)$$

Going over to the Hamiltonian formalism, the canonical momenta are 
given by 
$$p_{A_{\ell}}=-{1 \over g^2} F_{0\ell}~,~~~p_{\chi}=i\chi^T~~~, 
\eqno(3.11a)$$
and the axial gauge trace Hamiltonian 
is 
$${\bf H}={\bf H}_{A} +{\bf H}_{\chi}~~~,\eqno(3.11b)$$
with 
$$\eqalign{
{\bf H}_{A}=&\int d^3x {\rm Tr} \left( {-g^2 \over 2} \sum_{\ell=1}^2 
p^2_{A_{\ell}} -{1 \over 2 g^2} F_{03}^2 \right.\cr
&\left. -{1 \over 2g^2}(\partial_1A_2-\partial_2A_1
+ [A_1,A_2])^2 -{1 \over 2g^2}
[(\partial_3 A_1)^2 + (\partial_3 A_2)^2 ]\right)~~~, \cr
{}F_{03}=&{1\over2} g^2 \int_{-\infty}^{\infty}dz^{\prime} 
\epsilon(z-z^{\prime})
[-(p_{\chi}\chi+\chi^Tp^T_{\chi})+D_1p_{A_1}+D_2p_{A_2}]
\vert_{z^{\prime}}~~~,\cr
{\bf H}_{\chi}=&-i\int d^3x {\rm Tr}(p_{\chi} \hat \gamma^0\gamma_{\ell} 
D_{\ell}\chi) ~~~,\cr
}\eqno(3.11c)$$
where we have taken care to write {\bf H} in a form symmetric in the 
identical quantities $p_{\chi}$ and $i\chi^T$, and where $\epsilon(z)=1(-1)$  
for $z>0(z<0)$.
The trace three momentum is 
$$ {\bf P}_m=-\int d^3x {\rm Tr} (\sum_{\ell=1}^3 F_{m\ell}p_{A_{\ell}}
+p_{\chi}D_m\chi )~~~,\eqno(3.12)$$
and the conserved operator $\tilde C$ of Eq.~(2.6) is given by 
$$\tilde C=\int d^3x (\sum_{\ell=1}^2[A_{\ell},p_{A_{\ell}}]
-\{\chi,p_{\chi}\})
~~~,\eqno(3.13a)$$
with a contraction of the spinor indices in the final term of Eq.~(3.13a) 
understood.  By virtue of the constraint of Eq.~(3.10b), the conserved 
operator $\tilde C$ can also be written as 
$$\tilde C=-\int d^3x \sum_{\ell=1}^2  \partial_{\ell} p_{A_{\ell}}=
-\int_{\rm sphere~at~\infty} d^2 S_{\ell}~p_{A_{\ell}}~~~,\eqno(3.13b)$$
which vanishes when the surface integral in Eq.~(3.13b) is zero.  
The conserved trace quantity ${\bf N}$ of Eq.~(2.4e) has the same form 
as in the Wess-Zumino model, 
$${\bf N}=\int d^3x {\rm Tr} \chi^T \chi~~~.\eqno(3.14)$$

Making now the supersymmetry variations 
$$\eqalign{
\delta A_{\mu}=&ig \bar \epsilon \gamma_{\mu} \chi~~~, \cr
\delta \chi=&\big({i\over 8g} [\gamma_{\mu},\gamma_{\nu}]F^{\mu\nu} 
+{i \over 2} \gamma_5 D)\epsilon~~~, \cr 
\delta D=&i \bar \epsilon \gamma_5 \gamma^{\mu} D_{\mu} \chi~~~,  \cr
}\eqno(3.15)$$
in the trace Lagrangian, with $\epsilon$ again a Grassmann $c$-number, 
we find using cyclic invariance under the trace 
and the gamma matrix identities given in Appendix D 
that when $\epsilon$ is constant, the variation vanishes.  
When $\epsilon$ is not a constant, the variation of {\bf L} is given by 
$$\eqalign{
\delta {\bf L} =&\int d^3x {\rm Tr} (\bar J^{\mu} 
\partial_{\mu} \epsilon)~~~, \cr
\bar J^{\mu}=&-{i \over 4g} \bar \chi \gamma^{\mu} F_{\nu \sigma}
[\gamma^{\nu},\gamma^{\sigma}]~~~,\cr
}\eqno(3.16a)$$
from which we construct the trace supercharge ${\bf Q}_{\alpha}$ as 
$${\bf Q}_{\alpha}=\int d^3x {\rm Tr} {i \over 8g}(p_{\chi}+i\chi^T)
{}F_{\nu \sigma} [\gamma^{\nu},\gamma^{\sigma}]\alpha~~~,\eqno(3.16b)$$
again with $\alpha$ a  $c$-number Grassmann spinor.  
Again, it is straightforward to check, using the equations of motion and 
the cyclic identity, that $\bar{\bf J}_{\mu}\equiv {\rm Tr} \bar J_{\mu}$ 
is a conserved trace 
supercurrent, which implies that the trace supercharge is conserved. 

One can now verify the closure of the 
supersymmetry algebra under 
the generalized Poisson bracket of Eq.~(1.11a), which for the 
Hamiltonian dynamics of the supersymmetric Yang-Mills model gives
$$\eqalign{
\{ {\bf Q}_{\alpha},{\bf Q}_{\beta} \}=&{\rm Tr}\big[ \sum_{l=1}^2
{\delta {\bf Q}_{\alpha} \over \delta A_{\ell}} 
{\delta {\bf Q}_{\beta} \over \delta p_{A_{\ell}} }
-\sum_{d=1}^4 {\delta {\bf Q}_{\alpha} \over \delta \chi^d} 
{\delta {\bf Q}_{\beta} \over \delta p_{\chi^d}}
-\big(\alpha \leftrightarrow \beta\big) 
\big]\cr 
=&\bar\alpha \gamma^0 \beta {\bf H}
-\bar \alpha \vec \gamma \beta \cdot \vec {\bf P} ~~~,\cr
}\eqno(3.17)$$
with ${\bf H}$ and $\vec {\bf P}$ given by Eqs.~(3.11b,c) and Eq.~(3.12a) 
respectively.  
Examining the 
role of the supercharge as a generator of transformations, in analogy with 
Eq.~(3.8b),  the supercharge in the Yang-Mills case is found to generate 
the supersymmetry variations of Eq.~(3.15), plus an infinitesimal change 
of gauge.         

\medskip
\centerline{3C.~~~The Matrix Model for M Theory}
\medskip

As our third example of a trace dynamics model with global supersymmetry, 
we consider the matrix model [12]  that has been recently studied [13] in a
string-theory context under the name ``the matrix model for M-theory''.  
This model, formulated in zero spatial dimensions, has 
the trace Lagrangian {\bf L} given by 
$${\bf L}={\rm Tr}  \left( {1\over 2} D_t X_i D_t X^i
+ i \theta^T D_t \theta 
+{1\over 4} [X_i,X_j] [X^i,X^j] + \theta^T \gamma_i [\theta, X^i]\right)~~~,
\eqno(3.18)$$
with the covariant derivative defined now by 
$D_t {\cal O}=\partial_t {\cal O} -i [A_0, {\cal O}]$.  
In Eq. (3.18), 
a summation convention is understood on the indices $i,j$ which range 
from 1 to 9; $A_0$ and the $X_i$ are self-adjoint $N \times N$  
complex matrices, while $\theta$ is a 16-component fermionic spinor 
each element of 
which is a self-adjoint $N \times N$ complex Grassmann matrix, 
with the transpose $T$ acting 
only on the spinor structure but not on the $N \times N$ matrices, so that 
$\theta^T$ is simply the 16 component row spinor corresponding to the 
16 component column spinor $\theta$.  The potential $A_0$ has no kinetic 
term and so is a pure gauge degree of freedom.  Finally, 
the $\gamma_i$ are a set 
of nine $16 \times 16$ matrices, which are related to the standard 
$32 \times 32$ matrices $\Gamma^{\mu}$ as well as to the Dirac matrices of 
spin(8), as conveniently described in [14] and obeying identities    
summarized in Appendix D.  
 
Starting from the trace Lagrangian of Eq.~(3.18), using the 
definition of Eq.~(1.3b) to take operator variations, the operator
Euler-Lagrange equations of Eq.~(1.7c) give the equations of motion  
of the matrix model, 
$$\eqalign{
D_t^2X^i=&[[X^j,X^i],X_j]+2\theta^T \gamma^i \theta~~~,   \cr
D_t\theta^T=&i[\theta^T \gamma_i,X^i]~~ \Rightarrow~~ D_t\theta=
i[\gamma_i \theta, X^i]  ~~~,\cr
}\eqno(3.19a)$$
together with the constraint that the generic conserved operator $\tilde C$ 
of Eq.~(2.6) vanishes in this 
model, 
$$\tilde C = [X^i,D_t X_i]-2i\theta^T \theta=0~~~.\eqno(3.19b)$$
To transform the dynamics to trace Hamiltonian form, we define the 
canonical momenta $p_{X_i}$ and $p_{\theta}$ by 
$$\eqalign{
p_{X_i}=&{\delta {\bf L} \over \delta (\partial_t X_i) }=D_tX^i~~~, \cr
p_{\theta}=&{\delta {\bf L} \over \delta (\partial_t \theta) }=i \theta^T
~~~,\cr
}\eqno(3.20a)$$
so that the trace Hamiltonian is given by 
$${\bf H}={\rm Tr}(p_{X_i}\partial_t X_i + p_{\theta} \partial_t \theta )
-{\bf L}             
={\rm Tr}\left({1 \over 2} p_{X_i} p_{X^i} - {1 \over 4}[X_i,X_j][X^i,X^j]
+i p_{\theta} \gamma_i [\theta,X^i] +i A_0 \tilde C \right)~~~.
\eqno(3.20b)$$
Again, because $p_{\theta}=i\theta^T$, we have written the  
trace Hamiltonian in a form that is manifestly symmetric under the 
replacements $p_{\theta} \to i \theta^T,~~\theta \to -ip_{\theta}^T$.

Let us next consider the variation of the trace 
Lagrangian under the supersymmetry transformation defined by 
$$\eqalign{
\delta X^i=&-2 \epsilon^T \gamma^i \theta = 2 \theta^T \gamma^i 
\epsilon~~~, \cr
\delta \theta =&-\left(iD_tX^i \gamma_i +{1 \over 2}[X^i,X^j] \gamma_{ij}
\right) \epsilon + \epsilon^{\prime}~~~,  \cr
\delta A_0=&-2 \epsilon^T \theta=2 \theta^T \epsilon   ~~~.\cr
}\eqno(3.21)$$
Here $\epsilon$ and $\epsilon^{\prime}$ are 16 component Grassmann 
$c$-number spinors, that is, they are column vectors each of whose 16 
components is an independent $1 \times 1$ Grassmann matrix.  Using  
the cyclic trace identities of Sec.~1 and the gamma matrix properties 
summarized Appendix D, it is 
a matter of straightforward but lengthy calculation to verify that the 
trace Lagrangian is invariant under the transformation of Eq.~(3.21) 
when $\epsilon$ and $\epsilon^{\prime}$ are time independent.  
When $\epsilon$ and $\epsilon^{\prime}$ have a time dependence,  
$\delta {\bf L}$ 
is no longer zero, but instead is given by 
$$\eqalign{
\delta{\bf L}=&\partial_t {\rm Tr} \big[ -i \theta^T \epsilon^{\prime}
+\big( \theta^T \gamma_i D_t X^i + {1 \over 2}i\theta^T \gamma_{ij} [X^i,X^j]
\big) \epsilon \big] \cr    
+&{\rm Tr} \big[ 2i\theta^T\partial_t\epsilon^{\prime} 
+\big( 2 \theta^T\gamma_i D_t X^i  - i \theta^T\gamma_{ij} [X^i,X^j]  \big)
\partial_t \epsilon    \big] ~~~.\cr
}\eqno(3.22a)$$
This identifies the trace supercharges ${\bf Q}^{\prime}_{\alpha}$ and 
${\bf Q}_{\alpha}$ as
$$\eqalign{
{\bf Q}^{\prime}_{\alpha}=&{\rm Tr}  2i\theta^T \alpha~~~, \cr 
{\bf Q}_{\alpha} =&{\rm Tr}\big( 2\theta^T\gamma_iD_tX_i -i \theta^T 
\gamma_{ij} [X^i,X^j] \big) \alpha  ~~~,\cr
}\eqno(3.22b)$$
with $\alpha$ a 16-component $c$-number Grassmann spinor, 
and their conservation is easily checked using the equations of motion
and $\gamma$ matrix identities.    
To check the supersymmetry algebra, 
we must first write the supercharges of Eq.~(3.22b)  
in Hamiltonian form, symmetrized with respect to  
$p_{\theta}$ and $i \theta^T$, giving 
$$\eqalign{
{\bf Q}^{\prime}_{\alpha}=& {\rm Tr}(p_{\theta}+i\theta^T) \alpha~~~, \cr 
{\bf Q}_{\alpha}=&-{\rm Tr}(p_{\theta}+i\theta^T)\big( i\gamma_ip_{X_i} 
+{1\over 2} \gamma_{ij} [X^i,X^j] \big) \alpha  ~~~.\cr
}\eqno(3.22c)$$
Using the generalized Poisson bracket corresponding 
to the Hamiltonian structure of our model, defined now by  
$$\{{\bf A},{\bf B} \}={\rm Tr}\left(  {\delta{\bf A} \over \delta X_i}
{\delta{\bf B} \over \delta p_{X_i}}-{ \delta{\bf B} \over \delta X_i}
{\delta{\bf A} \over \delta p_{X_i}} -{\delta{\bf A}\over \delta \theta}
{\delta {\bf B} \over \delta p_{\theta} }+{\delta {\bf B} \over 
\delta \theta}
{\delta {\bf A} \over \delta p_{\theta} } \right)~~~,\eqno(3.23)$$
it is straightforward to evaluate the supercharge algebra, and to show  
that it has the expected form  [10].  

\bigskip
\centerline{3D.~~~Superspace Considerations and Remarks}
\bigskip

The derivations of Secs. 3A,B,C have all been carried out 
in the component 
formalism, which requires doing a separate computation for each Poincar\'e 
supersymmetry multiplet.  However, there is a simple and general superspace 
argument for the results we have obtained.  Recall that superspace is 
constructed by introducing four fermionic coordinates $\theta_{\alpha}$  
corresponding to the four space-time coordinates $x_{\mu}$.  The graded 
Poincar\'e algebra is then represented by differential operators 
constructed from the superspace coordinates, and superfields are represented 
by finite polynomials in the fermionic coordinates $\theta_{\alpha}$, with 
coefficient functions that depend on $x_{\mu}$.  To generalize the 
superspace formulation to give trace dynamics models, one simply replaces 
these coefficient functions by $N\times N$ matrices (or operators), and 
one inserts a trace Tr acting on the superspace integrals used to 
form the action.  Then the standard argument that the 
action is invariant under superspace translations still holds for the trace 
action formed this way from the matrix components of the superfields.  
We immediately see from this argument why it is essential for the 
supersymmetry parameter $\epsilon$ to be a 
Grassmann $c$-number and not also a matrix; 
this parameter appears as the magnitude of an infinitesimal superspace 
translation, and since the superspace coordinates $x_{\mu}$ and 
$\theta_{\alpha}$ are $c$-numbers, the parameter $\epsilon$ must be one also.  
The construction just given gives reducible supersymmetry representations, 
and various constraints must be applied to the superfields to pick out 
irreducible representations.  Since these constraints act linearly on the 
expansion coefficients, they can all be immediately generalized (with the 
usual replacement of complex conjugation for $c$-numbers by the adjoint) to 
the case in which the coefficient functions are matrices or operators.  

The simplicity of this argument suggests that generally, for rigid 
supersymmetry theories for which there exists a superspace construction,  
there will exist a corresponding trace dynamics generalization.  The 
superspace argument also suggests why it has not been possible [15] to 
construct trace dynamics generalizations of local supersymmetry theories, 
such as supergravity.  The commutator of two local supersymmetries 
with supersymmetry parameters $\epsilon_1$ and $\epsilon_2$ is a linear 
combination [11] of a local Lorentz transformation, a general coordinate 
transformation, and a supersymmetry transformation with supersymmetry 
parameter proportional to 
$$\bar \epsilon_2 \gamma_{\mu} \epsilon_1  \psi^{\mu}~~~,
\eqno(3.24)$$
with $\psi^{\mu}$ 
the Rarita-Schwinger gravitino field.  Even if we start with 
$\epsilon_{1,2}$ that are $c$-numbers, the new supersymmetry parameter 
given by Eq.~(3.24) will be matrix valued in a trace dynamics 
generalization where the gravitino field $\psi^{\mu}$ is matrix valued.  
Thus, an extension of the results of this section to local supersymmetries  
would appear to require a generalization of the results presented above  
to the case in which the supersymmetry   
parameter $\epsilon$ is matrix-valued, rather than a $c$-number as assumed 
throughout our discussion.
\bigskip                                                           
\vfill\eject

\twelvepoint
\doublespace
\pageno=42
\overfullrule=0pt
\centerline{4.~~~Statistical Mechanics of Matrix Models}
\bigskip
Up to this point we have discussed matrix models as classical dynamical 
systems.   We shall now start laying the groundwork for the emergence 
of quantum mechanical behavior from a matrix dynamics in which quantization   
is not assumed a priori.  We begin this discussion by emphasizing 
that we shall {\it not} follow the traditional 
route [16] of canonically quantizing a matrix model, in which working from   
Eqs.~(1.8) and (1.9a) one takes each classically 
conjugate matrix 
element pair $(q_r)_{ij}$ and $(p_r)_{ji}=
\partial {\bf L} / \partial (\dot q_r)_{ij}$, and elevates them to quantum  
operators that satisfy canonical commutation relations, such as (for bosonic 
degrees of freedom) 
$$\eqalign{ 
[(q_r)_{ij},(q_s)_{kl}]=&[(p_r)_{ij},(p_s)_{kl}]=0~~~,\cr
[(q_r)_{ij},(p_s)_{kl}]=&i\delta_{rs}\delta_{il}\delta_{jk}~~~.\cr
}\eqno(4.1)$$
In this approach, each classical matrix $q_r$, because it has $N^2$ matrix 
elements, ends up spawning $N^2$ quantum operators.  The canonical 
quantization approach is appropriate, for example, when dealing with a 
matrix model that arises as a discretized approximation to a continuum 
field system, or as an approximation to a many-body system with a large 
number of independent degrees of freedom.  In these cases, the matrix 
elements each represent a degree of freedom to which, assuming that one 
is dealing with a quantum field or a quantum many body system, the usual 
quantization rules apply.  
To repeat, this is {\it not} what we shall do, because we are not assuming   
that quantum theory applies at the underlying trace dynamics level.  

Instead, we shall require that the only matrix 
(or operator) structure present is that which is already present in 
the classical matrix model [17].   Thus, 
each $q_r$ and each $p_r$ corresponds 
to a single operator degree of freedom, and these degrees of freedom 
do not obey any simple 
commutation algebra:  in general for bosonic degrees of freedom, 
$[q_r,q_s]$, $[p_r,p_s]$, and $[q_r,p_s]$ will all be nonzero, and similarly   
when fermionic degrees of freedom are included, 
with anticommutators replacing commutators as appropriate.  However, 
we shall argue that in the statistical dynamics of matrix models with 
a global unitary invariance, within thermodynamic averages over polynomials 
of the $q$'s and $p$'s, the matrix variables $q_r$ and $p_s$ obey (when 
a specific approximation is made) an {\it effective} 
commutator/anticommutator  
algebra of the familiar canonical form.  
In other words, we shall show that quantum  
behavior is an emergent feature of the statistical mechanics of 
a particular class of matrix models, 
with each matrix variable pair $q_r$, $p_r$ corresponding to {\it one} 
quantum mechanical operator degree of freedom.  The first step 
in such a program is to set up the statistical mechanics of matrix models, 
and that is what we shall do in this section.  

\medskip
\centerline{4A.~~~The Liouville Theorem}

We begin our statistical treatment of matrix models by deriving [7] an 
analog of the Liouville theorem, which states that the matrix model trace 
dynamics leaves a suitably defined phase space volume element invariant.  
We shall actually derive a more general result, showing that the phase 
space volume element is invariant under the general canonical transformations 
introduced in Eqs.~(2.13a,b), of which the trace Hamiltonian dynamics 
is a special case.  

{}Following the notation of Sec.~1, we denote the general matrix 
element of the operator $x_r$ by $(x_r)_{mn}$, which can be decomposed 
into real and imaginary parts according to
$$(x_r)_{mn}=(x_r)_{mn}^0+i(x_r)_{mn}^1~~~,\eqno(4.2a)$$
where $(x_r)_{mn}^A$  with $A=0,1$ are real numbers.  
If for the moment we ignore adjointness restrictions, 
the natural phase space measure is defined by
$$\eqalign{
d\mu=&\prod_A d \mu^A~~~, \cr 
d\mu^A \equiv& \prod_{r,m,n} d(x_r)_{mn}^A~~~;\cr 
}\eqno(4.2b)$$
when adjointness restrictions are taken into account, certain factors in
Eq.~(4.2b) become redundant and are omitted.  Our strategy is first to 
ignore adjointness restrictions and to prove the canonical invariance of
each individual factor $d\mu^A$ in the first line of Eq.~(4.2b), and then   
to indicate how the argument is altered when adjointness restrictions are 
taken into account. 
 
Under the general canonical transformation of Eq.~(2.13b), the matrix 
elements
of the new variables $x_r^{\prime}$ are related to those of the original
variables $x_r$  by
$$(x_r^{\prime})_{mn}^A=(x_r)_{mn}^A+ \sum_s  \omega_{rs} 
\left( {\delta {\bf G} \over \delta x_s} \right)_{mn}^A~~~. \eqno(4.3a)$$
Inserting a complete set of intermediate states into the fundamental
definition
$$\delta {\bf G} = {\rm Tr} \sum_s {\delta {\bf G} \over \delta x_s} 
\delta x_s~~~,  \eqno(4.3b)$$
and using the reality of ${\bf G}$, we get
$$\delta {\bf G} =  \sum_{s,m,n,A} 
\epsilon^A \left( {\delta {\bf G} \over \delta x_s} \right)_{mn}^A 
(\delta x_s)_{nm}^A~~~,\eqno(4.3c)$$
where $\epsilon^0=1$ and $\epsilon^1=-1$. 
Thus, we see that
$$\left( {\delta {\bf G} \over \delta x_s} \right)_{mn}^A =
\epsilon^A {\partial {\bf G} \over \partial (x_s)_{nm}^A }
~~~,\eqno(4.3d)$$
a result that can also be obtained by decomposing Eq.~(1.8) into real and 
imaginary parts.  Equation  (4.3d) 
allows us to rewrite Eq.~(4.3a) in terms of conventional  
partial derivatives of
the total trace functional {\bf G},
$$(x_r^{\prime})_{mn}^A=(x_r)_{mn}^A+ \sum_s  \omega_{rs} 
\epsilon^A {\partial {\bf G} \over \partial (x_s)_{nm}^A } 
~~~. \eqno(4.3e)$$
Differentiating Eq.~(4.3e) with respect to $(x_{r^{\prime}})_{m^{\prime}
n^{\prime}}^A$, we get for the transformation matrix 
$${\partial (x_r^{\prime})_{mn}^A  \over \partial 
(x_{r^{\prime}})_{m^{\prime} n^{\prime}}^A  }=
\delta_{r r^{\prime}} \delta_{m m^{\prime}} \delta_{n n^{\prime}}
+ \sum_s \omega_{rs}  \epsilon^A
{ \partial^2 {\bf G} \over \partial (x_s)_{nm}^A
\partial (x_{r^{\prime}})_{m^{\prime} n^{\prime}}^A  }  ~~~. \eqno(4.4)$$

Since for an infinitesimal matrix $\delta X$ we have $\det(1+\delta X)
\approx 1+{\rm Tr} \delta X$, we learn from Eq.~(4.4) that 
the Jacobian of the transformation is
$$\eqalign{
J=&1+\Sigma~~~, \cr
\Sigma=&\sum_{r,s,m,n} 
\omega_{rs}  \epsilon^A
{ \partial^2 {\bf G} \over \partial (x_s)_{nm}^A \partial (x_r)_{mn}^A  } \cr
}~~~. \eqno(4.5a)$$
Interchanging in the expression for $\Sigma$ in Eq.~(4.5a) 
the summation indices $r$ and $s$,  and 
also interchanging
the summation indices $m$ and $n$, we get
$$
\Sigma=\sum_{r,s,m,n} 
\omega_{sr} \epsilon_n \epsilon^A
{ \partial^2 {\bf G} \over \partial (x_r)_{mn}^A \partial (x_s)_{nm}^A  } 
~~~. \eqno(4.5b)$$
However, now using the fact that for bosonic $r,s$ we have 
$$\eqalign{
\omega_{sr}=&-\omega_{rs}~~~,\cr
{ \partial^2 {\bf G} \over \partial (x_r)_{mn}^A \partial (x_s)_{nm}^A  } 
=&{ \partial^2 {\bf G} \over  \partial (x_s)_{nm}^A \partial (x_r)_{mn}^A } 
~~~,\cr
}\eqno(4.5c)$$
while for fermionic $r,s$ we have 
$$\eqalign{
\omega_{sr}=&\omega_{rs}~~~,\cr
{ \partial^2 {\bf G} \over \partial (x_r)_{mn}^A \partial (x_s)_{nm}^A  } 
=&-{ \partial^2 {\bf G} \over  \partial (x_s)_{nm}^A \partial (x_r)_{mn}^A } 
~~~,\cr
}\eqno(4.5d)$$
we see that Eqs.~(4.5a-d) imply that 
$\Sigma=-\Sigma$; hence $\Sigma$
vanishes and the Jacobian of the transformation is unity.  

Now let us see how this argument is modified when we take the adjointness 
restrictions on the phase space variables $x_r$ into account.  
Inspection of the argument just given  
shows that the diagonal $(m=n)$ and off--diagonal
$(m \neq n)$ terms in the sum $\Sigma$ vanish separately, and 
for each of these, 
the summed contribution from the canonical coordinate and momentum 
pair $q_r, p_r$ for each fixed $r$ also vanishes separately.  
This observation
permits us to readily take the adjointness restrictions into account; in the  
following discussion we shall write $d\mu = d\mu_B d\mu_F$, with $d\mu_B$ 
and $d\mu_F$ respectively the bosonic and fermionic integration measures. 

{}For a bosonic pair of phase space variables $q_r, p_r$, 
the $x_r$ variables are independent but are both self-adjoint, and thus
$$(x_r)_{mn}^A= \epsilon^A (x_r)_{nm}^A~~~. \eqno(4.6a) $$  
This means that the integration
measure must be redefined to include only the factors that are real 
diagonal in $m,n$ (the imaginary diagonal ones are identically zero), 
and only
the upper diagonal off-diagonal factors (since the lower diagonal ones 
are related to the upper diagonal ones by complex conjugation), 
so that the bosonic integration measure becomes
$$d\mu_B= \prod_{r,m} d(x_r)_{mm}^0 
\prod_{r,m<n,A} d(x_r)_{mn}^A~~~. 
\eqno(4.6b)$$
The argument for the diagonal terms in this product proceeds just as did that
for the diagonal terms in the unrestricted case, while the argument for the
off-diagonal terms differs from that in the unrestricted case only by 
inserting a factor of ${1\over 2}$ in front of $\Sigma$ in Eq.~(4.5a) and the subsequent 
equations that follow from it, which has no effect on the argument for   
the vanishing of $\Sigma$. 

{}For a fermionic pair of phase space variables constructed according to
the recipe $q_r=\psi_r,~~ p_r=i \psi_r^{\dagger} = iq_r^{\dagger}$ of   
Eq.~(2.4b), 
the $x_r$ variables are
no longer independent.  However, this construction implies that 
$$ (q_r)_{mn}^1= (p_r)_{nm}^0~,~~ (p_r)_{mn}^1=(q_r)_{nm}^0~~~, 
\eqno(4.7a)$$
and thus the fermionic integration measure 
must be redefined as
$$d\mu_F= \prod_{r, m ,n}  d(x_r)_{mn}^0~~~. \eqno(4.7b)$$
The argument, for both the diagonal and the off-diagonal 
factors in $m,n$ then 
proceeds just as in the unrestricted case, again apart from 
insertion of an irrelevant factor of ${1\over 2}$ in front of $\Sigma$ 
in Eq.~(4.5a) and 
the subsequent equations that follow from it.

To summarize, we have shown that the matrix operator phase space integration 
measure $d \mu$ is invariant under general canonical transformations. 
As noted at the beginning of this section, 
an important corollary of this result follows when {\bf G}
is taken as the generator $dt {\bf H}$ of an infinitesimal time translation,
since we then learn that $d \mu$ is invariant under the dynamical evolution
of the system, giving a trace dynamics analog of Liouville's
theorem of classical mechanics.  Since  
no restrictions on the form of the generator {\bf G} were needed in the 
above argument for the invariance of $d \mu$, the argument applies even when 
{\bf G} is formed from the operator phase space variables 
using {\it operator} coefficients.  Thus, 
the integration measure $d \mu$ is invariant under a
unitary transformation on the basis of states in Hilbert space, the effect of 
which on the variables $\{ x_r \}$ can be represented by Eqs.~(2.11a-c). 
(Note, however, that this transformation is not itself global unitary 
invariant (cf. Eq.~(2.12a)),  and so is only a covariance, rather than an 
invariance, of the conserved operator $\tilde C$.)

\medskip
\centerline{4B.~~~The Canonical Ensemble}

The matrix equations of motion of trace dynamics determine
the time evolution of the matrix $q$'s and $p$'s at all times,
given their values on an initial time slice.  However, these initial
values are themselves not determined.  We shall now make 
the assumption that for a large enough system, the {\it statistical 
distribution} of initial values can be treated by the methods of statistical
mechanics.  Specifically, we shall assume 
that the {\it a priori} distribution
of initial values is uniform over the unbounded matrix operator phase 
space, so that the 
equilibrium distribution is determined solely by maximizing the combinatoric 
probability subject to the
constraints imposed by the generic conservation laws.  Liouville's
theorem implies that if the assumption of a uniform {\it a priori}
probability distribution is made at one time, then it
is valid at all later times, assuring the consistency of the concept of an
equilibrium ensemble.  We do not propose to address the question of how the
randomness in the initial value distribution arises:  it could come from a
random initial condition, an ordered initial condition
followed by evolution under a chaotic and 
effectively ergodic dynamics, or some combination of
the two.

More specifically, let $d\mu = d\mu[\{x_r\}]$ denote the operator phase
space measure discussed in detail in the preceding section.  In what follows
we shall not need the specific form of this measure, but only the properties
that it obeys Liouville's theorem, and that the
measure is invariant under infinitesimal 
matrix operator shifts $\delta x_r$, 
that is
$$d\mu[\{x_r+ \delta x_r\}]= d\mu[\{x_r\}]~~~.\eqno(4.8)$$
(This property will be used later on, when we discuss the
equipartition or Ward identities.)
{}For a system in statistical equilibrium, there is an equilibrium 
distribution of matrix initial values $\rho[\{x_r\}]$, such that
$$ d P=d\mu[\{x_r\}]   \rho[\{x_r\}]  ~~~\eqno(4.9a)$$     
is the infinitesimal probability of finding the system in the
operator phase space volume element $d\mu$, with the total probability  
equal to unity,
$$1=\int  d P=\int d\mu[\{x_r\}]   \rho[\{x_r\}]  ~~~.\eqno(4.9b)$$     
The first task in a statistical
mechanical analysis is to determine the equilibrium distribution $\rho$.

Since equilibrium implies that $\dot \rho=0$, the equilibrium distribution
can depend only on conserved operators and total trace functionals.  
In the generic case for a matrix model that is global unitary invariant,  
we have seen in Secs.~1 and 2 that, in addition to the conserved  
trace Hamiltonian ${\bf H}$  (which we assume to be bounded from below),  
there is a conserved  operator  $\tilde C$, which is anti-self-adjoint when    
fermionic adjointness is assigned as in Eqs.~(2.4b) and (2.18d).  
If the model is assumed to be constructed in a way that 
balances fermionic $q$'s and $p$'s, there is additionally a conserved trace 
``fermion number'' ${\bf N}$.  When the discrete mode index $r$ labels 
infinitesimal boxes in a spatial manifold, and the model is Lorentz  
invariant on this manifold, there will also 
be a locally conserved trace stress-energy tensor ${\bf T}_{\mu\nu}$, 
from which one can obtain by spatial integration 
not only the conserved trace Hamiltonian, but also conserved trace generators 
for three-momentum ${\bf \vec P}$, total angular momentum ${\bf \vec J}$, 
and Lorentz boosts ${\bf \vec K}$.  

We shall assume henceforth a statistical ensemble that is at rest and is not 
rotating, so that ${\bf \vec P}={\bf \vec J}=0$.  Hence the distribution 
function has no dependence on ${\bf \vec P}$ and ${\bf \vec J}$, and the 
Lorentz invariant trace mass  $[{\bf P^0}^2-{\bf \vec P}^2]^{1 \over 2}$ 
reduces to its rest frame value ${\bf H}$.  Since the ensemble 
picks out a preferred frame, which we tentatively identify 
with the frame in which the cosmological black-body radiation is isotropic,    
it is clearly not Lorentz invariant, even when (as we shall always assume)   
the underlying trace dynamics action and equations of motion are Lorentz 
invariant.  

(We shall later argue that the Lorentz 
invariance of the emergent quantum field theory is a reflection of a very 
weak dependence of low energy physics on ${\bf H}$, together with  
the fact that $\tilde C$, by virtue of 
its invariance under global unitary invariant canonical transformations  
that generate Lorentz transformations of the trace action, 
is Lorentz invariant.)   For the time being we shall allow a possible 
dependence of the ensemble on ${\bf N}$.  With these assumptions, the  
general equilibrium  distribution has the form 
$$\rho=\rho(\tilde C, {\bf H}, {\bf N})~~~.  \eqno(4.9c)$$       

In addition to its dependence on the dynamical variables,     
$\rho$ can also depend on constant parameter values, with
the functional form of $\rho$ and the values of the parameters together 
defining the statistical ensemble.  Including a traceless 
anti-self-adjoint operator parameter $\tilde \lambda$  and  real number 
parameters $\tau$ and $\eta$, which correspond to the respective 
structures of $\tilde C$, ${\bf H}$, and ${\bf N}$, the 
general form of the equilibrium ensemble corresponding to 
Eq.~(4.9c) is
$$\rho=\rho(\tilde C, \tilde \lambda;{\bf H},\tau; 
{\bf N}, \eta)
~~~.\eqno(4.9d)$$
In the canonical ensemble, we shall see that the dependence on $\tilde C$ and
$\tilde \lambda$ is only through the single real number 
${\rm Tr} \tilde \lambda
\tilde C$, and so specializing to this case, Eq.~(4.9d) becomes
$$\rho=\rho({\rm Tr} \tilde \lambda \tilde C;{\bf H},\tau; 
{\bf N}, \eta)
~~~.\eqno(4.9e)$$

We shall now show that some significant consequences follow from the
general form of Eq.~(4.9e), together with the assumption 
that ${\bf H}$ is constructed from the operators $\{ x_r \}$ 
using only $c$--number coefficients (as needed to insure its global unitary 
invariance). 
{}For a general operator ${\cal O}$, let us define the ensemble average 
$\langle {\cal O} \rangle_{\rm AV}$ by
$$\langle {\cal O} \rangle_{\rm AV}= {\int d\mu \rho {\cal O}  
\over \int d\mu 
\rho}~~~.\eqno(4.10a)$$
Then when ${\cal O}$ is constructed from the $\{ x_r \}$  
using only $c$-number coefficients, 
the ensemble average $\langle  {\cal O} \rangle_{\rm AV}$ must have the form
$$\langle {\cal O} \rangle_{\rm AV}=F_{\cal O}(\tilde \lambda)~~~,
\eqno(4.10b)$$ 
with the function $F_{\cal O}$ constructed from its argument using
only $c$-number coefficients (in which we include the $\tau$ and $\eta $ 
dependence).   As a consequence, the ensemble 
parameter $\tilde \lambda$ commutes with $\langle {\cal O} \rangle_{\rm AV}$,  
$$[\tilde \lambda, \langle {\cal O} \rangle_{\rm AV}] =0~~~.\eqno(4.10c)$$
                                                  
Let us now exploit the fact that the anti-self-adjoint operator 
$\tilde \lambda$ can always be diagonalized by a unitary transformation on 
the basis of states in Hilbert space, which we have seen is 
also an invariance of the integration measure $d \mu$.  
Specializing to ${\cal O}=\tilde C$, 
the functional relationship of Eq.~(4.10b) between 
$\tilde \lambda$ and $\langle \tilde C \rangle_{\rm AV}$ then implies that 
$\langle \tilde C \rangle _{\rm AV}$ is diagonal in this basis as well. This
brings $\langle \tilde C \rangle_{\rm AV}$ into the following 
canonical form, written in terms of a ``magnitude'' operator  $D_{\rm eff}$ 
and a unitary ``phase'' operator $i_{\rm eff}$: 
$$\eqalign{
&\langle \tilde C\rangle_{\rm AV}=i_{\rm eff} D_{\rm eff}~,
~~~{\rm Tr}(i_{\rm eff} D_{\rm eff})=0~,~~~  
i_{\rm eff}=-i_{\rm eff}^{\dagger}~,~~~ 
i_{\rm eff}^2=-1~,~~~\cr
&[i_{\rm eff},D_{\rm eff}]=0~,~~~ 
D_{\rm eff}~~ {\rm real~diagonal~and~non-negative}~.~~~\cr
}\eqno(4.11a)$$
Although the case of general $D_{\rm eff}$, which corresponds to an ensemble  
that is asymmetrical in the Hilbert space basis, is interesting, 
we shall restrict
ourselves henceforth to the special case in which $D_{\rm eff}$ is a
real constant times the unit operator.  In other words, {\it we assume that
the ensemble does not favor any state in Hilbert space over any other}, as a 
result of initial conditions for the underlying dynamics (which presumably   
arise at the origin of the universe in the ``big bang''.)  
Since we shall see in Sec.~5 that this real constant, which has 
the dimensions of action, plays the role of Planck's constant in the 
emergent quantum mechanics derived from the canonical ensemble, we shall 
denote it by  $\hbar$, and so we have
$$\eqalign{
\langle \tilde C \rangle_{\rm AV}=&i_{\rm eff} \hbar~,~~~\cr
{\rm Tr}i_{\rm eff}=&0~.~~~\cr
}\eqno(4.11b)$$
Since the relations $i_{\rm eff}=-i_{\rm eff}^{\dagger}$ and 
$i_{\rm eff}^2=-1$ imply that $i_{\rm eff}$ can be diagonalized to take 
the form $i{\rm diag}(\pm1,\pm1,...., \pm1)$, Eq.~(4.11b) implies 
that the dimension 
$N$ of the underlying matrix Hilbert space must be even, say $N=2K$, 
and therefore 
$i_{\rm eff}$ diagonalizes to the form 
$$i_{\rm eff}=i\,{\rm diag}(1,-1,1,-1,...,1,-1)~~~.\eqno(4.11c)$$
The restriction to even $N$ is a direct result of our assumption 
that the magnitude matrix 
$D_{\rm eff}$ in Eq.~(4.11a) is a multiple of the unit matrix; 
if one were to start 
off with a matrix space with $N$ odd, then 
${\rm Tr}(i_{\rm eff}D_{\rm eff})=0$  
from Eq.~(4.11a) would require $D_{\rm eff}$ to have one null 
eigenvalue, since a one dimensional traceless matrix must vanish.

We turn now to the calculation of the functional form of $\rho$ in the
canonical ensemble, which is the ensemble relevant for describing the 
behavior of a large system that is a subsystem of a still larger system. 
The form of $\rho$ is determined [18] 
by maximizing the entropy,  
$$S=-\int d\mu \rho \log \rho ~,~~~\eqno(4.12a)$$
subject to the constraints
$$\eqalign{
\int d\mu \rho=&1~,~~~\cr
\int d\mu \rho \tilde C =&\langle \tilde C \rangle_{\rm AV}~,~~~\cr
\int d\mu \rho {\bf H}=&\langle {\bf H} \rangle_{\rm AV}~,~~~\cr     
\int d\mu \rho {\bf N}=&\langle {\bf N} \rangle_{\rm AV}~.~~~\cr     
}\eqno(4.12b)$$
The standard procedure is to impose the constraints with Lagrange 
multipliers $\theta,\tilde \lambda,\tau,\eta$ by writing
$${\cal F}=\int d\mu\rho \log \rho+\theta \int d\mu \rho + \int d\mu \rho 
{\rm Tr} \tilde \lambda \tilde C +\tau \int d\mu \rho {\bf H} 
+\eta \int d\mu \rho {\bf N} ~~~,  \eqno(4.13a)$$
and maximizing $-{\cal F}$ (or equivalently, minimizing ${\cal F}$), 
treating all variations of $\rho$ as independent.  
Varying Eq.~(4.13a) with respect to $\rho$ then gives
$$\rho=\exp(-1-\theta -{\rm Tr}\tilde \lambda \tilde C-\tau  
{\bf H} - \eta {\bf N})~,~~~
\eqno(4.13b)$$
which on imposing the condition that $\rho$ be normalized to unity gives 
finally 
$$\eqalign{
\rho=&Z^{-1}\exp(-{\rm Tr}\tilde \lambda\tilde C -\tau {\bf H}
-\eta {\bf N})~,~~~\cr
Z=&\int d\mu\exp(-{\rm Tr}\tilde \lambda\tilde C -\tau {\bf H}
-\eta {\bf N}) ~.~~~\cr
}\eqno(4.13c)$$

{}From Eq.~(4.13c) we can derive some 
elementary statistical properties
of the equilibrium ensemble.  For the entropy $S$, we find
$$S=-\langle \log \rho \rangle_{\rm AV}= \log Z + {\rm Tr} \tilde \lambda 
\langle \tilde C \rangle_{\rm AV} + \tau \langle {\bf  H} \rangle_{\rm AV}
+ \eta \langle {\bf N} \rangle_{\rm AV}~.~~~
\eqno(4.14a)$$
Since the ensemble averages which appear in Eq.~(4.14a) are given by 
$$\eqalign{
\langle \tilde C \rangle_{\rm AV}=&-{\delta \log Z 
\over \delta \tilde \lambda} ~,
~~~\cr
\langle {\bf H} \rangle_{\rm AV}=&-{\partial \log Z 
\over \partial \tau}~,~~~\cr 
\langle {\bf N} \rangle_{\rm AV}=&-{\partial \log Z \over \partial 
\eta}~,~~~\cr 
}\eqno(4.14b)$$
Eq.~(4.14a) takes the form
$$S=\log Z -{\rm Tr} \tilde \lambda {\delta 
\log Z \over \delta \tilde \lambda}
-\tau {\partial \log Z \over \partial \tau } 
-\eta {\partial \log Z \over \partial \eta  
}~.~~~\eqno(4.14c)$$
Thus the entropy is a thermodynamic quantity 
determined solely by the partition function.  
Taking second derivatives of the partition function, we can similarly derive 
the thermodynamic formulas 
for the averaged mean square fluctuations of the conserved quantities 
$\tilde C$, ${\bf H}$, and ${\bf N}$,
$$\eqalign{
\Delta_{{\rm Tr} \tilde P \tilde C}^2\equiv& \langle
( {\rm Tr} \tilde P \tilde C 
- \langle {\rm Tr} \tilde P \tilde C \rangle_{\rm AV})^2 \rangle_{\rm AV}
=\langle ({\rm Tr} \tilde P \tilde C)^2\rangle_{\rm AV}
-\langle {\rm Tr} \tilde P
\tilde C\rangle_{\rm AV}^2 = 
({\rm Tr} \tilde P {\delta \over \delta \tilde \lambda})^2 \log Z~,~~~\cr
\Delta_{\bf H}^2 \equiv&\langle 
({\bf H} -\langle {\bf  H} \rangle_{\rm AV})^2
\rangle_{\rm AV}= \langle {\bf  H}^2 \rangle_{\rm AV}-\langle {\bf  H} 
\rangle_{\rm AV}^2
={\partial^2 \log Z \over (\partial \tau)^2}~,~~~\cr
\Delta_{\bf N}^2 \equiv&\langle ({\bf N} -\langle {\bf N} \rangle_{\rm AV})^2
\rangle_{\rm AV}= \langle {\bf N}^2 \rangle_{\rm AV}
-\langle {\bf N} \rangle_{\rm AV}^2
={\partial^2 \log Z \over (\partial \eta)^2}~,~~~\cr
}\eqno(4.14d)$$
with $\tilde P$ an arbitrary fixed anti-self-adjoint operator.  
Similar expressions hold for the cross-correlations of ${\rm Tr}\tilde P 
\tilde C$, ${\bf H}$, and ${\bf N}$, for example 
$$\langle 
( {\rm Tr} \tilde P \tilde C    
- \langle {\rm Tr} \tilde P \tilde C \rangle_{\rm AV})  
({\bf H} -\langle {\bf  H} \rangle_{\rm AV}) \rangle_{\rm AV}
=\langle  {\rm Tr} \tilde P \tilde C  {\bf H}  \rangle_{\rm AV}
-\langle  {\rm Tr} \tilde P \tilde C \rangle_{\rm AV} 
\langle {\bf H} \rangle_{\rm AV}=
{\rm Tr} \tilde P {\delta \over \delta \tilde \lambda}
{\partial \over \tau} \log Z~~~.\eqno(4.14e)$$
The complete set of such relations, in a more compact notation, is given 
in the next section.  

As a final remark, we note that in subsequent sections we shall follow
the conventional practice of introducing for each matrix variable $x_r$  
a matrix source $j_r$, of the same bosonic or fermionic type and with the 
same adjointness properties as $x_r$, 
which can be varied and which is then set to zero
after all variations have been performed.  
With the sources included, the equilibrium distribution and 
partition function take the form
$$\eqalign{
\rho=&Z^{-1}
\exp(-{\rm Tr} \tilde \lambda \tilde C- \tau {\bf  H}-\eta 
{\bf N}  - \sum_r {\rm Tr} j_r x_r  )~~~, \cr 
Z=&\int d\mu
\exp(-{\rm Tr} \tilde \lambda \tilde C-\tau {\bf H}-\eta
{\bf N}  - \sum_r {\rm Tr} j_r x_r   )~~~. \cr 
}\eqno(4.15a)$$
Continuing to use the expression $\langle {\cal O} 
\rangle_{\rm AV}$ to denote 
the average of a general operator over the equilibrium distribution of 
Eq.~(4.15a) which includes sources, the variations of $\log Z$ 
with respect to 
its source arguments are related to the averages of the $x_r$ by 
$$\epsilon_r \langle x_r \rangle_{\rm AV}=-{\delta \log Z \over \delta j_r}
~.~~~\eqno(4.15b)$$
\vfill
\eject

\medskip
\centerline{4C.~~~The Microcanonical Ensemble}

In the previous section, we derived the canonical ensemble for trace 
dynamics by maximizing the entropy subject to the generic constraints, 
which were imposed in an averaged sense.  In this section, we shall give [5] 
an alternative and more fundamental 
derivation of the canonical ensemble, by starting from the 
microcanonical ensemble, in which the constraints are imposed in a sharp 
sense.  We shall see that the canonical ensemble then arises as the 
appropriate description of a large system in equilibrium with a much 
larger ``bath'', with the equilibrium conditions determining in an 
intrinsic manner the ensemble 
parameters, or generalized ``temperatures''  $\tilde \lambda$, $\tau$, 
and $\eta$.  Apart from using the microcanonical 
ensemble to derive the canonical ensemble in this section we shall not 
employ the microcanonical ensemble further in our subsequent analysis.  
The reason for our primary focus on the canonical ensemble is that 
the Ward identities that imply emergent quantum behavior, which are 
derived in Sec.~5, are properties of the canonical ensemble, but not of 
the microcanonical ensemble.  Thus, if the microcanonical ensemble is 
taken to represent the entire universe, our subsequent analysis suggests 
that emergent quantum mechanics is a property only of subsystems of the 
universe that are large but still appreciably smaller than the universe 
as a whole.  

It is convenient at this point to introduce a condensed notation for the 
exponent appearing in the canonical ensemble of Eq.~(4.13c), which takes 
the anti-self-adjointness of $\tilde C$ and $\tilde \lambda$ into account.  
Writing as in Eq.~(4.2a), 
$$\eqalign{
(\tilde C)_{mn}=& (\tilde C)_{mn}^0   +  i    (\tilde C)_{mn}^1 ~~~,\cr    
(\tilde \lambda)_{mn}=& (\tilde \lambda)_{mn}^0   
                      +  i    (\tilde \lambda)_{mn}^1 ~~~,\cr    
}\eqno(4.16a)$$                      
the anti-self-adjointness restrictions on $\tilde C$ and $\tilde \lambda$ 
take the form
$$(\tilde C)_{mn}^A=-\epsilon^A (\tilde C)_{nm}^A~,~~    
(\tilde \lambda)_{mn}^A=-\epsilon^A (\tilde \lambda)_{nm}^A~~~,    
\eqno(4.16b)$$
with $\epsilon^0=1$ and $\epsilon^1=-1$ as before.  Then a simple 
calculation shows that 
$${\rm Tr} \tilde \lambda \tilde C=
-\sum_n (\tilde \lambda)_{nn}^1 (\tilde C)_{nn}^1 
-2 \sum_{n<m}[ (\tilde \lambda)_{nm}^0 (\tilde C)_{nm}^0 
+(\tilde \lambda)_{nm}^1 (\tilde C)_{nm}^1    ]~~~,    \eqno(4.16c)$$ 
with all the terms on the right hand side independent.  It is now convenient 
to introduce a vector notation for the exponent in Eq.~(4.13c), by 
defining 
$$\eqalign{
\xi\equiv & (\xi^1,...,\xi^M)\cr 
\equiv& ({\bf H},{\bf N}, [(\tilde C)_{nn}^1, 
n=1,...,N], [(\tilde C)_{nm}^0, n<m=1,...,N],   
[(\tilde C)_{nm}^1, n<m=1,...,N] )~~~,\cr   
\sigma\equiv & (\sigma^1,...,\sigma^M)\cr \equiv& (\tau,\eta, 
-[(\tilde \lambda)_{nn}^1, 
n=1,...,N], -2[(\tilde \lambda)_{nm}^0, n<m=1,...,N],   
-2[(\tilde \lambda)_{nm}^1, n<m=1,...,N] )~~~,\cr   
}\eqno(4.17a)$$
which permits us to write 
$${\rm Tr}\tilde \lambda \tilde C + \tau {\bf H} + \eta {\bf N} 
= \vec \sigma \cdot \vec \xi~~~.\eqno(4.17b)$$
In other words, $\vec \xi$ is the vector of all the real number generic 
conserved quantities, and $\vec \sigma$ is the vector of the corresponding 
canonical ensemble parameters (which are analogs of the inverse temperature 
parameter $\beta$ and the chemical potential parameter $\mu$ of ordinary 
statistical mechanics).  The dimensionality $M$ of both vectors is 
$M=2+N^2$.  

We can now introduce the microcanonical ensemble $\Gamma(\vec \Xi)$, 
which is defined as the volume of the shell of phase space for 
the ``universe'' in which the 
conserved quantities take the sharp values $\vec \Xi$. In other words,  
we write 
$$\Gamma(\vec \Xi)=\int d\mu \prod_{a=1}^{M}\delta(\Xi^a-\xi^a)~~~,
\eqno(4.18a)$$
which has an associated entropy 
$$S(\vec \Xi)=\log \Gamma(\vec \Xi)~~~.\eqno(4.18b)$$
Let us now divide the universe into a ``system'' $s$, which is still large in 
a statistical sense but is much smaller than the universe, 
and a ``bath'' $b$ which is the complement of degrees of freedom in the 
universe not included in the system.   We now assume 
that the vector of conserved quantities $\vec \xi$ is to a good 
approximation additively decomposable over the system and the bath, when 
both are very large.   
In other words, we assume that  
$$\vec \xi \simeq \vec \xi_s + \vec \xi_b~~~, \eqno(4.19a)$$ 
with $\vec \xi_s$ and $\vec \xi_b$ the values of the conserved quantities 
appropriate to the system and to the bath, respectively.  
Note that if the system and bath are defined simply by a partitioning of the 
canonical degrees of freedom ${q_r,p_r}$, additivity is automatic for 
${\bf N}$ and $\tilde C$, which are additive sums over the degrees of 
freedom, but not for ${\bf H}$, which in general has nonlinear interactions 
between the degrees of freedom.      
Letting $d\mu_s$ and $d\mu_b$ be the phase space measures for the system 
and the bath, so that $d\mu\simeq d\mu_s d\mu_b$, and introducing a dummy 
variable of integration $\vec \Xi_s$, we can rewrite Eq.~(4.18a) as 
$$\Gamma(\vec \Xi)=\prod_a \int d\Xi_s^a 
\Gamma_b(\vec \Xi-\vec \Xi_s) \Gamma_s(\vec \Xi_s) 
~~~,\eqno(4.19b)$$
with the system and bath microcanonical subensembles defined by 
$$\eqalign{
\Gamma_s(\vec \Xi_s)\equiv & \int d\mu_s 
\prod_{a=1}^M  \delta(\Xi_s^a-\xi_s^a)~~~,\cr  
\Gamma_b(\vec \Xi-\vec \Xi_s)\equiv& \int d\mu_b 
\prod_{a=1}^M  \delta(\Xi^a -\Xi_s^a-\xi_b^a)~~~.\cr  
}\eqno(4.19c)$$

We now assume that the integrand in Eq.~(4.19b) has a maximum that 
dominates the integral when the number of degrees of freedom is large.  
Although we give no a priori justification of this assumption, we shall  
later on show that it is self-consistent.  The necessary condition for 
the integrand in Eq.~(4.19b) to have an extremum at $\vec \Xi_s =
\vec X_s$ is 
$${\partial \over \partial \Xi_s^a} 
[\Gamma_b(\vec \Xi-\vec \Xi_s) \Gamma_s(\vec \Xi_s)]\vert_{\vec X_s} 
=0~~~, \eqno(4.20a)$$
which can be rewritten as 
$$\sigma^a \equiv {\partial \over \partial \Xi_s^a}  
\log \Gamma_s(\vec \Xi_s) \vert_{\vec X_s}
={\partial \over \partial \Xi^a} \log \Gamma_b(\vec \Xi-\vec \Xi_s) 
\vert_{\vec X_s} ~~~.\eqno(4.20b)$$
Thus, at the assumed maximum, the logarithmic derivatives in Eq.~(4.20b) 
define a set of equilibrium parameters $\vec \sigma$ common to the bath 
and the system.   Recalling the entropy definition of Eq.~(4.18b), we  
can rewrite the bath phase space volume at the extremum as
$$\Gamma_b(\vec \Xi-\vec X_s) =\exp(S_b(\vec \Xi-\vec X_s))
\simeq \exp(S_b(\vec \Xi))\exp\left(-\sum_a X_s^a {\partial \over 
\partial \Xi^a} \log \Gamma_b (\vec \Xi) \right)~~~,\eqno(4.20c)$$
which neglecting a small shift from $\vec \Xi-\vec X$ to $\vec \Xi$ in the 
definition of the equilibrium parameters $\vec \sigma$, gives us 
$$\Gamma_b(\vec \Xi-\vec X_s) 
\simeq \exp(S_b(\vec \Xi))\exp(- \vec \sigma  \cdot \vec X )
~~~. \eqno(4.20d)$$
Renaming the free parameter $\vec X_s$ in Eq~(4.20d) as $\vec \Xi_s$, 
we deduce from this equation that 
$$\Gamma_b(\vec \Xi-\vec \Xi_s) 
\simeq \exp(S_b(\vec \Xi))\exp(-    \vec \sigma \cdot \vec \Xi_s )
~~~. \eqno(4.20e)$$

Returning now to Eq.~(4.19b) and substituting the approximate form   
of Eq.~(4.20e) for the bath phase space volume factor, we get 
$$\Gamma(\vec \Xi)\simeq  \exp(S_b(\vec \Xi)) Z_s~~~,\eqno(4.21a)$$   
with $Z_s$ the integral defined by 
$$Z_s=\prod_a \int d\Xi_s^a 
\exp(-\vec \sigma \cdot \vec \Xi_s )\Gamma_s(\vec \Xi_s) 
~~~.\eqno(4.21b)$$
On substituting Eq.~(4.19c) for the system phase space volume  
$\Gamma_s(\vec \Xi_s)$ and 
carrying out the integration over the dummy variables $\Xi_s^a$, we can  
rewrite $Z_s$ as
$$Z_s=\int d\mu_s  \exp(-\vec \sigma \cdot \vec \xi_s ) ~~~.\eqno(4.21c)$$    
We conclude that when the system and bath are in equilibrium, and the 
overall ``universe'' comprising the system and bath is in a microcanonical 
ensemble, the system variables are weighted in the phase space integral  
according to the normalized distribution 
$$\rho_s=Z_s^{-1}  \exp(-\vec \sigma \cdot \vec \xi_s )~~~,\eqno(4.22a)$$   
which defines the standard canonical ensemble.  Since all of the above 
manipulations go through if $d\mu_s$ is replaced by $d\mu_s f_s$, with 
$f_s$ any function of the system variables, we have shown that the 
average $\langle f_s \rangle_{\rm  AV} $ defined 
in the microcanonical ensemble, 
$$\langle f_s \rangle_{\rm AV} \equiv 
{\int d\mu \prod_{a=1}^{M}\delta(\Xi^a-\xi^a) f_s    \over
\int d\mu \prod_{a=1}^{M}\delta(\Xi^a-\xi^a) 1 } ~~~\eqno(4.23a)$$ 
can be equivalently calculated as 
$$\langle f_s \rangle_{\rm AV} \simeq \int d\mu_s \rho_s f_s 
~~~.\eqno(4.23b)$$ 
This justifies the use of the canonical ensemble in calculating 
thermodynamic averages of system quantities.

As a consistency check on the calculation, we must verify that within  
our approximations, the extremum of 
Eq.~(4.20a) is a maximum.  Using the approximated form of the 
integrand in Eq.~(4.21b), the condition for an extremum is 
$${\partial \over \Xi_s^a} [ -\vec \sigma \cdot \vec \Xi_s  +
\log \Gamma_s(\vec \Xi_s)  ]=0~~~.\eqno(4.24a)$$
In other words, 
$$\sigma^a={\partial \over \Xi_s^a} \log \Gamma_s(\vec \Xi_s) ~~~,
\eqno(4.24b)$$
in agreement (at $\vec \Xi_s =\vec X_s$)
with the definition of Eq.~(4.20b).  In order for the 
extremum to be a maximum, the matrix of second derivatives 
$$ {\partial \over \Xi_s^a} {\partial \over \Xi_s^b} 
\log \Gamma_s(\vec \Xi_s)  =  {\partial \over \Xi_s^a} \sigma^b~~~
\eqno(4.24c)$$ 
must be negative definite.   This will be true provided that the 
inverse matrix $\partial \Xi_s^a / \partial \sigma^b$ is negative 
definite (and bounded), with $\Xi_s^a$  the location of the maximum of the  
integrand in Eq.~(4.21b), which for a large system is closely 
approximated by the canonical ensemble average of $\xi_s^a$.   
But from Eq.~(4.21c), we see that 
$$-{\partial \log Z_s \over \partial \sigma^a} 
= \langle \xi_s^a \rangle_{\rm AV}~~~,
\eqno(4.25a)$$
and differentiating again, 
$$-{\partial^2 \log Z_s \over \partial \sigma^a \partial \sigma^b} = 
{\partial   \langle \xi_s^a \rangle_{\rm AV} \over \partial \sigma^b} 
=-\langle (\xi_s^a -\langle \xi_s^a \rangle _{\rm AV} ) 
(\xi_s^b -\langle \xi_s^b \rangle _{\rm AV} )  \rangle_{\rm AV}~~~, 
\eqno(4.25b)$$
which is negative definite and bounded.  Thus the assumption that the 
extremum 
in the phase space integral is a maximum is self-consistent.  Referring  
back to the correlation formulas of Eq.~(4.14e), we see that Eq.~(4.25b) 
gives the most general such formula in our condensed notation.

\medskip
\centerline{4D.~~~Gauge Fixing in the Partition Function $^*$}

Up to this point, in discussing the statistical mechanics of trace dynamics 
we have assumed that one is dealing with an unconstrained system, leading to 
the generic form of the canonical ensemble given in Eq.~(4.13c).
In order to apply Eq.~(4.13c) directly to a constrained system, one must 
first explicitly integrate out the constraints.  A simple example where
this is possible is provided 
by the trace dynamics transcription of supersymmetric Yang-Mills theory, 
discussed in Sec. 3B.  (Further examples of gauge invariant trace dynamics 
models are given in Appendix E.)  In axial gauge,  
where $A_3=0$, the covariant 
derivative $D_3$ simplifies to $D_3=\partial_3$, allowing the constraint    
of Eq.(3.10b) to be integrated out, giving the explicit 
expression for the trace Hamiltonian of Eqs.~(3.11b,c) and the  
corresponding expression for the conserved operator $\tilde C$ of  
Eq.~(3.13a).  
The axial gauge partition function is then given by 
Eq.~(4.13c), with the phase space measure $d \mu$ given by 
$$d \mu_{\rm axial}=\prod_{\vec x} \prod_{\ell=1}^2 dA_{\ell}(\vec x) 
dp_{A_{\ell}(\vec x)}~~~.\eqno(4.26)$$

The problem addressed in this section  is how to generalize the 
axial gauge partition function to other gauges in which it may not be 
possible to explicitly integrate out the constraint [19].  The 
problem of correctly incorporating a gauge invariance group with a continuous 
infinity of group parameters and an infinite invariant group volume is a 
familiar one
in the theory of path integrals, and we shall use methods 
similar to the ones employed there to give a solution.  However, since 
the partition function singles out a Lorentz frame, we will have to make 
a restriction not encountered in the Lorentz scalar path 
integral case, namely we will consider only nontemporal 
gauge conditions that do 
not involve the scalar potential $A_0$.  This still allows us to 
consider gauge transformations that rotate the axial gauge axis, or that 
transform to rotationally invariant gauges such as Coulomb gauge.  
We shall also make the further assumption that the allowed gauge 
transformations leave invariant the surface integral which, according  
to Eq.~(3.13b), determines $\tilde C$, thus placing a restriction on 
the gauge transformation at the point at infinity.   We proceed by 
developing an analog of the standard De Witt-Faddeev-Popov 
method to write the axial gauge partition function in a general nontemporal 
gauge, subject to the surface term restriction just stated.  Since 
we have seen in Sec.~3 that trace dynamics incorporates 
rigid supersymmetry, 
and since  BRST invariance is a particular rigid supersymmetry 
transformation, we shall find that the generalized expression for the 
partition function, when reexpressed in terms of ghost fermions,  
admits a BRST invariance.  As in our preceding discussion, we assume  
convergence of the partition function, which may well require 
restrictions on the class of trace Hamiltonians being considered.  

To express the partition function in a general nontemporal  
gauge, we follow closely the treatment of the De Witt-Faddeev-Popov 
construction in the familiar functional integral case, as given in 
the text of Weinberg [20].   Let us consider the integral
$$\eqalign{
Z_G=&\int d\mu B[f(A_{\ell})] \delta(Y)\det{\cal F}[A_{\ell}]
\exp(-\tau {\bf H}-{\rm Tr}\tilde \lambda \tilde C)~~~,\cr
d\mu=& \prod_{\vec x} \prod_{\ell=1}^3 dA_{\ell}(\vec x)
dp_{A_{\ell}(\vec x)}~~~,\cr
}\eqno(4.27a)$$
with $\tilde C$ given by Eq.~(3.13b) and 
with the constraint $Y$ given by 
$$Y\equiv \sum_{\ell=1}^3D_{\ell}p_{A_{\ell}}+2\bar \chi \gamma^0 \chi
~~~.\eqno(4.27b)$$
The trace Hamiltonian ${\bf H}$ in Eq.~(4.27a) is given by Eq.~(3.11b),  
with the gauge part ${\bf H}_A$ given by 
$${\bf H}_A=\int d^3x {\rm Tr}(-{g^2\over 2}\sum_{\ell=1}^3 p_{A_{\ell}}^2
-{1\over 4g^2}\sum_{\ell,m=1}^3 F_{\ell m}^2)~~~,\eqno(4.27c)$$  
which is valid in a general gauge on the constraint surface $Y=0$ selected by 
the delta function in Eq.~(4.27a). 
The delta function of the anti-self-adjoint matrix valued argument $Y$ 
appearing in Eq.~(4.27a) is given, in terms of 
ordinary delta functions of the real ($R$) and imaginary ($I$) parts of 
the matrix elements, by  
$$\delta(Y)=\prod_{m<n} \delta((Y_R)_{mn})\prod_{ m \leq n} 
\delta((Y_I)_{mn})~~~,\eqno(4.28a)$$
and the integration measure over the anti-self-adjoint matrix 
$A_{\ell}$ is defined by 
$$dA_{\ell}=\prod_{m<n} d(A_{\ell R})_{mn}
\prod_{m \leq n} d(A_{\ell I})_{mn}~~~,\eqno(4.28b)$$
and similarly for $dp_{A_{\ell}}$.  The function $B[f]$ is an arbitrary 
integrable scalar valued function of the matrix valued 
argument $f(A_{\ell})$, 
which is used to specify the gauge condition.  We shall treat $f$ as a  
column vector $f_{\alpha}$ with $\alpha$ a composite index formed from 
the matrix row and column indices $m,n$; the argument ${\cal F}$ 
of the De Witt-Faddeev-Popov determinant is then given in terms of $f$ by 
the expression  
$${\cal F}_{\alpha \, \vec x,\beta \, \vec y}[A_{\ell}] 
\equiv { \delta f_{\alpha}(A_{\ell}(\vec x)+D_{\ell} \Lambda(\vec x))
\over \delta \Lambda_{\beta}(\vec y)}\vert_{\Lambda=0}~~~,\eqno(4.28c)$$
where $\delta$ is the usual functional derivative and 
$\beta$ is the composite of the row and column indices of the 
infinitesimal gauge transformation matrix $\Lambda$.  

We now demonstrate two properties of the integral $Z_G$ defined 
in Eq.~(4.27a):
(i) first, we show that when the gauge fixing functions $B[f]$ and 
$f(A_{\ell})$ are chosen to correspond to the axial gauge condition, 
then Eq.~(4.27a) reduces 
(up to an overall constant) to the axial gauge partition function; 
(ii)  second, we show that $Z_G$ is in fact independent of the 
function $f(A_{\ell})$, and depends on the function $B[f]$ only through 
an overall constant.  These two properties together imply that 
$Z_G$ gives the wanted extension of the axial gauge partition function to 
general nontemporal gauges.  

To establish property (i), we make the conventional  axial gauge choice 
$$B[f(A_{\ell})]=\delta(A_3)=\prod_{m<n} \delta((A_{3R})_{mn})
\prod_{m \leq n}\delta((A_{3I})_{mn})
~~~,\eqno(4.29a)$$
so that 
$$\int dA_3 B[f(A_{\ell})]=\int dA_3 \delta(A_3)=1~~~.\eqno(4.29b)$$
With this gauge choice, 
$$D_3p_{A_3}=\partial_3p_{A_3}~~~,\eqno(4.29c)$$
which implies that 
$$\eqalign{
\delta(Y)=&\delta(\partial_3p_{A_3}+\sum_{\ell=1}^2 D_{\ell}p_{A_{\ell}}
+2 \bar \chi \gamma^0 \chi) \cr
=&|\partial_3|^{-1}\delta(p_{A_3}+
{1\over 2} \int dz^{\prime} \epsilon(z-z^{\prime}) 
(\sum_{\ell=1}^2 D_{\ell}p_{A_{\ell}}+ 2 \bar \chi \gamma^0 \chi) )~~~.\cr 
}\eqno(4.29d)$$
Hence the integral over $p_{A_3}$ in $Z_G$ can be done explicitly, giving 
(up to an overall constant factor coming from the Jacobian $|\partial_3|^
{-1}$) the expression 
$$Z_G=\int d \mu_{\rm axial}  
\exp(-\tau {\bf H}-{\rm Tr}\tilde \lambda \tilde C)|_{A_3=0;~p_{A_3}=
-{1\over 2} \int dz^{\prime} \epsilon(z-z^{\prime}) (\sum_{\ell=1}^2 
D_{\ell}p_{A_{\ell}}+ 2 \bar \chi \gamma^0 \chi) }
~~~,\eqno(4.30)$$
which agrees (recalling from Eq.~(3.3) that $2 \bar \chi \gamma^0 \chi
=-(p_{\chi} \chi+ \chi^T p_{\chi}^T)$)
with the axial gauge partition function constructed from   
${\bf H}_A$ of Eq.~(3.11c).

To establish property (ii), we first examine the gauge transformation 
properties of the various factors in the integral defining $Z_G$.  
We begin with the integration measure $d \mu$.  Under the infinitesimal gauge 
transformation (with $\Lambda$ anti-self-adjoint)
$$A_{\ell} \to A_{\ell}+D_{\ell}\Lambda=A_{\ell}+\partial_{\ell}\Lambda 
+[A_{\ell},\Lambda]~~~,\eqno(4.31a)$$ 
the inhomogeneous term $\partial_{\ell} \Lambda$ does not contribute 
to the transformation of the differential $dA_{\ell}$.  Therefore  
$dA_{\ell}$ obeys the homogeneous transformation law $dA_{\ell} 
\to dA_{\ell}+\Delta_{\ell}$, with 
$$\Delta_{\ell}\equiv [dA_{\ell},\Lambda]~~~.\eqno(4.31b)$$
Hence to first order in $\Lambda$, the Jacobian of the transformation 
(calculated by the same reasoning that led from Eq.~(4.4) to Eq.~(4.5a)) is 
$$\eqalign{
J=&1
+\sum_{m<n}{\partial(\Delta_{\ell R})_{mn} \over 
\partial (dA_{\ell R})_{mn} }  
+\sum_{m\leq n}{\partial(\Delta_{\ell I})_{mn} \over 
\partial (dA_{\ell I})_{mn} }  \cr
=&1
+(\sum_{m<n}+\sum_{m\leq n}) [(\Lambda_R)_{nn}-(\Lambda_R)_{mm}] \cr
=&1 ~~~,\cr
}\eqno(4.31c)$$
since the anti-self-adjointness of $\Lambda$ implies that $(\Lambda_R)_{nm}
=-(\Lambda_R)_{mn}$, and so the diagonal matrix elements 
$(\Lambda_R)_{nn}$ are all zero.  
Thus each factor $dA_{\ell}(\vec x)$ in the integration measure is 
gauge invariant. A similar argument applies to each factor 
$dp_{A_{\ell}(\vec x)}$ in the integration measure, and also to the factor 
$\delta(Y)$ in the integrand,  since $Y$ obeys the 
homogeneous gauge transformation law 
$Y \to Y +[Y, \Lambda]$.  Turning to the exponential, the terms 
${\rm Tr} p^2_{A_{\ell}}$ and ${\rm Tr}F^2_{\ell m}$ are gauge invariant, 
and so the trace Hamiltonian ${\bf H}$ is gauge invariant.  
By hypothesis, the surface term determining $\tilde C$ 
is left invariant by the class of gauge transformations under 
consideration.  To summarize, we see that the integral $Z_G$ has the form 
$$Z_G=\int d\mu {\cal G}[A_{\ell}] B[f(A_{\ell})]\det {\cal F}[A_{\ell}]
~~~,\eqno(4.32a)$$ 
with the integration measure $d\mu$ and the integrand factor 
$${\cal G}[A_{\ell}]=\delta(Y)\exp(-\tau{\bf H}-{\rm Tr} \tilde \lambda 
\tilde C)~~~\eqno(4.32b)$$
both gauge invariant.  Hence $Z_G$ has exactly the form assumed in the 
discussion of  Weinberg [20], and the proof given there 
completes the demonstration of property (ii).  

Continuing to follow the standard path integral analysis, let us  
represent the De Witt-Faddeev-Popov determinant $\det {\cal F}[A_{\ell}]$
as an integral over fermionic ghosts, by writing 
$$\det {\cal F}[A_{\ell}]=\int d\omega^*d\omega
\exp(\int d^3x d^3y \omega^*_{\alpha}(\vec x)
{\cal F}_{\alpha \, \vec x,\beta \, \vec y}[A_{\ell}] 
\omega_{\beta}(\vec y))~~~,\eqno(4.33a)$$
with
$$d\omega=\prod_{\vec x}\prod_{m,n}d\omega_{mn}(\vec x)~~~,~~~ 
  d\omega^*=\prod_{\vec x}\prod_{m,n}d\omega^*_{mn}(\vec x)~~~. 
  \eqno(4.33b)$$
Let us also take for $B[f]$ the usual Gaussian 
$$B[f]=\exp(-{1 \over 2 \xi} \int d^3x {\rm Tr} f(A_{\ell}(\vec x))^2)
~~~,\eqno(4.33c)$$
and for $f(A_{\ell})$ the linear gauge condition 
$$f(A_{\ell})=\sum_{\ell} L^{\ell} A_{\ell}~~~,\eqno(4.33d)$$
in which $L^{\ell}$ can be either a fixed vector (such as $\delta_{\ell3}$ 
in axial gauge) or a differential operator (such as $\partial_{\ell}$ in 
Coulomb gauge), and a summation of $\ell$ from 1 to 3 is understood.     
With this choice of $f(A_{\ell})$, we find from Eq.~(4.28c)  
that 
$$\eqalign{
{\cal F}_{nm \vec x,pq \vec y}[A_{\ell}]=&
{\delta f_{nm}(A_{\ell}(\vec x)+D_{\ell}\Lambda(\vec x)) \over 
\delta \Lambda_{pq}(\vec y) } \cr
=&\sum_{\ell} L^{\ell}_{\vec x}\left({\partial \delta(\vec x-\vec y) \over 
\partial x^{\ell} } \delta_{np}\delta_{mq} 
+\delta(\vec x-\vec y)[(A_{\ell})_{np} \delta_{mq}
-\delta_{np}(A_{\ell})_{qm}] \right)~~~,\cr
}\eqno(4.34a)$$
which when substituted into the exponent in Eq.~(4.33a) gives, after  
integrations by parts, 
$$\int d^3x d^3y \omega^*_{\alpha}(\vec x)
{\cal F}_{\alpha \, \vec x,\beta \, \vec y}[A_{\ell}] 
\omega_{\beta}(\vec y)
=\int d^3x {\rm Tr} \overline{\omega}(\vec x) \sum_{\ell}L^{\ell}D_{\ell}
\omega(\vec x)~~~,\eqno(4.34b) $$
where we have defined $\overline {\omega}_{mn}=\omega^*_{nm}$.  
Hence the expression of Eq.~(4.32a) for $Z_G$ becomes 
$$Z_G=\int d\mu d\overline{\omega} d\omega {\cal G}[A_{\ell}] 
\exp\left[-\int d^3x{\rm Tr} \left({1 \over 2 \xi} 
(\sum_{\ell}L^{\ell}A_{\ell})^2 -
\overline{\omega}(\vec x)\sum_{\ell}L^{\ell}D_{\ell}\omega(\vec x) 
\right)\right]
~~~~.\eqno(4.35a)$$
An alternative way of writing Eq.~(4.35a), that is 
convenient for exhibiting 
the BRST invariance, is to introduce an auxiliary self-adjoint 
matrix field $h$ and 
to reexpress Eq.~(4.35a) as 
$$Z_G=\int d\mu dh d\overline{\omega} d\omega {\cal G}[A_{\ell}] 
\exp\left[-\int d^3x{\rm Tr} 
\left({\xi \over 2} h^2 + i h \sum_{\ell}L^{\ell}A_{\ell}
-\overline{\omega}(\vec x)\sum_{\ell}L^{\ell}D_{\ell}\omega(\vec x) 
\right)\right]
~~~~.\eqno(4.35b)$$

Starting from Eq.~(4.35b), we can now show that $Z_G$ has a 
BRST invariance 
of the familiar form.  Let $\theta$ be an $\vec x$-independent  
$c$-number Grassmann parameter 
(i.e., a $1 \times 1$ Grassmann matrix), and consider the variations 
defined by 
$$\eqalign{
\delta \omega=&\omega^2 \theta~~~, \cr
\delta A_{\ell}=&D_{\ell}\omega \theta~~~, \cr 
\delta \overline{\omega}=&-i h \theta~~~, \cr
\delta h =& 0 ~~~. \cr
}\eqno(4.36)$$
We begin by showing that Eq.~(4.36) defines a nilpotent transformation, 
in the sense that the second variations of all quantities are zero.  
To verify this, we show that the variations of $\omega^2$ and 
$D_{\ell}\omega$ are zero (the variations of $h$ and of $0$ are trivially  
0), as follows:
$$\eqalign{
\delta \omega^2=&\{ \delta \omega, \omega \}=\{\omega^2 \theta,\omega\}  
=\omega^2 \{\omega, \theta \}=0~~~, \cr
\delta D_{\ell}\omega=&[\delta A_{\ell},\omega] + D_{\ell} \delta \omega 
= [D_{\ell}\omega \theta, \omega] +D_{\ell} \omega^2 \theta 
=-\{D_{\ell} \omega,\omega\} \theta + \{D_{\ell}\omega ,\omega\} \theta =0
~~~.\cr
}\eqno(4.37a)$$
To see that $Z_G$ is invariant, we note that the action on $A_{\ell}$ 
of the BRST 
transformation of Eq.~(4.36) is just a gauge transformation 
(albeit with a Grassmann valued parameter), and so 
the factors $d \mu$ and ${\cal G}[A_{\ell}]$ 
are  invariant.  The measure $dh$ is trivially invariant, and the measure 
$d \overline{\omega}$  is invariant because $\delta \overline{\omega}$ 
has no dependence on $\overline{\omega}$.  
Using  
$$\delta(d\omega)=d(\delta \omega)=d(\omega^2 \theta)=(\omega d\omega +
d\omega \omega)\theta~~~,\eqno(4.37b)$$
we have 
$$(\delta(d\omega))_{mn}=(\omega_{mm}d\omega_{mn}+d\omega_{mn} \omega_{nn})
\theta+...=d\omega_{mn}(\omega_{nn}-\omega_{mm})\theta+...,\eqno(4.37c) $$
with the ellipsis ``...'' denoting terms that 
contain matrix elements $d\omega_{m^{\prime}
n^{\prime}}$ with  $(m^{\prime},n^{\prime}) \not= (m,n)$.  Consequently the 
Jacobian of transformation for $d \omega$ 
differs from unity by a term proportional to 
$$\sum_{nm}(\omega_{nn}-\omega_{mm}) \theta =0~~~,\eqno(4.37d)$$
and so the measure $d\omega$ is also invariant.  To complete 
the demonstration that $Z_G$ 
is BRST invariant, we have to show that the gauge fixing part of the 
Hamiltonian, 
$${\bf H}_G\equiv \int d^3x {\rm Tr}({\xi \over 2} h^2 
+i h \sum_{\ell}L^{\ell}A_{\ell}- \overline{\omega}\sum_{\ell}L^{\ell}
D_{\ell} \omega)~~~,\eqno(4.38a)$$
is BRST invariant.  Since we have already seen that $D_{\ell} \omega$
is invariant, and since $h$ is trivially invariant, we have only to verify 
that 
$$0=\int d^3x {\rm Tr}[ih\sum_{\ell}L^{\ell}\delta A_{\ell}
-(\delta \overline{\omega}) \sum_{\ell} L^{\ell} D_{\ell} \omega]
=\int d^3x {\rm Tr}ih \sum_{\ell}L^{\ell}D_{\ell}\{\omega, \theta\}~~~,
\eqno(4.38b)$$
which checks, completing the demonstration of BRST invariance of the  
generalized partition function.

\medskip
\centerline{4E.~~~Reduction of the Hilbert Space Modulo $i_{\rm eff}$}

Our aim in this section is to further study the structure of averages 
of dynamical variables over the canonical ensemble, and more specifically,   
to study the implications of the fact that the canonical ensemble 
only partially breaks the originally assumed global unitary invariance 
group.  
We have seen in Sec.~4B that the canonical ensemble introduces an 
effective imaginary unit operator $i_{\rm eff}$  through  
$$\langle \tilde C \rangle_{\rm AV} =i_{\rm eff} 
D_{\rm eff}~~~,\eqno(4.39a)$$
where $D_{\rm eff}$ is assumed to be a real constant times the unit operator, 
and that the ensemble parameter $\tilde \lambda$ is functionally related 
to $\langle \tilde C \rangle_{\rm AV}$ using only c-number coefficients.  
This means that the traceless, anti-self-adjoint parameter 
$\tilde \lambda$ must have the form 
$$\tilde \lambda = \lambda i_{\rm eff}~~~,\eqno(4.39b)$$ 
with $\lambda$ a real c-number.  
Therefore if $U_{\rm eff}$ is a 
unitary matrix that commutes with $i_{\rm eff}$, 
$$U_{\rm eff}^{\dagger} U_{\rm eff}=U_{\rm eff} U_{\rm eff}^{\dagger} =
1~,~~~[U_{\rm eff},i_{\rm eff}]=0~~~,
\eqno(4.39c)$$
then $U_{\rm eff}$ also commutes with $\tilde \lambda$, 
$$[U_{\rm eff},\tilde \lambda]=0 \Rightarrow U_{\rm eff}
\tilde \lambda U_{\rm eff}^{\dagger}  = \tilde \lambda ~~~.\eqno(4.39d)$$
As a consequence, the canonical ensemble partially respects the assumed 
global unitary invariance of the dynamics:  the integration 
measure $d \mu$, the trace Hamiltonian ${\bf H}$, and the trace quantity  
${\bf N}$ are all invariant under general global unitary transformations 
of the matrix dynamical variables (cf. Eq.~(2.7b) and the discussion 
of Sec. 4A), but as we shall see in detail,   
the term in the exponent in the canonical ensemble ${\rm Tr} \tilde \lambda 
\tilde C$ is invariant only under the subset $U_{\rm eff}$ 
of global unitary transformations that commute with $i_{\rm eff}$. This 
has important consequences that we shall explore in this section.   We 
shall develop a formalism for isolating the effects of the residual 
global unitary invariance, and after establishing that it is necessary  to 
break this invariance in order to extract the full implications of the 
canonical ensemble, we shall give an explicit method for  breaking 
the residual invariance by modifying the  operator phase space measure.

Introducing the standard $2 \times 2$ Pauli matrices $\tau_1,\tau_2,\tau_3$, 
it is convenient to rewrite Eq.~(4.11c) in the form 
$$i_{\rm eff}=i \tau_3 1_K~~~,\eqno(4.40)$$
with $1_K$ a $K \times K $ unit matrix, where we recall that we have taken   
the dimension of the underlying matrix Hilbert space to be $N=2K$. 
Letting $\tau_0=1_2$ denote the 
$2 \times 2$ unit matrix corresponding to the Pauli 
matrices $\tau_{1,2,3}$, a general $N \times N$ matrix $M$ can be 
decomposed in the form 
$$M={1 \over 2} (\tau_0+\tau_3) M_{+} + {1\over 2} (\tau_0-\tau_3) M_{-} 
+\tau_1 M_1 + \tau_2 M_2 ~~~,\eqno(4.41a)$$
with $M_{+,-,1,2}$ four $K\times K$ matrices that commute with all of the 
Pauli matrices $\tau_{1,2,3}$.  Thus, corresponding to $M=i_{\rm eff}$, 
we would have $M_{+}=-M_{-}=i1_K~,~~M_{1,2}=0$.  For 
general $M$, let us define 
$M_{\rm eff}$ and $M_{12}$ by 
$$\eqalign{
M_{\rm eff}=& {1 \over 2} (M-i_{\rm eff} M i_{\rm eff})
={1 \over 2} (\tau_0+\tau_3) M_{+} + {1\over 2} (\tau_0-\tau_3) M_{-} 
~~~,\cr
M_{12} =& M-M_{\rm eff}=\tau_1 M_1 + \tau_2 M_2~~~,\cr
}\eqno(4.41b)$$
so that $M_{\rm eff}$ and $M_{12}$ give, respectively, the parts 
of $M$ that commute and anti-commute with $i_{\rm eff}$, 
$$\eqalign{
i_{\rm eff} M_{\rm eff}=&M_{\rm eff} i_{\rm eff}~~~,\cr
i_{\rm eff} M_{12}=&-M_{12} i_{\rm eff}~~~.\cr
}\eqno(4.41c)$$
Combining Eqs.~(4.41b,c), we get the useful relation  
$$2 i_{\rm eff} M_{\rm eff}=\{i_{\rm eff},M\}~~~.\eqno(4.41d)$$
We see that for the subset of matrix operators $M_{\rm eff}$ that commute 
with $i_{\rm eff}$, the original $N$ dimensional Hilbert space diagonalizes 
into two subspaces of dimension $K$, on the first of 
which $i_{\rm eff}$ 
acts as $i1_K$ and $M_{\rm eff}$ acts as $M_{+}$, and on the second 
of which $i_{\rm eff}$ acts as $-i1_K$ and $M_{\rm eff}$ acts as 
$M_{-}$.  

Using this notation, let us examine the unitary 
transformation behavior of the 
term ${\rm Tr} \tilde \lambda \tilde C$ in the partition function.  
Substituting Eq.~(4.39b), we have 
$${\rm Tr} \tilde \lambda \tilde C   
= \lambda {\rm Tr} i_{\rm eff} \tilde C~~~.\eqno(4.42a)$$  
Under a general unitary transformation of the dynamical variables, 
we have 
$$q_r \to U^{\dagger} q_r U~,~~p_r \to U^{\dagger} p_r U~,~~
\tilde C \to U^{\dagger} \tilde C U~~~,\eqno(4.42b)$$ 
and so the right hand side of Eq.~(4.42a) becomes 
$$ \lambda {\rm Tr}U i_{\rm eff} U^{\dagger} \tilde C~~~,\eqno(4.42c)$$  
which in general differs from Eq.~(4.42a) because a general $U$ does not 
commute with $i_{\rm eff}$.  Thus the 
${\rm Tr} \tilde \lambda \tilde C$ term in the canonical 
ensemble breaks the global unitary invariance of the underlying dynamics. 
However, when $U$ in Eq.~(4.42b) is restricted to have the 
structure $U_{\rm eff}$ that commutes with 
$i_{\rm eff}$, the transformation of Eq.~(4.42b) is still an invariance 
of Eq.~(4.42a), and hence is an invariance of the canonical ensemble.  

In group theoretic terms [8], the original $U(N)$ global unitary invariance 
of ${\bf H}$ is broken, by the term 
${\rm Tr} \tilde \lambda \tilde C$ in the canonical ensemble, 
to $U(K) \times U(K) \times R$, 
with $R$ the discrete reflection symmetry that interchanges the eigenvalues 
$\pm i$ of $i_{\rm eff}$.  This is clearly the largest symmetry group of 
the ensemble for which one can have $\langle \tilde C \rangle_{AV} \not=0$.  
If one were to attempt to preserve the full $U(N)$ symmetry by  
taking an ensemble 
with $\tilde \lambda=i\lambda$, with $\lambda$ a $c$-number, 
then in the canonical ensemble the term 
${\rm Tr}\tilde \lambda \tilde C$ would vanish by 
virtue of the tracelessness 
of $\tilde C$, and the resulting ensemble would have 
$\langle \tilde C \rangle_{AV} =0$. Requiring the largest 
possible nontrivial  
symmetry group plays the role in our derivation of 
making the emergent Planck 
constant a $c$-number; if on the other hand, we 
were to sacrifice all of the $U(N)$ symmetry by allowing a generic $\tilde 
\lambda$, then the emergent canonical commutation relations derived 
in Sec. 5C below 
would generically yield a matrix  $\hbar$  acting non-trivially on the 
states of Hilbert space, which  would be inconsistent with an 
emergent Heisenberg dynamics.  
It would clearly be desirable to have a deeper 
justification from first principles of our choice of canonical ensemble, 
perhaps based on a more detailed understanding 
of the underlying trace dynamics, 
but at present we must simply introduce it as a postulate.  

The residual unitary invariance of the canonical ensemble 
has the following consequence.  Let us write the integration 
measure $d \mu$ as 
$$d \mu= d[U_{\rm eff}] d\hat \mu~~~,\eqno(4.43a)$$
with $d[U_{\rm eff}]$ the Haar measure for integration over the subgroup of 
global unitary transformations $U_{\rm eff}$ that commute with $i_{\rm eff}$, 
and with $d\hat  \mu$ the integration measure over the operator phase 
space subject to the restriction that an overall global unitary 
transformation $U_{\rm eff}$ is kept fixed.  Let us consider the canonical 
ensemble average of a polynomial operator $R_{\rm eff}$, 
that is a function (with $c$-number coefficients) of 
$i_{\rm eff}$ and of the underlying 
dynamical variables $q_r,p_r$ and 
that commutes with $i_{\rm eff}$, 
$$R_{\rm eff\,AV} \equiv \langle R_{\rm eff} \rangle_{\rm AV}
={\int d[U_{\rm eff}] d\hat  \mu \rho R_{\rm eff} 
\over \int d[U_{\rm eff}] d\hat  \mu \rho 1}~~~.\eqno(4.43b)$$
We can relate the general operator variables $q_r,p_r$ to operator variables 
$\hat q_r,\hat p_r$ that have an overall $U_{\rm eff}$ rotation frozen, by 
writing 
$$q_r=U_{\rm eff}^{\dagger} \hat q_r U_{\rm eff}~,~~
p_r=U_{\rm eff}^{\dagger} \hat p_r U_{\rm eff}~~~,\eqno(4.43c)$$
and so correspondingly we have 
$$R_{\rm eff}= U_{\rm eff}^{\dagger} \hat R_{\rm eff} U_{\rm eff} 
~~~,\eqno(4.43d)$$
with $\hat R_{\rm eff}$ obtained from $R_{\rm eff}$ by the replacements 
$q_r,p_r \to \hat q_r, \hat p_r$.  
Since the canonical ensemble is invariant under unitary transformations 
that commute with $i_{\rm eff}$, we have $\rho=\hat \rho$, with $\hat \rho$
constructed in the same manner as $\rho$, but using the variables 
$\hat q_r,\hat p_r$ in place of $q_r,p_r$.  Putting all these ingredients 
together, we can rewrite Eq.~(4.43b) in the form 
$$R_{\rm eff\,AV} = {\int d[U_{\rm eff}] U_{\rm eff}^{\dagger}
 R_{\rm {eff}\,\hat{\rm AV}} 
U_{\rm eff} \over \int d[U_{\rm eff}]  1}~~~,\eqno(4.44a)$$
with $ R_{\rm {eff}\,\hat{\rm AV}}$ given by 
$$ R_{\rm{eff}\,\hat {\rm AV}} 
\equiv {\int d\hat  \mu \hat \rho \hat R_{\rm eff} 
\over \int  d\hat  \mu \hat \rho 1}~~~.\eqno(4.44b)$$
Writing all matrix quantities in terms of $+$ and $-$ components according 
to Eq.~(4.41b), Eqs.~(4.44a,b) separate into the independent 
$\pm$ components 
$$R_{{\rm eff}\,{\rm AV}\pm} = 
{\int d[U_{{\rm eff}\pm} ] U_{{\rm eff}\pm}^{\dagger}
 R_{{\rm eff}\,\hat {\rm AV}\pm} 
U_{{\rm eff}\pm} \over \int d[U_{{\rm eff}\pm}]  1}~~~,\eqno(4.44c)$$
with $ R_{{\rm eff} \, \hat {\rm AV}\pm}$ given by 
$$ R_{{\rm eff} \, \hat {\rm AV}\pm} 
\equiv{\int d\hat  \mu \hat \rho \hat R_{{\rm eff} \pm} 
\over \int  d\hat  \mu \hat \rho 1}~~~.\eqno(4.44d)$$

In both the $\pm$ cases, the integral on the right hand side of  
Eq.~(4.44c) has the general form 
$$I[M_K]=
{\int d[U_K] U_K^{\dagger} M_K U_K \over \int d[U_K]}~~~,\eqno(4.45a)$$
with $M_K$ a $K \times K$ matrix and with $U_K$ a $K\times K$ unitary 
matrix.   But replacing $M_K$ by $V_K^{\dagger} M_K V_K$, with $V_K$ 
unitary, and using the invariance property 
$d[V_KU_K] =d[U_K]$ of the Haar measure, we see 
that $I[M_K]=I[V_K^{\dagger} M_K V_K]$ for arbitrary unitary $V_K$.  Thus 
$I[M_K]$ is a linear, unitary invariant function of $M_K$, which on the 
unit matrix takes the value $I[1_K]=1_K$. These properties imply that  
$I[M_K]$ is given by the trace 
$$I[M_K]={1 \over K} 1_K {\rm Tr}_K M_K~~~.\eqno(4.45b)$$
We learn from this that if we take an unrestricted average of $R_{\rm eff}$  
over the canonical  
ensemble, the interesting matrix operator structure is averaged out.  To  
preserve this structure, we must restrict the integration in the canonical 
ensemble to leave an overall  global unitary transformation $U_{\rm eff}$ 
fixed, as in $R_{\rm eff\,\hat{\rm AV}}$ of Eq.~(4.44b).  

Let us now give an explicit recipe for fixing an overall global unitary 
transformation that commutes with $i_{\rm eff}$.  We begin by assuming 
that the structure of the trace Hamiltonian is such that we cannot split 
the dynamical variables $\{x_r\}$ into two disjoint sets $\{x_r^I\}$ 
and $\{x_r^{II}\}$, such that the trace Hamiltonian {\it exactly} 
separates into disjoint pieces, 
$${\bf H}={\bf H}^I[\{x_r^I\}] + {\bf H}^{II}[\{x_r^{II}\}]~~~.
\eqno(4.46)$$
If Eq.~(4.46) were to hold, the fact that $\tilde C$ and ${\bf N}$ are 
additive over the dynamical degrees of freedom would then 
imply exact factorization of the partition function $Z$ according 
to $Z=Z^IZ^{II}$, 
and we would then have to address the same problem of fixing a global  
unitary invariance at the level of both $Z^I$ and $Z^{II}$.  Put another way, we 
begin by assuming that our trace dynamics is irreducible, in the sense that 
it cannot be exactly reduced to two or more independent trace dynamics 
systems.  

Once this assumption has been made, it suffices to fix a global 
unitary rotation of any {\it one} dynamical variable $x_R$, which we shall 
assume to be a self-adjoint bosonic variable, and which we denote for further 
discussion by $M$.  Let $d\mu(M)$ be the factor contributed by $M$ to the 
integration measure $d\mu$ of Eq.~(4.2b), that is   
$$d\mu(M)=\prod_{m\leq n,A}dM_{mn}^A~~,\eqno(4.47a)$$
with self-adjointness implying that the diagonal imaginary parts $M_{mm}^1$ 
are zero (cf. Eq. (4.6b)).  Making the Pauli matrix decomposition of $M$ 
given in Eq.~(4.41a), with self-adjointness of $M$ 
implying that $M_{+,-,1,2}$ 
are all self-adjoint $K \times K$ matrices, we have 
$$d\mu(M)={\rm constant} \times d\mu(M_+)d\mu(M_-)d\mu(M_1)d\mu(M_2)~~~, 
\eqno(4.47b)$$ 
with the overall constant not relevant for computing averages over the 
canonical ensemble since it cancels out between numerator and normalizing  
denominator in $\langle {\cal O} \rangle$  
(cf. Eq.~(4.10a)).  In order to fix an overall unitary rotation 
$U_{\rm eff}$, it suffices to fix independent overall unitary rotations 
in the submatrices $M_{+,-}$ that determine $M_{\rm eff}$.  This can be 
done explicitly by writing $M_{+,-}$ as unitary rotations from diagonal  
matrices $D_{+,-}$,  
$$\eqalign{
M_+=&U_{{\rm eff} +}^{\dagger} D_+ U_{{\rm eff} +}~~~,\cr  
M_-=&U_{{\rm eff} -}^{\dagger} D_- U_{{\rm eff} -}~~~,\cr  
}\eqno(4.48a)$$                        
and then using the formula [21] 
$$ d\mu(M_{\pm} )={\rm constant} \times  d[U_{{\rm eff} \pm}] 
\Delta(\lambda_{\pm})^2 \prod_i d\lambda_{\pm i}~~~. \eqno(4.48b)$$
Here  $\lambda_{\pm i}$ are the real eigenvalues of $M_{\pm}$,  
$\Delta(\lambda_{\pm})$ is the Vandermonde determinant 
$$\Delta(\lambda)=\prod_{i>j} (\lambda_i-\lambda_j)~~~,\eqno(4.48c)$$
and $d[U_{\rm eff \pm} ]$ are the Haar measures for the unitary groups 
generated by $U_{\rm eff \pm}$.   Simply omitting the integrations over 
these unitary groups then gives the necessary freezing of the global 
unitary invariance $U_{\rm eff}$. In other words, the restricted measure 
$d\hat   \mu$ is explicitly defined by replacing Eq.~(4.48b) by 
$$ d\hat  \mu(M_{\pm} )={\rm constant} \times   
\Delta(\lambda_{\pm})^2 \prod_i d\lambda_{\pm i}~~~,  \eqno(4.48d)$$
for the one chosen self-adjoint variable $M$.  

In field theory applications, the index $r$ for degrees of freedom will  
contain the specification of a spatial point $\vec x$, and the unitary 
invariance fixing discussed above is implemented by taking $M$ to be 
a bosonic variable at a specific spatial point $\vec X$, on the specified 
time slice $t=T$ used in the canonical ensemble.  As a result of this global 
unitary fixing, if we form the average $ R_{{\rm eff}\,\hat{\rm AV}}$ 
of a polynomial 
function of $i_{\rm eff}$ and of operators at various space-time 
coordinates $x^{\nu}$, and 
then shift all space time coordinates by a common amount $a^{\nu}$, 
the resulting average will be changed, even when the dynamics is space-time 
translation invariant, because the variable that is frozen is changed from 
$M(\vec x,T)$ to $M(\vec x+\vec a, T+a^0)$.  (Even when $\vec a=0$ and there 
is only a time translation $a^0$, dynamical evolution of the 
system results in  $M(\vec x, T+a^0)$ differing 
from $M(\vec x,T)$, and so the 
global unitary fixings based on them will differ.)  In a theory 
with space-time translation invariant dynamics, the shifted 
result will be related to the unshifted one by a global unitary 
transformation that commutes with $i_{\rm eff}$, 
$$R(a^{\nu})_{{\rm eff}\,\hat{\rm AV}}= U(a^{\nu})_{\rm eff}^{\dagger} 
 R(0)_{{\rm eff}\,\hat{\rm AV}} U(a^{\nu}_{\rm eff}) ~~~,\eqno(4.49a)$$
since this is the general structure of the 
transformation that relates two inequivalent global unitary invariance 
fixings.  When substituted into Eq.~(4.44a), the transformation 
$U(a^\nu)_{\rm eff}$ can be absorbed into the integration measure, leading to the 
expected result that the unfixed average is space-time translation invariant. 
(Formally, this invariance results from the facts that space-time 
translations 
are a canonical transformation generated by the trace four-momentum 
${\bf P}^{\nu}$, and that this canonical transformation is an invariance of 
the unfixed integration measure $d\mu$. Hence the effect of an overall 
space-time translation can simply be absorbed 
into the integration measure in the unfixed case.)
Because the shifts by $a^{\nu}$ form an Abelian group, invoking Stone's 
theorem tells us that $U(a^{\nu})_{\rm eff}$ must have the generator 
form 
$$U(a^{\nu})_{\rm eff}=\exp(i_{\rm eff} \hbar^{-1} a_{\nu} 
P_{\rm eff}^{\nu})~~~,\eqno(4.49b)$$
with $\hbar^{-1} P_{\rm eff}^{\nu}$ a self-adjoint matrix 
that commutes with $i_{\rm eff}$ (and with $a_{\nu}=\eta_{\mu\nu}a^{\nu}$,   
where $\eta_{\mu\nu}=(1,1,1,-1)$ is the Minkowski metric and we have used 
the usual relativistic summation convention.)
Equations (4.49a,b) thus 
have the same structure as the space time translation 
properties of a polynomial function of field operators in quantum field 
theory,  giving a hint already 
that under certain conditions a quantum field theory structure can emerge
from the statistical mechanics of trace dynamics with a global unitary 
invariance.  Developing this idea further is the subject to which 
we turn in the next section.  
\vfill
\eject

\twelvepoint
\doublespace
\pageno=78
\overfullrule=0pt
\centerline{5.~~~Ward Identities and the Emergence of Quantum Field 
Dynamics}
\bigskip
In Sec. 2, we have seen that a generic feature of matrix models with 
a global unitary invariance is the existence of a conserved   
operator $\tilde C$,  
$$\tilde C \equiv \sum_{r,B}[q_r,p_r]-\sum_{r,F}
\{q_r, p_r\}=\sum_{r,u}x_r \omega_{ru}x_u~~,\eqno(5.1)$$
which is anti-self-adjoint when we adopt the fermion adjointness 
assignment of Eqs.~(2.4b) and (2.18d).  
The operator $\tilde C$, which is given by the sum of bosonic $q_r,p_r$
commutators minus the corresponding sum
of fermionic anticommutators,  together with the trace quantities 
${\bf H}$ and ${\bf N}$, plays a role in the equilibrium 
statistical mechanics of trace dynamics closely
analogous to that played by the energy in 
classical statistical physics.  This analogy suggests the idea [7] 
that the canonical
commutation relations of quantum field theory may arise from a 
trace dynamics analog of the classical theorem of equipartition of 
energy.  To pursue this thought, let us begin by reviewing a 
simple derivation [22] of the classical equipartition theorem.  
Let $H(\{x_r\})$ be the classical Hamiltonian
as a function of classical phase space variables $\{x_r\}$, and let 
$d\mu(\{x_r\})$ be the classical phase space integration measure.  We
consider the integral
$$\eqalign{
&\int d\mu {\partial [x_r \exp(-\beta H)] \over \partial x_s}\cr 
=&\int d\mu \delta_{rs} \exp(-\beta H) \cr      
-&\int d\mu x_r {\partial [\beta H] \over \partial x_s} 
\exp(-\beta H)~~~, \cr
}\eqno(5.2a)$$
the left hand side of which  
is the integral of a total derivative and vanishes when
the integrand is sufficiently rapidly vanishing at infinity.  Assuming this,
we get 
$$\delta_{rs}={ \int d\mu x_r \beta (\partial H/ \partial x_s) 
\exp(-\beta H) \over \int d\mu \exp(-\beta H) } ~~~, \eqno(5.2b)$$
which is the classical theorem of equipartition of energy.  The method of
derivation is similar to that used to derive Ward identities from functional
integrals in quantum field theory (see, e.g. [23]), and the equipartition
theorem can be viewed as a Ward identity application in classical statistical
mechanics.
\bigskip
\centerline{5A.~~~The General Ward Identity}
\bigskip
We now apply a similar procedure to the statistical mechanics of matrix 
models.  We begin by specifying some notation.  
Let $d\hat  \mu$ be the measure introduced in Eqs.~(4.43a) and (4.48d), in 
which the integration over {\it one} dynamical variable $x_R$ is restricted 
so as to break the subgroup of the global unitary group that commutes with 
$i_{\rm eff}$.  Also, let $\langle {\cal O}\rangle_{\hat{\rm AV}}$ be the 
average over the canonical ensemble, using this restricted measure,    
of a general operator depending on the $\{x_r\}$.  This average is 
formed as in Eq.~(4.10a), 
$$\langle {\cal O}\rangle_{\hat{\rm AV}} 
={ \int d\hat  \mu \rho {\cal O} \over \int d\hat  \mu \rho}~~~,\eqno(5.3a)$$
with $\rho$ and the partition function $Z$ given by Eq.~(4.13c), as modified 
by the replacement of $d\mu$ by the restricted integration 
measure $d\hat  \mu$.       
Since we shall wish to include sources, let 
$\langle {\cal O}\rangle_{\hat{\rm AV},j}$ be the corresponding average 
in the presence of a complete set $\{j_r\}$ of external sources, given by 
$$\langle {\cal O}\rangle_{\hat{\rm AV},j} 
={ \int d\hat  \mu \rho {\cal O} \over \int d\hat  \mu \rho}~~~,\eqno(5.3b)$$
with $\rho$ and $Z$  given by the expressions (cf.  
Eq.~(4.15a)) that include sources, now with restricted integration measure,   
$$\eqalign{
\rho=&Z^{-1}
\exp(-{\rm Tr} \tilde \lambda \tilde C- \tau {\bf  H}-\eta 
{\bf N}  - \sum_r {\rm Tr} j_r x_r  )~~~, \cr 
Z=&\int d\hat \mu
\exp(-{\rm Tr} \tilde \lambda \tilde C-\tau {\bf H}-\eta
{\bf N}  - \sum_r {\rm Tr} j_r x_r   )~~~. \cr 
}\eqno(5.3c)$$
Thus, in this notation, we have
$$\langle {\cal O}\rangle_{\hat{\rm AV}}= 
\langle {\cal O}\rangle_{\hat{\rm AV},0}~~~,\eqno(5.3d)$$ 
with the subscript $0$ on the right hand side denoting the average in 
which all sources are zero, that is, with $\{j_r\}=\{0\}$.  
In the deriving the Ward identity, we shall employ the fact that the 
integration measure $d \mu$ is invariant under a constant shift 
$x_r \to x_r + \delta x_r$ of any of the dynamical variables $x_r$.  When 
we use the restricted integration measure $d\hat  \mu$, this invariance 
still holds, provided $r$ is not equal to the index $R$ of the one variable 
$x_R$ for which the phase space 
integration is restricted.  In other 
words, with respect to the restricted measure we have 
$$0=\int d\hat  \mu \delta_{x_r} (\rho {\cal O})~,~~r \neq R~~~,\eqno(5.4a)$$
with 
$$\delta_{x_r} {\cal A}={\cal A}|_{x_r +\delta x_r} - {\cal A}|_{x_r}~~~.
\eqno(5.4b)$$

With this notation established, we begin our derivation by considering 
$$Z \langle {\rm Tr} \{ \tilde C,i_{\rm eff}\} W \rangle_{\hat{\rm AV}} 
= \int d\hat  \mu 
\exp(-{\rm Tr} \tilde \lambda \tilde C- \tau {\bf  H}-\eta 
{\bf N}  - \sum_r {\rm Tr} j_r x_r  ) {\rm Tr} \{ \tilde C,i_{\rm eff}\} W 
~~~,\eqno(5.5a)$$
with $W$ any bosonic polynomial function of the dynamical variables, and   
where the trace ${\rm Tr}$ is understood to act on the product of all factors
standing to its right.  
Using Eqs.~(5.4a,b), when we make a shift of $x_s,~s \neq R$  in the 
integrand of Eq.~(5.5a), we have 
$$0= \int d\hat  \mu  \delta_{x_s} [
\exp(-{\rm Tr} \tilde \lambda \tilde C- \tau {\bf  H}-\eta 
{\bf N}  - \sum_r {\rm Tr} j_r x_r  ) {\rm Tr} \{ \tilde C,i_{\rm eff}\} W ]
~~~,\eqno(5.5b)$$
which on applying the chain rule for differentiation becomes 
$$\eqalign{
0=& \int d\hat  \mu  
\exp(-{\rm Tr} \tilde \lambda \tilde C - \tau {\bf  H}- \eta 
{\bf N}  - \sum_r {\rm Tr} j_r x_r  ) \cr 
&\times [(
-{\rm Tr} \tilde \lambda \delta_{x_s} \tilde C- \tau \delta_{x_s} {\bf  H} 
-\eta \delta_{x_s} {\bf N}  -  {\rm Tr} j_s \delta x_s  )  
{\rm Tr} \{ \tilde C,i_{\rm eff}\} W
+ \delta_{x_s}
{\rm Tr} \{ \tilde C,i_{\rm eff}\} W ]  ~~~.\cr
}\eqno(5.5c)$$

We now have to evaluate the variations with respect to $x_s$ appearing in 
the various terms on the right hand side of Eq.~(5.5c).  From Eq.~(2.6) 
for $\tilde C$, we have 
$${\rm Tr} \tilde \lambda \delta_{x_s} \tilde C 
={\rm Tr} \tilde \lambda \sum_r( \delta x_s \omega_{sr} x_r + 
x_r \omega_{rs} \delta x_s )~~~,\eqno(5.6a)$$
which by Eq.~(1.17) can be rewritten as 
$${\rm Tr} \tilde \lambda \delta_{x_s} \tilde C  
={\rm Tr}  \tilde \lambda \sum_r \omega_{rs} 
(- \epsilon_r \delta x_s x_r +  x_r  \delta x_s )~~~,\eqno(5.6b)$$
and by Eqs.~(1.1a,b) simplifies to 
$${\rm Tr} \tilde \lambda \delta_{x_s} \tilde C  
={\rm Tr}  [\tilde \lambda, \sum_r \omega_{rs} x_r] \delta x_s~~~.
\eqno(5.6c)$$
We next consider $ \delta_{x_s} {\bf  H}$, which by Eqs.~(1.3b) and (1.15b)  
through (1.17) is given by 
$$ \delta_{x_s} {\bf  H} ={\rm Tr} {\delta {\bf H} \over \delta x_s} 
\delta x_s=\sum_r \omega_{rs} {\rm Tr} \dot x_r \delta x_s~~~.\eqno(5.7)$$
Turning next to the evaluation of $\delta_{x_s} {\bf N}$, 
let us define $\tilde \omega_{rs}$ 
by 
$$\tilde \omega_{rs}={\rm diag}(0,...,0,\Omega_B,...,\Omega_B)~~~,
\eqno(5.8a)$$
where the $2 \times 2$ matrix $\Omega_B$ is defined in Eq.~(1.16b).  
Recalling our convention that we list all bosonic variables 
before all fermionic 
ones in the $2D$-dimensional phase space vector $x_r$, the definition of 
Eq.~(5.8a) states that $\tilde \omega_{rs}$ acts as $0$ on any bosonic 
phase space pair $q_r,p_r$ and acts as $\Omega_B$ on any fermionic phase 
space pair $q_r,p_r$, so that 
$$\sum_{r,u} \tilde \omega_{ru} x_r x_u = \sum_{r,F} [q_r,p_r] 
~~~,\eqno(5.8b)$$
giving the quantity appearing in Eq.~(2.2a) defining ${\bf N}$.  We then 
find 
$$\delta_{x_s} {\bf N} ={1\over 2} i \sum_r {\rm Tr} 
(\tilde \omega_{sr} \delta x_s x_r  + \tilde \omega_{rs} x_r \delta x_s)
=i \sum_r \tilde \omega_{rs} {\rm Tr} x_r \delta x_s~~~.\eqno(5.8c)$$
This completes the calculation of variations of terms that come from the 
exponent in the canonical ensemble $\rho$. 

The remaining terms come from 
$$\delta_{x_s} {\rm Tr} \{\tilde C,i_{\rm eff}\} W 
={\rm Tr}(\{\delta_{x_s} \tilde C, i_{\rm eff} \} W    
+ \{\tilde C,i_{\rm eff}\} \delta_{x_s} W )~~~.\eqno(5.9a)$$
{}For the first term on the right hand side of Eq.~(5.9a), we use Eq.~(5.6c), 
with $\tilde \lambda$ replaced by 
$\{i_{\rm eff},W\}=2i_{\rm eff} W_{\rm eff}$ (where we have simplified using  
Eq.~(4.41d)), giving 
$${\rm Tr}\{\delta_{x_s} \tilde C, i_{\rm eff} \} W   
={\rm Tr} \{i_{\rm eff}, W\}  \delta_{x_s} \tilde C
={\rm Tr}  [2 i_{\rm eff} W_{\rm eff}, \sum_r \omega_{rs} x_r] \delta x_s~~~.
\eqno(5.9b)$$
To evaluate the second term on the right hand side of Eq.~(5.9a), we write  
the operator structure of $W$ in the form 
$$W=\sum_{\ell} W_s^{L\ell} x_s W_s^{R\ell}~~~,\eqno(5.10a)$$ 
where $\ell$ indexes each occurrence of $x_s$ in $W$, so that in this 
notation we have 
$${\delta {\bf W} \over \delta x_s} =\sum_{\ell} \epsilon_{\ell} W_s^{R\ell} 
W_s^{L\ell}~~~,\eqno(5.10b)$$
with $\epsilon_{\ell}$ the grading factor 
appropriate to $W_s^{R\ell}$ and to 
$W_s^{L\ell}x_s$ (which must both 
be of the same grade since we have defined $W$ to 
be bosonic).  With this definition, we find 
$$\delta_{x_s} W= \sum_{\ell} W_s^{L\ell}\delta x_s W_s^{R\ell}~~~,
\eqno(5.10c)$$
giving finally 
$${\rm Tr}  \{\tilde C,i_{\rm eff}\} \delta_{x_s} W
= \sum_{\ell} \epsilon_{\ell}  {\rm Tr}
W_s^{R\ell} \{\tilde C,i_{\rm eff}\} W_s^{L\ell}  
\delta x_s~~~.\eqno(5.10d)$$

When these results are collected and substituted back into Eq.~(5.5c), this 
equation takes the form (after multiplication by $Z^{-1}$)
$$0=\langle {\rm Tr} \Sigma_s \delta x_s\rangle_{\hat{\rm AV},j} 
~~~,\eqno(5.11a)$$
with $\Sigma_s$ a shorthand for the sum of the contributions coming from 
Eqs.~(5.6a) through (5.10d),   
$$\eqalign{
\Sigma_s=&(-[\tilde \lambda,\sum_r \omega_{rs} x_r] -\tau \sum_r\omega_{rs} 
\dot x_r -i\eta \sum_r \tilde \omega_{rs} x_r -j_s)2 {\rm Tr} \tilde C 
i_{\rm eff} W_{\rm eff}\cr  
+&2 [i_{\rm eff} W_{\rm eff}, \sum_r \omega_{rs}  x_r]
+\sum_{\ell} \epsilon_{\ell} W_s^{R\ell} \{\tilde C,i_{\rm eff} \} 
W_s^{L\ell}
~~~.\cr
}\eqno(5.11b)$$
Since the variation $\delta x_s$ is arbitrary, 
subject to the adjointness restrictions on $x_s$, and since both the 
real and imaginary parts of Eq.~(5.11a) must vanish separately, 
we can conclude 
from Eq.~(5.11a) that 
$$\langle \Sigma_s \rangle_{\hat{\rm AV},j}=0~~~.\eqno(5.11c)$$    

To get the final form of the  Ward identity, we perform 
several algebraic manipulations on Eq.~(5.11c).  First of all, we form  
the effective projection by taking one half of the anticommutator of 
Eq.~(5.11c) with $i_{\rm eff}$ (cf. Eq.(4.41d)).  Since $\tilde \lambda 
=\lambda i_{\rm eff}$ (cf. Eq.~(4.39b)), and since $i_{\rm eff}$ commutes 
with $x_{r {\rm eff}}$, the first term on the right hand side of Eq.~(5.11b) 
drops out and we are left with 
$$\langle \Sigma_{s{\rm eff}} \rangle_{\hat{\rm AV},j}=0~~~,\eqno(5.12a)$$    
with $ \Sigma_{s{\rm eff}} $ given by 
$$\eqalign{
\Sigma_{s{\rm eff}}=&(-\tau \sum_r\omega_{rs} 
\dot x_{r{\rm eff}} -i\eta \sum_r \tilde \omega_{rs} x_{r{\rm eff}} 
-j_{s {\rm eff}} )
2 {\rm Tr} \tilde C i_{\rm eff} 
W_{\rm eff} \cr  
+&2 [i_{\rm eff} W_{\rm eff}, \sum_r \omega_{rs}  x_{r{\rm eff}}]
+\sum_{\ell} \epsilon_{\ell} (W_s^{R\ell} \{\tilde C,i_{\rm eff} \} 
W_s^{L\ell})_{\rm eff}
~~~.\cr
}\eqno(5.12b)$$
(This step of the derivation requires use of the canonical ensemble: 
in the microcanonical ensemble, the analogous term arising from 
the $\tilde C$ dependence of the ensemble $\Gamma$ does not 
have the form of a commutator $[\tilde \lambda,...]$, and so does not 
vanish on taking the effective projection.)  

Next, we multiply Eqs.~(5.12a,b) by 
${1 \over 2} \omega_{us}$ and sum on $s$, evaluating the sums using 
Eq.~(1.17), which gives  
$$\sum_s \omega_{us}\omega_{rs} =\delta_{ur}~~~,\eqno(5.13a)$$
together with the the formula
$$\sum_{rs} \omega_{us} \tilde \omega_{rs}x_r=-\xi_ux_u~~~,\eqno(5.13b)$$
with $\xi_u=0$ for any bosonic $x_u$ and with  $\xi_u=1(-1)$ for $x_u$ 
a fermionic $q(p)$.  This gives the final result  
$$\langle \Lambda_{u{\rm eff}} \rangle_{\hat{\rm AV},j}=0~~~,\eqno(5.14a)$$ 
with $ \Lambda_{u{\rm eff}} $ given by 
$$\eqalign{
\Lambda_{u{\rm eff}}=&{1\over 2} \sum_s \omega_{us} \Sigma_{s{\rm eff}}\cr
=&(-\tau \dot x_{u{\rm eff}} +i\eta \xi_u x_{u{\rm eff}}
-\sum_s \omega_{us} j_{s {\rm eff}}) 
 {\rm Tr} \tilde C i_{\rm eff} W_{\rm eff} \cr  
+& [i_{\rm eff} W_{\rm eff},  x_{u{\rm eff}}]
+\sum_{s,\ell} \omega_{us} \epsilon_{\ell} \left(W_s^{R\ell} 
{1\over 2}  \{\tilde C,i_{\rm eff} \} 
W_s^{L\ell}\right)_{\rm eff}
~~~.\cr
}\eqno(5.14b)$$
\bigskip
\centerline{5B.~~~Variation of the Source Terms}
\bigskip
The Ward identity of Eqs.~(5.14a,b) is exact, and still includes the full 
source term structure.  Our next step is to show how, through variation 
of the source terms, we can generate similar Ward identities involving 
general polynomials in the effective projections 
$x_{r {\rm eff}}$ of the dynamical variables.  To see how such polynomials 
can be generated, we recall that in the notation  
introduced in Eq.~(4.41b), 
when the source term in the canonical ensemble
is rewritten using the decompositions $x_r=x_{r {\rm eff}}+ 
x_{r12}$ and $j_r=j_{r {\rm eff}}+j_{r12}$, we
get 
${\rm Tr}j_r x_r={\rm Tr}(j_{r{\rm eff}} x_{r{\rm eff}} +
j_{r12} x_{r12})$, 
and so varying with respect to $j_{r{\rm eff}}$ brings down a factor 
of ${\rm Tr}\delta j_{r{\rm eff}}  x_{r{\rm eff}}$.  Stripping away 
the general variation $\delta  j_{r{\rm eff}}$ from the left  
then leaves us with a 
matrix factor $ x_{r{\rm eff}}$.  As we shall see, doing this repeatedly  
allows one to build up general polynomials.  
We can now state the result to be demonstrated in 
this section as follows:  defining ${\cal D} x_{u {\rm eff}}$ by 
$$\eqalign{
{\cal D} x_{u {\rm eff}} 
=&(-\tau \dot x_{u{\rm eff}} +i\eta \xi_u x_{u{\rm eff}})
{\rm Tr} \tilde C i_{\rm eff} W_{\rm eff} \cr  
+& [i_{\rm eff} W_{\rm eff},  x_{u{\rm eff}}]
+\sum_{s,\ell} \omega_{us} \epsilon_{\ell} \left(W_s^{R\ell} 
{1\over 2}  \{\tilde C,i_{\rm eff} \} 
W_s^{L\ell}\right)_{\rm eff} ~~~,\cr
}\eqno(5.15a)$$
we can rewrite Eq.~(5.14a) as 
$$\langle {\cal D} x_{u {\rm eff}}  -\sum_s \omega_{us} j_{s {\rm eff}} 
{\rm Tr} \tilde C i_{\rm eff} W_{\rm eff}  \rangle_{\hat{\rm AV},j} =0~~~.
\eqno(5.15b)$$ 
Then we shall show that Eq.~(5.15b) also implies the 
relations at zero sources, 
$$\langle S_L(x_{t{\rm eff}}) ({\cal D} S(x_{r{\rm eff}})) 
S_R(x_{t{\rm eff}})  
\rangle_{\hat {\rm AV},0} =0~~~,
\eqno(5.15c)$$ 
with $S$ a general polynomial (with $c$-number coefficients) 
in the effective variables 
$x_{r {\rm eff}}$ on which ${\cal D}$ acts by the Leibnitz product rule
$${\cal D}(x_{r{\rm eff}}x_{s{\rm eff}})=({\cal D}x_{r{\rm eff}}) 
x_{s{\rm eff}} + x_{r{\rm eff}} ({\cal D} x_{s{\rm eff}})~~~,\eqno(5.15d)$$
and with $S_L$ and $S_R$ polynomials (also with $c$-number coefficients) 
with the property that for all $x_t$ in 
$S_{L,R}$  and all $x_r$ in $S$, the structure constant $\omega_{tr}$ is 
zero.  
An equivalent way of stating the result of Eq.~(5.15c) is obtained by noting  
that for a ${\cal D}$ 
that acts by the Leibnitz rule, we have 
$$ {\cal D} S(x_{r{\rm eff}}) =
\sum_u S(x_{r{\rm eff}}, \, r\neq u; {\cal D} x_{u{\rm eff}} ) 
~~~,\eqno(5.16a)$$
with all factor orderings left undisturbed when ${\cal D}$ acts on monomials  
in $x_{r{\rm eff}}$.  Since the case of a polynomial $S$ with $c$-number 
coefficients can be obtained by linearity from the case of a monomial $S$, 
to establish Eq.~(5.15c) it suffices  to prove  
that for monomials $S$, $S_L$, 
and $S_R$, we have 
$$\langle   S_L(x_{t{\rm eff}}) 
\sum_u S(x_{r{\rm eff}} \, r\neq u; {\cal D} x_{u{\rm eff}} ) 
 S_R(x_{t{\rm eff}})       \rangle_{\hat {\rm AV},0} =0~~~.
\eqno(5.16b)$$ 

To prove this assertion, we begin by multiplying Eq.~(5.15b) by 
$Z\delta j_{u {\rm eff}}$ and taking the trace, giving 
$$Z\langle {\rm Tr} \delta j_{u {\rm eff}}{\cal D} x_{u {\rm eff}}  
-\sum_s \omega_{us}{\rm Tr} (\delta j_{u{\rm eff}} j_{s {\rm eff}} )  
{\rm Tr} \tilde C i_{\rm eff} W_{\rm eff}  \rangle_{\hat{\rm AV},j} =0~~~.
\eqno(5.17a)$$ 
We now make sequential, independent variations of the  sources 
$j_{r{\rm eff}}$ associated 
with all $x_{r{\rm eff}}$ in the monomial $S$ other than the 
one $x_{u{\rm eff}}$  that is explicitly exhibited in Eq.~(5.17a), as well as 
variations of the sources associated with all   
$x_{t{\rm eff}}$ in the monomials $S_{L,R}$.  To     
get repeated factors of some particular $x_{T{\rm eff}}$, we make independent source 
variations for each, 
and after all variations have been performed, we set all sources equal 
to zero.  In the second term in Eq.~(5.17a), if $j_{s{\rm eff}}$ is not 
varied in this process, it makes a vanishing contribution after the 
sources are set to zero. The case when  $j_{s{\rm eff}}$ is varied can only  
arise when we are varying with respect to the source for a variable 
$x_{v{\rm eff}}$ contained in $S$, since by hypothesis the coefficient 
tensor $\omega_{ut}$ vanishes for all $x_{t{\rm eff}}$ in $S_{L,R}$.   
In general, 
if we focus on two variables $x_{v{\rm eff}}$ and $x_{u{\rm eff}}$ 
for which $\omega_{uv} \neq 0$, 
the polynomial $S$ will have the form 
$...x_{u{\rm eff}}....x_{v{\rm eff}}....$ (or a similar 
expression with the roles of $u$ and $v$ interchanged; we do not assume 
symmetrization of this structure over $u$ and $v$), with $...$ denoting 
factors that are not explicitly exhibited.  Correspondingly, the derivative 
${\cal D}S$ formed by use of the Leibnitz rule will have the form 
$$...({\cal D}x_{u{\rm eff}})...x_{v{\rm eff}}... + 
...x_{u{\rm eff}}... ({\cal D}x_{v{\rm eff}})...~~~~.\eqno(5.17b)$$  
In forming the first term in Eq.~(5.17b) by source variations, the second 
term of Eq.~(5.17a) will contribute the second variation expression 
$$...\omega_{uv} {\rm Tr} (\delta j_{u{\rm eff}} j_{v {\rm eff}} )... ~~~, 
\eqno(5.17c)$$ 
while in forming the second term in Eq.~(5.17b), there will be a 
corresponding second variation expression 
$$...\omega_{vu} {\rm Tr} (\delta j_{v{\rm eff}} j_{u {\rm eff}} )... ~~~. 
\eqno(5.17d)$$ 
Since both second variation expressions multiply identical factors, they 
contribute through their sum 
$$...[\omega_{uv} {\rm Tr} (\delta j_{u{\rm eff}} j_{v {\rm eff}} )
+ \omega_{vu} {\rm Tr} (\delta j_{v{\rm eff}} j_{u {\rm eff}})]... ~~~.
\eqno(5.17e)$$
However, this expression vanishes, because in the bosonic (fermionic) 
case $\omega_{uv}$ is antisymmetric (symmetric) in $u,v$ 
while the  trace multiplying it is symmetric (antisymmetric) in $u,v$.

Thus, in forming Eq.~(5.16b) by source variation,  the explicit 
source term in Eq.~(5.15b) does not contribute after all sources are set 
equal to zero, and so the effect of multiple source variations is to    
lead to an expression of the form 
$$Z\langle  \prod_{r \neq u} {\rm Tr}(\delta j_{r{\rm eff}} x_{r{\rm eff}})
{\rm Tr}(\delta j_{u{\rm eff}} {\cal D} x_{u{\rm eff}} ) \
\rangle_{\hat{\rm AV},0} ~~~.\eqno(5.17f)$$
Since the sources variations 
in Eq.~(5.17f) are all independent, we can use them to ``break open'' the 
traces as in the discussion of Eqs.~(1.4a-d), giving the expression  
$$Z\langle  \prod_{r \neq u}  (x_{r{\rm eff}})_{n_r m_r}
( {\cal D} x_{u{\rm eff}} )_{n_u m_u} \rangle_{\hat{\rm AV},0} 
~~~,\eqno(5.17g)$$
with the average now involving a product of matrix elements. 
By linking the matrix elements in 
Eq.~(5.17g) in the appropriate order, we can form the general matrix element 
of ${\cal D}$ acting on a general monomial $S$,  
multiplied on the left and right by monomials $S_{L,R}$ containing 
variables  $x_{t{\rm eff}}$ that are not 
linked by $\omega_{rt}$ to any variable $x_{r{\rm eff}}$ in $S$.  
This then gives the identity of Eq.~(5.16b), completing the proof of 
Eq.~(5.15c).  
\vfill\eject
\centerline{5C.~~~Approximations/Assumptions Leading to the Emergence}
\centerline{of Quantum Theory}
\bigskip

Starting from Eqs. (5.15a--c), we proceed now to show that 
with a few plausible assumptions and approximations, the general 
formalism of quantum field theory emerges.  The assumptions and 
approximations that we make all have the effect of simplifying 
the structure of ${\cal D} x_{u {\rm eff}}$.  They are:

\item{(1)}  We assume that the support properties of $\dot x_{u {\rm eff}}$ 
and of $\tilde C_{\rm eff}=-{1\over 2} i_{\rm eff} \{i_{\rm eff},\tilde C\}$ 
in the operator phase space 
are such that the term 
$$\eqalign{
-&\tau \dot x_{u{\rm eff}} {\rm Tr} \tilde C i_{\rm eff} W_{\rm eff}~~~\cr
=&-\tau \dot x_{u{\rm eff}} 
{\rm Tr}\tilde C_{\rm eff} i_{\rm eff} W_{\rm eff}  ~~~ \cr
}\eqno(5.18a)$$ 
in Eq. (5.15a) can be neglected.  
Specifically, we shall assume that this 
comes about in the following way.  We identify the time scale $\tau$ and 
mass ${\tau}^{-1}$ with the ``fast'' or ``high'' physical scale given by the  
Planck scale,  and we assume 
that the underlying theory develops a mass hierarchy, so that observed 
physics corresponds to ``slow'' components of $x_{u {\rm eff}}$  
that  are very slowly varying in comparison to the time $\tau$.  We also  
assume that the ``fast'' components of $x_{u {\rm eff}}$ have disjoint 
support from the support of $\tilde C_{\rm eff}$.  
Then for the ``slow'' components 
of  $x_{u {\rm eff}}$, the contribution to 
Eq.~(5.18a) will be small because it 
is suppressed by one power of the mass hierarchy,  
while for the ``fast'' components of  $x_{u{\rm eff}}$, the contribution  
to Eq.~(5.18a) will be small by virtue of the assumed support properties.  
A more detailed discussion of this assumption 
and its implications will be given in the next section.  (This assumption 
could perhaps be weakened to allow for the possibility that the ``fast'' 
components of $x_{u{\rm eff}}$ have regions of common support with 
$\tilde C_{\rm eff}$, but that rapid relative phase oscillations make the 
contribution of Eq.~(5.18a) to the Ward identities very small.  However, 
we shall not pursue this possibility further in what follows.)  

\item{(2)}  We assume that the ``chemical potential'' $\eta$ is very 
small, so that the term
$$i\eta \xi_ux_{u {\rm eff}} {\rm Tr} \tilde C i_{\rm eff} W_{\rm eff}~~~
\eqno(5.18b)$$
can be neglected.  Since this term is identically zero for $u$ bosonic (cf. 
the line following Eq.~(5.13b)), this assumption is only operative in 
the fermionic sector of the theory.

\item{(3)}  We assume that in the final term of Eq.~(5.15a), we can replace 
the conserved quantity $\tilde C_{\rm eff}$ by its 
zero source ensemble average 
$\langle \tilde C_{\rm eff} \rangle_{\hat{\rm AV}}=
\langle \tilde C_{\rm eff} \rangle_{\rm AV}=i_{\rm eff}\hbar$. 
In other words, 
we assume that in this term, fluctuations in $\tilde C_{\rm eff}$ 
over the canonical 
ensemble are unimportant.  (As discussed in the next section, the 
corresponding replacement cannot be made in the $\tau$ term in Eq.~(5.15a), 
since it would change the support properties in a significant way.)  
With this 
assumption, the final term of Eq.~(5.15a) becomes 
$$-\hbar 
\sum_{s\ell} \omega_{us} \epsilon_{\ell} (W_s^{R\ell} 
W_s^{L\ell})_{\rm eff} 
=-\hbar \sum_s \omega_{us} \left( 
{\delta {\bf W} \over \delta x_s} \right)_{\rm eff}~~~,\eqno(5.18c)$$
where in simplifying we have made use of Eq.~(5.10b).

With these assumptions, Eq.~(5.15a) simplifies dramatically to take the 
form 
$${\cal D} x_{u {\rm eff}} 
= i_{\rm eff} [W_{\rm eff},  x_{u{\rm eff}}]
-\hbar \sum_s \omega_{us} \left( 
{\delta {\bf W} \over \delta x_s} \right)_{\rm eff}~~~,\eqno(5.19a)$$
where we have used the fact that $i_{\rm eff}$ commutes with all effective 
quantities to pull it outside the commutator in the first term.  
We can now use this simplified form of ${\cal D}$ in the Ward 
identity with sources 
given in Eq.~(5.15b), and in the Ward identity obtained after 
source variations given in Eq.~(5.15c).  As discussed in [7], specialization 
of the operator polynomial $W$  gives a number of important results.  

\noindent{(A)}  First, let us take $W$  in Eq.~(5.19a) 
to be the operator Hamiltonian 
$H=\sum_r p_r \dot q_r -L$,  with ${\bf H} = {\rm Tr} H$  
the corresponding trace Hamiltonian, as in Eq.~(1.9b). Then by using    
the trace dynamics equation of motion of Eq.~(1.15b), Eq.~(5.19a) 
simplifies to 
$${\cal D} x_{u {\rm eff}} 
= i_{\rm eff} [H_{\rm eff},  x_{u{\rm eff}}]
-\hbar \dot x_{u {\rm eff}}~~~.\eqno(5.19b)$$
Using Eq.~(5.19b) in Eq.~(5.15c), we learn that $x_{u{\rm eff}}$ 
obeys an effective Heisenberg picture equation of motion, which holds when  
sandwiched between  polynomials $S_{L,R}$  that do not  
contain $x_u$ and averaged over the zero source canonical ensemble.  
If we now take  $S$ in Eq.~(5.15c) as the Hamiltonian 
$H_{\rm eff}$, and substitute  ${\cal D}$ from 
Eq.~(5.19b) and use the chain rule, we learn that 
$$\eqalign{ 
&\langle S_L(x_{t{\rm eff}}) \dot H_{\rm eff} S_R(x_{t{\rm eff}})  
\rangle_{\hat {\rm AV},0}\cr 
=&
\langle S_L(x_{t{\rm eff}}) i_{\rm eff} \hbar^{-1} 
[H_{\rm eff}, H_{\rm eff}] 
S_R(x_{t{\rm eff}})  \rangle_{\hat {\rm AV},0}=0~~~,\cr 
}\eqno(5.19c)$$ 
showing that within our approximations $H_{\rm eff}$ behaves as a constant  
of the motion, as required for consistency of the interpretation of 
Eqs.~(5.15c) and (5.19b) as an effective Heisenberg dynamics.  

Integrating now with respect to time, we learn from Eq.~(5.19b) that 
$$ x_{u {\rm eff}}(t) = \exp(i_{\rm eff} \hbar^{-1}  H_{\rm eff} t)  
x_{u {\rm eff}}(0) 
\exp(-i_{\rm eff} \hbar^{-1}  H_{\rm eff} t) ~~~,\eqno(5.19d)$$ 
when sandwiched between  polynomials $S_{L,R}$ that do not 
contain $x_u$, and averaged over the zero source ensemble.  If we 
consider the ensemble average of an arbitrary polynomial $S$ formed  
from dynamical variables, 
taken at different times, and advance them all in time by the same 
time increment $a^0$, we correspondingly get 
$$\langle S|_{\Delta t=a^0} \rangle_{\hat{\rm AV},0} = 
\langle  \exp(i_{\rm eff} \hbar^{-1}  H_{\rm eff} a^0)  S|_{\Delta t=0}
  \exp(-i_{\rm eff} \hbar^{-1} H_{\rm eff} a^0) \rangle_{\hat{\rm AV},0} ~~~.
\eqno(5.20a)$$
Since  $H_{\rm eff}$ is a constant of motion, we expect it to be 
approximately constant over the peak of the canonical ensemble.  Making  
the approximation of taking the exponentials in  $H_{\rm eff}$ 
outside the ensemble 
average,  Eq.~(5.20a) becomes 
$$\langle S|_{\Delta t=a^0} \rangle_{\hat{\rm AV},0} = 
\exp(i_{\rm eff} \hbar^{-1} H_{\rm eff} a^0) 
\langle  S|_{\Delta t=0} \rangle_{\hat{\rm AV},0}
\exp(-i_{\rm eff} \hbar^{-1} H_{\rm eff} a^0)  ~~~.\eqno(5.20b)$$ 
When we make the identification $P^0_{\rm eff}= H_{\rm eff}$,  
this equation agrees with the exact relations of Eqs.~(4.49a,b),  
which were obtained from general considerations of fixing a global unitary 
transformation $U_{\rm eff}$ in the integration defining the canonical  
ensemble.

\noindent{(B)}  Next, let us take $W$ in Eq.~(5.19a) 
to be $\sigma_t x_t$, with 
$\sigma_t$ an auxiliary $c$-number parameter which is 
a real or complex number 
for $t$ bosonic, and is a real or complex 
Grassmann number for $t$ fermionic, so that   
$W$ in both cases is bosonic.  We then have 
$\delta {\bf W} / \delta x_s =\sigma_t \delta_{st}$, and 
Eq.~(5.19a) becomes (after multiplication through by $i_{\rm eff}$)
$$i_{\rm eff}{\cal D} x_{u {\rm eff}} 
=  [x_{u{\rm eff}},    \sigma_t x_{t{\rm eff}}]
-i_{\rm eff}\hbar \omega_{ut}  \sigma_t
~~~.\eqno(5.21a)$$
Equation (5.15c) then tells us that this expression vanishes when 
sandwiched between general polynomials $S_{L,R}$ (which do not contain 
$x_u$) inside the zero source canonical ensemble.   Translating from 
the compact symplectic notation used in Eq.~(5.21a) by 
recalling the definition of $\omega_{ut}$ given in 
Eqs.~(1.16a,b), and factoring 
away the parameter $\sigma_t$ with attention to the fact that it 
is Grassmann for fermionic $t$, we learn from Eqs.~(5.15c) and 
(5.21a) that 
within ensemble averages, we have the effective canonical commutators 
$$ [q_{u{\rm eff}},q_{t{\rm eff}}]=[p_{u{\rm eff}},p_{t{\rm eff}}]=0~,~~~
[q_{u{\rm eff}},p_{t{\rm eff}}]=i_{\rm eff} \hbar \delta_{ut}
~~~\eqno(5.21b)$$    
for $u,t$ bosonic, and the effective canonical anticommutators 
$$\{q_{u{\rm eff}},q_{t{\rm eff}}\}=\{p_{u{\rm eff}},p_{t{\rm eff}}\}=0~,~~~ 
\{q_{u{\rm eff}},p_{t{\rm eff}}\}=i_{\rm eff} \hbar \delta_{ut}
~~~\eqno(5.21c)$$    
for $u,t$ fermionic, 
with the corresponding commutators of bosonic with fermionic quantities 
all vanishing.  (In other words, the entire canonical algebra for boson  
and fermion degrees of freedom is compactly encoded in the statement 
that the right hand side of Eq.~(5.21a) vanishes.)  
Thus, as suggested at the beginning of this section, the 
Ward identities do in fact lead to an equipartitioning of $\tilde C$, 
giving rise in an emergent fashion to the canonical commutator/anticommutator 
structure that is the basis for field quantization.  Note that the emergence 
of the canonical algebra implicitly requires a limiting process.  
As is well known, in a complex Hilbert space the canonical 
algebra $[q,p]=i,~ [q,i]=[p,i]=0$ cannot have finite dimensional (or 
more generally, trace class) 
representations, since this algebra implies, for example,  the relation
$q^2p^2+p^2q^2-2qp^2q=-2$, which in a finite dimensional Hilbert space  
would have a left hand side with trace zero and a right hand side with 
trace nonzero.  However, it is consistent for the canonical algebra 
to emerge as the limit $N \to \infty$ of a 
matrix algebra in an $N$ dimensional  
Hilbert space, or as an idealized approximation to a matrix  
algebra in a Hilbert space with $N$ large but finite (or as an approximation  
to a more general trace class algebra).  

\noindent{(C)}  Finally, let us take $W$ in Eq.~(5.19a) to be a general 
self-adjoint 
polynomial $G$, so that ${\bf G}$ is the generator of general 
canonical transformations 
of the trace dynamics as discussed in Eqs.~(2.13a,b).  Then combining 
Eq.~(5.19a) with Eq.~(2.13b), and dividing by $\hbar$, we get 
$$\hbar^{-1} {\cal D} x_{u {\rm eff}} 
= i_{\rm eff}\hbar^{-1}  [G_{\rm eff},  x_{u{\rm eff}}]
- \delta x_{u{\rm eff}}~~~,\eqno(5.22a)$$ 
which tells us that sandwiched between polynomials $S_{L,R}$ in the 
zero source canonical ensemble, general infinitesimal 
canonical transformations are 
effectively generated by unitary transformations of the form  
$$ U_{{\rm can~eff}} = \exp(  i_{\rm eff}\hbar^{-1}  G_{\rm eff})~~~. 
\eqno(5.22b)$$
In trace dynamics models defined as continuum theories, such as the 
examples of Sec. 3, this result states that the canonical 
trace Poincar\'e generators are represented at the effective level 
by operator generators.  In particular, in Poincar\'e invariant 
theories the trace three-momentum will 
have a corresponding operator three momentum $\vec P_{\rm eff}$ that is 
time independent (by similar arguments to those used to show that 
$H_{\rm eff}$ is time independent), and that 
serves to generate translations through a unitary transformation 
$$ U_{{\rm trans~eff}}(\vec a) = \exp(  i_{\rm eff}\hbar^{-1}\vec a \cdot   
\vec P_{\rm eff}  )~~~. 
\eqno(5.22c)$$

To sum up, we have shown that the basic structure of quantum field theory -- 
the canonical commutator/anticommutator structure, 
time evolution in the Heisenberg picture, 
and the unitary generation of canonical transformations -- emerges 
from the statistical thermodynamics of matrix models with a global unitary 
invariance.  The results that we have obtained suggest a direct 
correspondence between averages of operator polynomials in the zero source 
canonical ensemble and the corresponding operator polynomials in  complex 
quantum field theory, 
$$ \langle S(\{x_{r{\rm eff}}\}) \rangle_{\hat{\rm AV}} \Leftrightarrow 
S(\{X_{r {\rm eff}}\})~~~,\eqno(5.23a)$$
with the $X_{r{\rm eff}}$ on the right quantized operators 
in a quantum theory 
with the role of $i$ played by the matrix $i_{\rm eff}$, which commutes  
with all of the $X_{r{\rm eff}}$.  Since $i_{\rm eff}$ diagonalizes into 
two $K\times K$ blocks as indicated in Eq.~(4.40), the correspondence 
of Eq.~(5.23a) actually gives two uncoupled copies of a complex quantum 
field dynamics, on one of which $i_{\rm eff}$ acts as $i$ and on the 
other of which $i_{\rm eff}$ acts as the complex conjugate $-i$.  Let 
us now focus on the copy on which $i_{\rm eff}$ acts as $i$; similar   
considerations, apart from sign changes in the fermion sector that will 
be discussed below, apply to the copy on which  $i_{\rm eff}$ acts as $-i$. 

We now make two further structural assumptions:  

\item{(4)}  In the continuum limit, with the indices $r$ labeling 
infinitesimal spatial boxes as well as internal symmetry structure, the  
underlying trace Lagrangian ${\bf L}$ is Poincar\'e invariant.   

\item{(5)}  The underlying dynamics leads to an 
$H_{\rm eff}$ that is  
bounded from below by the magnitude of the corresponding 
effective three-momentum operator 
$\vec P_{\rm eff}$, with $P^{\mu}_{\rm eff}$ the generator   
defined in Eq.~(4.49b),    
and there is a unique eigenvector  
$\psi_0$ with lowest eigenvalue of $H_{\rm eff}$.

With these two assumptions we can make contact with quantum field theory.  
According to assumption (5), the eigenvector $\psi_0$ acts as the 
conventional vacuum state, and sandwiching Eq.~(5.23a) between 
$\psi_0^{\dagger}$ and $\psi_0$, we get the correspondence between  
trace dynamics 
canonical ensemble averages and Wightman functions in the emergent  
quantum field theory,
$$\psi_0^{\dagger} \langle S(\{x_{r{\rm eff}}\}) 
\rangle_{\hat{\rm AV}} \psi_0
\Leftrightarrow 
\langle {\rm vac}|S(\{X_{r {\rm eff}}\})|{\rm vac} \rangle~~~.\eqno(5.23b)$$
(Equation (5.23b) applies in the sector where $i_{\rm eff}$ acts as $i$; 
in the sector where $i_{\rm eff}$ acts as $-i$, it is necessary to 
identify the fermionic $\psi^{\dagger}$ on the left of Eq.~(5.23b) with 
the field operator $-\Psi^{\dagger}$ on the right, in order to obtain 
consistency with the positivity properties of the fermionic field 
anticommutator.  This will be discussed in further detail in Sec. 5E).  
When the underlying trace dynamics is Poincar\'e invariant, the fact that 
Poincar\'e transformations are 
canonical transformations with global unitary invariant generators implies,  
by the discussion of Sec. 2, that $\tilde C$ is a Poincar\'e invariant.  
Hence the term ${\rm Tr} \tilde \lambda \tilde C$ 
in the canonical ensemble is a Poincar\'e invariant, and in particular is   
a Lorentz scalar.  Although 
the term $\tau {\bf H}$ in the canonical ensemble is 
{\it not} a Lorentz scalar, 
in the low energy regime where approximation (1) 
is valid, so that variations in the $\tau$ term can be neglected in the 
Ward identities, the Lorentz non-invariance associated with this term 
decouples, giving a Lorentz invariant effective field theory.  Hence 
with the assumptions and approximations that we have made, the 
correspondence of Eq.~(5.23b) defines Wightman functions of a Lorentz 
invariant effective field theory.  Although our Ward identity derivations  
required the exclusion of certain index values (the $x_s$ that is 
varied cannot  
be the $x_R$ for which the phase space integral is restricted, and the 
polynomial $S$ in Eq.~(5.15c) cannot contain variables $x_t$ that appear 
in $S_{L,R}$), in the continuum limit with $s$ labeling infinitesimal 
spatial boxes, these exclusions amount to a set of measure zero and have 
negligible effect.  

Let us briefly enumerate the basic properties of 
Wightman functions [24] (see Appendix F for further details)  and 
indicate why, with 
the correspondence of Eq.~(5.23b), it is plausible that they are obeyed.  
(i)  By constructing the Wightman functions from thermodynamic averages, 
they are expected to have the requisite smoothness properties, i.e., they   
should be tempered distributions.  
(ii)  Lorentz covariance of the Wightman functions follows from Lorentz 
invariance of $\tilde C$ and our assumption that the noninvariant 
$\tau$ terms 
effectively decouple.  (iii)  The spectral condition for the Wightman 
functions follows from the formula of Eqs.~(4.49a,b) for the effect of 
a uniform space-time translation, together with the positivity and 
boundedness assumption (5).  (iv)  Locality of the Wightman 
functions follows from  the canonical 
commutation/anticommutation relations derived, in the 
approximation of neglecting 
the $\tau$ terms, in Eqs.~(5.21a--c).  
(v)  The hermiticity and positivity properties of 
the Wightman functions follow   
directly from the correspondence with thermodynamics averages of Eq.~(5.23b).
(vi) The cluster property of the Wightman functions corresponds to the  
assumptions about ${\bf H}$ made in the 
clustering argument used in the discussion of the microcanonical ensemble 
in Eqs.~(4.19a--c).  

To conclude this section, we note that there is a natural hierarchy of matrix 
structures leading from the underlying trace dynamics, to the emergent 
effective complex quantum field theory, to the classical limit.  In the 
underlying theory, the matrices $x_r$ are of completely general structure.  
No commutation properties of the $x_r$ are assumed at the trace dynamics 
level, and since all degrees of freedom communicate with one another, the 
dynamics is completely nonlocal.  (As a consequence of this nonlocality, 
the Bell theorem arguments [25] against local hidden variables do not 
apply to trace dynamics.)   In the effective quantum theory, the $x_r$ are  
still matrices, but with a restricted structure that obeys the canonical 
commutator/anticommutator algebra.  Thus, locality is an emergent property 
of the effective theory, even though it is not a property of the underlying   
trace dynamics.  Finally, in the limit in which the matrices $x_r$ are 
dominated by their $c$-number or classical parts defined in Eqs.~(2.16a,b),   
the effective quantum field dynamics becomes an effective classical 
dynamics.  Thus, both classical mechanics and quantum mechanics are 
subsumed in the more general trace dynamics, as reflected in the following 
hierarchy of matrix structures, corresponding to increasing specialization:  
general $\to$ canonical quantum $\to$ $c$-number, classical.  The 
indeterminacy characteristic of quantum mechanics appears only at the 
middle level of this hierarchy of matrix structures.  At the bottom or 
trace dynamics level, there is no indeterminacy (at least in principle, 
given the initial conditions) because no dynamical information has been 
discarded, while at the top or classical level, quantum indeterminacies 
are masked by large system size.  
\vfill\eject

\centerline{5D.~~~Restrictions on the Underlying Theory Implied}
\centerline{by Further Ward Identities}
\bigskip
In this section we shall take a critical look at the assumptions and 
approximations made in the previous section in the course of our argument  
for the emergence of quantum behavior.   We shall do this by deriving 
further Ward identities, and showing that their consistency with the 
Ward identity derived in Sec. 5A, and with the approximations 
made in Sec. 5C, 
places nontrivial constraints on the structure of the underlying trace 
dynamics.  

The first additional Ward identity is obtained by proceeding as we did 
starting from 
Eq.~(5.5a), but with $\{ \tilde C, i_{\rm eff} \}
=2i_{\rm eff} \tilde C_{\rm eff}$ in the factor 
${\rm Tr} \{ \tilde C, i_{\rm eff} \}W$ 
replaced at the outset by its ensemble average $-2\hbar$.  
Thus, factoring away the constant $-2\hbar$,  
we start now from 
$$ Z \langle {\rm Tr}  W \rangle_{\hat{\rm AV}} 
= \int d\hat  \mu 
\exp(-{\rm Tr} \tilde \lambda \tilde C- \tau {\bf  H}-\eta 
{\bf N}  - \sum_r {\rm Tr} j_r x_r  ) {\rm Tr}  W 
~~~,\eqno(5.24a)$$
and proceed as we did in Sec. 5A.  
The derivation completely parallels 
that leading to Eqs.~(5.14a,b), except that the term 
corresponding to Eq.~(5.9b), 
which comes from the variation of the factor $\tilde C$ in 
${\rm Tr} \{\tilde C,i_{\rm eff} \}W$, is absent.  Dropping this term 
from Eq.~(5.14b), replacing $\{ \tilde C, i_{\rm eff} \}$ on 
the right hand side by 
$-2 \hbar$ and factoring away $-\hbar$, and 
using Eq.~(5.10b) to simplify the final term on  
the right, we get 
$$\langle \Lambda_{u{\rm eff}}^{\prime} \rangle_{\hat{\rm AV},j}=0~~~,
\eqno(5.24b)$$ 
with $ \Lambda_{u{\rm eff}}^{\prime}$ given by 
$$\eqalign{
\Lambda_{u{\rm eff}}^{\prime}
=&(-\tau \dot x_{u{\rm eff}} +i\eta \xi_u x_{u{\rm eff}}
-\sum_s \omega_{us} j_{s {\rm eff}}) 
 {\bf W}  \cr  
+& \sum_s \omega_{us} 
\left({\delta {\bf W} \over \delta x_s}\right)_{\rm eff}
~~~.\cr
}\eqno(5.24c)$$
Correspondingly, after variation of source terms, instead of Eq.~(5.16b) 
we now get 
$$\langle   S_L(x_{t{\rm eff}}) 
\sum_u S(x_{r{\rm eff}}, \, r\neq u; {\cal D}^{\prime} x_{u{\rm eff}} ) 
 S_R(x_{t{\rm eff}})       \rangle_{\hat {\rm AV},0} =0~~~,
\eqno(5.25a)$$ 
with ${\cal D}^{\prime} x_{u {\rm eff}}$ given by 
$$\eqalign{
{\cal D}^{\prime} x_{u {\rm eff}} 
=&(-\tau \dot x_{u{\rm eff}} +i\eta \xi_u x_{u{\rm eff}})
{\bf W}  \cr  
+& \sum_{s} \omega_{us} 
\left({\delta {\bf W} \over \delta x_s}\right)_{\rm eff} ~~~.\cr
}\eqno(5.25b)$$

It is apparent from Eq.~(5.25b) that we {\it cannot} make an 
approximation of neglecting both the $\eta$ term and the 
$\tau$ term on the right hand 
side, since this would lead to the incorrect conclusion, for example, that 
$$\langle \left({\delta {\bf W} \over \delta x_s}\right)_{\rm eff}
 \rangle_{\hat {\rm AV},0} =0~~~\eqno(5.26a)$$
for general $W$.  
In other words,  assumption (1) of Sec. 5C, which we recall states that 
$$ \tau \dot x_{u{\rm eff}} {\rm Tr} \tilde C_{\rm eff} i_{\rm eff} 
W_{\rm eff}= 
\tau \dot x_{u{\rm eff}} {\rm Tr} \tilde C_{\rm eff} i_{\rm eff} W 
 ~~~\eqno(5.26b)$$
can be neglected, cannot be extended to the 
assumption that the corresponding expression obtained by 
replacing $ \tilde C_{\rm eff}$ 
by its ensemble average $i_{\rm eff}\hbar$ can also be neglected. This 
does not invalidate the reasoning of Sec. 5C, but does place constraints 
on the support structure of the underlying trace dynamics.  To give 
a simple illustration, if $f$ and $g$ are non-negative functions, the 
vanishing of the the average $(fg)_{\rm AV}$ over a domain including 
the supports of $f$ and $g$ does not 
contradict the fact that 
$((f)_{\rm AV} g)_{\rm AV} = (f)_{\rm AV}(g)_{\rm AV}$ is nonzero and 
positive; what is necessary and sufficient 
to achieve the vanishing of the former 
is for $f$ and $g$ to have nonintersecting domains of support, so that   
$f=0$ where $g>0$ and vice versa.  We are faced with a similar situation 
with respect to $\tilde C_{\rm eff}$, which 
although not of definite sign has, by 
assumption, a nonvanishing canonical ensemble average $ \hbar$. 
A sufficient condition for us to be able to neglect Eq.~(5.26b), without 
contradicting Eq.~(5.25b), is for $\dot x_{u{\rm eff}}$  
and $\tilde C_{\rm eff}$ 
to have disjoint domains of support, which is why we have phrased  
assumption (1) in terms of support properties.   

At the same time, we 
cannot impose this property by requiring the stronger condition that 
$x_{u{\rm eff}}$ and $\tilde C_{\rm eff}$ have disjoint support, 
since this would 
contradict assumption (3), which states that $\{ \tilde C, i_{\rm eff} \}$ 
can be replaced 
by its ensemble average in the final term in Eq.~(5.15a).   
Thus, what is required is that the ``slow'' components 
of  $x_{u{\rm eff}}$, for which $\tau \dot x_{u{\rm eff}}$ is effectively 
zero, share a common domain of support with $\tilde C_{\rm eff}$, 
and be slowly 
varying with respect to the scale of variations of $\tilde C_{\rm eff}$, 
so that 
in ``slow'' terms in the Ward identity $\tilde C_{\rm eff}$ is 
effectively equal to 
its ensemble average.  At the same time, we must 
require that the ``fast'' components 
of  $x_{u{\rm eff}}$, for which $\dot x_{u{\rm eff}}$ is significant, 
have disjoint support from $\tilde C_{\rm eff}$, so that 
the ensemble average of 
their product is effectively zero.  To see that these requirements 
are compatible, and suffice to do what is needed, let us write 
$$x_{r{\rm eff}}=  x_{r{\rm eff}}^{\rm slow}    
+     x_{r{\rm eff}}^{\rm fast}~~~,\eqno(5.27a)$$
and postulate that $\tilde C_{\rm eff}$ has disjoint support from 
$ x_{r{\rm eff}}^{\rm fast}$.  Then in the final term of Eq.~(5.15a) we have 
$$
\left(W_s^{R\ell} 
{1\over 2}  \{\tilde C,i_{\rm eff} \} 
W_s^{L\ell}\right)_{\rm eff}=
\left(W_s^{R\ell \,{\rm slow}} 
{1\over 2}  \{\tilde C,i_{\rm eff} \} 
W_s^{L\ell \, {\rm slow}}   \right)_{\rm eff}~~~,\eqno(5.27b)$$
so that the ``fast'' terms do not appear in this expression, and we can 
then apply assumption (3) to replace $\tilde C_{\rm eff}$ 
by $i_{\rm eff} \hbar$ in 
this term.  Similarly, by the assumed support properties,
$$\tau \dot x_{r {\rm eff}} 
{\rm Tr} i_{\rm eff} \tilde C_{\rm eff} W_{\rm eff}  
= \tau \dot x_{r{\rm eff}}^{\rm slow}
{\rm Tr} i_{\rm eff} \tilde C_{\rm eff} W_{\rm eff}^{\rm slow}  
\simeq 0~~~,\eqno(5.27c)$$
but on the other hand 
$$\eqalign{
\tau \dot x_{r {\rm eff}} {\rm Tr} W  
=\tau \dot x_{r {\rm eff}}  {\rm Tr}  W_{\rm eff}  
=& \tau  (\dot x_{r{\rm eff}}^{\rm slow} + \dot x_{r{\rm eff}}^{\rm fast})
{\rm Tr} (W_{\rm eff}^{\rm slow} +  W_{\rm eff}^{\rm fast})  \cr
\simeq & \tau   \dot x_{r{\rm eff}}^{\rm fast}
{\rm Tr} (W_{\rm eff}^{\rm slow} +  W_{\rm eff}^{\rm fast}) 
\neq 0~~~,\cr
}\eqno(5.27d)$$
so the additional Ward identity derived in Eq.~(5.25b) can be satisfied.    

More specific statements of our assumptions (1) and (3) can be given if  
we specialize the underlying trace dynamics (i) to the supersymmetric 
case, where the numbers $n_B$ and $n_F$ of bosonic and fermionic degrees 
of freedom are equal, and (ii) to the specific case of supersymmetric 
Yang-Mills theory, where we have seen in Sec. 3B  that 
$\tilde C$ reduces to a surface integral at spatial infinity.  Following 
the notation of Eq.~(4.41b), let us write $x_r=x_{r {\rm eff}}+x_{r\,12}$, 
which when substituted into Eq.~(2.6) for $\tilde C$ gives 
$$\tilde C
=\sum_{r,s}\omega_{rs}( x_{r{\rm eff}} x_{s{\rm eff}} + 
 x_{r{\rm eff}} x_{s\,12} +   x_{r\,12} x_{s{\rm eff}} + 
 x_{r\,12} x_{s\,12} )~~~.\eqno(5.28a)$$ 
Taking the effective projection, and using Eq.~(4.41c), which implies that 
$(x_{s\,12})_{\rm eff}= (x_{r\,12})_{\rm eff}=0$, 
we get 
$$\tilde C_{\rm eff}
=\sum_{r,s}\omega_{rs}(x_{r{\rm eff}} x_{s{\rm eff}} + 
(x_{r\,12} x_{s\,12})_{\rm eff} )~~~.\eqno(5.28b)$$ 
The first term on the right hand side in Eq.~(5.28b) can be rewritten as 
$$\sum_{r,s}\omega_{rs}x_{r{\rm eff}} x_{s{\rm eff}}
=\sum_{r,B} [q_{r {\rm eff}},p_{r {\rm eff}}] 
-\sum_{r,F} \{ q_{r {\rm eff}},p_{r {\rm eff}}\}~~~.
\eqno(5.28c)$$
Within ensemble averages, we have seen that the commutators and 
anticommutators  
in Eq.~(5.28c) have the effective canonical values given in Eqs.~(5.21b,c), 
and so within averages Eq.~(5.28c) is effectively $(n_B-n_F) i_{\rm eff} 
\hbar$, and we have   
$$\tilde C_{\rm eff}\simeq
(n_B-n_F) i_{\rm eff}
+\sum_{r,s}\omega_{rs}(x_{r\,12} x_{s\,12})_{\rm eff} ~~~,\eqno(5.28d)$$ 
with $n_B$ and $n_F$ respectively the number of bosonic and fermionic 
matrix degrees of freedom.  

We can now apply this equation in two ways.  First of all, taking its 
ensemble average we get 
$$i_{\rm eff} \hbar \simeq 
(n_B-n_F) i_{\rm eff}
+\langle \sum_{r,s}\omega_{rs}(x_{r\,12} x_{s\,12})_{\rm eff} 
\rangle_{\hat {\rm AV},0} ~~~.\eqno(5.29a)$$ 
This shows that if the difference $n_B-n_F$ becomes infinite, then there 
is an inconsistency unless the second term in Eq.~(5.29a)  becomes large 
in such a way as to cancel the infinite part of the first term.  This 
is implausible, and so we conclude that consistency of our approximations 
requires that $n_B-n_F$ should be finite, even when $n_B$ becomes large. 
(A more precise version of this argument for boson--fermion balance will 
be given shortly.)  Specializing now to the case of supersymmetric theories,
for which $n_B$ is exactly equal to $n_F$, the first term on the right 
in Eq.~(5.28d) vanishes, and we get 
$$\tilde C_{\rm eff}\simeq
\sum_{r,s}\omega_{rs}(x_{r\,12} x_{s\,12})_{\rm eff} ~~~.\eqno(5.29b)$$ 
{}From this equation, we see that a sufficient requirement for the assumed 
support properties is that $ x_{u{\rm eff}}^{\rm fast} $ should have 
disjoint support from $x_{s\,12}$  for all $u,s$, while 
$ x_{u{\rm eff}}^{\rm slow} $ should have a common support with $x_{s\,12}$, 
in such a way that $\tilde C_{\rm eff}$ can be replaced by its ensemble 
average in expressions involving ``slow'' quantities.  From Eq.~(5.29a) 
with $n_B-n_F=0$, we see that 
$$i_{\rm eff} \hbar \simeq 
\langle \sum_{r,s}\omega_{rs}(x_{r\,12} x_{s\,12})_{\rm eff} 
\rangle_{\hat {\rm AV},0} ~~~.\eqno(5.29c)$$ 
Since $i_{\rm eff}$ anticommutes with $x_{r\,12}$ and with $x_{s\,12}$, 
it commutes with the right hand side of Eq.~(5.29c), as required by the 
structure of the left hand side.

We have learned from this that the quantities for which disjoint support 
is required are distinct components under the separation into ``eff'' 
and ``12'' components of Eq.~(4.41c).  A further distinction is present 
in supersymmetric Yang-Mills theories, in which we have seen that 
$\tilde C$ reduces to a surface term at spatial infinity, giving the 
emergence of quantum mechanics in these theories a ``holographic'' flavor. 
(For a review of recent ideas on a possible holographic structure of 
physical theories, see [26].)  The  degrees of freedom $x_r$ which  
describe observable physics are then degrees of freedom 
residing in the interior volume, whereas the 
components $x_{r\,12}$ entering into Eq.~(5.29b) reside on the surface 
at infinity.  For such theories, the required support properties state  
that for volume degrees of freedom, $ x_{r{\rm eff}}^{\rm fast} $  should 
have disjoint phase space support from degrees of freedom $x_{r\,12}$
residing on the surface at spatial infinity, while the volume degrees 
of freedom $ x_{r{\rm eff}}^{\rm slow} $  should have a common support 
with the degrees $x_{r\,12}$ residing on the surface at infinity and should 
be slowly varying relative to $\tilde C_{\rm eff}$.  (Recall that the 
distinction between volume and surface here refers to operator labels $r$, 
and does not directly translate into support properties in the operator phase 
space.)   These statements are 
as far as we have been able to carry a general  
analysis of the needed support properties.   A further understanding of  
whether they can be realized will require a study of specific models for 
the underlying trace dynamics; constructing realistic candidates 
requires a solution to the hierarchy problem, and so will not be    
a simple matter.  At a minimum, what our analysis has 
accomplished is to show that the support properties needed for 
assumptions (1) and (3) are not contradictory, and so cannot be used 
in any obvious way to 
construct an argument falsifying our program.  

The support properties required for the emergence of quantum behavior 
can also be characterized in physical terms as the requirement that the 
canonical ensemble should possess a certain ``rigidity'', in the sense 
that the contribution of the ensemble variation 
$(\delta \rho/\delta x_s)_{\rm eff}$ can be neglected in deriving the 
Ward identities.  The 
need for a rigid statistical ensemble in our context suggests a possible 
analogy with the concept of London rigidity in the theory of 
superconductivity.  In the presence of an applied vector potential 
$\vec A$, the 
induced current density $\vec j$ in a metal is given by 
$$\langle \vec j ~\rangle = -{n e \over m} \langle \vec p + e 
\vec A
~\rangle ~~~,\eqno(5.29d))$$
with $n,m, e,\vec p$ respectively the electron 
density, mass, charge, and three momentum operator.  
In a normal metal the two terms on the right hand side of Eq.~(5.29d) nearly 
cancel, leaving a small residual diamagnetism.  However, in a superconductor 
the rigidity of the wave 
function leads to the vanishing of $\langle \vec p~ \rangle$, giving perfect 
diamagnetism and the Meissner effect.   An analogy with the  
analysis of this section would equate normal metal behavior with the case in 
which $\tilde C$ can be replaced by its ensemble average in all 
terms, including the $\tau$ term, 
in the Ward identities; in this case $-\hbar$ times 
the right hand side of Eq.~(5.24c) is  equal to the 
non-commutator part of the 
right hand side of Eq.~(5.15a) when the sources $j_s$ are 
set equal to zero, leading to vanishing of the emergent 
canonical commutator/anticommutator and to an effective 
classical dynamics.   Similarly,  
the analogy would equate superconducting behavior 
with the case in which the $\tau$ term containing $\tilde C$ can be dropped 
because of ``rigidity'' of $(\delta \rho/\delta x_s)_{\rm eff}$, leading as 
seen in Sec.~5C to an emergent canonical 
commutator/anticommutator as an analog of the 
superconductive Meissner effect.  
In this analogy, the Planck energy and the associated energy scale 
hierarchy would play the role of the superconductive energy gap.  We 
suggest that this analogy may be useful in identifying the particular 
underlying trace dynamics for which the assumptions needed for emergent 
quantum mechanics are realized.  

We turn now to deriving a more precise statement [8] of the requirement 
of boson -- fermion balance, which we achieve by deriving yet another   
Ward identity.  This is obtained by proceeding as we did 
starting from 
Eq.~(5.5a), but with $\{ \tilde C, i_{\rm eff} \}$ 
replaced by $\tilde C$ in the trace factor 
${\rm Tr} \{ \tilde C, i_{\rm eff} \}W$, so that this factor is taken now as  
${\rm Tr} \tilde C W$.  
The derivation  parallels 
that leading to Eqs.~(5.11b,c), except that the term 
corresponding to Eq.~(5.9b), 
which comes from the variation of the factor $\tilde C$ in 
${\rm Tr} \{ \tilde C, i_{\rm eff} \} W$, becomes  now
${\rm Tr}  [W, \sum_r \omega_{rs} x_r] \delta x_s$.  
Making the corresponding replacements in Eq.~(5.11b), we get the Ward 
identity 
$$\langle \Sigma_s^{\prime\prime} \rangle_{\hat{\rm AV},j}=0
~~~,\eqno(5.30a)$$ 
with $\Sigma_s^{\prime\prime}$  given by 
$$\eqalign{
\Sigma_s^{\prime \prime}=
&(-[\tilde \lambda,\sum_r \omega_{rs} x_r] -\tau \sum_r\omega_{rs} 
\dot x_r -i\eta \sum_r \tilde \omega_{rs} x_r -j_s) {\rm Tr} \tilde C W\cr  
+& [W, \sum_r \omega_{rs}  x_r]
+\sum_{\ell} \epsilon_{\ell} W_s^{R\ell} \tilde C W_s^{L\ell}
~~~.\cr
}\eqno(5.30b)$$
We now follow a different procedure from that used in Sec. 5A, by      
immediately taking the sources $j$ to vanish, and by not taking an overall  
effective projection.  The contribution to the Ward identity of the first 
term on the right hand side of Eq.~(5.30b) can be rewritten as  
$$-[\tilde \lambda, \langle \sum_r \omega_{rs} x_r {\rm Tr} \tilde C W   
 \rangle_{\hat{\rm AV},0}]=0~~~,\eqno(5.30c)$$
since the zero source expectation inside the commutator in Eq.~(5.30c) 
is a function of the operator $\tilde \lambda$ and no 
other operators, and so its commutator with $\tilde \lambda$ vanishes.  
Multiplying by $\omega_{us}$ and summing over $s$, and evaluating 
the sums using Eqs.~(5.13a,b), in place of Eqs.~(5.14a,b) we now get the 
Ward identity
$$\langle \Lambda_{u{\rm eff}}^{\prime \prime} 
\rangle_{\hat{\rm AV},0}=0~~~,\eqno(5.31a)$$ 
with $ \Lambda_{u}^{\prime\prime} $ given by 
$$\eqalign{
\Lambda_{u}^{\prime\prime}
=&(-\tau \dot x_{u} +i\eta \xi_u x_{u})
{\rm Tr} \tilde C  W \cr  
+& [ W,  x_{u}]
+\sum_{s,\ell} \omega_{us} \epsilon_{\ell} W_s^{R\ell} 
  \tilde C W_s^{L\ell}
~~~.\cr
}\eqno(5.31b)$$

Let us now apply this to the particular choice $W=\sigma_t x_t$, with 
$\sigma_t$ again an auxiliary $c$-number parameter which is a complex 
number for $t$ bosonic, and a complex Grassmann number for $t$ fermionic.  
Then (cf. Eqs.~(5.10b,c)) we have only one term in the sum over $\ell$, with 
$\epsilon_{\ell=1}=1$,  $W_s^{L\ell=1} =\sigma_t \delta_{st}$, 
and $W_s^{R\ell=1}=\delta_{st}$, 
so that the final term in Eq.~(5.31b) reduces to 
$$ \omega_{ut} \tilde C \sigma_t~~~,\eqno(5.31c)$$ 
and has the ensemble average 
$$ \omega_{ut} i_{\rm eff} \hbar \sigma_t~~~.\eqno(5.31d)$$
Hence in this special case the Ward identity  of Eqs.~(5.31a,b) reduces to 
$$0=\langle (-\tau \dot x_u +i\eta \xi_u x_u) 
{\rm Tr} \tilde C  \sigma_t x_t \rangle_{\hat{\rm AV},0}  
+ \langle [ \sigma_t x_t,  x_u] \rangle_{\hat{\rm AV},0}
+ \omega_{ut} i_{\rm eff}\hbar \sigma_t~~~.
\eqno(5.32a)$$
Since the final two terms in Eq~(5.32a) are manifestly traceless, and 
since $\tilde C$ is traceless, 
projecting out the traceless part of the first term using the notation 
of Eq.~(2.16a), and rearranging terms, we arrive at [8]
$$\langle [x_u, \sigma_t x_t] \rangle_{\hat{\rm AV},0} =
 i_{\rm eff}\hbar \omega_{ut} \sigma_t
+\langle (-\tau \dot x_u^{\prime} +i\eta \xi_u x_u^{\prime}) 
{\rm Tr} \tilde C  \sigma_t x_t^{\prime}  \rangle_{\hat{\rm AV},0}~~~.  
\eqno(5.32b)$$

Letting the indices $t$ and $u$ in Eq.~(5.32b) be either both 
bosonic or both fermionic, and in the fermionic case, for simplicity,  
setting the ``chemical potential'' $\eta$  
equal to zero (it is easy to extend the analysis to $\eta \not= 0$),  
we get the respective relations 
$$\eqalign{
\langle [q_r,p_r]\rangle_{\hat{\rm AV},0}=&i_{\rm eff}\hbar 
-\tau\langle{\dot q}_r^{\prime}
{\rm Tr}\tilde C p_r^{\prime}\rangle_{\hat{\rm AV},0}~~~~~r~{\rm bosonic}\cr
\langle\{q_r,p_r\}\rangle_{\hat{\rm AV},0}=&i_{\rm eff}\hbar
-\tau\langle{\dot q}_r^{\prime}
{\rm Tr}\tilde C p_r^{\prime}\rangle_{\hat{\rm AV},0}
~~~~~r~{\rm fermonic}~~~.\cr
}\eqno(5.33a)$$
Substituting this into Eq.~(2.6) for $\tilde C$, taking the ensemble average, 
and using Eq.~(4.11b), we get 
$$\eqalign{                                        
i_{\rm eff}\hbar =& \langle \tilde C \rangle_{\hat{\rm AV},0}
=\langle  \sum_{r,B}[q_r,p_r]-\sum_{r,F}\{q_r, p_r\} 
\rangle_{\hat{\rm AV},0}\cr
=&(\sum_{r,B}-\sum_{r,F} ) 
i_{\rm eff}\hbar
-\tau (\sum_{r,B}-\sum_{r,F}) 
\langle{\dot q}_r^{\prime}{\rm Tr}\tilde 
C p_r^{\prime}\rangle_{\hat{\rm AV},0} ~~~.\cr
}\eqno(5.33b)$$
After division by $\hbar$, transposition of terms, and use of  
$\Sigma_{r,B} 1=n_B~,~~  \Sigma_{r,F} 1=n_F$, this gives 
$$ (\sum_{r,B}-\sum_{r,F}) 
\hbar^{-1} \tau\langle{\dot q}_r^{\prime}{\rm Tr}\tilde C 
p_r^{\prime}\rangle_{\hat{\rm AV},0} \break
=i_{\rm eff} (n_B-n_F-1 )~~~.
\eqno(5.33c)$$
When the condition of approximation (1) of Sec. 5C is 
satisfied, the left hand side of Eq.~(5.33c) is a sum of very small terms. 
Assuming that this sum yields at most a finite, bounded total, let us  
consider the case in which $r$ includes the spatial label 
of a translation invariant field theory.  Then 
the number of bosonic and fermionic modes per unit volume contributing on 
the right hand side of Eq.~(5.33c) must be equal, since if not, 
the right hand side of Eq.~(5.33c) 
would become infinite as the spatial volume grows to infinity, contradicting 
the boundedness of the left hand side.  Therefore, a trace dynamics 
that is a candidate 
pre-quantum mechanics must have equal numbers 
of bosonic and fermionic degrees of freedom (up to a finite residue). This  
is a much weaker requirement than supersymmetry, but of course is always 
satisfied by supersymmetric theories.  When the numbers of 
bosonic and fermionic modes are in balance, Eq.~(5.33c) simplifies to 
$$ (\sum_{r,B}-\sum_{r,F}) 
\hbar^{-1} \tau\langle{\dot q}_r^{\prime}{\rm Tr}\tilde C 
p_r^{\prime}\rangle_{\hat{\rm AV},0} \break
=-i_{\rm eff}~~~, \eqno(5.33d)$$
showing that the $\tau \dot x_r$ terms neglected in making approximation (1) 
sum in Eq.~(5.33c) to give a total of unit magnitude.  

The above analysis of boson--fermion balance has implications for the 
behavior of $\tilde C$ in the thermodynamic limit of large system size.  
Although 
the bosonic and fermionic contributions to $\tilde C$ each grow linearly 
with the size of the system, the near cancellation of their  
contributions to  $\tilde C$ suggests that the total  
rate of growth of $\tilde C$ could be much smaller than those of the   
bosonic or fermionic parts taken separately.  Hence, even 
though $\tilde C$ is formally an extensive thermodynamic quantity (it 
is additive for disjoint subsystems), it may remain bounded, or have 
much smaller than a linear rate of growth, as the system size gets large.  
In the first case $\hbar$ would remain bounded in the limit of large 
system size, while in the second case $\hbar$ could still be a weakly 
increasing function of system size; our general 
analysis does not determine the expected behavior.    
                
\bigskip
\centerline{5E.~~~Derivation of the Schr\"odinger Equation}
\bigskip
In Sec. 5C we have argued that the statistical thermodynamics of matrix 
models with a global unitary invariance leads to an emergent Heisenberg 
picture quantum mechanics.  In this section we shall make the transition 
to the corresponding Schr\"odinger picture formulation.  We take as our 
starting point the correspondence of canonical ensemble averages to 
operators in an effective quantum theory given in Eq.~(5.23a).  Transcribing 
the canonical commutators inside averages of Eqs.~(5.21a--c) into operator 
statements, we get 
$$[X_{u{\rm eff}},    \sigma_t X_{t{\rm eff}}]
=i_{\rm eff}\hbar \omega_{ut}  \sigma_t ~~~,\eqno(5.34a)$$
which encodes the canonical commutators 
$$ [Q_{u{\rm eff}},Q_{t{\rm eff}}]=[P_{u{\rm eff}},P_{t{\rm eff}}]=0~,~~~
[Q_{u{\rm eff}},P_{t{\rm eff}}]=i_{\rm eff} \hbar \delta_{ut}
~~~\eqno(5.34b)$$    
for $u,t$ bosonic, and the effective canonical anticommutators 
$$\{Q_{u{\rm eff}},Q_{t{\rm eff}}\}=\{P_{u{\rm eff}},P_{t{\rm eff}}\}=0~,~~~ 
\{Q_{u{\rm eff}},P_{t{\rm eff}}\}=i_{\rm eff} \hbar \delta_{ut}  
~~~\eqno(5.34c)$$    
for $u,t$ fermionic, with all boson-fermion commutators vanishing.  
Similarly, the operator transcription of Eqs.~(5.19b) and (5.15c) 
for the time evolution of $x_{u{\rm eff}}$ becomes 
$$\dot X_{u{\rm eff}}= i_{\rm eff} \hbar^{-1}[H_{\rm eff}, 
X_{u{\rm eff}}]~~~,
\eqno(5.34d)$$
which extends by the chain rule to 
$$\dot S_{\rm eff} 
=i_{\rm eff} \hbar^{-1}[H_{\rm eff}, S_{\rm eff}]~~~,\eqno(5.34e)$$
with $S_{\rm eff}$ any polynomial function of the 
operators $\{X_{r{\rm eff}}\}$.  

Before turning to the transition to the Schr\"odinger picture, we first 
discuss consistency issues raised by treating the fermionic anticommutators 
as operator equations.  If we assign fermionic adjoint properties as in 
Eq.~(2.4b), which has the operator transcription  
$$Q_{r{\rm eff}}=\Psi_{r{\rm eff}}~,
~~~P_{r{\rm eff}}=i\Psi_{r{\rm eff}}^{\dagger}~~~,\eqno(5.35a)$$
then the nonvanishing anticommutator in Eq.~(5.34c) takes the form 
$$\{ \Psi_{u{\rm eff}}, \Psi^{\dagger}_{t{\rm eff} }\}=
-i  i_{\rm eff} \hbar \delta_{ut} ~~~.\eqno(5.35b)$$ 
In the $K$ dimensional subspace of Hilbert space on which $i_{\rm eff}$ 
acts as $\pm i$,  Eq.~(5.35b) takes the form 
$$\{ \Psi_{u{\rm eff}}, \Psi^{\dagger}_{t{\rm eff} }\} =
\pm \hbar \delta_{ut} ~~~.\eqno(5.35c)$$ 
The $+$ sign case of Eq.~(5.35c) corresponds to the normal field theoretic 
fermionic anticommutator, but the $-$ sign case is inconsistent:  Setting  
$u=t$, the $-$ sign case gives 
$$\{ \Psi_{t{\rm eff}},\Psi^{\dagger}_{t{\rm eff}} \} = -\hbar~~~,
\eqno(5.35d)$$
which is not possible because the left hand side of this relation is the 
sum of two positive semidefinite operators.  Therefore the operator 
transcription of Eq.~(5.23b) must be modified to include an extra $-$ sign 
in $\Psi_{t{\rm eff}}^{\dagger}$ in the $i_{\rm eff}=-i$ sector; in other 
words, a consistent form of the operator transcription for fermions 
is given by 
$$q_{r{\rm eff}} \leftrightarrow Q_{r{\rm eff}}=\Psi_{r{\rm eff}}~,~~~
p_{r{\rm eff}} \leftrightarrow \tau_3 P_{r{\rm eff}}
=i\tau_3 \Psi_{r{\rm eff}}^{\dagger}=i_{\rm eff} \Psi_{r{\rm eff}}^{\dagger}
~~~,\eqno(5.36a)$$
with $\tau_3$ the $2 \times 2$ matrix introduced in Eq.~(4.40).  
Correspondingly, the nonvanishing anticommutator in Eq.~(5.35b) is changed to 
$$\{ \Psi_{u{\rm eff}}, \Psi^{\dagger}_{t{\rm eff} }\} =
 \hbar \delta_{ut} ~~~,\eqno(5.36b)$$ 
which has the correct positive sign on the right in both the 
$i_{\rm eff}=\pm i$ sectors of Hilbert space.  
The presence of an extra factor $\tau_3$ with eigenvalues $\pm 1$ in the
consistent operator transcription of Eq.~(5.36a) can be viewed as a 
reflection of the fact that the use of $i$ instead of $-i$ in the 
adjointness assignment of Eq.~(2.4b) was completely arbitrary:  with 
either $p_r=i\psi_r$ of $p_r=-i\psi_r$, a self-adjoint Lagrangian and an 
anti-self-adjoint $\tilde C$ are obtained.  

Turning to the boson sector, let us now introduce 
effective bosonic creation and 
annihilation operators in the effective quantum theory, denoted 
by $A_{r{\rm eff}}$ and $A_{r{\rm eff}}^{\dagger}$,  
by writing 
$$Q_{r{\rm eff}}={1 \over \surd 2}(A_{r{\rm eff}}+ 
A_{r{\rm eff}}^{\dagger})~,~~~
P_{r{\rm eff}}={1 \over  i_{\rm eff}\surd 2  } 
(A_{r{\rm eff}} - A_{r{\rm eff}}^{\dagger})~~~.\eqno(5.37a)$$
Since $i_{\rm eff}$ commutes with all effective operators, the definition  
of Eq.~(5.37a) 
is clearly consistent with the self-adjointness of $Q_r$ and $P_r$.    
Rewriting the commutator algebra of Eq.~(5.34b) in terms of $A_{r{\rm eff}}$ 
and its adjoint, we get 
$$[A_{u{\rm eff}},A_{t{\rm eff}}]=
[A_{u{\rm eff}}^{\dagger},A_{t{\rm eff}}^{\dagger}]=0~,~~~
[A_{u{\rm eff}},A_{t{\rm eff}}^{\dagger}] = \hbar \delta_{ut}~~~,
\eqno(5.37b)$$
which, as was the case in Eq.~(5.36c), 
has the correct positive sign on the right in both the 
$i_{\rm eff}=\pm i$ sectors of Hilbert space.  Note that again a factor of 
$i_{\rm eff}$ appears in the transformation from $Q_{r{\rm eff}}~,~~
P_{r{\rm eff}}$ to the corresponding creation and annihilation operators   
(which  for both fermions and bosons, as defined above, differ by a 
factor of $\hbar^{1 \over 2}$ from the customary ones).  

We are now ready to discuss the transition from our emergent Heisenberg 
picture quantum mechanics to the Schr\"odinger picture, and to derive the 
usual nonrelativistic Schr\"odinger equation.  Since the Heisenberg 
equation of motion of Eq.~(5.34e) and the commutation relations in the 
form given in Eqs.~(5.36b) and (5.37b) have the standard quantum mechanics  
form, what we do now is standard quantum mechanics, and makes no explicit 
reference to emergent origins of the quantum equations.  Restricting 
ourselves to the case in which the effective Hamiltonian has no 
intrinsic time dependence, we define $U_{\rm eff}(t)$ by 
$$U_{\rm eff}(t) =\exp(-i_{\rm eff} \hbar^{-1} t  H_{\rm eff})
~~~,\eqno(5.38a)$$
so that 
$$\eqalign{
{d \over dt}  U_{\rm eff}(t)=&-i_{\rm eff} 
\hbar^{-1} H_{\rm eff} U_{\rm eff}(t)~~~,\cr  
{d\over dt} U_{\rm eff}(t)^{\dagger} =&i_{\rm eff} \hbar^{-1} 
U_{\rm eff}(t)^{\dagger} H_{\rm eff}~~~.\cr  
}\eqno(5.38b)$$
Then from the time-independent Heisenberg 
picture state vector $\psi$ and an operator 
$S_{\rm eff}(t)$ with no intrinsic time dependence, 
we can form a Schr\"odinger picture state vector 
$\psi_{\rm Schr}$ and operator $S_{\rm eff~Schr}$ by 
the usual construction, 
$$\eqalign{
\psi_{\rm Schr}(t)=& U_{\rm eff}(t) \psi~~~,\cr
S_{\rm eff~Schr} =& U_{\rm eff}(t) S_{\rm eff}(t) 
U_{\rm eff}(t)^{\dagger}~~~,\cr
}\eqno(5.39a)$$
giving 
$$\eqalign{
i_{\rm eff} \hbar {d \over dt} \psi_{\rm Schr}(t)=&   
H_{\rm eff} \psi_{\rm Schr}(t) ~~~,\cr
{d \over dt} S_{\rm eff~Schr} =& 0~~~.\cr
}\eqno(5.39b)$$

To derive the nonrelativistic Schr\"odinger equation, let us consider 
the spacetime continuum case in which $r$ is the label $\vec x$, so that 
the fermionic anticommutation relations of Eq.~(5.36b) take the form 
$$\{ \Psi_{\rm eff}(\vec x), \Psi^{\dagger}_{\rm eff}(\vec y) \} =
 \hbar \delta^3(\vec x-\vec y) ~~~.\eqno(5.40a)$$ 
Assuming that the nonrelativistic operator $\Psi_{\rm eff}$ 
annihilates the vacuum state $|{\rm vac} \rangle$, so that 
$$   \Psi_{\rm eff}(\vec x) |{\rm vac} \rangle =0~~~, \eqno(5.40b)$$
and sandwiching Eq.~(5.40a) between $\langle {\rm vac} |$ 
and $| {\rm vac} \rangle $ 
gives 
$$\langle {\rm vac}|   \Psi_{\rm eff}(\vec x) 
\Psi^{\dagger}_{\rm eff}(\vec y)|{\rm vac} \rangle = 
 \hbar \delta^3(\vec x-\vec y) ~~~.\eqno(5.40c)$$ 
In the bosonic case, we start from the bosonic commutation relation 
of Eq.~(5.37b), and assuming 
that the nonrelativistic operator $A_{\rm eff}$ annihilates 
$|{\rm vac}\rangle$, we end up with 
$$\langle {\rm vac}|   A_{\rm eff}(\vec x) 
A^{\dagger}_{\rm eff}(\vec y) |{\rm vac} \rangle = 
 \hbar \delta^3(\vec x-\vec y) ~~~,\eqno(5.40d)$$ 
which has the same form as in the fermionic case.  So  
it suffices to restrict ourselves henceforth to the fermionic 
case.

Let us now introduce a complete set of single fermion intermediate 
states into Eq.~(5.40c), by inserting $1=\sum_n |n\rangle \langle n|$,   
giving 
$$\sum_n \langle {\rm vac}|  \Psi_{\rm eff}(\vec x)  |n\rangle
 \langle n| \Psi^{\dagger}_{\rm eff}(\vec y) | {\rm vac} \rangle = 
 \hbar \delta^3(\vec x-\vec y) ~~~.\eqno(5.41a)$$ 
If we now define a wave function $\Psi_n(\vec x)$ by 
$$\eqalign{
\hbar^{1\over 2} \Psi_n(\vec x)=& 
\langle {\rm vac}| \Psi_{\rm eff}(\vec x) | n \rangle ~~~,\cr
\hbar^{1\over 2} \Psi_n^*(\vec x)=& 
\langle n| \Psi^{\dagger}_{\rm eff}(\vec x) | {\rm vac} \rangle ~~~,\cr
}\eqno(5.41b)$$ 
then after dividing by $\hbar$ Eq.~(5.41a) can be rewritten as 
$$\sum_n \Psi_n^*(\vec x) \Psi_n(\vec y)=  \delta^3(\vec x-\vec y) ~~~,
\eqno(5.41c)$$
which is the usual completeness relation in coordinate representation. 
Multiplying by $\int d^3y \Psi_m^*(\vec y)$, we get 
$$\sum_n \Psi_n^*(\vec x) \int d^3y \Psi_m^*(\vec y)  \Psi_n(\vec y) 
=  \Psi_m(\vec x)    ~~~,\eqno(5.41d)$$
which by linear independence of the $\Psi_n$ tells us that 
$$ \int d^3y \Psi_m^*(\vec y)  \Psi_n(\vec y)=\delta_{mn}~~~,\eqno(5.41e)$$ 
which is the orthonormality condition in coordinate representation.  
Taking the time derivative of the first line of 
Eq.~(5.41b) and using the Heisenberg 
equation of motion of Eq.~(5.34d), we get 
$$\eqalign{
\hbar^{1\over 2}{d\over dt} \Psi_n(\vec x) 
=&\langle {\rm vac}| {d\over dt} \Psi_{\rm eff}(\vec x)| n \rangle \cr
=&\langle {\rm vac} | i_{\rm eff} \hbar^{-1} 
[H_{\rm eff},\Psi_{\rm eff}(\vec x)]| n \rangle~~~. \cr 
}\eqno(5.42a)$$ 
If we take $H_{\rm eff}$ to be a one body operator of the form 
$$H_{\rm eff}=\int d^3 y \Psi_{\rm eff}^{\dagger}(\vec y) 
{\cal H}_{\rm eff}(\vec y) \Psi_{\rm eff}(\vec y)~~~,\eqno(5.42b)$$ 
then the commutator appearing in Eq.~(5.42a) is given by 
$$ [H_{\rm eff},\Psi_{\rm eff}(\vec x)]
=-  {\cal H}_{\rm eff}(\vec x) \Psi_{\rm eff}(\vec x) ~~~,\eqno(5.42c)$$
and so the right hand side of Eq.~(5.42a) becomes 
$$-i_{\rm eff} \hbar^{-1}  {\cal H}_{\rm eff}(\vec x)  
\langle {\rm vac} | \Psi_{\rm eff}(\vec x)| n \rangle
=  -i_{\rm eff} \hbar^{-{1\over 2}}  
{\cal H}_{\rm eff}(\vec x)\Psi_n(\vec x)~~~.\eqno(5.42d)$$      
Multiplying through by $i_{\rm eff}\hbar^{1\over 2}$, Eq.~(5.42a) then 
yields the standard nonrelativistic Schr\"odinger equation 
in coordinate representation, 
$$i_{\rm eff} \hbar {d\over dt} \Psi_n(\vec x)  =
{\cal H}_{\rm eff}(\vec x)\Psi_n(\vec x)~~~.\eqno(5.43)$$      

This is of course all standard quantum mechanics and quantum field theory. 
The point of going through it in detail is to emphasize that once we have 
obtained emergent canonical commutation relations and an emergent 
Heisenberg equation of motion for operators, the Schr\"odinger picture 
and Schr\"odinger equation of quantum mechanics follow in a straightforward 
way. To complete the argument for an emergent quantum mechanics, we must  
address the issue of how the probability interpretation (the Born rule) 
follows from our framework; this is the topic of the next section.  
\bigskip
\centerline{5F.~~~Brownian Motion Corrections to Schr\"odinger Dynamics}
\centerline{and the Emergence of the Probability Interpretation}
\bigskip

Up to this point we have worked in the thermodynamic limit, with  
our reasoning based on the study of averages of dynamical variables 
in the canonical ensemble, with all fine grained structure averaged out.    
However, as in classical statistical mechanics, there are contexts 
in which fluctuations around the averages, which can be modeled in a  
natural way by a 
generalized Brownian motion, are important.   We shall argue in this section 
that Brownian motion corrections to emergent quantum mechanics provide 
the 
mechanism responsible both for reduction of the state vector, and for the  
emergence of the Born and L\"uders probability rules.    

To do this, we shall return to the general Ward identity of Eqs.~(5.15a--c),  
in which the source terms have been varied and then set equal to zero. 
We continue to make approximations (1) and (2) of Sec. 5C, that is, 
we neglect the $\tau$ and $\eta$ terms in Eq.~(5.15a),   
but now we do not make approximation (3), so that Eq.~(5.15a) 
takes the form  
$${\cal D} x_{u {\rm eff}} 
= i_{\rm eff}[ W_{\rm eff},  x_{u{\rm eff}}]   
+\sum_{s,\ell} \omega_{us} \epsilon_{\ell} \left(W_s^{R\ell} 
{1\over 2}  \{\tilde C,i_{\rm eff} \} 
W_s^{L\ell}\right)_{\rm eff} ~~~,\eqno(5.44a)$$
and Eq.~(5.15c) states that this expression vanishes inside suitable 
canonical ensemble averages.  (To recapitulate some terminology, 
$i_{\rm eff}$ is defined, together with $\hbar$, from the ensemble average 
of $\tilde C$ in Eq.~(4.11b); $W$ is a general bosonic polynomial in the 
matrix dynamical variables $\{x_r\}$; the effective of ``eff'' projection 
is defined in Eqs.~(4.41a-d); the structure coefficients $\omega_{us}$ and 
the grading factor $\epsilon_{\ell}$ are defined in Eqs.~(1.16a,b) and the 
line following Eq.~(1.10b), respectively; and finally, the decomposition 
of $W$ into left and right factors $W_s^{L\ell}$ and $W_s^{R\ell}$ with 
respect to $x_s$ is defined in Eq.~(5.10a).)
We shall now proceed in two steps:  first we study the implications of 
Eq.~(5.44a) for the normalization and completeness of 
wave functions at a fixed initial time, and then we 
study its implications for the time development 
of wave functions, that is, for the Schr\"odinger equation derived in 
Eq.~(5.43).  

{}For the  first step  we observe, 
recalling the discussion preceding Eq.~(5.31c), that 
if we take $W=\sigma_t x_t$, then there is only one term in the sum over 
$\ell$, with 
$\epsilon_{\ell=1}=1$,  $W_s^{L\ell=1} =\sigma_t \delta_{st}$, 
and $W_s^{R\ell=1}=\delta_{st}$, 
so that the final term in Eq.~(5.44a) reduces to 
$$ \omega_{ut} {1\over 2}  \{\tilde C,i_{\rm eff} \}   
\sigma_t~~~.\eqno(5.44b)$$ 
Without making further approximations, this 
term has zero source ensemble average 
$$- \omega_{ut} \hbar \sigma_t~~~,\eqno(5.44c)$$
and so setting the sources equal 
to zero in Eq.~(5.15b), we get 
$$\langle   i_{\rm eff}[ \sigma_t x_{t{\rm eff}},  x_{u{\rm eff}}]   
- \omega_{ut} \hbar \sigma_t  \rangle_{\hat{\rm AV},0}=0~~~.\eqno(5.45a)$$
Multiplying Eq.~(5.45a) on the left by $\psi_0^{\dagger}$ and on the right 
by $\psi_0$, and assuming the correspondence between canonical ensemble 
averages and Wightman functions given in Eq.~(5.23b), we see that 
$$\langle {\rm vac} |
  i_{\rm eff}[ \sigma_t X_{t{\rm eff}},  X_{u{\rm eff}}]   
 - \omega_{ut} \hbar \sigma_t  |{\rm vac}\rangle =0~~~.\eqno(5.45b)$$

Using Eq.~(5.45b) in place of Eq.~(5.34a) as the starting point for the 
analysis of Sec.~5E, we learn that the vacuum expectation of Eq.~(5.40a), 
$$\langle {\rm vac} |    
\{ \Psi_{\rm eff}(\vec x), \Psi^{\dagger}_{\rm eff}(\vec y) \} 
| {\rm vac} \rangle 
= \hbar \delta^3(\vec x-\vec y) ~~~,\eqno(5.45c)$$ 
does not make use of approximation (3), in other words, Eq.~(5.45c) 
holds even when fluctuations of $\tilde C$ about its ensemble average are 
taken into account.  Thus, the orthonormalization and completeness of  
Schr\"odinger wave functions, 
derived in Eqs.~(5.41a-e), also does not make use of
approximation (3), as a result of the insensitivity of the vacuum expectation of the 
canonical algebra to fluctuations in $\tilde C$. 

{}For the second step, we take $W$ in Eq.~(5.44a) to be the operator 
Hamiltonian $H$, as we did in Eq.~(5.19b) of Sec.~5C, but we now take     
the fluctuations of $\tilde C$ about its ensemble average into account.  
We denote this fluctuating term by  
$$\Delta \tilde C =\tilde C-\langle \tilde C \rangle_{{\rm AV},0} 
~~~,\eqno(5.46a)$$
and following the notation of Eq.~(4.41b), we write 
$$\Delta \tilde C=  \Delta \tilde C_{\rm eff} + \Delta \tilde C_{12}~~~.
\eqno(5.46b)$$
We now make an Ansatz for the structure of the first term in 
Eq.~(5.46b), 
$$\Delta \tilde C_{\rm eff}=i_{\rm eff} \hbar {\cal K}~~~, \eqno(5.46c)$$
where ${\cal K}={\cal K}_0+i{\cal K}_1$ is taken to be a 
complex $c$-number that is rapidly fluctuating, so 
that it has vanishing zero source ensemble average, 
$$\langle {\cal K} \rangle_{{\rm AV},0}=0~~~.\eqno(5.46d)$$
Note that if ${\cal K}$ is a $c$-number, then Eq.~(5.46c) is automatically 
consistent with the general 
condition ${\rm Tr} \tilde C=0$, and because we have seen in Sec.~2 
that $\tilde C$  can have a self-adjoint part, which can arise from 
time reversal noninvariance, the 
imaginary part ${\cal K}_1$ of 
the fluctuating term can be nonzero.  
When we form $\{i_{\rm eff},\tilde C\}$, the second term of Eq.~(5.46b) 
drops out, and so with the Ansatz of Eq.~(5.46c) we get 
$$\{\tilde C,i_{\rm eff}\}=-2 \hbar (1+{\cal K})~~~,\eqno(5.47a)$$
with the term ${\cal K}$, which we assume to be 
much smaller than unity, giving 
the leading fluctuation correction to our earlier analysis.  There will 
in general be higher order terms in the fluctuations, which can have a 
nontrivial operator structure, and so we shall 
include an $O({\cal K}^2)$ error 
term in the subsequent formulas.  However, we shall see that general physical
requirements determine much of the 
structure of these 
higher order terms, given the structure of the linear term.  

There are two possible dynamical scenarios for having a nonzero imaginary 
part ${\cal K}_1$.  In the first scenario, $\tilde C$ has 
an explicit self-adjoint 
part, as in the examples discussed in Eqs.~(2.18a) through (2.22d).  In this 
case, there will be an additional Lagrange multiplier term in the exponent 
of the canonical ensemble $\rho$, associated with conservation of the 
self-adjoint part of $\tilde C$, and the variation of this term will lead 
to an additional term in the Ward identity, which must be retained for   
consistency.  In the second scenario, $\tilde C$ 
remains anti-self-adjoint, so that the structure of the Ward identity 
is unaltered, and the imaginary part ${\cal K}_1$ in the fluctuation 
term appears because of spontaneous breaking of time reversal symmetry.  
In this case, consistency requires  the canonical ensemble to  
contain 
an implicit boundary condition or contour prescription $i\epsilon~,~~
\epsilon \to 0^{+(-)}$, which selects the sign $+(-)i$ of the imaginary  
unit multiplying ${\cal K}_1$.  

We shall assume that the second scenario is the relevant one, and so 
shall continue to use the Ward identity of Eqs.~(5.15a-c), and the 
consequences derived from it in Sec.~5C.  
Since ${\cal K}$ is a $c$-number, when Eq.~(5.47a) is 
substituted into Eq.~(5.44a), 
we can still use Eq.~(5.10b) to evaluate the sum over $\ell$, just as 
we did in Eq.~(5.18c), giving 
$${\cal D} x_{u {\rm eff}} 
= i_{\rm eff} [W_{\rm eff},  x_{u{\rm eff}}]
-\hbar(1+{\cal K}) \sum_s \omega_{us} \left( 
{\delta {\bf W} \over \delta x_s} \right)_{\rm eff} +O({\cal K}^2)
~~~.\eqno(5.47b)$$
Taking $W$ to be the operator Hamiltonian $H$, we now find that Eq.~(5.19b) 
is replaced by 
$${\cal D} x_{u {\rm eff}} 
= i_{\rm eff} [H_{\rm eff},  x_{u{\rm eff}}]
-\hbar(1+{\cal K}) \dot x_{u {\rm eff}}+O({\cal K}^2)~~~,\eqno(5.47c)$$
which is still to be interpreted as an equality holding when sandwiched 
between polynomials $S_{L,R}$ that do not contain $x_u$, and averaged over 
the zero source canonical ensemble.  

We now wish to reinterpret the 
vanishing of Eq.~(5.47c) as an operator statement in the 
effective quantum field theory, using the correspondence of Eq.~(5.23a),  
and restricting ourselves henceforth to the $i_{\rm eff}=i$ sector.  
Since ${\cal K}$ is a rapidly fluctuating quantity in the underlying matrix 
operator phase space, it does not have a direct transcription to the 
effective field theory.  However, consistent with the idealization involved 
in describing an ergodic, time dependent matrix dynamics by the static 
canonical ensemble, it is natural to model ${\cal K}$ in the effective field 
theory transcription of Eq.~(5.47c) as a time dependent, rapidly fluctuating 
complex $c$-number noise term ${\cal K}(t)$.  Hence we provisionally 
reinterpret 
Eq.~(5.47c) as a field theory equation of motion
$$i_{\rm eff} [H_{\rm eff},  X_{u{\rm eff}}]
-\hbar(1+{\cal K}(t)) \dot X_{u{\rm eff}}+O({\cal K}^2)=0~~~.\eqno(5.47d)$$
Rewriting this as 
$$\eqalign{
\dot X_{u{\rm eff}}=& 
i_{\rm eff} \hbar^{-1} (1-{\cal K}(t)) 
[H_{\rm eff}, X_{u{\rm eff}}] +O({\cal K}^2)  \cr
=&
i_{\rm eff} \hbar^{-1} (1-{\cal K}_0(t)) 
[H_{\rm eff}, X_{u{\rm eff}}] 
-i i_{\rm eff} \hbar^{-1} {\cal K}_1(t) 
[H_{\rm eff}, X_{u{\rm eff}}] 
+O({\cal K}^2)~~~,  \cr
}\eqno(5.47e)$$
we see that the ${\cal K}_1(t)$ term in $\dot X_{u{\rm eff}}$ has the 
opposite adjointness from $X_{u{\rm eff}}$, that is, if $X_{u{\rm eff}}$ is 
self-adjoint then the ${\cal K}_1(t)
$ term in $\dot X_{u{\rm eff}}$ is anti-self-adjoint, and vice versa.  
As a result, there is an inconsistency if we regard Eq.~(5.47e) as an 
operator equation of motion that holds for all operators 
$X_{u{\rm eff}}$, since if 
we replace $X_{u{\rm eff}}$ in Eq.~(5.47e) by its 
adjoint $X_{u{\rm eff}}^{\dagger}$, 
and then take the adjoint of Eq.~(5.47e), we obtain an equation for 
$\dot X_{u{\rm eff}}$ in which the sign of the ${\cal K}_1$ term is reversed.  
(A further discussion of this problem is given in Appendix H.)

We can avoid this inconsistency in the same manner that the problems with 
imposing a covariant Lorentz gauge constraint are avoided in quantum 
electrodynamics, by regarding Eq.~(5.47e) not as an operator equation of 
motion, but as a constraint on the vacuum state $|{\rm vac}\rangle$, 
so that we have 
$$\dot X_{u{\rm eff}}|{\rm vac}\rangle= 
i_{\rm eff} \hbar^{-1} (1-{\cal K}(t)) 
[H_{\rm eff}, X_{u{\rm eff}}]|{\rm vac}\rangle +O({\cal K}^2) 
~~~.\eqno(5.47f)$$
Since we are assuming that $H_{\rm eff}|{\rm vac}\rangle=0$, this 
equation simplifies to 
$$\dot X_{u{\rm eff}}|{\rm vac}\rangle= 
i_{\rm eff} \hbar^{-1} (1-{\cal K}(t)) 
H_{\rm eff} X_{u{\rm eff}}|{\rm vac}\rangle +O({\cal K}^2) 
~~~.\eqno(5.47g)$$
When $ X_{u{\rm eff}}$ is an annihilation operator $\Psi_{u{\rm eff}}$ 
or $A_{u{\rm eff}}$, the left and right hand sides of Eq.~(5.47g) are 
equal to zero and the equation becomes trivial.  However, it has nontrivial 
content when $ X_{u{\rm eff}}$ is a creation operator 
$\Psi_{u{\rm eff}}^{\dagger}$ or $A_{u{\rm eff}}^{\dagger}$. Similarly, 
taking the adjoint of Eq.~(5.47g), we get 
$$\langle {\rm vac}|\dot X_{u{\rm eff}}^{\dagger}= 
\langle {\rm vac}|  X_{u{\rm eff}}^{\dagger} H_{\rm eff}  
(- i_{\rm eff}) \hbar^{-1} (1-{\cal K^*}(t)) 
  +O({\cal K}^2) 
~~~,\eqno(5.47h)$$
which is trivial when $X_{u{\rm eff}}^{\dagger}$ is a creation operator 
and nontrivial when it is an annihilation operator.  

Using Eq.~(5.47h), projected on a general state $|...\rangle$,  
in the Schr\"odinger  
equation derivation of Eqs.~(5.42a-5.43), we obtain
$$ {d\over dt} \Phi(\vec x)  =
-i \hbar^{-1}(1-{\cal K}^*(t)) {\cal H}_{\rm eff}(\vec x)\Phi(\vec x)
+O({\cal K}^2) \Phi(\vec x)
~~~.\eqno(5.48a)$$      
We have here denoted a generic state vector by $\Phi$ rather 
than by $\Psi_n$.  We have dropped the basis label $n$ because it is 
irrelevant for the present discussion, and have used the notation 
$\Phi$ rather than $\Psi$ to emphasize the fact that 
when the imaginary part  of ${\cal K}(t)$ is nonzero, Eq.~(5.48a) does 
not preserve the norm of the state vector $\Phi$.  The fact that the norm 
of $\Phi$ is not preserved, we suggest, is a reflection of the 
approximation involved in transcribing the time independent ${\cal K}$ that
fluctuates over the underlying operator phase space into a time fluctuating 
process ${\cal K}(t)$ in the effective field theory.  

Since we have seen that the orthonormality structure of state vectors 
at an initial time $t=0$ is preserved in the presence of fluctuations 
in $\tilde C$, and moreover, since in the nonrelativistic limit all particle 
species are conserved in number, we must restore conservation of the norm 
under time evolution in order to obtain the physical state vector $\Psi$.  
A norm preserving Schr\"odinger 
equation, incorporating the fluctuation corrections, can be obtained by  
making a suitable choice of the $O({\cal K}^2)$ term in Eq.~(5.48a) and then 
identifying the physical state $\Psi$ with the renormalized $\Phi$, 
$$\Psi= {\Phi \over 
[\int d^3x \Phi^*(\vec x) \Phi(\vec x)]^{1\over 2} }~~~. \eqno(5.48b)$$
To complete the specification of the stochastic dynamics for 
$\Phi$,  we must also specify the nonvanishing time averages 
(which we denote by $E[...]$) of the 
rapidly varying noise term, by writing 
$$E[{\cal K}_A(t)]=0~,~~~E[ {\cal K}_A(t_1) {\cal K}_B(t_2)]  
=\delta_{AB} D_A(t_1-t_2)~,~~~A,B=0,1~~~,\eqno(5.48c)$$
where the  $D_A$ are certain specified functions, which in the 
case of white noise are constants times delta functions.  

Equations (5.48a-c) summarize our basic results for the fluctuation 
modified Schr\"odinger equation.  In essence, what we have found is 
that the fluctuations of $\tilde C$ have the effect of replacing the
Planck constant $\hbar$ in the Schr\"odinger equation by a new Planck 
``constant'' $\hbar (1+{\cal K})$ that has small complex 
random fluctuations.  
Although we have not specified the $O({\cal K}^2)$ term, once the part of 
Eq.~(5.48c) linear in ${\cal K}$ is given, the  $O({\cal K}^2)$ 
term in the white noise case is completely 
determined by general structural considerations.  The easiest 
way to see this is to go over to a standard  
It\^o calculus representation of the 
fluctuation term, by writing 
$$i \hbar^{-1} {\cal K}^*(t)dt = \beta_R dW_t^R + i \beta_I dW_t^I
~~~,\eqno(5.49a)$$
where $\beta_{R,I}$ are real constants and where $dW_t^R$ and 
$dW_t^I$ are stochastic differentials that, together with $dt$,  
obey the usual It\^o calculus rules (see also Appendix G) 
$$\eqalign{
(dW_t^R)^2=&(dW_t^I)^2=dt~~~,\cr
dW_t^RdW_t^I=&dW_t^Rdt=dW_t^Idt=dt^2=0~~~.\cr   
}\eqno(5.49b)$$
Writing $\Phi(\vec x)=\langle \vec x|\Phi\rangle$, Eq.~(5.48a) then  
takes the form 
$$ d| \Phi\rangle  =
-i \hbar^{-1}  H_{\rm eff}|\Phi \rangle dt
+O(\beta^2)| \Phi\rangle dt + \beta_R  H_{\rm eff}|\Phi \rangle dW_t^R 
+ i\beta_I  H_{\rm eff}|\Phi \rangle dW_t^I 
~~~.\eqno(5.49c)$$
Using the standard It\^o product rule (or stochastic integration by  
parts formula) given by 
$$d(FG)=(dF) G + F dG + dF dG~~~, \eqno(5.49d)$$
it is then straightforward to show [27] that 
up to an overall constant phase, the normalized state vector 
$|\Psi\rangle$ corresponding to Eq.~(5.49d) satisfies the 
stochastic differential equation 
$$\eqalign{ 
d|\Psi\rangle  =&
-i \hbar^{-1} H_{\rm eff}|\Psi\rangle dt  
+[Q-\langle Q \rangle - {1\over 2}  \beta_R^2 
(   H_{\rm eff}-\langle   H_{\rm eff} \rangle )^2  -{1\over 2} \beta_I^2 
H_{\rm eff}^2]  |\Psi\rangle dt \cr
+&\beta_R (   H_{\rm eff}-\langle  H_{\rm eff} \rangle ) 
|\Psi\rangle dW_t^R  
+i\beta_I H_{\rm eff}|\Psi\rangle dW_t^I  ~~~,\cr
}\eqno(5.50a)$$
where $\langle  {\cal O} \rangle$ denotes the expectation of
the operator ${\cal O}$ in the normalized state $|\Psi \rangle$, 
$$\langle  {\cal O} \rangle =\langle \Psi | {\cal O} | \Psi \rangle
~~~.\eqno(5.50b)$$
In Eq.~(5.50a), $Q$ is an arbitrary self-adjoint operator of order 
$O(\beta^2)$ which is not determined by the requirement of norm preservation.     
However, a simple calculation [28] using Eq.~(5.49d) 
shows that the  evolution of the density 
matrix $\hat \rho=|\Psi\rangle \langle \Psi|$ 
corresponding to Eq.~(5.50a) is 
$$\eqalign{
d \hat \rho =& i \hbar^{-1} [\hat \rho,  H_{\rm eff} ] dt
-{1\over 2} |\beta|^2[ H_{\rm eff} ,[ H_{\rm eff},\hat \rho]]dt
+[\hat \rho,[\hat \rho,Q]]dt\cr 
+ &\beta_R [\hat \rho,[\hat \rho,H_{\rm eff}]]dW_t^R
+i\beta_I[H_{\rm eff},\hat \rho] dW_t^I~~~,\cr
}\eqno(5.51a)$$
which when we take its expectation (i.e., averaging over the fluctuations)  
gives the ordinary differential equation 
$${dE[\hat\rho]\over dt} = i \hbar^{-1} [E[\hat \rho],  H_{\rm eff} ]
-{1\over 2} |\beta|^2[ H_{\rm eff} ,[ H_{\rm eff},E[\hat\rho]]]
+E[[\hat \rho,[\hat \rho,Q]]] ~~~.\eqno(5.51b)$$

There are now two independent structural arguments for imposing the 
vanishing of $Q$.  First, when $Q$ is not zero, Eq.~(5.51b) implies that  
the expected value of the energy $E[{\rm Tr}\hat \rho H_{\rm eff}]$  is 
not conserved, but instead obeys 
$${d\over dt} E[{\rm Tr}\hat \rho H_{\rm eff}]
=E[{\rm Tr}[\hat \rho,[\hat \rho,Q]] H_{\rm eff} ] \not= 0~~~.\eqno(5.51c)$$
Thus, to achieve energy conservation in the mean we must take $Q=0$.  
Second, when $Q$ is not zero, 
the evolution of Eq.~(5.51b) is nonlinear, which opens up the possibility of
instantaneous (faster than light) signaling [29]. If we assume that   
the underlying matrix dynamics has a structure that forbids superluminal 
signaling in the emergent quantum theory, then we must again set $Q=0$.     

With $Q=0$, the stochastic expectation $E[\hat\rho]$ then obeys a linear 
master equation of the Lindblad [30] type characteristic quite generally of 
open system dynamics.  Thus, although we have not 
explicitly calculated the $O(\beta^2)$ term 
in Eq.~(5.49c), general structural 
requirements lead to the unique norm-preserving stochastic equation 
$$\eqalign{ 
d|\Psi\rangle  =&
-i \hbar^{-1} H_{\rm eff}|\Psi\rangle dt  
 - {1\over 2}  [\beta_R^2 
(   H_{\rm eff}-\langle   H_{\rm eff} \rangle )^2  
+\beta_I^2 H_{\rm eff}^2]|\Psi\rangle dt \cr
+&\beta_R (   H_{\rm eff}-\langle  H_{\rm eff} \rangle ) 
|\Psi\rangle dW_t^R + i \beta_I H_{\rm eff} |\Psi\rangle dW_t^I ~~~, \cr
}\eqno(5.51d)$$
with the corresponding density matrix evolution
$$d \hat\rho = i \hbar^{-1} [\hat\rho,  H_{\rm eff} ] dt
-{1\over 2} |\beta|^2[ H_{\rm eff} ,[ H_{\rm eff},\hat\rho]]dt
 + \beta_R [\hat\rho,[\hat\rho,H_{\rm eff}]]dW_t^R
 +i \beta_I [H_{\rm eff},\hat \rho] dW_t^I~~~.\eqno(5.51e)$$

Through Eqs.~(5.48a-c) and Eqs.~(5.51d,e), we establish a connection  
between the quantum dynamics 
emergent from matrix model dynamics, and a large body of literature 
dealing with stochastic modifications to the Schr\"odinger equation.  
We shall not attempt to review the stochastic Schr\"odinger equation   
literature here.  For recent reviews focusing on the  
spontaneous localization 
approach and references, 
see [31], and for 
detailed mathematical and phenomenological 
studies of the case in which the stochasticity 
is driven by the Hamiltonian, as in the equations derived above, 
see respectively [28,32] and [33].  The connection between   
general Gaussian noise, and the simpler case of white noise, is 
discussed in [34].  The seminal ideas in the stochastic reduction 
program arose from work over the last    
twenty five years by Pearle [35], Ghirardi, Rimini, and Weber [36],  
Ghirardi, Pearle, and Rimini [37], 
Gisin [38],  Di\'osi  [39], and Percival [40], supplemented and extended in 
more recent work by many 
others.  The main result coming from the stochastic Schr\"odinger program,   
as applied to Eqs.~(5.51d,e), 
is that when the imaginary part ${\cal K}_1$ of the noise term is nonzero, 
corresponding in Eqs.~(5.51d,e) to a nonzero real part $\beta_R$, 
then as it evolves  
the state vector reduces to energy eigenstates with the correct 
Born rule probabilities.   An outline of the
proof of this, focusing on the case of a nondegenerate Hamiltonian,  
is given in Appendix G. 
More generally, in the case of energy degeneracies, one finds [32] reduction 
in the generalized sense of the L\"uders projection postulate.  
The phenomenological analysis of [33] 
(for a discussion see Appendix I) shows that reduction to energy 
eigenstates or degenerate energy manifolds, 
assuming a Planckian magnitude for the coefficient of the stochastic 
term and taking into account environmental energy fluctuations, is compatible 
with all known experiments, both for cases in which 
the maintenance of coherence is 
observed, and those in which a measurement is made and 
state vector reduction results.  These considerations strongly suggest 
that the statistical mechanics of matrix models with a global 
unitary invariance lead not only to an emergent complex quantum mechanics, 
but also to the emergence of the usual probabilistic framework needed 
for the application of quantum theory.  

Returning to Eq.~(5.44a), which was the starting point for our analysis 
of stochastic reduction, we note that there is no obvious route for 
obtaining an analog of Eqs.~(5.51d,e) in which $H_{\rm eff}$ is replaced 
by a more general operator, such as one leading to spatial localization.  
The reason is that if we attempt to generalize beyond the Ansatz of 
Eq.~(5.46c), by taking $\Delta \tilde C_{\rm eff}$ to be a general 
effective operator (rather than a $c$-number multiple of $i_{\rm eff}$), 
then $\{ \tilde C_{\rm eff},i_{\rm eff} \}$ becomes an operator rather 
than a $c$-number, and it is no longer possible to use Eq.~(5.10b) to 
evaluate the sum over $\ell$ in Eq.~(5.44a) so as to give Eq.~(5.47b).  
Consequently, when $W$ in Eq.~(5.44a) is taken as $H_{\rm eff}$, in the 
case of a general operator fluctuation $\Delta \tilde C_{\rm eff}$ one no 
longer recovers the time derivative $\dot x_{u{\rm eff}}$ from the 
second term in Eq.~(5.44a), and the route we employed 
to derive a stochastic time evolution equation is blocked.  

One feature of stochastic Schr\"odinger equations that has led to much 
discussion in the literature is the fact that the usual formulations 
are nonrelativistic, 
and attempts to construct  relativistic 
generalizations have encountered serious obstacles [41].  Within the 
framework given here, this is not surprising, since the canonical ensemble 
that we have used to derive emergent quantum mechanics picks 
out a preferred rest 
frame, which we have tentatively identified with the rest frame of the 
cosmic blackbody radiation.  In the decoupling limit in which the $\tau$ 
terms are neglected and in which fluctuations in $\tilde C$ are neglected, 
we have argued that a Lorentz invariant effective quantum theory results  
when the underlying trace Lagrangian is Lorentz invariant.  
However, in order for fluctuations in the canonical ensemble to have 
a finite magnitude, the convergence factor $\exp(-\tau {\bf H})$ in the 
canonical ensemble is needed, and so fluctuation processes in the ensemble 
are necessarily frame-dependent.  From this point of view, the 
frame-dependent structure of the stochastically modified Schr\"odinger 
equation is a natural feature.  

\vfill\eject

\overfullrule=0pt
\twelvepoint
\doublespace
\pageno=125
\bigskip    
\centerline{6.~~~Discussion and Outlook}
\bigskip
In the preceding sections we have developed a new approach to quantum 
mechanics, based on the idea that quantum theory is an emergent phenomenon 
arising from the statistical dynamics of an underlying matrix model with 
a global unitary invariance.  What we have done is to establish a general 
framework, and to formulate specific assumptions and approximations, 
that lead to the emergence of quantum mechanics as an effective dynamics 
for low frequency or ``soft'' degrees of freedom, which are separated 
by a large hierarchy of scale and effectively decoupled, from very  
high frequency or ``hard'' degrees of freedom  characterizing the 
underlying dynamics.  However, we have not identified a candidate for 
{\it the} specific matrix model that realizes our assumptions: this is 
a task for the future.  

Since, as noted in the Introduction, quantum theory is our most successful 
physical theory, one can ask why try to replace it with something else?  
To conclude, we respond to this question by listing a number of problems 
with conventional quantum mechanics, that  
are solved, or may be solved in the future, by an approach along the lines   
of the one developed here.  
\bigskip
\noindent
\centerline{(1)~~ What is the origin of ``canonical quantization''?}
\medskip
The standard approach to constructing a quantum theory consists in  
first writing down the corresponding classical theory, and 
then ``quantizing''
it by reinterpreting the classical quantities as operators, and replacing 
the classical Poisson brackets by commutators/anticommutators.  However, 
since quantum theory is more fundamental than classical theory, it seems 
odd that one has to construct it by starting from the classical limit; the 
canonical quantization approach has very much the flavor of an algorithm 
for inverting the classical limit of quantum mechanics.  In the trace 
dynamics approach developed here, one works with operators from the outset, 
and the full structure of the canonical commutation/anticommutation 
relations is derived 
as a reflection of the structure of the conserved operator $\tilde C$.
\bigskip
\noindent
\centerline{(2)~~The quantum measurement problem.}  
\medskip
The unitary evolution of standard quantum mechanics does not 
describe what happens when measurements are made, but 
conventionally has to be 
supplemented by an additional postulate of nonunitary  state vector reduction  
when a ``measurement'' is performed by a ``classical'' apparatus.  
In the emergent approach developed here, both unitary evolution and 
state vector reduction are seen to be different aspects of the underlying 
nonunitary matrix model dynamics.  Unitary evolution of the 
emergent quantum mechanics reflects properties  
of the statistical thermodynamics of the underlying matrix dynamics, 
while the phenomenon of state vector reduction together with the Born 
probability rules arise as a consequence of Brownian motion corrections to 
this thermodynamics.  
\bigskip
\noindent
\centerline{(3)~~Infinities and nonlocality}
\medskip
An outstanding problem in quantum mechanics (or more specifically, in  
quantum field theory) is the presence of  
infinities, and an outstanding puzzle is the nonlocality of quantum 
mechanics seen, for example, in Einstein-Podolsky-Rosen type experiments.  
In the emergent approach developed here, these are both aspects of 
an underlying dynamics that is totally nonlocal, as manifested in the 
fact that the underlying matrix variables have no assumed commutativity 
properties.  These matrix variables are postulated to lead to convergent 
traces obeying the usual cyclic properties, and since the associated 
trace dynamics is an extension of classical dynamics, it should involve no 
inconsistencies or infinities.  The canonical local 
commutation/anticommutation relations of quantum field theory, with their 
corresponding short distance singularities, emerge only as an idealization 
approximating the thermodynamic limit of the underlying trace dynamics as  
it applies to the ``soft'' degrees of freedom.  As already noted in Sec. 5C, 
because the underlying trace dynamics is a nonlocal ``hidden variables'' 
theory, it is not subject to the Bell inequalities that rule out attempts 
at local hidden variables extensions of quantum mechanics.  We also remark 
that the nonlocal structure of the underlying dynamics may have significant 
implications for aspects of early universe cosmology, such as the 
horizon problem.  
\bigskip 
\noindent
\centerline{(4)~~Unification of quantum theory with gravitation} 
\medskip
There are a number of indications that conventional quantum mechanics 
must be modified in a profound fashion in order for it to be  
successfully combined with gravitational 
physics.  In generic curved spacetimes, it is 
not possible to give a precise formulation of the particle production 
rate, nor is there necessarily a well-defined concept of conserved energy.     
In trace dynamics, there 
is no conserved energy operator, but only a conserved trace energy ${\bf H}$,
suggesting that it may give an arena in which the properties of 
quantum mechanics and gravity can be reconciled.  (We note that recent work 
on the string theory approach to quantum gravity has also 
suggested certain classes of matrix models as an  
underlying dynamics [42].)  A second indication that quantum mechanics must 
be modified when combined with gravitational physics is provided by recent 
ideas on ``holography'', which suggest that the association of degrees of 
freedom with volume subdivisions must break down near the 
Planck energy [26,43]. 
As we have noted, the fact that $\tilde C$ becomes a surface integral in   
supersymmetric Yang-Mills theories suggests a possible connection between 
trace dynamics and holographic ideas.  

A third indication of possible new 
quantum physics in the gravitational context comes from the fact that 
in addition to ordinary matter,  cosmology appears to require two additional 
forms of matter, the mysterious ``dark matter'' and ``dark energy'' 
contributions to the closure of the universe.  These of course may have 
more conventional explanations, but we note that in the theory of emergent 
quantum mechanics developed here, there are three distinct sectors: 
the $i_{\rm eff}= i$ and $i_{\rm eff}=-i$ sectors in the emergent quantum 
theory, which are not directly coupled to one another but which still 
both interact with the same spacetime metric, and the high energy 
degrees of freedom (given the subscript ``12'' in 
Sec.~4E) that anticommute with $i_{\rm eff}$.  Could these correspond, 
respectively, to the ordinary matter, dark matter, and dark energy sectors?
Settling such speculations will require understanding precisely how 
gravitation, and the standard model particle forces, fit into our framework. 
\medskip  
\noindent
\centerline{(5)~~The cosmological constant}  
\medskip
Another indication that quantum mechanics may have to be modified to 
deal with gravitational phenomena is provided by the problem of the 
cosmological constant. 
In conventional quantum field theory it is very hard to understand 
why the observed cosmological constant is 120 orders magnitude than the 
natural scale provided by the Planck energy.  Either unbroken scale 
invariance or unbroken supersymmetry would forbid the appearance of a 
cosmological constant, but they also forbid the appearance of a realistic 
particle mass spectrum, and so in conventional quantum 
theory they do not provide a basis for 
solving the cosmological constant problem.  The difficulty that arises here 
can be formulated as a mismatch between the single constraint needed -- a 
sum rule dictating the vanishing of the cosmological constant -- and the 
infinite number of constraints arising from having  conserved operator scale 
or conformal transformation generators 
or a conserved operator supercurrent.  
In trace dynamics, we have seen that scale invariance is manifested 
by the vanishing of the matrix trace of the Lorentz 
trace of the stress-energy 
tensor ${\rm Tr} T_{\mu}^{~\mu}={\bf T}_{\mu}^{~\mu}$, providing 
a single number condition.  Similarly, we have seen that supersymmetry 
implies only the conservation of a trace supercurrent ${\bf J}^{\mu}$ 
(and its conjugate $\bar {\bf J}^{\mu}$), again providing a single number 
condition.  Neither the vanishing of ${\bf T}_{\mu}^{~\mu}$ nor the 
conservation of ${\bf J}^{\mu}$ imply corresponding conditions on the  
operators $T_{\mu}^{~\mu}$ or $ J^{\mu}$.  Thus in a suitably constructed 
unified trace dynamics theory of the forces, it is possible that either  
scale invariance [44] or supersymmetry 
could provide a single number constraint forcing the vanishing of the 
cosmological constant, without simultaneously forcing the emergent 
quantum field theory to be either massless or exactly supersymmetric.  
\vfill\eject

\bigskip
\centerline{Acknowledgements}
\bigskip    
I have many people to thank for their assistance in aspects of this work.  
The discovery by my thesis student Andrew Millard   
of the conservation of $\tilde C$ 
provided the underpinning for the entire project.  I  am
greatly indebted to him, and to my other collaborators in the course of 
parts of this work, Gyan Bhanot, Dorje Brody, Todd Brun, Larry Horwitz, 
Lane Hughston, Achim Kempf, Indrajit Mitra, John Weckel, and Yong-Shi Wu.  
I am grateful to Angelo Bassi, 
Todd Brun, and Lane Hughston for many insightful comments 
on the first draft of this book.  
I have benefited 
from conversations and/or email correspondence with a great many people; 
a list (undoubtedly incomplete) includes:  Jeeva Anandan, Philip Anderson, 
John Bahcall, 
Vijay Balasubramanian, 
Angelo Bassi, 
Lowell Brown, Tian-Yu Cao, Sudip Chakravarty, Laslo Di\'osi, Freeman Dyson, 
Sheldon Goldstein, 
GianCarlo Ghirardi, Siyuan Han, William Happer, 
Abraham Klein, John Klauder, Pawan Kumar, Joel Lebowitz, James Lukens, 
G. Mangano, 
Ian Percival, Michael Ramalis, Soo-Jong  
Rey, Lee Smolin, Yuri Suhov, Leo Stodolsky, Terry Tao, Charles Thorn,   
Sam Treiman, Walter Troost, Frank Wilczek, David Wineland, and Edward Witten.  

I also wish to acknowledge the hospitality of the Aspen Center for Physics,  
and of both the Department of Applied Mathematics and Theoretical Physics 
and Clare Hall at  
Cambridge  University, as well as my home base at the Institute for 
Advanced Study in Princeton.  My Albert Einstein Professorship 
there is partially funded by 
the State of New Jersey, and my work is also supported in part by the 
Department of Energy under Grant No. DE-FG02-90ER40542.

\bigskip
\vfill
\eject
\overfullrule=0pt
\twelvepoint
\doublespace
\pageno=131
\centerline{Appendices}
\bigskip    
To keep the discussion of this book self-contained, a number of topics 
that are briefly mentioned in the text are treated in more detail in the 
appendices that follow.  Appendices A through F deal with issues related to 
trace dynamics and our argument that it leads to an emergent quantum theory. 
Appendices G, H, and I give a survey of the mathematical and phenomenological 
aspects of the energy-driven stochastic Schr\"odinger equation.

The notation of the appendices follows that of the text, except in 
Appendix I, where some changes of notation are introduced in order to 
make contact with the conventions used in the relevant literature.  
Throughout this book, we indicate sums explicitly, {\it except} that 
the usual Einstein summation convention is used for sums over Greek letter 
four-vector and tensor indices.  Our Minkowski metric convention is 
$\eta_{\mu\nu}={\rm diag}(1,1,1,-1)$, and we have taken the velocity of 
light to be unity, so that $c$ does not appear in the equations.  However, 
Planck's constant $\hbar$ is retained throughout (because our approach 
implies that it has a dynamical origin), except that in the formulas 
of Appendix I we set $\hbar$ equal to unity.  
\bigskip
\centerline{Appendix A:  Modifications in Real 
and Quaternionic Hilbert Space}
\bigskip
In a complex Hilbert space the scalars that are used to 
form superpositions of  
Hilbert space vectors are complex numbers.  In real Hilbert space the  
scalars take only real number values, while in quaternionic Hilbert 
space the scalars can be quaternions of the form $r_0 + r_1 i 
+ r_2 j + r_3 k$, with $r_{0,1,2,3}$ real, and with $i,j,k$ the quaternion 
imaginary units obeying the noncommutative algebra  
$i^2=j^2=k^2=-1$ and $ij=-ji=k, \, jk=-kj=i, \,
ki=-ik=j$.  The distinguishing feature of complex Hilbert space is that 
there is an  anti-self-adjoint $c$-number $i1$ that commutes with all 
operators on Hilbert space, and the trace can 
take complex values.  By contrast, in real and in quaternionic Hilbert 
space, there exists no anti-self-adjoint $c$-number, since  
the only operators that 
commute with all operators on Hilbert space 
are of the form $r1$, with $r$ real, and hence are self-adjoint.  Also, 
in real and in quaternionic Hilbert space, the trace is real.  
In real Hilbert
space the reality of the trace is self-evident. In quaternionic Hilbert 
space it follows from the fact that the diagonal sum $\sum_n (B_1B_2)_{nn}$ 
does not obey the cyclic property of Eq.~(1.1a) as a result of the 
noncommutativity of the matrix elements $(B_1)_{mn}$ and $(B_2)_{nm}$ 
in the quaternion algebra; so the trace in the quaternionic case must be 
defined [45] as ${\rm Tr} {\cal O}={\rm Re} \sum_n {\cal O}_{nn}$, which 
does obey the cyclic properties of Eqs.~(1.1a--c).  

Because there is no anti-self-adjoint $c$-number in real and quaternionic 
Hilbert space, in these cases one cannot make the choice of Eq.~(2.18d) 
for the matrix $A_{rs}$ that appears in the fermion kinetic term. Instead, 
this matrix must have a nontrivial structure in its indices $rs$; 
the simplest 
case, which corresponds to doubling the number of fermion species, arises 
from taking $A_{rs}=(i\tau_2)_{rs}$, with $\tau_2$ the standard Pauli matrix. 
The main results of Secs.~2-6 generalize to the real and quaternionic 
cases, except for those that depend on the fact that the complex trace can 
have a nonzero imaginary part, or on the fact that in complex Hilbert 
space $i$ acts as an anti-self-adjoint $c$-number.  An example of the 
former is the derivation leading from Eq.~(5.30a) to Eq.~(5.33a), while 
examples of the latter are the discussions in Secs.~4E and 5E that use the 
diagonalization of $i_{\rm eff}$ into $\pm i 1_K$ sectors, and the 
Brownian motion discussion of Sec. 5F, which depends on the existence 
of a complex $c$-number ${\cal K}$.  Further details of the results that 
do generalize, including the Liouville theorem discussion of Sec.4A,  
are found in the Appendices of Ref. [7].  

\vfill\eject 
\centerline{Appendix B: Algebraic Proof of the Jacobi Identity for the} 
\centerline{Generalized Poisson Bracket}
\bigskip
We give here a basis-independent, algebraic proof of the Jacobi 
identity for the generalized Poisson bracket [2].    For ease 
of exposition, we shall use a more compact notation 
than was employed in Sec.~1.  Derivatives with respect to 
$q_r$ and $p_r$ of a total trace functional 
${\bf A}$ will be denoted by ${\bf A}_r$ and ${\bf A}^r$ respectively.  The
operation ${\rm Tr}$ will be implied by a parenthesis $(~)$; this 
means that we can cyclically permute the factors within a parenthesis, 
if we include a factor $\epsilon_r$ every time a $q_r$ or $p_r$ is moved 
from the front of a parenthesis to the back, with $\epsilon_r=1 (-1)$ for 
bosonic (fermionic) degrees of freedom.  Thus, in our shorthand 
notation, $(q_r{\cal O})=\epsilon_r({\cal O}q_r)$, and
the generalized Poisson bracket is given by
$$
{\bf \{A,B\}}= \sum_r~ \epsilon_r\left( {\bf A}_r {\bf B}^r -
                 {\bf B}_r {\bf A}^r \right).
\eqno(B.1)
$$
 
It is useful to illustrate with an example how derivatives are computed.
Consider the case where we have two kinds of matrix variables
$q_1,p_1$ and $q_2,p_2$.  Given the total trace functional
${\bf A} =  \left( q_1 p_1 q_2 q_1 p_2 q_1 \right)$, its
derivative with respect to $q_1$ is denoted by
${\bf A}_1$ and is given by
$$
{\bf A}_1 = q_1 p_1 q_2 q_1 p_2 + \epsilon_1 \epsilon_2 p_2 q_1 q_1 p_1 q_2
+ \epsilon_1 p_1 q_2 q_1 p_2 q_1~.
\eqno(B.2)
$$
The three terms result from the three possible $q_1$ factors to 
differentiate, and the $\epsilon$ factors come from 
cyclically permuting the matrix factors to
bring the particular $q_1$ which is to be differentiated to the right.
 
The first term on the left hand side of the Jacobi identity of 
Eq.~(1.13a), expanded out in this
notation, is 
$$ {\bf\{A,\{B,C\}\}} = \sum_r~\{{\bf A},\epsilon_r
\left( {\bf B}_r {\bf C}^r -
                 {\bf C}_r {\bf B}^r \right)\}~,
\eqno(B.3a)
$$
which can be expanded further to
$$
{\bf\{A,\{B,C\}\}} = \sum_{r,s}~\epsilon_r\epsilon_s \left(
{\bf A}_s \left({\bf B}_r{\bf C}^r\right)^s-
{\bf A}_s\left({\bf C}_r{\bf B}^r\right)^s-
\left({\bf B}_r{\bf C}^r\right)_s{\bf A}^s
+\left({\bf C}_r{\bf B}^r\right)_s{\bf A}^s
\right)~.
\eqno(B.3b)
$$
Cyclic permutations of ${\bf A, B}$, and ${\bf C}$ give the other two terms 
in Eq.~(1.13a).  Thus, the left hand side of Eq.~(1.13a) is
$$\eqalign{
\sum_{r,s}\epsilon_r\epsilon_s[&\left(
{\bf A}_s \left({\bf B}_r{\bf C}^r\right)^s
-{\bf A}_s\left({\bf C}_r{\bf B}^r\right)^s-
\left({\bf B}_r{\bf C}^r\right)_s{\bf A}^s
+\left({\bf C}_r{\bf B}^r\right)_s{\bf A}^s
\right)  \cr
+&\left(
{\bf B}_s \left({\bf C}_r{\bf A}^r\right)^s
-{\bf B}_s\left({\bf A}_r{\bf C}^r\right)^s-
\left({\bf C}_r{\bf A}^r\right)_s{\bf B}^s
+\left({\bf A}_r{\bf C}^r\right)_s{\bf B}^s
\right)\cr
+&\left(
{\bf C}_s \left({\bf A}_r{\bf B}^r\right)^s
-{\bf C}_s\left({\bf B}_r{\bf A}^r\right)^s-
\left({\bf A}_r{\bf B}^r\right)_s{\bf C}^s
+\left({\bf B}_r{\bf A}^r\right)_s{\bf C}^s
\right)]~~~.\cr
}\eqno(B.4)$$
 
Let us first consider how the terms in Eq.~(B.4) cancel in the classical,
$c$-number case.  A similar cancellation mechanism will 
also apply in the more
general matrix operator case.  For $c$-numbers, the trace 
operation is trivial,
derivatives of functionals commute, and one can apply the 
Leibnitz product rule
to expand the terms.  For instance, 
$$\left({\bf B}_r{\bf C}^r\right)^s = {\bf B}_r^{~s}{\bf C}^r+
{\bf B}_r{\bf C}^{rs}~.
\eqno(B.5)
$$
Note that ${\bf B}_r^{~s}$ means that the $q_r$ derivative 
is applied before the
$p_s$ derivative.  ${\bf B}_{~r}^s$ would mean that the same derivatives 
are applied in the opposite order.  This distinction is
meaningless for $c$-number fields, where derivatives commute, but it is
crucial for noncommutative operators $\{q_r\}$ and $\{p_r\}$.
 
Equation~(B.5) implies that each summand term in Eq.~(B.4) will generate 
two terms.  These terms cancel in pairs in the $c$-number case.  
{}For example, in the first term in Eq.~(B.4),
consider the derivative with respect to $p_s$ applied to ${\bf B}_r$.
This generates the term $+{\bf A}_s{\bf B}_r^{~s}{\bf C}^r$.  This
cancels against the term $-{\bf A}_r{\bf B}_{~s}^{r}{\bf C}^s$
obtained by applying the derivative with respect to $p_s$ on ${\bf B}_r$ in
the eleventh term (the dummy indices $r$ and $s$ need to be interchanged 
for the terms
to be the same).  The other half of the eleventh term will in turn be
cancelled by a part of the eighth term, and so on.  After twelve such
double terms have been computed, we come back to the beginning and all
terms have been cancelled.
 
The order in which these cancellations occur classically in the summand 
of Eq.~(B.4) is as follows,
$$
\eqalign{
\longleftrightarrow
&({\bf A}_s \left( {\bf B}_r{\bf C}^r \right)^s)
\longleftrightarrow
(\left( {\bf A}_s{\bf B}^s \right)_r {\bf C}^r)
\longleftrightarrow
(\left( {\bf A}_r{\bf C}^r \right)_s {\bf B}^s)
\longleftrightarrow
({\bf A}_r \left( {\bf C}_s{\bf B}^s \right)^r)
\longleftrightarrow
\cr &({\bf C}_s \left( {\bf A}_r{\bf B}^r \right)^s)
\longleftrightarrow
(\left( {\bf C}_s{\bf A}^s \right)_r {\bf B}^r)
\longleftrightarrow
(\left( {\bf C}_r{\bf B}^r \right)_s {\bf A}^s)
\longleftrightarrow
({\bf C}_r \left( {\bf B}_s{\bf A}^s \right)^r)
\longleftrightarrow
\cr &({\bf B}_s \left( {\bf C}_r{\bf A}^r \right)^s)
\longleftrightarrow
(\left( {\bf B}_s{\bf C}^s \right)_r {\bf A}^r)
\longleftrightarrow
(\left( {\bf B}_r{\bf A}^r \right)_s {\bf C}^s)
\longleftrightarrow
({\bf B}_r \left( {\bf A}_s{\bf C}^s \right)^r)
\longleftrightarrow~~~, \cr
}\eqno(B.6)$$
where we have used the fact that $r$ and $s$ are dummy indices and
have interchanged them in some of the terms, and where the lower right 
of Eq.~(B.6) links back to the upper left.  By Eq.~(B.5), each entry in 
Eq.~(B.6) generates two terms; one of these 
cancels against a term from the entry to the immediate left in the 
chain, and the other cancels against a term from the entry to the 
immediate right.
 
We will now proceed to show that in the general operator case, the
cancellations occur in a similar way.  However, the absence of both
commutativity and the Leibnitz product rule for 
operators makes the proof a little less
trivial.  For the rest of this discussion, we focus, as in Eq.~(B.6), on 
the summands which appear, 
summed over $r$ and $s$, in the Jacobi identity.  Also, we will assume 
that ${\bf A}$, ${\bf B}$,
${\bf C}$ are monomials in $\{q_r\}$ and $\{p_r\}$.  The proof for 
the general
case of polynomial functionals follows from expanding out the
generalized Poisson brackets in Eq.~(1.13a) in terms of monomials.  
 
When one computes the derivative of some monomial
with respect to $q_r$ (say), each particular occurrence of $q_r$ generates
one term in the result.  Consider the expression
$$
\left({\bf B}_r{\bf C}^r\right)^s~,
\eqno(B.7)
$$ 
which appears in the first entry of Eq.~(B.6).  In this
expression, there are three derivatives, and there is a sum over the set 
of choices of which
occurrence of $q_r$, $p_r$, and $p_s$ is differentiated in the
appropriate factors.  Each one of the set of choices will 
produce a particular
monomial term in the result.  If $q_r$ appears $N({\bf B},q_r)$ times
in the monomial ${\bf B}$, and $p_r$ appears $N({\bf C},p_r)$ times in 
${\bf C}$,
and so on, then the number of terms produced by Eq.~(B.7) is at most $N({\bf
B},q_r)N({\bf C},p_r)[N({\bf B},p_s)+N({\bf C},p_s)]$.
 
We will show that in Eq.~(B.4), each such monomial term in the result, for 
fixed $r$,
$s$ (i.e., for a fixed choice of $q_r, p_r, q_s, p_s$), will cancel
with its counterpart in the order defined by Eq.~(B.6).  Consider the 
case where the $p_s$ derivative is applied to ${\bf B}$
in the first entry and the $q_r$ derivative is applied to ${\bf B}$ in the
second entry of Eq.~(B.6).  For these to give nonvanishing 
contributions, ${\bf B}$
must contain at least one instance of both $q_r$ and $p_s$.  Therefore
the most general form for ${\bf B}$ is
$$
{\bf B} = (\alpha q_r\beta p_s)~,
\eqno(B.8)
$$ where $\alpha$ and $\beta$ are arbitrary monomials (and could
possibly contain $q_r$ and $p_s$).
The displayed $q_r$ and $p_s$
are the particular instances of these coordinates in ${\bf B}$ upon which
the derivatives will act.
 
We have
$$
\eqalignno{({\bf A}_s({\bf B}_r{\bf C}^r)^s)
&=({\bf A}_s((\alpha q_r\beta p_s)_r{\bf C}^r)^s)\cr
&= \epsilon_{\alpha}\epsilon_r ({\bf A}_s(\beta p_s\alpha {\bf 
C}^r)^s)\cr
&= \epsilon_{\alpha}\epsilon_r \epsilon_{\beta} \epsilon_s
({\bf A}_s\alpha {\bf C}^r\beta)~,&(B.9)\cr}
$$
and
$$
\eqalignno{(({\bf A}_s{\bf B}^s)_r{\bf C}^r)
&=(({\bf A}_s(\alpha q_r\beta p_s)^s)_r {\bf C}^r)\cr
&= (({\bf A}_s\alpha q_r \beta)_r{\bf C}^r )\cr
&= \epsilon_{\beta}(\beta {\bf A}_s \alpha {\bf C}^r)\cr
&=({\bf A}_s\alpha {\bf C}^r\beta)~. &(B.10)\cr}
$$
If ${\bf B}$ is not identically zero [in which case the 
equality of Eqs.~(B.9) and 
(B.10) is trivial], it
must have an even number of fermion factors.  Therefore,
$\epsilon_{\alpha}
\epsilon_r\epsilon_{\beta}\epsilon_s=1$,
and so the right--hand sides of Eqs.~(B.9) and (B.10) are always the same.  
{}Finally, these same cancellations can be
shown to occur for every summand term in Eq.~(B.4) in the order indicated by 
Eq.~(B.6), and apply both to the summands with $r\not= s$ and to those
with $r=s$, including the parts of the summands with $r=s$ in which 
there are two derivatives with respect to the same variable $q_r$ (or 
$p_r$).  This proves that the
Jacobi identity is true for arbitrary bosonic and fermionic matrix 
operator variables $\{q_r\}$ and $\{p_r\}$.
\bigskip
\centerline{Appendix C: Symplectic Structures in Trace Dynamics}
\bigskip
We shall demonstrate here that there is a close correspondence [4] between 
the tangent vector field and symplectic structures of trace dynamics and 
of classical mechanics [46].  Let $X_{\bf A}$ be the tangent vector field 
associated with a trace functional ${\bf A}$, defined as a formal 
derivative operator by 
$$X_{\bf A}\equiv 
{\rm Tr}\left[ \sum_r \left( \epsilon_r
{\delta {\bf A} \over \delta q_r}{\delta  \over \delta p_r}
-{\delta {\bf A} \over \delta p_r}{\delta  \over \delta q_r} \right) \right]
~~~,\eqno(C.1)$$ 
which by definition acts on any trace functional ${\bf B}$ as 
$$X_{\bf A} {\bf B}= {\bf B}X_{\bf A} + (X_{\bf A} {\bf B})~~~,\eqno(C.2)$$
with $(X_{\bf A}{\bf B})$ given by (cf. Eq.~(1.11a))
$$\eqalign{
(X_{\bf A}{\bf B})=&
{\rm Tr}\left[ \sum_r \left( \epsilon_r
{\delta {\bf A} \over \delta q_r}{\delta{\bf B} \over \delta p_r}
-{\delta {\bf A} \over \delta p_r}{\delta{\bf B} \over \delta q_r} 
\right) \right]   \cr
=&
{\rm Tr} \sum_r \epsilon_r \left(
{\delta {\bf A} \over \delta q_r}{\delta {\bf B} \over \delta p_r}
-{\delta {\bf B} \over \delta q_r} {\delta {\bf A} \over \delta p_r} \right)
=\{{\bf A}, {\bf B} \} ~~~.
}\eqno(C.3)$$
In terms of this operator, the time development of a general trace 
functional ${\bf B}[\{q_r\},\{p_r\}]$ with no 
intrinsic time dependence, under the dynamics 
governed by ${\bf A}$ as trace Hamiltonian, can be written as (cf. 
Eq.~(1.11b)) 
$${d {\bf B} \over dt}=-(X_{\bf A}{\bf B})~~~.\eqno(C.4)$$
Thus the tangent vector field $X_{\bf A}$ can be viewed as (minus) the 
directional derivative along the time evolution orbit  (called the phase 
flow in [46]) of the phase space point $(\{q_r\},\{p_r\})$, which is  
determined by the Hamiltonian equations of motion 
$$\dot p_r=-{\delta {\bf A} \over \delta q_r}~,~~~
\dot q_r=\epsilon_r{\delta {\bf A} \over \delta p_r}~,~~~\eqno(C.5)$$
with ${\bf A}$ acting as the total trace Hamiltonian and with the dot 
denoting a time derivative. Following the terminology of classical  
mechanics [46], we call a tangent vector field of the form of 
Eq.~(C.1) a Hamiltonian vector field.  It is easily verified that the 
directional derivative $X_{\bf A}$ obeys the Leibnitz product rule 
when applied to the generalized Poisson bracket, 
$$(X_{\bf A}\{{\bf B},{\bf C}\})=\{(X_{\bf A}{\bf B}),{\bf C}\}
+\{{\bf B},(X_{\bf A} {\bf C})\}~~~,\eqno(C.6)$$
because this equation is equivalent to the Jacobi identity of Eq.~(1.13a).

Let us now study the algebraic structure of Hamiltonian vector fields, by 
computing the action of the commutator of two tangent vector fields 
$X_{\bf A}$ and $X_{\bf B}$ on a third trace functional ${\bf C}$, 
$$\eqalign{
([X_{\bf A},X_{\bf B}] {\bf C})=&(X_{\bf A}(X_{\bf B}{\bf C})) 
- (X_{\bf B}(X_{\bf A}{\bf C}))  \cr
=&\{ {\bf A}, \{ {\bf B},{\bf C}\} \}-\{ {\bf B}, \{ {\bf A},{\bf C}\} \} \cr
=&\{ {\bf A}, \{ {\bf B},{\bf C}\} \}+\{ {\bf B}, \{ {\bf C},{\bf A}\} \}
~~~.\cr}\eqno(C.7)$$
Using Eq.~(C.3) with ${\bf A}$ replaced by $\{ {\bf A},{\bf B} \}$ and 
${\bf B}$ replaced by ${\bf C}$, we also get
$$(X_{ \{ {\bf A},{\bf B} \} }{\bf C})=\{\{ {\bf A},{\bf B}\},{\bf C}\}\}~~~,
\eqno(C.8)$$
and subtracting Eq.~(C.8) from Eq.~(C.7) gives finally 
$$\eqalign{
&(([X_{\bf A},X_{\bf B}]-X_{ \{ {\bf A},{\bf B} \} } ){\bf C})\cr
=&~~~~~~\{ {\bf A},\{ {\bf B},{\bf C} \}\}    
+\{ {\bf C},\{ {\bf A},{\bf B} \}\}    
+\{ {\bf B},\{ {\bf C},{\bf A} \}\} =0~~~.\cr
}\eqno(C.9)$$   
Hence the validity of the Jacobi identity for the generalized Poisson 
bracket implies that the Hamiltonian vector fields $X_{\bf A}$ obey the 
commutator algebra
$$[X_{\bf A},X_{\bf B}]=X_{ \{ {\bf A},{\bf B} \} }~~~,\eqno(C.10)$$
and thus form a Lie algebra that is isomorphic to the Lie algebra of 
trace functionals under the generalized Poisson bracket.  This gives 
a trace dynamics analog of a standard result [46] in classical mechanics.  

We next show that the symplectic geometry of classical mechanics extends 
to trace dynamics, and, as in the classical case, it is preserved by 
phase space flows produced by Hamiltonian time evolutions.  Symplectic 
geometry is defined by an antisymmetric metric in the tangent or 
cotangent spaces of a phase space.  (This contrasts with with Riemannian 
geometry, which is defined by a symmetric metric in the tangent or 
cotangent spaces of a manifold.)  To avoid differential forms, let us 
work in the cotangent space, which is spanned by covariant vectors the 
components of which form the gradient of a function on phase space.  The 
standard symplectic metric, or inner product, between two 
classical functions 
on phase space is provided by their classical Poisson bracket.  The analogs 
of classical functions in trace dynamics are trace functionals, with 
differentials given by the phase space version of Eq.~(1.3b), 
$$\delta {\bf A}={\rm Tr} \sum_r \left( {\delta {\bf A} \over \delta q_r} 
\delta q_r      + {\delta {\bf A} \over \delta p_r} \delta  p_r \right)~~~.
\eqno(C.11)$$
We can then use the generalized Poisson bracket to define a generalized 
symplectic structure $\Omega$ on the operator phase space, by defining 
the inner product between two cotangent vectors $\delta{\bf A}$ 
and $\delta{\bf B}$ by 
$$\Omega(\delta{\bf A},\delta{\bf B})\equiv\{ {\bf A}, {\bf B} \} 
~~~.\eqno(C.12)$$

To see that this symplectic structure is preserved by the Hamiltonian  
dynamics given by Eq.~(C.4), we observe that the time derivative of the 
inner product along the phase flow is 
$${d \over dt} \Omega(\delta {\bf B}, \delta {\bf C} )= 
{d \over dt} \{ {\bf B}, {\bf C} \}= \{ \{ {\bf B}, 
{\bf C} \} ,{\bf A} \}~~~,
\eqno(C.13a),$$ 
while that of the differential $\delta {\bf B}$ along the same flow is 
$${d \over dt} \delta {\bf B} =\delta \dot {\bf B}~~~.\eqno(C.13b)$$
Therefore we have 
$$\eqalign{
\Omega(\delta \dot {\bf B}, \delta {\bf C})
+&\Omega(\delta {\bf B}, \delta \dot {\bf C})
=\{ \dot {\bf B}, {\bf C} \} + \{ {\bf B}, \dot {\bf C} \} \cr
=&\{\{ {\bf B},{\bf A}\}, {\bf C} \} +\{ {\bf B},\{ {\bf C}, {\bf A} \} \}
~~~,\cr}\eqno(C.13c)$$
which comparing with Eq.~(C.13a) and using the Jacobi identity of Eq.~(1.13a) 
implies that 
$$ {d \over dt} \Omega(\delta {\bf B}, \delta {\bf C} )= 
\Omega(\delta \dot {\bf B}, \delta {\bf C})
+\Omega(\delta {\bf B}, \delta \dot {\bf C})  ~~~.\eqno(C.14)$$
In other words, the symplectic structure is invariant under Hamiltonian 
phase flow.  This statement can be viewed as a dual form of the Liouville 
theorem for trace dynamics.  

Thus, trace dynamics,  with noncommuting operator phase space 
variables, nonetheless has an underlying symplectic geometry which is 
preserved by the time evolution generated by any trace Hamiltonian, or 
equivalently, by the flow corresponding to  
any general canonical transformation as defined in Eq.~(2.13a).  
This is due to the existence of a graded trace ${\rm Tr}$, that permits 
cyclic permutation of the noncommuting operator variables, which implies 
the validity of the Jacobi identity for the generalized Poisson bracket.  
Hence, in analogy with classical mechanics,  the basic concepts 
and theorems of trace dynamics will be invariant under the group of 
transformations that preserve its generalized symplectic structure.  
\bigskip
\centerline{Appendix D: Gamma Matrix Identities for Supersymmetric} 
\centerline{Trace Dynamics Models}
\bigskip
We give here the gamma matrix identities needed for carrying out the 
calculations involving supersymmetric trace dynamics models 
sketched in Secs. 3A, 3B, and 3C.  

{}For the calculations 
of Secs. 3A and 3B, it is convenient to use Majorana representation 
$\gamma$ matrices constructed explicitly as follows.  
Let $\sigma_{1,2,3}$ and $\tau_{1,2,3}$ be two 
independent sets of Pauli spin matrices; then we take 
$$\eqalign{
\gamma^0=&-\gamma_0=-i\sigma_2 \tau_1 ~~~,\cr
\hat \gamma^0=&i \gamma^0=\sigma_2 \tau_1 ~~~,\cr
\gamma^1=&\gamma_1=\sigma_3 ~~~,\cr 
\gamma^2=&\gamma_2=-\sigma_2\tau_2  ~~~,\cr
\gamma^3=&\gamma_3=-\sigma_1 ~~~,\cr
\gamma_5=&i\gamma^1\gamma^2\gamma^3\gamma^0=-\sigma_2\tau_3 ~~~,\cr
\hat \gamma^0 \gamma_5=& i\tau_2 ~~~, \cr
}\eqno(D.1a)$$
so that $\hat\gamma^0,\gamma_5,\hat\gamma^0\gamma_5$ are skew symmetric 
and $\gamma^1,\gamma^2,\gamma^3$ are symmetric, and 
$$  \hat \gamma^0 \gamma^{\mu T} \hat \gamma^0=-\gamma^{\mu}  ~~~.
\eqno(D.1b)$$
{}For this choice of $\gamma$ matrices, the four matrices $\gamma^{\mu}$ are 
real.  

To prove supersymmetry of the trace dynamics version of the 
Wess-Zumino model, one uses cyclic invariance of the trace 
together with the cyclic identity valid for Majorana representation 
$\gamma$ matrices, 
$$\sum_{{\rm cycle}~ a \rightarrow b \rightarrow d \rightarrow a} 
[\hat\gamma^0_{ab} \hat\gamma^0_{cd} + (\hat \gamma^0 \gamma_5)_{ab}
(\hat \gamma^0 \gamma_5)_{cd}]=0~~~.\eqno(D.2)$$
To prove supersymmetry of the trace dynamics version of the 
supersymmetric Yang-Mills model, one uses cyclic invariance of the trace 
together with another cyclic identity valid for Majorana representation 
$\gamma$ matrices, 
$$\sum_{{\rm cycle}~ a \rightarrow b \rightarrow d \rightarrow a} 
(\hat\gamma^0 \gamma^{\mu})_{ab} (\hat\gamma^0 \gamma_{\mu})_{cd} 
=0~~~.\eqno(D.3)$$
To verify closure of the supersymmetry algebra under the generalized 
Poisson bracket, one can proceed in either of two ways. The first is to  
directly rearrange into the expected form, verifying 
along the way the various 
$\gamma$ matrix identities that are needed; for example, in the case of 
the Wess-Zumino model, one needs the cyclic identity of Eq.~(D.2) 
together with the additional identity 
(with $\ell,m,n$ spatial indices, and $\epsilon_{\ell m n}$ 
the three index antisymmetric tensor with $\epsilon_{123}=1)$
$$\eqalign{
&\gamma^{\ell}_{ab}\hat \gamma^0_{cd}+\gamma^{\ell}_{db}\hat \gamma^0_{ca}
-(\gamma^{\ell}\gamma_5)_{ab}(\hat \gamma^0 \gamma_5)_{cd} 
-(\gamma^{\ell} \gamma_5)_{db}(\hat \gamma^0 \gamma_5)_{ca}  \cr
&=\delta_{ad}(\hat \gamma^0 \gamma_{\ell})_{bc}
-(\hat \gamma^0 \gamma_{\ell})_{ad} \delta_{bc} 
+\epsilon_{\ell mn}(\gamma_{\ell}\gamma_m\gamma_5)_{ad}
(\gamma_{\ell}\gamma_n)_{cb}  ~~~.\cr
}\eqno(D.4)$$
An alternative method for verifying the closure of the supersymmetry 
algebra is to first Fierz transform using the standard Fierz identity 
given in Eq.~(A.80) of the book of West [11], so as to isolate expressions of 
the form $\alpha^T \Gamma \beta$, and then to show that the coefficients 
of the various terms of this type, with different Dirac matrix 
structures $\Gamma$, have the form required by closure.  

The identities of Eqs. (D.2-4) are representation 
covariant, in that they do not take the same form in representations in 
which the Dirac gamma matrices are complex rather than real.  To see this, 
we note that the matrices in a general representation $\gamma^{\mu}_G$ 
are related to the Majorana representation matrices $\gamma^{\mu}$ 
given above by
$$\gamma^{\mu}_G=U^{\dagger} \gamma^{\mu} U=U^{T*} \gamma^{\mu} U 
~~~,\eqno(D.5)$$
with $U$ a unitary matrix which in general is complex, as a result of   
which the row and column indices transform with different matrices.  
However, the identities of Eqs.~(D.2-4) mix row and column indices; for  
example, in Eq.~(D.2) there is one term in the cyclic sum in which $a$ is 
a row index, and two terms in which $a$ is a column index.  (By way of 
contrast, the more familiar Fierz identities only interchange two row 
indices, and so do not mix row and column indices.)  Hence we cannot get 
a representation invariant form of the identity by two applications of 
Eq.~(D.5), since in the second and third terms of the cyclic sum, we 
will have a row index contracted with a $U$ and a column index contracted 
with a $U^*$, which does not correspond with Eq.~(D.5).  However,   
we can easily get a representation covariant form of Eq.~(D.2) by contracting 
all indices with a $U^*$, and wherever $U^*$ contracts with a column 
index using the identity 
$$U^*=UU^{*T}U^* =U\gamma^*~~~,\gamma \equiv U^TU ~~~,\eqno(D.6)$$
with $\gamma$ a matrix that appears on pp. 341-342 of the book of Adler  
[1] (which is introduced there because it plays a role in the 
transformation properties of the 
Dirac equation in quaternionic quantum mechanics).  
We can then apply Eq.~(D.5) to all the gamma matrices, giving 
for Eq.~(D.2), for example, the representation covariant form 
$$\sum_{{\rm cycle}~a \rightarrow b \rightarrow d \rightarrow a} 
[(\hat \gamma^0\gamma^*)_{ab}  (\hat \gamma^0\gamma^*)_{cd}
+(\hat \gamma^0\gamma_5\gamma^*)_{ab}(\hat \gamma^0\gamma_5\gamma^*)_{cd}]
=0~~~.\eqno(D.7)$$
{}For a change of representation which preserves reality of the $\gamma$ 
matrices, we have $U^*=U,~~~\gamma=U^TU=U^{*T}U=1$, and 
Eq.~(D.7) is identical to Eq.~(D.2), but for general changes of 
representation the identity is form covariant but not form invariant.  

{}For the calculations in the ``M Theory'' model of Sec.~3C, one uses 
a set $\gamma_i$ of nine 
$16 \times 16$ matrices, that are 
related [14] to the standard 
$32 \times 32$ matrices $\Gamma^{\mu}$ as well as to the Dirac matrices of 
spin(8). A number of properties of the real, symmetric 
matrices $\gamma_i$ play a role in the calculation.   These 
matrices satisfy the anticommutator algebra 
$$\{ \gamma_i, \gamma_j \} =2 \delta_{ij} ~~~,\eqno(D.8)$$
as well as the cyclic identity  
$$\sum_{{\rm cycle}~~ p \rightarrow q \rightarrow n \rightarrow p} 
(\delta^{mn}\delta^{pq}-
\gamma_i^{mn}\gamma_i^{pq})=0~~~,\eqno(D.9)$$
with $i$ again summed over and with the indices $m,n,p,q$ spinorial indices 
ranging from 1 to 16. (The identity of Eq.~(D.9)  also has 
the same spinor index appearing both as a row and as a column index, 
and so is only form covariant under changes of gamma matrix 
representation, and  
is obtained  by chiral projection with ${1 \over 2}(1-i\Gamma^9)$ 
from Eq.~(4.A.6) of [14].)   Defining 
$$\gamma_{ij}={1 \over 2} [\gamma_i, \gamma_j] ~~~,\eqno(D.10a)$$
so that 
$$\gamma_i \gamma_j = \delta_{ij}+\gamma_{ij}~~~,\eqno(D.10b)$$
one readily derives from  Eq.~(D.9) an identity given in  [47], 
$$\gamma_{ij}^{mn}\gamma_i^{pq}+\gamma_{ij}^{pq}\gamma_i^{mn}
+ (m \leftrightarrow p )=2(\gamma_j^{nq} \delta^{mp}-
\gamma_j^{mp}\delta^{nq})
~~~.\eqno(D.10c)$$
By standard gamma matrix manipulations using Eq.~(D.8), one also derives 
the the fact that the matrix 
$$A_{ijk}=\gamma_i\gamma_j\gamma_k-\delta_{ij}\gamma_k+\delta_{ik}
\gamma_j- \delta_{jk} \gamma_i \eqno(D.11)$$
is totally antisymmetric in the indices $i,j,k$ (it is just the 
antisymmetrized product $\gamma_{[i}\gamma_j\gamma_{k]}$ with normalization 
factor ${1\over 6})$, as well as the identity 
$${1\over 2}\{\gamma_{\ell m},\gamma_{ij} \}= \gamma_{[\ell}\gamma_m\gamma_i
\gamma_{j]}
+\delta_{\ell j} \delta_{im}- \delta_{mj}\delta_{i\ell}~~~,\eqno(D.12)$$  
with the first term on the right the antisymmetrized  
product including normalization factor ${1 \over 24}$.  

\bigskip
\centerline{Appendix E: Trace Dynamics Models With Operator Gauge Invariance}
\bigskip

In  Sec. 3B, we studied the trace dynamics version of the 
supersymmetric Yang-Mills model, which is 
an example of a general class of trace dynamics models with 
a local  operator gauge invariance, and a corresponding operator 
constraint.  In Sec. 4D, we discussed  methods for taking this constraint 
into account in forming the canonical ensemble and the partition function.
Lest it appear that operator gauge invariance is linked to supersymmetry, 
we give here 
further examples [1] of trace dynamics models, now non-supersymmetric, 
which admit an operator gauge invariance.   

As our first example, we consider a matrix scalar field 
$\phi$, which is not restricted to be self-adjoint (or 
anti-self-adjoint), and which is subjected to the general local gauging
$$
\phi\to U\phi U^{\prime\dagger}~,~~~~~~UU^\dagger=U^\dagger 
U=U^\prime U^{\prime\dagger}=U^{\prime\dagger}U^\prime =1~~~,
\eqno(E.1a)
$$
with $U$ and $U^{\prime}$ independent unitary matrices.  (Thus, the 
superscript ${}^{\prime}$ (prime) in this Appendix does {\it not} have 
the significance of ``noncommutative part'' as in the discussion of 
Eqs.~(2.16a-c) of Sec.~2.) 
Let us introduce independent anti-self-adjoint 
gauge potentials $B_\mu,B_\mu^\prime$ 
which transform as
$$
B_\mu\to UB_\mu U^\dagger -(\partial_\mu 
U)U^\dagger~,~~~~~~B_\mu^\prime\to U^\prime B_\mu^\prime 
U^{\prime\dagger}-(\partial_\mu U^\prime)U^{\prime\dagger}~~~,
\eqno(E.1b)
$$
and the covariant derivative and field strengths
$$
\eqalign{
D_\mu\phi 
&= \partial_\mu\phi+B_\mu\phi-\phi B_\mu^\prime~~~,\cr
{}F_{\mu\nu} &=\partial_\mu B_\nu-\partial_\nu B_\mu+[B_\mu,B_\nu]~~~,\cr
{}F_{\mu\nu}^\prime &=\partial_\mu B_\nu^\prime - \partial_\nu 
B_\mu^\prime+[B_\mu^\prime,B_\nu^\prime]~~~,
}\eqno(E.1c)
$$
which correspondingly transform as
$$
D_\mu\phi\to UD_\mu\phi U^{\prime\dagger}~,~~~~~~F_{\mu\nu}\to 
UF_{\mu\nu} U^\dagger~,~~~~~~F_{\mu\nu}^\prime\to U^\prime 
{}F_{\mu\nu}^\prime U^{\prime\dagger}~~~.
\eqno(E.1d)
$$
Then the trace Lagrangian density given by 
$$
\eqalign{{\bf {\cal L}}
&={\bf {\cal L}}_\phi +{\bf {\cal L}}_B+{\bf {\cal L}}_{B^\prime}~~~,\cr
{\bf {\cal L}}_\phi &={\rm Tr}\,\left\{{1\over 
2}~[-(D_\mu\phi)^\dagger D^\mu\phi-m^2\phi^\dagger\phi]-{\lambda\over 
4}~(\phi^\dagger\phi)^2\right\}~~~,\cr
{\bf {\cal L}}_B &={\rm Tr}\,\left({1\over 
4g^2}~F_{\nu\mu}F^{\nu\mu}\right)~,~~~~~~{\bf {\cal L}}_{B^\prime}={\rm 
Tr}\,\left({1\over 
4(g^\prime)^2}~F_{\nu\mu}^\prime F^{\prime\nu\mu}\right)~~~,\cr}
\eqno(E.2a)
$$
is gauge invariant, as may be verified by substituting Eqs.~(E.1a-d) and 
using cyclic invariance under the trace.   
The  trace Lagrangian ${\bf L}$ and action ${\bf S}$ are formed 
from ${\bf{\cal L}}$ by the usual recipe
$$
{\bf L}=\int~d^3x{\bf{\cal L}}~,~~~~~~{\bf S}=\int~dt\,{\bf L}~.
\eqno(E.2b)
$$
When we form the Euler-Lagrange equations by varying  ${\bf S}$, 
through $\delta F_{\mu\nu}$ and $\delta F_{\mu\nu}^\prime$ we encounter 
new covariant derivatives $\hat D_\mu$ and $\hat D_\mu^\prime$ defined by
$$
\eqalign{
\hat D_\mu{\cal O}
&=\partial_\mu{\cal O}+[B_\mu,{\cal O}]~,~~~~~~\hat D_\mu^\prime,{\cal 
O}=\partial_\mu{\cal O}+[B_\mu^\prime,{\cal 
O}]~~~,\cr
\delta F_{\mu\nu} &=\hat D_\mu\delta B_\nu-\hat D_\nu\delta 
B_\mu~,~~~~~~\delta F_{\mu\nu}^\prime =\hat D_\mu^\prime\delta 
B_\nu^\prime-\hat D_\nu^\prime\delta B_\mu^\prime~~~,}
\eqno(E.3a)$$
and in integrating by parts we use the following ``intertwining identities'' 
that are easily derived from Eqs.~(E.1c) and (E.3a), 
$$
\eqalign{\hat D_\mu(\rho\eta^\dagger)
&= (D_\mu\rho)\eta^\dagger+\rho(D_\mu\eta)^\dagger~~~,\cr
\hat D_\mu^\prime(\rho^\dagger\eta) &= 
(D_\mu\rho)^\dagger\eta+\rho^\dagger D_\mu\eta~~~,\cr
\partial_\mu {\rm Tr}\,(\rho\eta^\dagger) &={\rm 
Tr}\,[(D_\mu\rho)\eta^\dagger+\rho(D_\mu\eta)^\dagger]~~~,\cr
\partial_\mu{\rm Tr}\,(\rho^\dagger\eta) &={\rm 
Tr}\,[(D_\mu\rho)^\dagger\eta+\rho^\dagger D_\mu\eta]~~~,\cr 
}\eqno(E.3b)
$$
which apply when $\rho$ and $\eta$ are both bosonic 
or both fermionic in type.  
We then get the operator equations of motion 
$$
\eqalign{
~~~~~~~~~~~~~&D_\mu D^\mu\phi-(m^2+\lambda\phi\phi^\dagger)\phi=0~~~,\cr
&\hat D^\mu F_{\nu\mu}=g^2{\cal J}_\nu~,~~~~~~{\cal J}_\nu={1\over 
2}~[\phi(D_\nu\phi)^\dagger-(D_\nu\phi)\phi^\dagger]~~~,\cr
&\hat D^{\prime\mu}F_{\nu\mu}^\prime=g^{\prime 2}{\cal 
J}_\nu^\prime~,~~~~~~{\cal J}_\nu^\prime ={1\over 2}~[\phi^\dagger 
D_\nu\phi-(D_\nu\phi)^\dagger\phi]~~~,\cr
}\eqno(E.3c)$$
in which the $\nu=0$ components of the gauge field equations are 
constraints.

We turn next to the case of  fermion fields, starting again 
with the operator gauging in which there is a fermion  
$\psi$ transforming as
$$
\psi \to U \psi U^{\prime\dagger}~~~.
\eqno(E.4a)
$$
The total trace Lagrangian density analogous to Eqs.~(E.2a) is
$$
{\bf {\cal L}}={\bf{\cal L}}_{\psi}+{\bf{\cal L}}_B+
{\bf{\cal L}}_{B^\prime}~,
\eqno(E.4b)
$$
with ${\bf {\cal L}}_B$ and ${\bf{\cal L}}_{B^\prime}$ as in 
Eq.~(E.2a), and with ${\bf{\cal L}}_{\psi}$ given by
$$\eqalign{
{\bf {\cal L}}_{\psi}=&{\rm Tr}
(-i\psi^\dagger\gamma^0\gamma^\mu D_\mu\psi
+i m \psi^\dagger \gamma^0 \psi )~~~,\cr
D_\mu\psi 
=& \partial_\mu\psi+B_\mu\psi-\psi B_\mu^\prime~~~,\cr
}\eqno(E.4c)$$
This Lagrangian density is again gauge invariant, and varying the 
trace action ${\bf S}$ 
to get the corresponding Euler-Lagrange equations, we find  
the operator equations of motion
$$
\eqalign{
&~~(-\gamma^\mu D_\mu+m)\psi=0~~~,\cr
&\hat D^{\mu} F_{\mu\nu}=g^2{\cal J}_\nu~,~~~~~~{\cal 
J}_\nu=i\psi^T\gamma_\nu^T\gamma^{0T}\psi^{\dagger 
T}~~~,\cr
& \hat D^{\prime\mu}F_{\mu\nu}^\prime=(g^\prime)^2{\cal 
J}_\nu^\prime~,~~~~~~{\cal J}_\nu^\prime=i\psi^\dagger
\gamma^0\gamma_\nu\psi~~~,\cr 
}\eqno(E.4d)$$
with $T$ indicating Dirac index (but {\it not} operator) transposition. 
Since $\psi$ and $\psi^{\dagger}$ are noncommutative matrix operators, 
the current ${\cal J}_\nu$ is not equal to $-{\cal J}_\nu^\prime$, as it 
would be if $\psi,\,\psi^{\dagger}$ were $c$-number Grassmann spinors.  
Again, the $\nu=0$ 
components of the gauge field equations are constraints.  

A further discussion of the bosonic and fermionic models briefly described 
here, including their Hamiltonian form and their discrete symmetries, 
can be found in [1].  
\bigskip
\centerline{Appendix F:  Properties of Wightman Functions Needed}
\centerline{for Reconstruction of Local Quantum Field Theory}
\bigskip
We review here, following [24], the properties of Wightman functions 
that are needed for the reconstruction from them of local quantum field 
theory.  For simplicity, we consider the case of a single self-adjoint 
scalar field $\phi(x)$.  Letting $|{\rm vac}\rangle$ denote the vacuum state, 
which is assumed unique, the Wightman functions are defined by 
$${\cal W}(x_1,x_2,...,x_n)=\langle {\rm vac}|
\phi(x_1)\phi(x_2)...\phi(x_n)
| {\rm vac} \rangle~~~,\eqno(F.1)$$
for all $n$ ranging from 0 to $\infty$.  Starting from the axioms of 
local quantum field theory, a number of properties of these functions 
can be derived.  Conversely, given Wightman functions satisfying the 
following properties, one can reconstruct a local quantum field theory:  
\medskip
\leftline{(i) ~~~Smoothness properties} 
The functions ${\cal W}(\{x\})$ must be tempered distributions. 
\medskip
\leftline{(ii)~~~Covariance} 
The functions ${\cal W}(\{x\})$ must satisfy the requirements of Poincar\'e 
invariance.  Translation invariance requires that 
$${\cal W}(x_1,x_2,...,x_n)=W(x_1-x_2,x_2-x_3,...,x_{n-1}-x_n)
\equiv W(\xi_1,\xi_2,...,\xi_{n-1})\equiv W(\{\xi\})~~~.\eqno(F.2)$$
Lorentz invariance for a scalar field $\phi$ requires that 
$$W(\{\xi\})=W(\{\Lambda \xi\})~~~,\eqno(F.3)$$ 
with $\Lambda_{\mu}^{~\nu}$ a proper orthochronous Lorentz transformation. 
When fields with spin appear in the Wightman functions, Eq.~(F.3) must 
be modified to include the appropriate Wigner rotations acting on the 
spin indices.  
\medskip
\leftline{(iii)~~~Spectral Condition} 
Let $\tilde {\cal W}(p_1,p_2,...,p_n)$ and $\tilde W(q_1,q_2,...,q_{n-1})$ 
be the Fourier transforms of the Wightman functions defined by 
$$\eqalign{
\tilde {\cal W}(p_1,p_2,...,p_n)=&\int dx_1...dx_n 
\exp(-i\sum_{j=1}^n p_j \cdot x_j) {\cal W}(x_1,x_2,...,x_n)~~~,\cr
\tilde W(q_1,q_2,...,q_{n-1})=&\int d\xi_1...d\xi_{n-1} 
\exp(-i\sum_{j=1}^{n-1} q_j \cdot \xi_j ) W(\xi_1,\xi_2,...,\xi_{n-1})~~~.\cr
}\eqno(F.4)$$
These must be related by 
$$\tilde {\cal W}(p_1,p_2,...,p_n)=(2 \pi)^4 \delta(\sum_{j=1}^n p_j)
\tilde W(p_1,p_1+p_2,...,p_1+p_2+...+p_{n-1})~~~,\eqno(F.5a)$$
and we must have
$$\tilde W(q_1,q_2,...,q_{n-1})=0~~~,\eqno(F.5b)$$ 
for any $q_j$ not in the forward light cone defined by $q^0\geq|\vec q\,|$.  
\medskip
\leftline{(iv)~~~Local Commutativity} 
The Wightman functions must obey 
$${\cal W}(x_1,...,x_j,x_{j+1},...,x_n)={\cal W}(x_1,...,x_{j+1},x_j,...,x_n)
~~,~~~~j=1,2,...,n-1~~~,\eqno(F.6)$$
whenever $x_j$ and $x_{j+1}$ are spacelike separated.
\medskip
\leftline{(v)~~~Hermiticity and Positivity Conditions} 
The Wightman functions must obey the Hermiticity condition 
$${\cal W} (x-1,...,x_n)={\cal W}(x_n,...,x_1)^*~~~,\eqno(F.7a)$$ 
where ${}^*$ denotes the complex conjugate.  They must also obey the 
positivity condition 
$$\sum \int...\int dx_1...dx_j dy_1...dy_k f_j(x_1,...,x_j)^* 
{\cal W}(x_j,...,x_1,y_1,...,y_k) f_k(y_1,...,y_k)\geq 0~~~,\eqno(F.7b)$$
for all finite sequences $f_0,f_1(x_1),f_2(x_1,x_2),...$ of test functions.  
\medskip
\leftline{(vi)~~~Cluster Property} 
The Wightman functions must cluster, in the sense that 
$$\lim_{S \to \infty} [{\cal W}(x_1,...,x_j,x_{j+1}+Sa,...,x_n+Sa)
-{\cal W}(x_1,...,x_j)  {\cal W}(x_{j+1},...,x_n)]=0~~~,\eqno(F.8)$$
when the unit four-vector direction $a$ of increasing 
separation $S$ is spacelike,

The proof of the reconstruction theorem, assuming Wightman functions 
obeying the conditions enumerated above, is given in [24], and a recent 
discussion is given in [48].
\vfill\eject
\centerline{Appendix G:  Proof of Reduction with Born Rule Probabilities}
\bigskip
We give here the proof, following ideas in [37,49,28], that Eqs.~(5.51d,e) 
imply state vector reduction to energy eigenstates with the Born rule 
probabilities.  We begin by reviewing 
the derivation of Eq.~(5.51e) from Eq.~(5.51d).   This calculation 
uses the It\^o calculus rules given in Eqs.~(5.49b) and (5.49d), to which 
an excellent expository introduction can be found in [50]. We start from   
the stochastic differential equation for the state vector 
given in Eq.~(5.51d), together with its adjoint, 
$$\eqalign{ 
d|\Psi\rangle  =&
-i \hbar^{-1} H_{\rm eff}|\Psi\rangle dt  
 - {1\over 2}  [\beta_R^2 
(   H_{\rm eff}-\langle   H_{\rm eff} \rangle )^2  
+\beta_I^2 H_{\rm eff}^2]|\Psi\rangle dt \cr
+&\beta_R (   H_{\rm eff}-\langle  H_{\rm eff} \rangle ) 
|\Psi\rangle dW_t^R + i \beta_I H_{\rm eff} |\Psi\rangle dW_t^I ~~~, \cr
&~~~~~~~~~\cr
d\langle \Psi|  =&
\langle \Psi| i \hbar^{-1} H_{\rm eff} dt
 -\langle \Psi| {1\over 2}  [\beta_R^2 
(   H_{\rm eff}-\langle   H_{\rm eff} \rangle )^2  
+\beta_I^2 H_{\rm eff}^2] dt \cr
+&\langle  \Psi| \beta_R (   H_{\rm eff}-\langle  H_{\rm eff} \rangle ) 
 dW_t^R -\langle \Psi| i \beta_I H_{\rm eff}  dW_t^I ~~~. \cr
}\eqno(G.1)$$
Substituting these into 
$$d \hat \rho = (d |\Psi \rangle ) \langle \Psi | 
+|\Psi \rangle (d \langle \Psi |) + (d |\Psi \rangle ) (d \langle \Psi |)
~~~,\eqno(G.2)$$
which follows from Eq.~(5.49d) as applied to the definition $\hat \rho=
|\Psi \rangle \langle \Psi|$,  and using the stochastic calculus rules 
of Eq.~(5.49b), we obtain after some straightforward algebra Eq.~(5.51e), 
which we repeat here for convenience, 
$$d \hat\rho 
= i \hbar^{-1} [\hat \rho,  H_{\rm eff} ] dt
-{1\over 2} |\beta|^2[ H_{\rm eff} ,[ H_{\rm eff},\hat \rho]]dt
 + \beta_R [\hat \rho,[\hat \rho,H_{\rm eff}]]dW_t^R
 +i\beta_I [H_{\rm eff},\hat \rho] dW_t^I~~~.\eqno(G.3)$$

We begin by remarking that for any operator 
$G$ commuting with $H_{\rm eff}$, we 
have 
$$\eqalign{
E[d\langle G \rangle]=&E[{\rm Tr} G d\hat \rho]
={\rm Tr} G E[d\hat \rho]  \cr 
=&{\rm Tr} G  (i \hbar^{-1} [E[\hat \rho],  H_{\rm eff} ]
-{1\over 2} |\beta|^2[ H_{\rm eff} ,[ H_{\rm eff},E[\hat \rho]]])dt     \cr
=&{\rm Tr} [G,H_{\rm eff}] (-i \hbar^{-1} E[\hat \rho] 
-{1\over 2} |\beta|^2[ H_{\rm eff},E[\hat \rho]])dt =0~~~,  \cr
}\eqno(G.4)$$
where $E[...]$ denotes the expectation with respect to the stochastic 
process, with $E[dW_t^R]=E[dW_t^I]=0$.  
Consider now the energy variance (the squared energy uncertainty), 
defined by  
$$V=\langle (H_{\rm eff} - \langle H_{\rm eff} \rangle)^2 \rangle  
={\rm Tr}\hat \rho H_{\rm eff}^2-({\rm Tr}\hat \rho H_{\rm eff})^2~~~.
\eqno(G.5)$$  
Using the It\^o product rule of 
Eq.~(5.49d),  together with the result of Eq.~(G.3), we have 
$$\eqalign{
dE[V]=&E[dV]=-E[{\rm Tr} (d\hat \rho H_{\rm eff})]^2~~~\cr 
=&- \beta_R^2 E[{\rm Tr} ([\hat \rho,H_{\rm eff}])^2 ]^2 dt
=-4\beta_R^2 E[V^2] dt~~~.\cr
}\eqno(G.6a)$$
Integrating with respect to time, we see that 
the expectation $E[V]$ satisfies the integral equation  
$$E[V(t)]=E[V(0)]-4 \beta_R^2 \int_0^t ds E[V(s)^2]~~~,\eqno(G.6b)$$
which using the inequality $0\leq E[(V-E[V])^2]=E[V^2]-E[V]^2$ 
gives the inequality 
$$E[V(t)] \leq E[V(0)] -4 \beta_R^2 \int_0^t ds E[V(s)]^2~~~.\eqno(G.6c)$$
Since the variance $V$ is necessarily non-negative, Eq.~(G.6c) 
implies that $E[V(\infty)]=0$, 
and again using non-negativity of $V$ this implies that $V(s)$ 
vanishes as $s \to \infty$, apart from a set of outcomes occurring with 
probability  zero.  Thus the stochastic Schr\"odinger 
equation of Eq.~(G.1) drives $|\Psi\rangle$,  as $t \to 
\infty$, to a definite energy eigenstate when the energy eigenvalues   
are nondegenerate, which for the time being we assume.  (We shall consider  
the degenerate case shortly.)

To see that Born rule probabilities emerge, we apply Eq.~(G.4) to 
the projectors $\Pi_e\equiv |e\rangle \langle e| $ on a complete set of 
energy eigenstates $|e \rangle$.  By definition, these projectors all 
commute with $H_{\rm eff}$, and so by Eq.~(G.4) the expectations   
$E[\langle\Pi_e\rangle]$ are time independent; 
additionally, by completeness of   
the states $|e\rangle$, we have $\sum_e \langle\Pi_e\rangle=1$.  
But these are just the conditions for   
Pearle's [35] gambler's ruin or martingale 
argument to apply.  At time zero, when the 
stochastic evolution has just started, 
$E[\langle\Pi_e\rangle]=\langle\Pi_e\rangle\equiv p_e$
is the absolute value squared of the quantum mechanical amplitude  
to find the initial state in energy eigenstate $|e \rangle$.  At $t=\infty$, 
the system always evolves to an energy eigenstate, with the eigenstate  
$|f\rangle $ occurring with some probability $P_f$.  The expectation 
$E[\langle\Pi_e\rangle]$, evaluated at infinite time, is then  
$$E[\langle\Pi_e\rangle]=1 \times P_e + 
\sum_{f \neq e} 0 \times P_f = P_e~~~;\eqno(G.7)$$
hence $p_e=P_e$ for each  $e$ and the state collapses into energy eigenstates 
at $t=\infty$ with probabilities given by the usual quantum mechanical 
Born rule applied to the initial wave function.  

Let us now consider the case in which the Hamiltonian $H_{\rm eff}$ is 
degenerate.  In this case, let us choose a basis of energy eigenstates 
so that, within each degenerate manifold, one basis element coincides (after 
normalization) with the projection of the initial state vector into that 
manifold, and the others are orthogonal to it.  (If the projection of 
the initial state vector into the manifold vanishes, any orthonormal basis 
for that manifold suffices.)  We can then apply the argument just given for 
the nondegenerate case, using this specially chosen energy eigenstate basis.  
We learn that the state vector reduces to one of the members of this 
basis, with a probability equal to the modulus squared of the projection 
of the initial state vector on this basis.  Thus, the state vector 
reduces into one or another of the degenerate energy manifolds, with 
the result of reduction being the normalized 
projection of the initial state vector into that manifold, and with the 
probability of obtaining this outcome equal to the squared modulus of the 
projection of the initial state into the manifold [51].     
This is precisely the result expected from  the L\"uders projection
postulate, which generalizes the Born rule to the degenerate case.     
Heuristically, the reason the L\"uders rule arises from Eq.~(G.1) is 
that this equation has the form $d|\Psi\rangle = {\cal O} |\Psi \rangle$, 
with ${\cal O}$ diagonal on an energy basis. Thus any energy eigenstate 
component that has coefficient zero in the eigenstate expansion of the 
initial state vector cannot obtain a nonzero coefficient through the 
subsequent stochastic evolution. 

{}Finally, referring to Eqs.~(G.6a-c), we see that state vector 
reduction occurs 
only when $\beta_R \not=0$, since the $\beta_I$ term in Eq.~(G.1) does 
not contribute to the evolution of the variance.  When $\beta_R=0$ 
and $\beta_I\not=0$, the initial state vector still evolves stochastically, 
but the energy variance remains constant in time.   The necessity for 
having $\beta_R\not= 0$ to achieve reduction motivates the detailed 
discussion in Sec. 5F of how to achieve a 
nonzero ${\cal K}_1$, which is related to $\beta_R$ by Eq.~(5.49a).  

\bigskip
\centerline{Appendix H:  ``No Go'' Theorem for  Heisenberg Picture Reduction}
\bigskip
In Sec.~5F we encountered difficulties in giving an operator formulation 
of state vector reduction, which were surmounted by projecting Eq.~(5.47e) 
on the vacuum state.  Here we shall work backwards from the state vector 
reduction equations given in Eq.~(G.1), and will 
prove the following: \hfill\break
``No go'' theorem.  When $\beta_R\not=0$, the linearized form of Eqs.~(G.1),  
$$\eqalign{ 
d|\Psi\rangle  =&
\beta_R   H_{\rm eff}|\Psi\rangle dW_t^R    
+ i \beta_I H_{\rm eff} |\Psi\rangle dW_t^I +O(dt)~~~, \cr
&~~~~~~~~~\cr
d\langle \Psi|  =&
\langle  \Psi| \beta_R   H_{\rm eff}  dW_t^R 
-\langle \Psi| i \beta_I H_{\rm eff}  dW_t^I +O(dt)~~~, \cr
}\eqno(H.1)$$
cannot be interpreted as the vacuum projection of a general operator 
equation that satisfies the Leibnitz product (or ``chain'') rule to 
order $O(dt)$, and also has a stochastic differential $d$ that commutes 
with the operator adjoint $\dagger$.  

To prove this, we write $|\Psi\rangle =\Psi^{\dagger} |{\rm vac}\rangle$, 
and look for an evolution equation for a general operator ${\cal O}$ that, 
when applied to ${\cal O}=\Psi^{\dagger}$, gives Eq.~(H.1) when acting on 
the vacuum state.  Since we need a structure that is linear in $H_{\rm eff}$, 
there are only two irreducible possibilities for giving the $\beta_R$  
term an operator interpretation.  In the first we write 
$$d_1{\cal O} = \beta_R [  H_{\rm eff},{\cal O} ] dW_t^R    
+i\beta_I [  H_{\rm eff},{\cal O} ] dW_t^I+O(dt)~~~,\eqno(H.2a)$$  
and in the second we write 
$$d_2{\cal O} = \beta_R \{  H_{\rm eff},{\cal O} \} dW_t^R    
+i\beta_I [  H_{\rm eff},{\cal O} ] dW_t^I+O(dt)~~~,\eqno(H.2a)$$  
both of which give the first equation in Eq.~(H.1) when acting from the 
left on $|{\rm vac}\rangle$.  
However, although Eq.~(H.2a) obeys the chain rule for differentiation,    
$$d_1({\cal O}_1{\cal O}_2)=(d_1{\cal O}_1){\cal O}_2 
+ {\cal O}_1 (d_1{\cal O}_2)~~~,
\eqno(H.3a)$$
it defines a differential that does not commute with 
the adjoint $\dagger$, since according to Eq.~(H.2a) we have 
$$\eqalign{
&(d_1{\cal O})^{\dagger} =-\beta_R [  H_{\rm eff},{\cal O} ] dW_t^R    
+i\beta_I [  H_{\rm eff},{\cal O} ] dW_t^I+O(dt)~~~\cr 
\not = &d_1({\cal O}^{\dagger}) 
= \beta_R [  H_{\rm eff},{\cal O}^{\dagger} ] dW_t^R    
+i\beta_I [  H_{\rm eff},{\cal O}^{\dagger} ] dW_t^I+O(dt)~~~. \cr
}\eqno(H.3b)$$  
Similarly, although Eq.~(H.2b) defines a differential that commutes with the 
adjoint $\dagger$, 
$$ (d_2{\cal O})^{\dagger} =d_2({\cal O}^{\dagger})~~~,\eqno(H.4a)$$
it does not obey the chain rule for differentiation, since the anticommutator 
structure multiplying $\beta_R dW_t^R$ behaves as 
$$  \{  H_{\rm eff},{\cal O}_1 {\cal O}_2 \}
\not= {\cal O}_1  \{  H_{\rm eff},{\cal O}_2 \}
+{\cal O}_1  \{  H_{\rm eff},{\cal O}_2 \}   ~~~.\eqno(H.4b)$$
Because the most general differential $d$ with the required action on the  
vacuum state is a linear combination of $d_1$ and $d_2$, 
$$d=c_1 d_1+ c_2  d_2~~,~~~  c_1+c_2=1~~~,\eqno(H.5)$$
there is no $d$ that both commutes with the adjoint and obeys the chain rule,   
completing the proof.  
\bigskip
\centerline{Appendix I: Phenomenology of Energy Driven Stochastic Reduction}
\bigskip
We discuss here phenomenological aspects of the energy driven  
stochastic reduction equation given in Eqs.~(5.51d,e).  We shall address 
the following issues:  clustering,  
bounds on the stochastic term implied 
by the maintenance of coherence where that is observed, 
the role of environmental interactions in reduction, 
and estimates of the reduction rate in measurement situations (assuming a   
Planckian magnitude for the coefficient $\beta_R$).  

We begin with some changes in notation, that are helpful in making 
contact with the relevant literature.  Since we have seen in Appendix G 
that the $\beta_I$ term in Eqs.~(5.51d,e) does not lead to state vector 
reduction, we shall set $\beta_I=0$, and shall write $\beta_R={1\over 2} 
\sigma$, with $\sigma$ the notation used for the stochastic parameter 
in [49], [28], [32], and [33].  We shall also omit: the caret 
on the density 
matrix $\hat \rho$, which we shall write simply as $\rho$, the subscript 
``eff'' on $H_{\rm eff}$, which we shall write simply as $H$, 
and the superscript $R$ on $dW_t^R$, which we shall write simply as $dW_t$. 
{}Finally, we shall set Planck's constant $\hbar$ equal to unity.  
Note that $H$, $\sigma$ and $\rho$ will now have different meanings 
from those assigned to  
these symbols in the trace dynamics discussion of the text!  

With these changes of notation, Eq.~(5.51d) of the text and Eq.~(G.1) 
of Appendix G take the form 
$$
d|\Psi\rangle  =
-i H|\Psi\rangle dt  
 - {1\over 8}  \sigma^2 (   H-\langle   H \rangle )^2  |\Psi\rangle dt 
+{1\over 2} \sigma (   H- \langle  H \rangle ) 
|\Psi\rangle dW_t  ~~~. 
\eqno(I.1)$$
Similarly, the density matrix evolution of Eq.~(5.51e) of the text and 
Eq.~(G.3) of Appendix G becomes 
$$
d\rho 
= i [\rho,  H ] dt      
-{1\over 8}\sigma^2  [ H ,[ H, \rho]]dt
 + {1\over 2}\sigma  N(\rho, H) dW_t~~~,
\eqno(I.2)$$
where the coefficient $N(\rho,H)$ of the It\^o noise term $dW_t$  
is  
$$N(\rho,H)=\{\rho,H\}-2\rho {\rm Tr}\rho H~~~~,\eqno(I.3a)$$ 
which by use of the pure  state condition $\rho^2=\rho$ is equivalent to   
$$N(\rho,H)= [\rho,[\rho,H]]~~~~.\eqno(I.3b)$$   
Both of these forms have the property that $\rho^2=\rho$ implies 
that $\{\rho,d\rho\}+(d\rho)^2=d\rho$, which can be rewritten as 
$(\rho+d\rho)^2=\rho+d\rho$, and so they preserve the pure state condition.

We begin our discussion with the issue of clustering.  
The Hamiltonian $H$ 
appearing in Eqs.~(I.1)-(I.3a,b) is the total world Hamiltonian.  In order  
for these equations to give a sensible phenomenology of reduction, they  
must separate under appropriate conditions 
into independent equations for isolated, noninteracting subsystems. 
{}Following [28] and [33], we study this question by considering the case of 
two independent subsystems, so that 
$H$ is the sum of two Hamiltonians $H_1,\, H_2$ which depend on disjoint 
sets of variables, and investigate the conditions under which Eqs.~(I.2) 
and (I.3a,b) 
admit factorized solutions $\rho=\rho_1\rho_2$,  with $\rho_{1,2}$ 
obeying equations of similar form driven by the respective Hamiltonians 
$H_{1,2}$.  Substituting $H=H_1+H_2$ and $\rho=\rho_1\rho_2$ 
into Eqs.~(I.3a), (I.3b),  
and using the facts that all variables in set 1 commute with all 
variables in set 2, and that ${\rm Tr}={\rm Tr}_1 {\rm Tr}_2$, we 
find from Eq.~(I.3a) that
$$N(\rho_1\rho_2,H_1+H_2)= 
\rho_2[\{\rho_1,H_1\}-2\rho_1 {\rm Tr_2} \rho_2 {\rm Tr_1}\rho_1 H_1] 
+\rho_1[\{\rho_2,H_2\}-2\rho_2 {\rm Tr_1} \rho_1 {\rm Tr_2}\rho_2 H_2] 
~~~,\eqno(I.4a)$$
while from Eq.~(I.3b) we find that 
$$N(\rho_1\rho_2,H_1+H_2)= 
 \rho_2^2 [\rho_1,[\rho_1,H_1]]  + \rho_1^2 [\rho_2,[\rho_2,H_2]] 
~~~.\eqno(I.4b)$$  
Clustering requires that 
$$N(\rho_1\rho_2,H_1+H_2)=\rho_2 N_1(\rho_1,H_1) + \rho_1 N_2(\rho_2,H_2)
~~~,\eqno(I.5)$$
with $N_{1,2}$ the restrictions of $N$ to the 1,2 subspaces.  
We see that Eq.~(I.4a) obeys the clustering property by virtue of the 
trace conditions ${\rm Tr_1}\rho_1=1$, ${\rm Tr_2}\rho_2=1$, while 
Eq.~(I.4b) satisfies the clustering property by virtue of the pure state 
conditions $\rho_1^2=\rho_1$, $\rho_2^2=\rho_2$.   

Let us now examine the clustering properties of the remaining terms in 
Eq.~(I.2).  For the left hand side, we find by use of the It\^o extension 
of the chain rule that
$$d(\rho_1\rho_2)=\rho_2 d\rho_1 +\rho_1 d\rho_2 +d\rho_1d\rho_2
~~~.\eqno(I.6a)$$
Thus, in order to have $d\rho_1$ and $d\rho_2$ obeying equations of the 
same form as $d\rho$ but restricted to the $1,2$ subspaces, the left hand 
side should take the form  
$$d(\rho_1\rho_2)=\rho_2 d\rho_1 +\rho_1 d\rho_2 + {1\over 4} \sigma^2 
N_1(\rho_1,H_1)
N_2(\rho_2,H_2) dt
~~~.\eqno(I.6b)$$
{}For the $dt$ terms on the right hand side of Eq.~(I.2), we have   
$$\eqalign{
i[\rho_1\rho_2,H_1+H_2] dt -&{1\over 8}\sigma^2 [H_1+H_2,[H_1+H_2,
\rho_1\rho_2]]dt\cr   
=&\rho_2\{i[\rho_1,H_1]dt -{1\over 8}\sigma^2 [H_1,[H_1,\rho_1]] dt \}\cr
+&\rho_1\{i[\rho_2,H_2]dt -{1\over 8}\sigma^2 [H_2,[H_2,\rho_2]] dt \}\cr
-&{1\over 4}\sigma^2 [H_1,\rho_1][H_2,\rho_2] dt~~~.\cr 
}\eqno(I.6c)$$
Assuming the conditions for the clustering property of Eq.~(I.5) to hold for 
the It\^o noise term, comparing Eqs.~(I.6a-c) we see that 
the complete density matrix evolution equation will 
cluster if and only if 
$$N_1(\rho_1,H_1) N_2(\rho_2,H_2)
= -[H_1,\rho_1][H_2,\rho_2] ~~~.\eqno(I.7)$$  
This condition does not hold as in identity for either of the two possible 
forms for $N(\rho,H)$ given in Eqs.~(I.3a), (I.3b), and 
so the $\sigma^2 dt$ or   
drift term in the stochastic evolution equation  does couple disjoint  
systems.  

However, there are two important special cases in which disjoint 
systems decouple asymptotically.  The first of these cases corresponds 
[28] to taking $N(\rho,H)$ as in Eq.~(I.3b), so that 
Eq.~(I.7) becomes
$$ [\rho_1,[\rho_1,H_1]] [\rho_2,[\rho_2,H_2]]
= -[H_1,\rho_1][H_2,\rho_2] ~~~.\eqno(I.8a)$$  
This equation is satisfied, by virtue of both the left and right hand sides 
vanishing, whenever either $[\rho_1,H_1]=0$ or $[\rho_2,H_2]=0$, conditions  
that are respectively obeyed when system 1 or system 2 is at the 
endpoint of the state vector reduction process.  In particular, if system 
1 represents a measurement process, and system 2 represents a pure state 
environment at the endpoint of its reduction process, then the stochastic 
dynamics of system 1 is completely independent of the dynamics of its 
environment.   

A more general case [33] in which disjoint systems decouple 
asymptotically corresponds to taking $N(\rho,H)$ as in Eq.~(I.3a), but not 
assuming the pure state condition so that this cannot be transformed to 
Eq.~(I.3b).  Equation (I.7) now becomes 
$$[\{\rho_1,H_1\}-2\rho_1{\rm Tr_1}\rho_1 H_1] 
[\{\rho_2,H_2\}-2\rho_2{\rm Tr_2}\rho_2 H_2]= 
-[H_1,\rho_1][H_2,\rho_2]~~~~.\eqno(I.8b)$$
This equation is satisfied, again by virtue of both the left and right hand 
sides vanishing, whenever either $\rho_1$ is a linear combination of 
projectors on a degenerate 
submanifold of $H_1$, or $\rho_2$ is a linear combination of  
projectors on a degenerate submanifold 
of $H_2$.  For example, in the latter case we would have $\rho_2 H_2=
H_2 \rho_2=E_2 \rho_2$ for some degenerate submanifold energy $E_2$, together 
with ${\rm Tr_2}\rho_2=1$, which imply the simultaneous vanishing of 
$\{\rho_2,H_2\}-2\rho_2{\rm Tr_2}\rho_2 H_2 $ and of $[H_2,\rho_2]$.  
Thus, 
if one were to adopt Eqs.~(I.2) and (I.3a) as a generalization of the density 
matrix evolution equation to the case of non-pure state density matrices, 
a pure state measurement process decouples from a mixed state environment 
whenever the density matrix for this environment is a linear combination of 
projectors on a degenerate submanifold of its Hamiltonian.  An application  
of these ideas to the case of thermal mixed state environments is given 
in [33].  

We note that the conclusions we have reached about clustering  do 
not extend [28] to  more general stochastic 
evolutions in which the stochastic process 
is driven by an operator $A$ differing from the Hamiltonian $H$, with $A$ 
taken to be additive over subsystems.  The reason is that there is 
now a competition between the stochastic terms, which 
in Eqs.~(I.2) and (I.3b)  
are constructed  
from double commutators with an innermost commutator $[A,\rho]$, and  
the Schr\"odinger evolution term, which involves 
the commutator $[H,\rho]$; 
the stochastic terms tend to drive the system to $A$ eigenstates, while 
the Schr\"odinger term  coherently mixes $A$ eigenstates, leading to   
evolution away from $A$ eigenstates.  Thus, a subsystem cannot 
remain indefinitely in an $A$ eigenstate, and 
as a result does not persist indefinitely 
as an unentangled independent subsystem in a larger system.  

Now that we are assured that  Eqs.(I.1)-(I.3a,b)  
can be applied to the evolution of an isolated system decoupled from its 
environment, with the Hamiltonian $H$ referring only to the system, we 
can embark on a discussion of phenomenological implications of these  
equations for measurements.   We first need a quantitative estimate of 
the reduction time given by Hughston [49].  Rewriting Eq.~(G.6a) 
in terms of the parameter 
$\sigma$, and approximating $E[V^2]$ by $E[V]^2$, we get a differential 
equation for $E[V]$, 
$${dE[V] \over dt}=-\sigma^2E[V]^2~~~,\eqno(I.9a)$$
which can be integrated to give 
$$E[V(t)]={E[V(0)] \over 1+\sigma^2 E[V(0)]t}~~~.\eqno(I.9b)$$
Thus, given an initial energy variance $E[V(0)]\equiv (\Delta E)^2$ 
and the parameter $\sigma$, 
state vector reduction will be completed for times significantly larger 
that $t_R$, with 
$$t_R={1 \over (\sigma \Delta E)^2 }~~~.\eqno(I.10a)$$
{}From this equation (or directly from the stochastic Schr\"odinger equation) 
we see that $\sigma$ has units $({\rm mass})^{-{1\over 2}}$. So writing 
$\sigma = M^{-{1\over 2}}$, with $M$ the characteristic mass scale for the 
fluctuations that give rise to the stochasticity in the Schr\"odinger 
equation, Eq.~(I.10a) takes the form 
$$t_R={M\over (\Delta E)^2}~~~.\eqno(I.10b)$$ 
If one assumes that $M$ is of order 
the Planck mass $M_P \sim 10^{19}$ GeV, then 
one gets the estimate [49] 
$$t_R \sim \left( {2.8 {\rm MeV} \over \Delta E}\right)^2 {\rm sec}~~~.
\eqno(I.10c)$$
Thus, for $\Delta E$ equal to a proton mass, $t_R \sim 10^{-5} {\rm sec}$, 
while for $\Delta E$ equal to the mass of a nitrogen molecule, 
one has $t_R \sim 10^{-8} {\rm sec}$.  

In order for stochastic energy-driven state vector reduction to give a 
viable phenomenology, it must satisfy the twin constraints of predicting 
the maintenance of coherence when this is observed, while 
predicting a rapid enough state vector reduction when  a  
probabilistic choice between alternative outcomes is observed. A detailed  
analysis of these issues is given in [33]; we give here only a brief
discussion.  

We first 
discuss the constraints imposed by the maintenance of coherence.  
According to Eqs.~(I.10a-c), the sole criterion 
governing how rapidly the state vector reduces is the energy variance; 
whether the system is microscopic or macroscopic plays no role.  
Coherent superpositions of macroscopic states, involving large numbers 
of particles, will persist in time if the energy spread between the 
superimposed states is small enough.  For example, consider the 
recent superconducting quantum interference device (SQUID) experiments  
[52] that observe the existence of coherent superpositions of macroscopic 
states consisting of oppositely circulating supercurrents.  The variance
$\Delta E$ in the Friedman et. al. experiment [52] is roughly 
$8.6 \times 10^{-6} {\rm eV}$, and the circulating currents each correspond 
to the collective motion of $\sim 10^9$ Cooper pairs.  
According to Eq.~(I.10c), 
despite the macroscopic structure of the state vector, 
the state vector reduction time $t_R$ for this experiment should be about 
$10^{23}~{\rm s} \sim 3 \times 10^{15}~{\rm yr}$, and so maintenance of 
coherence is expected over the measurement time of order a millisecond.  
Similarly, in atomic quantum intermittency experiments [53], which involve 
transitions between a metastable atomic energy 
level (with a lifetime of around a second) and the ground state, with 
typical energy separations of a few eV, 
we expect a state vector reduction 
time for Planckian $M$ of order $10^{12}$ s $\sim 3 \times 10^4$ yr.  
Hence in this case also, the maintenance of coherence is expected.  

We can turn this calculation around, and estimate a lower bound on the 
mass $M$ appearing in Eq.~(1.10b) in terms of the time $t_C$ over which 
a superposition of energy states differing by $\Delta E$ is observed to 
remain coherent, 
$$M>t_C (\Delta E)^2~~~.\eqno(I.11a)$$
The most straightforward cases to consider are those involving 
oscillations of neutrinos, $K$-mesons, or $B$-mesons, since these are   
can be treated as two-state systems with negligible interaction with 
the electromagnetic field, and so Eq.~(I.11a) can be directly applied.  
{}For a two state system with mass splitting $\Delta m$, and mean energy 
$E$ for the components, one has $\Delta E= \Delta m^2/(2E)$, and so we 
get from Eq.~(I.11a) the estimate 
$$M>{t_C (\Delta m^2)^2  \over 4 E^2}~~~.\eqno(I.11b)$$
{}For neutrinos [54], taking the coherence time $t_C$ to be the oscillation  
time  $2 \pi/(\Delta E)=4 \pi E/ (\Delta m^2)$,  Eq.~(I.11b) becomes
$$M  >2 \pi \Delta E =   {\pi \Delta m^2 \over E}~~~,\eqno(I.11c)$$ 
which for the parameters appropriate to both the atmospheric 
($\Delta m^2 \sim 3 \times 10^{-3} {\rm eV}^2$, $E \sim 1 {\rm GeV}$)  
and solar 
($\Delta m^2 \sim 5 \times 10^{-5} {\rm eV}^2$, $E \sim 8 {\rm MeV}$)  
neutrino oscillation observations, gives to within a factor of two the   
estimate $M > 10^{-20} {\rm GeV}$.  
{}For $K$- and $B$-mesons at rest in the lab 
frame, taking the coherence time $t_C$ to be the lifetime $\tau_S$ of the 
shorter-lived component 
(which is similar in magnitude to the oscillation time), Eq.~(I.11a) becomes  
$$M>\tau_s (\Delta m)^2~~~,\eqno(I.11d)$$
which for the parameters appropriate to the $K$-meson system 
($\Delta m \sim 4 \times 10^{-6} {\rm eV}$, $\tau_S \sim .9 \times 10^{-10}
{\rm s}$)  
and to the $B$-meson system 
($\Delta m \sim 3 \times 10^{-4} {\rm eV}$, $\tau_S \sim 1.6 \times 10^{-12}
{\rm s}$)  
gives respective bounds of $M> 2\times 10^{-15}\, {\rm GeV}$ and 
$M >2\times 10^{-13}\, {\rm GeV}$.  The $K$- and $B$-meson 
systems give better bounds than are obtained from neutrinos because
for the mesons $m/E$ is of order unity, whereas for neutrinos $m/E$ is 
very small.  However, even the meson system bounds are far removed from 
the Planck scale of $10^{19}$ GeV $\sim 10^{13} {\rm eV}^2 \, {\rm s}$.

Better limits may be obtainable in atomic and nuclear systems.  
Returning to the SQUID experiment discussed earlier, using 
the value $\Delta E \sim 8.6 \times 10^{-6} {\rm eV}$, and making the  
{\it assumption} that a lower limit for the coherence time $t_C$ 
is given by the 
millisecond duration of the microwave pulse used as a probe, one gets 
from Eq.~(I.11c) the bound $M > 7 \times 10^{-14} 
{\rm eV}^2$ s $\sim 10^{-7}$ GeV.   The existence of nuclear 
isomer states [55] 
lying 100-200 keV above the ground state and with very long lifetimes 
offers a possibility of greatly improved bounds, but this will require 
further theoretical work.  In the absence of coupling to the electromagnetic 
field, these isomers are energy eigenstates, and so are stable under the 
stochastic Schr\"odinger evolution.  When the electromagnetic coupling 
is added as a perturbation, the isomers are no longer eigenstates of the 
total Hamiltonian, but to understand what happens the standard theories 
of decay rates, line width, laser action, etc. will have to be generalized 
to take take the stochastic terms in Eq.~(I.1) into account.  

We turn now to the second requirement that must be satisfied by a 
phenomenology of state vector reduction, which is 
that it should lead to rapid reduction in experimental situations where a 
probabilistic outcome is observed.  According to the von Neumann model 
for measurement [56], a measurement sets up a  correlation  
between states $|f_{\ell}\rangle$ of a quantum  system being measured, and 
macroscopically distinguishable states $|{\cal M}_{\ell}\rangle$ of the 
measuring apparatus ${\cal M}$, in such a way that an initial state  
$$|f\rangle |{\cal M}_{\rm initial}\rangle=
\sum_{\ell}c_{\ell}|f_{\ell}\rangle
|{\cal M}_{\rm initial}\rangle~~~\eqno(I.12a)$$ 
evolves unitarily to 
$$\sum_{\ell} c_{\ell} |f_{\ell}\rangle  |{\cal M}_{\ell}\rangle~~~.
\eqno(I.12b)$$
An objective state vector reduction model must then account for the selection 
of {\it one} of the alternatives 
$|f_{\ell}\rangle  |{\cal M}_{\ell}\rangle$ from this superposition, 
with a probability given by $|c_{\ell}|^2$.  
If the energy spread among the states $|f_{\ell}\rangle$ has a typical 
atomic magnitude of a few eV, then as we have seen above,  for a Planckian 
magnitude of $M$, reduction times in the energy driven model are of 
order $10^4$ years, and cannot by themselves quantitatively account 
for state vector reduction.  The only way for reduction to occur within  
typical measurement times is for the energy spreads among the 
alternative {\it apparatus} states in the superposition 
to be much larger than   a few eV.  
Since in the ideal measurement model there is no energy transfer from the 
microscopic system to the apparatus, such an energy spread in the 
measurement apparatus states can only be present if induced by 
environmental interactions, which are ignored in the von Neumann analysis.  
{}For environmental interactions 
to be effective in producing state 
vector reduction, they must lead to energy fluctuations $\Delta E$ 
of the apparatus  
in the course of a measurement, that are large enough for Eq.~(I.10c) to 
predict a reduction time $t_R$ that is less than the time it takes 
to make the measurement.  

A detailed analysis of this issue is given in [33] (see also [28]), 
following on an 
initial suggestion in [49].  We summarize only a few of the results here.  
Although different measuring devices have different 
response times, let us assume that the 
relevant measurement time is of order $10^{-8}$ seconds,   
which requires for reduction a $\Delta E$ ranging up to $\sim 30$GeV,  
roughly the mass of a nitrogen molecule.  
The analysis of [33] considered three  possible sources of 
energy fluctuations:  thermal energy fluctuations, fluctuations in apparatus 
mass from particle accretion processes, and fluctuations in 
apparatus mass from amplified fluctuations in the currents that 
actuate the indicator devices.  Thermal energy fluctuations were found 
to be unable to produce the needed energy fluctuation within the 
measurement time.  On the other hand, both energy fluctuations from 
mass accretion, and from amplified current fluctuations, are relevant.  

As illustrations of accretion-induced fluctuations, assuming  
room temperature and 
atmospheric pressure (760 Torr),   
the time for one molecule to be accreted onto an area of 
$1 {\rm cm}^2$ is $3\times 10^{-24}~{\rm sec}$, while
at an ultrahigh vacuum of $10^{-13}$ Torr it is 
$3 \times 10^{-8} ~{\rm sec}$. 
Thus, for an apparatus in the atmosphere at standard temperature and  
pressure, 
where the bulk of the accreting atoms are nitrogen molecules, the minimum 
apparatus area required for one molecule to accrete in a reduction time 
of $10^{-8} {\rm sec}$ (corresponding to a  $\Delta E$ equal to the mass of 
a nitrogen molecule) is $3 \times 10^{-16} {\rm cm}^2$, with the   
corresponding minimum area needed at a pressure of $10^{-13}$ Torr equal to 
$3 {\rm cm}^2$.  Further estimates of this type are given in [33].  
{}For example, even in the sparsely populated environment of intergalactic 
space, one concludes that in a typical high precision molecular beam 
experiment, the reduction time induced by particle accretion 
on a capsule large enough 
to enclose the apparatus would be smaller, by at least an order  
of magnitude, than the measurement time.   

Estimates can also be made [33] of energy fluctuations arising 
from the amplified fluctuations in the currents which actuate 
experimental 
indicating or recording devices.  
Of course, if power sources are included, 
there are no overall current fluctuations, but power supplies are typically 
large in area and so when included in the system the 
accretion analysis just given   
indicates rapid reduction times.   In a typical electrically amplified  
measurement, a final total charge transfer $Ne$ (with $e$ the charge of 
an electron) actuates an 
indicator or recording device. 
Assuming that the fluctuation in the current is the amplified fluctuation 
in the initially detected signal, for amplification gain $G$ we have 
$\Delta N \sim G \times (N/G)^{1\over 2} =(NG)^{1 \over 2}$. 
Let us take  $N$ to correspond to a charge 
transfer of 1 milliampere (a  voltage change of 10 volts at 10 k$\Omega$ 
impedance) 
over a $10^{-8}$ second pulse, so that   
$N \sim 6 \times 10^7$, and assume a gain $G \sim 
10^4$, giving $\Delta N \sim 8 \times 10^5$.  Multiplying by the 
electron mass of 
$ .5 \times 10^{-3} {\rm GeV}$, we find that the corresponding energy  
fluctuation is $\Delta E \sim 4 \times 10^2 {\rm GeV}$, which leads to 
state vector reduction in $5 \times 10^{-11} {\rm sec}$.  Thus, electric 
current fluctuations play a significant role in state vector reduction 
when the ``apparatus'' is defined to exclude power sources.  

Our overall conclusion is that conditions under which laboratory 
experiments are performed, as well as conditions under which space 
capsule experiments might be performed 
in the foreseeable future, are consistent 
with state vector reduction times as estimated by Eq.~(I.10c) that are well 
within experimental measurement times.  As long as experimental outcomes 
are macroscopically distinguishable (which effectively defines an apparatus), 
the energy spread between different outcomes is sufficient to cause 
the state vector to reduce.  

We note finally that in the analysis of future experiments to improve 
the phenomenological bounds on $M$, one will either 
have to solve the stochastic 
Schr\"odinger equation analytically, or simulate it numerically [57],  
for typical experimental configurations. Powerful new techniques for 
performing such 
simulations, along with relevant analytical methods, are given in [58].  
\vfill\eject
\overfullrule=0pt
\twelvepoint
\doublespace  
\pageno=168
\centerline{References}
\bigskip
\noindent
\item{[1]}  S. L. Adler, Nucl. Phys. B 415, 195 (1994); S. L. Adler, 
``Quaternionic Quantum Mechanics and Quantum Fields'', Secs. 13.5-13.7 and 
App. A (Oxford Univ. Press, New York, 1995). In these papers, and others 
in the trace dynamics program before 1997, fermions were introduced through 
a $(-1)^F$ operator insertion in the trace, rather than by use of a Grassmann 
algebra as done in Ref. [8] and here.  The principal results of the 
earlier work are unaffected by this change, but certain details are altered. 
The idea of using a trace variational principle to generate operator 
equations goes back to the inception of quantum mechanics; see   
M. Born and P. Jordan, Zeit. f. Phys. 34, 858 (1925), who in Sec. 2 of their   
paper introduce a symbolic differentiation of operator 
monomials under a trace that is identical to the bosonic case of the one 
used here.  A Hamiltonian variational principle based on this idea has 
been used by A. Kerman and A. Klein, Phys. Rev. 132, 1326 (1963), Appendix B,
to generate equations of motion for many-body physics.   
(I am indebted to A. Klein for bringing these references 
to my attention; see W. R. Greenberg, A. Klein, I. Zlatev, 
and C.-T. Li, chem-ph/9603006, for further references to many-body theory 
applications.)  Our own first use of trace dynamics ideas appears in  
S. L. Adler, Phys. Lett. B86, 203 (1979).  
\bigskip
\noindent 
\item{[2]} S. L. Adler, G. V. Bhanot, and J. D. Weckel, J. Math. Phys. 
35, 531 (1994).  See also  Appendix A of the book of S. L. Adler, Ref. [1].  
\bigskip
\noindent
\item{[3]} S. L. Adler, Nucl. Phys. B415, 195 (1994), Appendix A. 
\bigskip
\noindent
\item{[4]} S. L. Adler and Y.-S. Wu, Phys. Rev. D49, 6705 (1994).  
\bigskip
\noindent
\item{[5]}  S. L. Adler and   L. P. Horwitz, J. Math. Phys. 37, 5429 (1996).
\bigskip
\noindent
\item{[6]}  A. C. Millard, private email  
communication to S. L. Adler, June, 1996; 
A. C. Millard, ``Non-Commutative Methods in Quantum 
Mechanics'', Princeton University PhD thesis, April, 1997. 
\bigskip
\noindent
\item{[7]}  S. L. Adler and A. C. Millard, Nucl. Phys. B 473, 199 (1996). 
\bigskip
\noindent
\item{[8]}  S. L. Adler and A. Kempf, J. Math. Phys. 39, 5083 (1998).
\bigskip
\noindent
\item{[9]}  S. L. Adler, Nucl. Phys. B499, 569 (1997).
\bigskip 
\noindent
\item{[10]}  S. L. Adler, Phys. Lett. B407, 229 (1997).
\bigskip
\noindent
\item{[11]}  P. West, ``Introduction to Supersymmetry and Supergravity'', 
extended second ed. (World Scientific, Singapore, 1990).
\bigskip
\noindent
\item{[12]}  E. Bergshoeff, E. Sezgin, and P. K. Townsend, Phys. Lett. 
B189, 75 (1987) and Ann. Phys. 185, 330 (1988);  
M. Claudson and M. B. Halpern, Nucl. Phys. B250, 689 (1985); 
V. Rittenberg and S. Yankielowicz, Ann. Phys. 162, 273 (1985); R. Flume, 
Ann. Phys. 164, 189 (1985).  For a recent survey, see B. de Wit, 
``Supersymmetric quantum mechanics, supermembranes and Dirichlet 
particles'', hep-th/9701169.  See also T. Banks, 
N. Seiberg, and S. Shenker, Ref. [47].  
\bigskip
\noindent
\item{[13]} B. de Wit, J. Hoppe, and H. Nicolai, Nucl. Phys. B305, 545 
(1988); P. K. Townsend, Phys. Lett. B373, 68 (1996); 
T. Banks, W. Fischler, S. H. Shenker, and L. Susskind, 
``M Theory as A Matrix Model: A Conjecture'', hep-th/9610043.  A large 
bibliography of related earlier work can be found here. 
\bigskip 
\noindent
\item{[14]} M. B. Green, J. H. Schwarz, and E. Witten, ``Superstring 
Theory'', Vol. 1, pp. 220, 246, and 288 
(Cambridge Univ. Press, Cambridge, 1987); 
see also E. Cremmer and B. Julia, Nucl. Phys. B159, 141 (1979).
\bigskip
\noindent
\item{[15]}  S. L. Adler and Y.-S. Wu, unpublished.   
\bigskip
\noindent
\item{[16]}  The standard quantum mechanical treatment of matrix models 
is used in the papers reprinted in E. Br\'ezin and S. R. Wadia, eds., 
``The Large N Expansion in Quantum Field Theory and Statistical Mechanics'' 
(World Scientific, Singapore, 1993).  
\bigskip
\noindent
\item{[17]}  A similar strategy towards the interpretation of matrix models, 
with a different detailed execution, has been proposed by Smolin.  See 
L. Smolin,  ``Derivation of quantum mechanics from a deterministic non-local 
hidden variables theory, I.  The two dimensional theory'', Institute for 
Advanced Study preprint, 1983; ``Stochastic mechanics, hidden variables and 
gravity'' in C. J. Isham and R. Penrose, eds., ``Quantum Concepts in Space 
and Time'' (Oxford Univ. Press,  New York, 1986); ``Matrix models 
as non-local hidden variables theories'', hep-th/020103.  
\bigskip
\noindent
\item{[18]}  D. ter Haar, ``Elements of Statistical Mechanics'', Third ed., 
Sec. 5.13 (Butterworth Heinemann, Oxford and Boston, 1995); A. Sommerfeld, 
``Thermodynamics and Statistical Mechanics'', Secs. 28, 29, 36, and 40 
(Academic Press, New York, 1956).  
\bigskip
\noindent
\item{[19]}  S. L. Adler, J. Math. Phys. 39, 1723 (1998).  
\bigskip
\noindent
\item{[20]}  S. Weinberg, ``The Quantum Theory of Fields'', Vol II, Sec. 
15.5 (Cambridge University Press, Cambridge, 1996).  
\bigskip
\noindent
\item{[21]}  M. L. Mehta, ``Random Matrices'', Chapt. 3 (Academic Press, 
New York, 1967).  See also E. Br\`ezin and S. R. Wadia, Ref. [16], Sec. 8. 
\bigskip
\noindent
\item{[22]}  F. Mohling, ``Statistical mechanics: methods and applications'', 
pp. 270-272. (Halsted Press/John Wiley, New York, 1982).  
\bigskip
\noindent
\item{[23]} M. Kaku, ``Quantum Field Theory'', pp. 407-410. (Oxford,  
New York, 1993).  
\bigskip
\noindent
\item{[24]}  R. F. Streater and A. S. Wightman, ``PCT, Spin \& Statistics, 
and All That'' (Benjamin, New York, 1968).
\bigskip
\noindent
\item{[25]}  J. S. Bell, Physics 1, 195 (1965).  See also J. S. Bell, 
``Speakable and Unspeakable in Quantum Mechanics'' (Cambridge University 
Press, Cambridge, 1987).
\bigskip
\noindent
\item{[26]}  R. Bousso, ``The holographic principle'', 
 hep-th/0203101 (Rev. Mod. Phys., in press).
\bigskip
\noindent
\item{[27]} S. L. Adler and T. Brun, J. Phys. A: Math. Gen. 34, 1 (2001). 
\bigskip
\noindent
\item{[28]} S. L. Adler and L. P. Horwitz, J. Math. Phys. 41, 2485 (2000).
The $Q$ term is not explicitly included in this paper, but 
it has an analogous 
structure to the coefficient of the stochastic term, which is evaluated 
in this reference.  
\bigskip
\noindent
\item{[29]}  N. Gisin, J. Phys. A: Math. Gen. 28, 7375 (1995).
\bigskip
\noindent
\item{[30]}  G. Lindblad,  Commun. Math. Phys. 48, 119 (1976); V. Gorini,   
A. Kossakowski, and E. C. G. Sudarshan, J. Math. Phys. 17, 821 (1976).  
\bigskip
\noindent
\item{[31]}  A. Bassi and G. C. Ghirardi, ``Dynamical Reduction Models'', 
Physics Reports (in press); P. Pearle, ``Collapse Models'', in H.-P. Breuer  
and F. Pettrucione, eds., ``Open Systems and Measurements in Relativistic 
Quantum Field Theory'' (Lecture Notes in Physics 526) (Springer, Berlin, 
1999).  
\bigskip
\noindent
\item{[32]}  S. L. Adler, D. C. Brody, T. A. Brun, and L. P. Hughston, 
J. Phys. A: Math. Gen: 34, 8795 (2001).
\bigskip
\noindent
\item{[33]}  S. L. Adler, J. Phys. A: Math. Gen. 35, 841 (2002).  See also 
L. P. Hughston, Ref. [49], which first suggested 
the relevance of environmental accretion effects, and Ref. [28], which 
gives related phenomenological estimates.  
\bigskip
\noindent
\item{[34]}  A. Bassi and G. C. Ghirardi, Phys. Rev. A65, 42144 (2002).    
\bigskip
\noindent
\item{[35]}  P. Pearle, Phys. Rev. D13, 857 (1976); Int. Journ. Theor. Phys.
18, 489 (1979); Phys. Rev. D29, 235 (1984); Phys. Rev. A 39, 2277 (1989).  
\bigskip
\noindent
\item{[36]}  G. C. Ghirardi, A. Rimini, and T. Weber, 
Phys. Rev. D34, 470 (1986).   
\bigskip
\noindent
\item{[37]} G. C. Ghirardi, P. Pearle, and A. 
Rimini, Phys. Rev. A42, 78 (1990).
\bigskip
\noindent
\item{[38]}  N. Gisin, Phys. Rev. Lett. 52, 1657 (1984); Helv. Phys. Acta 
62, 363 (1989).  
\bigskip
\noindent
\item{[39]}  L. Di\'osi, Phys. Lett. A 129, 419 (1988). 
\bigskip
\noindent
\item{[40]}  I. Percival, Proc. Roy. Soc. London A 447, 189 (1994).
\bigskip
\noindent
\item{[41]}  P. Pearle, ``Toward a relativistic theory of statevector 
reduction'', in ``Sixty-Two Years of Uncertainty: Historical, Philosphical, 
and Physical Inquiries into the Foundations of Quantum Mechanics'', A. I. 
Miller, ed.  (Plenum Press, New York, 1990).  For a recent discussion and 
extensive references, see Ref. [27].  
\bigskip
\noindent
\item{[42]}  W. Taylor, Rev. Mod. Phys. 73, 419 (2001).
\bigskip
\noindent
\item{[43]}  Y. J. Ng, ``Spacetime Foam'', gr-qc/0201022.
\bigskip
\noindent  
\item{[44]}  S. L. Adler, Gen. Rel. and Gravitation 29, 1357 (1997).  
The attempt in this paper at a concrete 
calculation of the induced cosmological 
constant is based on an interpretation of the trace dynamics 
canonical ensemble 
averages that we now know to be incorrect (as shown in Sec.~4E, an  
unrestricted ensemble average corresponds to a trace in the quantum 
theory interpretation, not to a vacuum expectation), and does not go through  
when the identification of canonical ensemble 
averages with Wightman functions is revised to take the form given in Sec.
5C of this book.    
\bigskip
\noindent
\item{[45]}  D. Finkelstein, J. M. Jauch, and D. Speiser, ``Notes on 
quaternion quantum mechanics'', CERN Report 59-7 (1959); reprinted in 
C. Hooker, ed., ``Logico-Algebraic Approach to Quantum Mechanics II'' 
(Reidel, Dordrecht, 1979).  
\bigskip
\noindent
\item{[46]} V. I. Arnold, ``Mathematical Methods of Classical Mechanics'', 
p. 211 (Springer-Verlag, New York, 1978); R. Abraham and J. E. Marsden, 
``Foundations of Mechanics'', 2nd edition, p. 194 (Benjamin/Cummings, 
Reading, MA, 1980).  
\bigskip
\noindent
\item{[47]} T. Banks, N. Seiberg, and S. Shenker, Nucl. Phys.  B490, 91   
(1997).
\bigskip
\noindent
\item{[48]}  F. Strocchi, ``Causal Properties of Quantum Field Theory'', 
Lecture Notes in Physics Vol. 51 (World Scientific, Singapore, 1993).  
\bigskip
\noindent
\item{[49]}  L. P. Hughston, Proc. Roy. Soc. Lond. A452, 953 (1996).  
\bigskip
\noindent
\item{[50]}  C. W. Gardiner, ``Handbook of Stochastic Methods'', Chapt. 4 
(Springer-Verlag, Berlin, 1990).
\bigskip
\noindent
\item{[51]}  S. L. Adler, D. C. Brody, T. A. Brun, and L. P. Hughston, 
J. Phys. A: Math. Gen. 34, 8795 (2001).
\bigskip
\noindent
\item{[52]} J. R. Friedman, V. Patel, W. Chen, S. K. Tolpygo, and J. E. 
Lukens, Nature 406, 43 (2000); 
C. H.  van der Wal, A. C. J. ter Haar, F. K. Wilhelm, 
R. N. Schouten, C. J. P. M. Harmans, T.P. Orlando, S. Lloyd, and  
J. E. Mooij, Science  290, 773 (2000).    
\bigskip 
\noindent
\item{[53]}  H. Dehmelt, in ``Laser Spectroscopy V''  
(Springer Series in Optical Sciences, Vol. 3)  
 A. R. W. McKellar, T. Oka, and 
B. P. Stoicheff, eds. (Springer, Berlin, 1981);  
H. Dehmelt, IEEE Trans. Instrum. Meas.  2, 83 (1982);
H. Dehmelt,  `` Advances in Laser  
Spectroscopy''  (Nato Advanced Study Institute, Vol. 95)  F. T. 
Arecchi, F. Strumia, and H. Walther, eds. (Plenum, New York, 1983); 
M. Porrati and S. Putterman, Phys. Rev. A 36, 929 (1987); 
C. Cohen-Tannoudji and J. Dalibard,  
Europhys. Lett.  1, 441 (1986); 
M. Porrati  and S. Putterman, Phys. Rev.  A  39,  3010 (1989).  
\bigskip
\noindent 
\item{[54]}  E. Lisi, A. Marrone, and D. Montanino, Phys. Rev. Lett. 85, 
1166 (2000); S. L. Adler, Phys. Rev. D62, 117901 (2000).  
\bigskip
\noindent
\item{[55]}  P. Walker and G. Dracoulis, Nature 399, 35 (1999); G. J. Perlow, 
W. Potzel, R. M. Kash, and H. De Waard, J. de Physique, Colloque C6, Suppl. 
No. 12, p. C6-197 (1974). 
\bigskip
\noindent
\item{[56]} J. von Neumann, ``Mathematische Grundlagen der  
Quantenmechanik'', Chap. VI (Springer, Berlin, 1932).     
[Engl. Transl.: R. T. Beyer,  
``Mathematical Foundations of Quantum Mechanics'', 
(Princeton University Press, Princeton, 1971).]  
{}For a recent pedagogical exposition, see the book of Adler cited in [1].    
\bigskip
\noindent
\item{[57]} R. Shack, T. A. Brun, and I. C. Percival, J. Phys. A: Math. Gen. 
28, 5401 (1995); R. Shack and T. A. Brun, Comp. Phys. 
Commun. 102, 210 (1997); 
I. Percival, ``Quantum State Diffusion'' (Cambridge University  
Press, Cambridge, 1998).
\bigskip
\noindent
\item{[58]} D. C. Brody and L. P. Hughston, ``Efficient Simulation of 
Quantum State Reduction'',  quant-ph/0203035.  
\bigskip

\vfill
\eject
\bigskip
\bye